\documentclass[epj]{svjour}
\usepackage{graphics}
\usepackage{amsmath,amssymb}
\usepackage{graphicx}
\usepackage[dvips]{epsfig}
\usepackage{color}

\newcommand{\pol}[1]{\mathaccent"017E{#1}}
\newcommand{\Szero}{\mbox{$^{1\!}S_0$}}
\newcommand{\dd}{\mbox{$\textrm{d}$}}
\newcommand{\half}{\mbox{${\textstyle \frac{1}{2}}$}}           
\newcommand{\fmn}[2]{\mbox{${\textstyle \frac{#1}{#2}}$}}

\begin{document}

\title{The legacy of the experimental hadron physics programme at COSY}
\author{C.~Wilkin
\thanks{Email: c.wilkin@ucl.ac.uk}}
\institute{Physics and Astronomy Department, UCL, Gower Street, London, WC1E
6BT, UK }
\date{Received: \today / Revised version:}
\abstract{The experimental hadronic physics programme at the COoler
SYnchrotron of the Forschungszentrum J\"{u}lich terminated at the end of
2014. After describing the accelerator and the associated facilities, a
review is presented of the major achievements in the field realized over the
twenty years of intense research activity.
}

\maketitle
\setcounter{tocdepth}{3} \tableofcontents

%
%
\section{Introduction} 
\label{introduction}\setcounter{equation}{0}%

At the end of 2014 the experimental priorities of the Institut f\"ur
Kernphysik (IKP) J\"ulich switched from the study of hadronic reactions to
precision measurements that are more in keeping with current particle
physics. Since many interesting results had been found in the field of
hadronic physics over the twenty years of operation of the laboratory's
COoler SYnchrotron COSY, it is clearly appropriate to try to describe some of
these phenomena in the form of a review article.

Much of the programme at COSY was influenced by that of the SATURNE machine
at the Saclay laboratory but, though the machines had similar maximum proton
or deuteron beam energies, the accelerators had very different
characteristics and so, before describing the physics programme at COSY, it
is necessary to discuss in some detail the machine and the associated
facilities that were available for experiments. The main difference is, of
course, that COSY acts as a storage ring so that the various detectors
described in sect.~\ref{Facilities} are divided between those used for
experiments inside the ring and those designed for use at external target
stations. In contrast, SATURNE concentrated on the development of a whole
series of state-of-the-art magnetic spectrometers for use at external beam
lines.

The review is not intended merely to provide a synthesis of IKP Annual
Reports but rather it hopefully gives a critical evaluation, while trying to
make links between experiments carried out using some of the different
facilities available around COSY or, indeed, at other laboratories. For this
reason it was decided that the review should be prepared by a single person
rather than follow the practice of the multi-author volume that described in
1998 the legacy of the SATURNE programme~\cite{SAT2000}.

It is certainly impossible or even undesirable in this review to go into the
details of all the several hundred research papers that have emerged from
COSY over twenty years. In all cases of interest the reader is advised to go
back to the original sources where, for example, systematic uncertainties and
limitations or approximations are discussed at length. Since the aim is to
present the COSY experimental programme, we have been rather cavalier in the
discussion of the theoretical motivation for an experiment or its analysis.
In general, in order to keep the length under control, we have confined
ourselves to presenting only the phenomenology required to understand the
experimental results at a rather basic level.

This review is concerned with the hadronic physics programme at COSY and so
it omits any discussion of the extensive studies of spallation and nuclear
breakup studies carried out by the Nessi~\cite{LET2002},
Jessica~\cite{NUN2002}, and PISA~\cite{BUB2007} collaborations, especially in
the first few years of COSY. The results in these early stages are summarized
by an internal report in 2003 ``10 years of COSY'' and this shows the
dominance of the experiments that were launched quickly, most notably
COSY-11. In contrast, only simple experiments from ANKE were described and,
of course, WASA had not even arrived in J\"{u}lich by then. There were, of
course, conference proceedings, such as those of
Refs.~\cite{MAC2005,BRI2007}, but these only gave partial snapshots of the
research that was current at the time. This review aims to present a more
balanced picture over the twenty years. Also omitted is any description of
tests of equipment for use at other facilities, in particular the extensive
developments for the PANDA detector at the future FAIR complex.

In the space available, the brief descriptions of the machine and the
facilities available at COSY are necessarily incomplete and biased. Thus
there is no serious discussion of pellet or cluster-jet targets but, in
contrast, space is devoted to the polarized targets that allow many refined
experiments to be carried out at internal target stations. Technical
experiments are then discussed which show, for example, how the beam momentum
and the luminosity can be determined in a storage ring environment. Though by
themselves not giving immediate hadronic physics results, they facilitated
such experiments and are potentially important elements in the future
precision physics programme at COSY.

The subsequent sections deal with the COSY experiments in the order given in
the Table of Contents but it must be remembered that the final analysis of an
experiment may come several years after the data had been taken. In one,
hopefully extreme, example a paper was submitted for publication in 2015
based upon data that were taken at COSY seven years earlier. Hence some of
the unpublished results presented in this review must be considered as being
\emph{preliminary}. Only results available by October 1$^{\rm st}$ 2016 will
be reported on, though some analyses currently being worked on might be
indicated.

Although the hadronic physics programme has finished at COSY, the machine
itself lives on as the basis for a challenging programme of precision
physics. The most important element here is its use in putting constraints on
the electric dipole moments of the deuteron and proton. The results obtained
at COSY will be vital for the design of a dedicated ring that will lower the
limits even further and possibly even find non-zero values. The TRIC
experiment will search for the violation of time-reversal invariance in
proton-deuteron collisions. It would therefore be remiss if we did not
describe some of these exciting developments for the future, and this we do
in the Conclusions.

Since we are interested in the ``legacy of the experimental hadron physics
programme at COSY'', in the Conclusions we also try to pick out ten
experiments that we presently believe will have an influence on the field
long after the termination of the hadron physics programme at COSY. This is
necessarily a very speculative choice and we invite the readers to draw up
their own lists of alternative experiments.

%
%
\section{Facilities} 
\label{Facilities}
\setcounter{equation}{0}%
\subsection{The COSY machine} 
\label{COSY}

The first detailed description of the COoler SYnchrotron and storage ring
COSY is almost twenty years old~\cite{MAI1997} but, despite several
modifications, the underlying structure remains the same. The machine, which
is capable of accelerating polarized or unpolarized protons or deuterons up
to momenta of about 3700~MeV/$c$, is equipped with both electron cooling and
stochastic cooling to provide quality beams.

\begin{figure}[htb]
\begin{center}
\includegraphics[width=0.95\columnwidth]{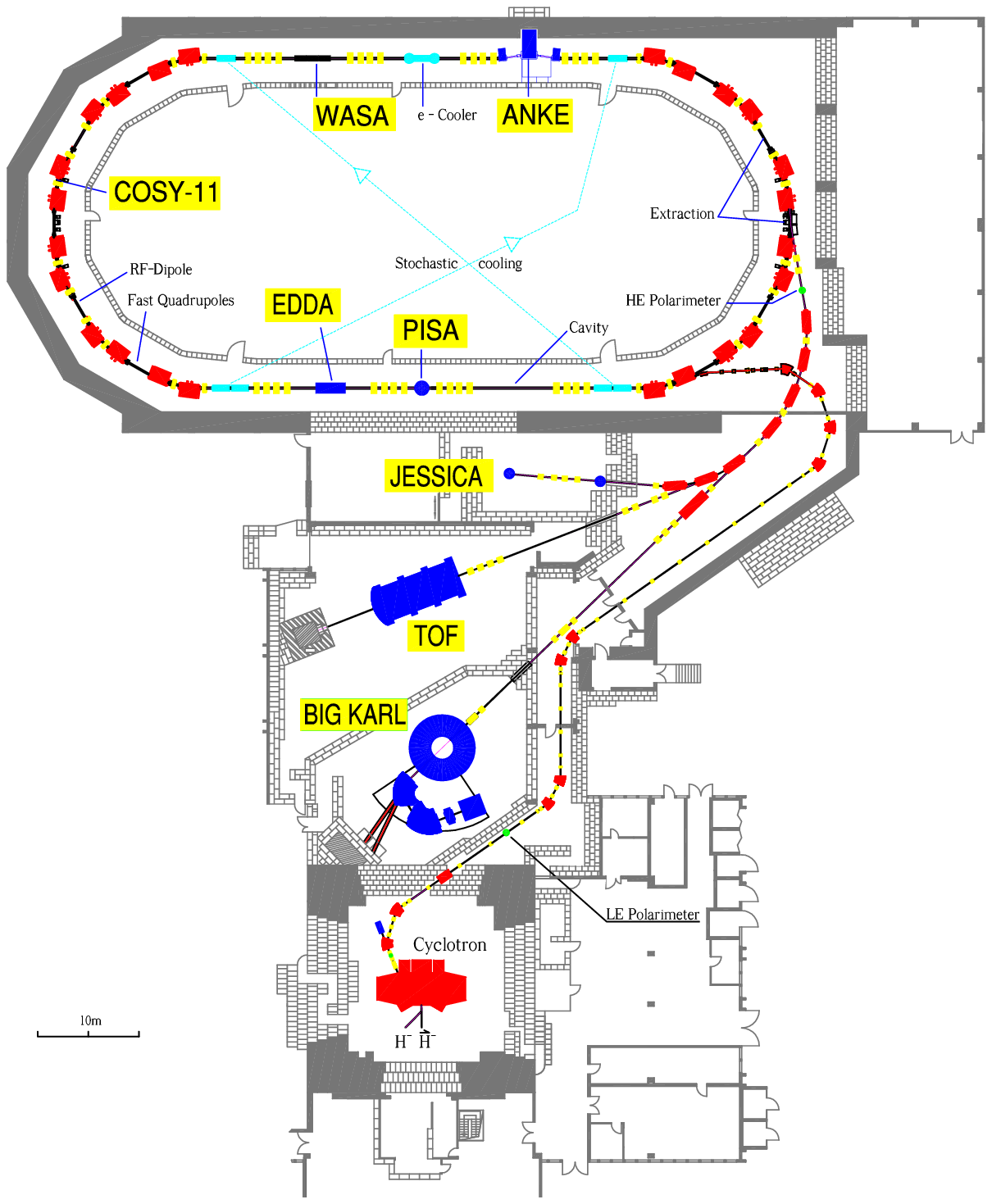}
\caption{\label{fig:COSY} Floor plan of the COSY facility, showing the
injection cyclotron JULIC and the principal internal and external detectors.
The PISA, Jessica, and Nessi (situated after TOF) detectors were used purely
for nuclear reaction studies and will not be discussed in this review. The
PISA location is now used for the PAX detector and the TRIC experiment that
is discussed in sect.~12.}
\end{center}
\end{figure}

A sketch of the overall layout of the facility is given in
Fig.~\ref{fig:COSY}. This shows the 100~m long transfer beam line from the
injector isochronous cyclotron to the ring and the cooler synchrotron itself,
which has a circumference of about 184~m. This racetrack is made up of two
arcs, each 52~m long, and two 40~m straight sections, where some of the
larger experimental equipments are installed. In addition to the experimental
detectors indicated\footnote{The PISA, Jessica, and Nessi (situated after
TOF) detectors were used purely for nuclear reaction studies and will not be
discussed in this review.}, the ring also contain accelerator-specific
components, such as the accelerating \emph{rf}-cavity, the electron cooler,
scrapers, the stochastic pick-up and kicker tanks, Schottky pick-ups, and
beam current monitors. There are also three extracted beam lines serving
external experimental areas.

COSY has two ion sources, one for polarized and another for unpolarized
beams, each of which yields H$^-$ and D$^-$ ion beams. By using different
combinations of the \emph{rf} transitions in the source, it is possible to
produce deuteron beams with different mixtures of vector and tensor
polarizations~\cite{WEI1996,FEL2002}. Unlike the SATURNE accelerator, no
attempt was made to create beams of heavier nuclei, such as $^4$He or $^6$Li.
The ions are pre-accelerated in the cyclotron JULIC, up to 295 MeV/$c$ for
H$^-$ and 539~MeV/$c$ for D$^-$, before being injected into the storage ring
via charge exchange, using a carbon foil stripper. The low energy polarimeter
in the injection line, which uses a carbon target, can determine the proton
and deuteron vector polarizations but is insensitive to the deuteron tensor
polarization.

Two different cooling techniques to shrink the beam phase space are
implemented at COSY. Electron cooling~\cite{STE2003a}\footnote{A more
powerful electron cooler is now installed at COSY but this was not available
for the hadron physics programme described in this review.} is successful up
to momenta of $600$~MeV/$c$ and this is complemented by a stochastic cooling
system that covers the upper momentum range from $1.5$~GeV/$c$ to
$3.3$~GeV/$c$~\cite{PRA2000}. These cooling techniques significantly reduce
the momentum spread of the COSY beam, such that a momentum resolution down to
$\Delta p/p = 10^{-3} - 10^{-5}$ has been achieved. The space charge limit on
the number of stored protons or deuterons in the ring is about
$2\times10^{11}$ and, by using stacking injection, values as high as
$6\times10^{10}$ have been obtained in practice. Since the beam revolution
frequency is of order 1~MHz, this would correspond to close to $10^{17}$
particles per second passing an internal target.

In a strong-focusing synchrotron, such as COSY, intrinsic or imperfection
resonances can lead to losses of polarization of a proton beam during
acceleration. In order to compensate for these effects, adiabatic spin-flip
has been used to overcome the imperfection resonances and tune-jumping to
deal with the intrinsic ones~\cite{LEH2003}. The situation is much simpler
for deuteron beams since, because of the much smaller gyromagnetic anomaly,
there are no resonances for deuterons throughout the whole of the COSY
momentum range.

Both resonant (slow) and stochastic extraction have been used at COSY to
populate the beam lines that serve the external experimental areas, whose
locations are shown in Fig.~\ref{fig:COSY}. The maximum extracted proton beam
momentum achieved was 3300~MeV/$c$, which is somewhat below the maximum
circulating momentum of 3700~MeV/$c$.

%
%

\subsection{Principal installations} 
\label{main}
\subsubsection{The COSY-11 spectrometer} 
\label{COSY-11}

COSY-11~\cite{BRA1996} was one of the simpler facilities to be implemented at
COSY and this allowed the collaboration to carry out quite rapidly many
near-threshold measurements. Its brilliant simplicity was that it used one of
the existing dipoles of the COSY ring as an analyzing magnet of a
spectrometer. Although this idea was also exploited at the CELSIUS storage
ring~\cite{CAL1998,BAR2006}, the COSY-11 installation was far more ambitious.

\begin{figure}[htb]
\begin{center}
\includegraphics[width=0.95\columnwidth]{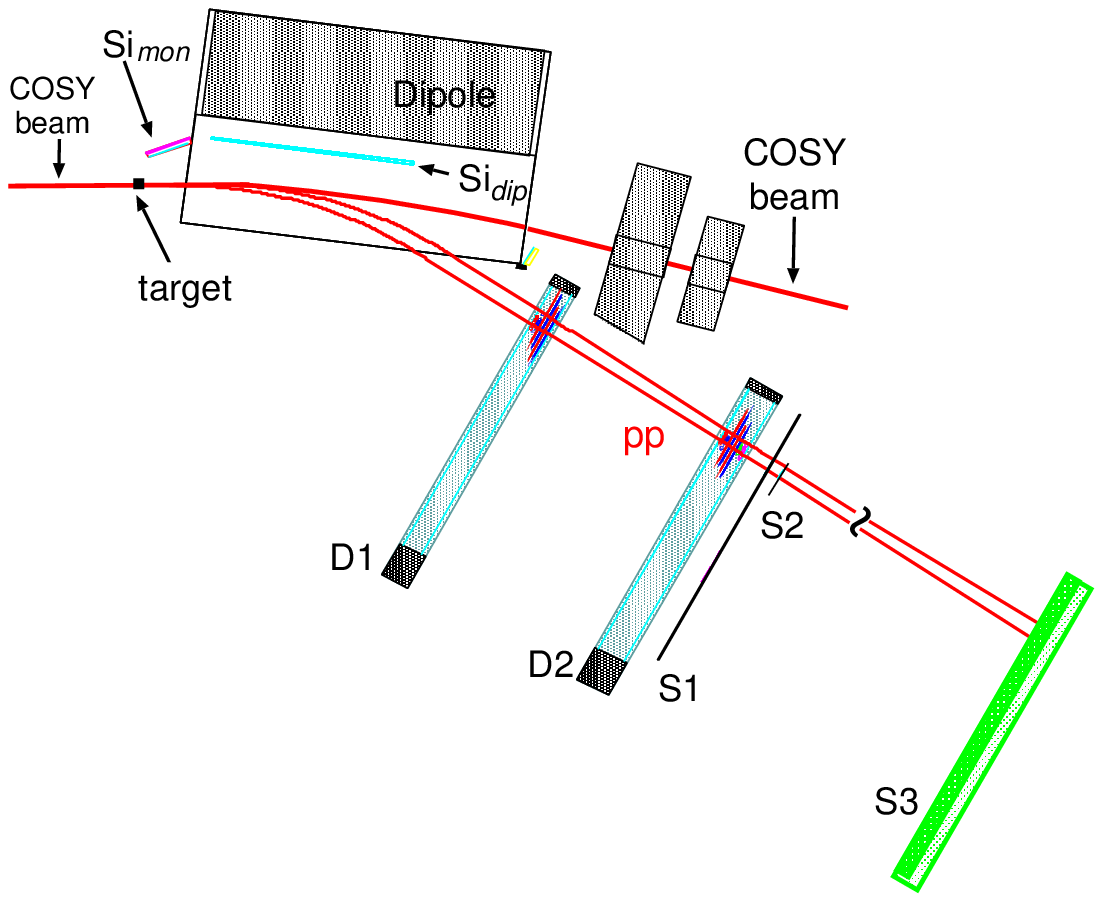}
\caption{\label{fig:COSY11} Schematic view of the COSY-11 facility, as used
for the measurement of the $pp\to ppX^0$ reactions near
threshold~\cite{SMY2000}. The proton trajectories are measured by means of
hits in two sets of drift chambers D1 and D2. The scintillation hodoscopes S1
and S2 are used as start detectors and S3 as the corresponding stop detector
for time-of-flight measurements. Si$_{\rm mon}$ is a silicon pad monitor
detector that is used to measure the recoil proton from $pp$ elastic
scattering for the normalization of cross sections and the calibration of the
detection system. Si$_{\rm dip}$ could be used to detect negatively charged
particles.}
\end{center}
\end{figure}

The position of the detector inside COSY, which is shown in
Fig.~\ref{fig:COSY}, is in a bending section of the ring in a dispersive
region, so that the effective beam momentum spread \emph{seen} by the target
is much reduced. The basic principles of the facility are illustrated for the
most straightforward $pp\to ppX^0$ reaction~\cite{SMY2000} in
Fig.~\ref{fig:COSY11}. This shows the location of the hydrogen cluster target
which, crucially, perturbs very little the circulating COSY beam. Due to
their lower momenta, the two outgoing protons from a meson production
reaction are separated from the beam in the magnetic field of the C--shaped
dipole and are bent towards the centre of the COSY ring, where they can be
seen by the COSY-11 detectors.

The proton trajectories are measured by means of hits in a set of two drift
chambers (marked D1 and D2 in Fig.~\ref{fig:COSY11}), which allow the momenta
to be determined by ray tracing back through the precisely known magnetic
field to the target position. Identification of the particles as protons is
ensured by measuring also the times of flight over a distance of
$\approx$~9.4~m between the start and stop scintillator hodoscopes (S1 and
S3). The neutral mesons, $X^0 = \eta\ \textrm{or}\ \eta^{\prime}$, are not
registered directly but are identified by peaks in the missing-mass
distributions. The isolation of these peaks is helped by looking at
background data taken just below the threshold for the production of that
meson. The geometrical acceptance of the COSY-11 detection system is limited,
especially in the vertical direction, due to the narrow opening of the dipole
gap with an internal height of 60~mm.

The beam and target parameters could be monitored and the cross sections
normalized by measuring proton-proton elastic scattering in
parallel~\cite{MOS2001}. This was achieved with the help of the silicon pad
monitor detector Si$_{\rm mon}$, which measured the recoil proton.

Close to threshold the two protons from the $pp\to pp\eta$ reaction must
emerge with very similar momenta aligned close to the beam direction. In this
case the geometric acceptance of COSY-11 is high but it falls quickly with
increasing excess energy. It must be stressed that, unlike detectors such as
WASA or Big Karl, where there is a hole that allows the passage of the beam,
the COSY-11 coverage is essentially the highest near the forward direction.
The detector is therefore well adapted to making measurements near threshold.

For other reactions involving three-body final states, such as
$K^+p\Lambda/K^+p\Sigma^0$, the proton and kaon are registered in D1/D2 and
S1/S2 and the hyperon isolated using the missing-mass method.

Though for certain experiments the facility was expanded by, for example, the
addition of a neutron wall, COSY-11 was most successful when it was kept
(comparatively) \emph{simple} and the collaboration had a remarkable record
in the measurement of near-threshold meson production that will be discussed
in later sections

\subsubsection{The ANKE spectrometer} 
\label{ANKE}

\begin{figure}[htb]
\begin{center}
\includegraphics[width=0.95\columnwidth]{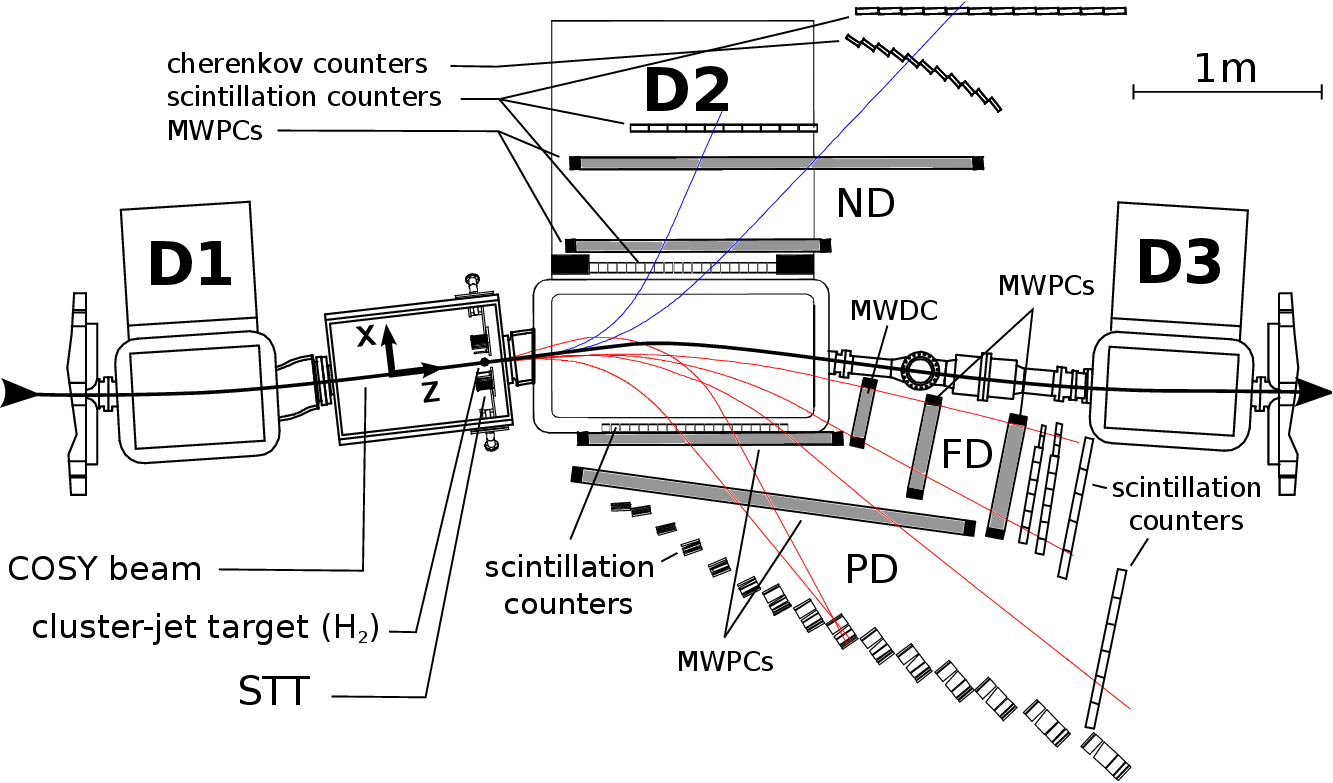}
\caption{\label{fig:ANKE} The ANKE spectrometer setup~\cite{BAR2001}, showing
the positions of the three dipole magnets (D1,D2,D3), the hydrogen
cluster-jet target, the Silicon Tracking Telescope (STT), and the Positive
(PD), Negative (ND), and Forward (FD) detectors. }
\end{center}
\end{figure}

The motivations for the COSY-11 and ANKE spectrometers have much in common
since they were both designed as Zero Degree Facilities or, more accurately,
as small angle facilities. However, in contrast to COSY-11, whose development
was ``straightforward'', ANKE has evolved into possibly the most complex
detector at COSY. The basic design is described in Ref.~\cite{BAR2001} and
the final layout shown in Fig.~\ref{fig:ANKE} allows much more space than is
available at COSY-11.

As shown in Fig.~\ref{fig:COSY}, ANKE is placed in one of the two straight
sections of COSY which means that, unlike COSY-11, it requires its own
dedicated analyzing magnet. ANKE's basic structure is built around three
dipoles. D1 deflects the circulating COSY beam through an angle $\alpha$ onto
the target. A spectrometer dipole magnet D2, is used to analyze the momentum
of the reaction products that originate from a beam-target interaction. D2
deflects the residual beam by an angle $-2\alpha$, which is then compensated
by D3, which returns the beam back to the nominal orbit. For each value of
$\alpha$, which is chosen optimally for a given experiment, D2 has to be
moved perpendicularly to the COSY straight section and, for this purpose, D2
is installed on rails. The whole system thus forms a kind of chicane in the
COSY racetrack.

Although Fig.~\ref{fig:ANKE} indicates a H$_2$ cluster-jet target, which may
also be D$_2$, one can alternatively use a strip target, where the beam is
steered onto it after injection and acceleration. For experiments requiring
single- and double-polarization, there is also space in the ANKE target
chamber for a polarized storage-cell gas target fed by an atomic beam source,
which are discussed in sects.~\ref{cells} and \ref{ABS}, respectively.

The silicon tracking telescopes (STT), described in sect.~\ref{STT}, which
can be placed in the target chamber, are particularly helpful in ensuring
some left-right symmetry and also in defining the vertex when using a storage
cell.

The ANKE detection system consists of three distinct parts, viz.\
\begin{itemize}
\item The forward detector (FD) measures high momentum particles.
    $0.8<p<3.7$~GeV/$c$, close to the COSY beam orbit.
\item The positive detector (PD) measures positive projectiles with $0.3
    <p <0.8$~GeV/$c$ and covers much larger angles than the FD.
\item The negative detector (ND), which is located partially inside the
    D2 magnet frame, is used to measure negatively charged pions and
    kaons. Its momentum coverage is similar to that of the PD.
\end{itemize}
All three systems employ multiwire proportional chambers for track
reconstruction and plastic scintillator counters to obtain time information.

In concept the forward detector is very similar to that used at COSY-11,
being optimized to measure charged particles emitted near the forward
direction, using tracking and time-of-flight information, Its angular
acceptance is about $12^{\circ}$ in the horizontal plane but only about
$3.5^{\circ}$ in the vertical.

Especial mention should be made of the 15 focal-surface telescopes placed
after the PD, that are used to unambiguously identify $K^+$. Each of these
telescopes is made up of a stop counter, an energy-loss counter, a
delayed-veto counter, and two passive degraders chosen such that a $K^+$
stops either at the edge of the first or in the second degrader. The products
from the $K^+$ decay are registered in the delayed-veto counter, with the
characteristic decay time of 12.4~ns. The delayed veto criterion leads to a
suppression of better than $10^{-5}$ in the non-kaon background for both
inclusive and coincidence measurements~\cite{BUS2002}.

\subsubsection{The WASA detector} 
\label{WASA_detector}	

In its first decade of its operation, COSY was only equipped with detectors
for charged particles. Although eventually some of the facilities, such as
COSY-11 or TOF, were enhanced through the addition of neutron walls, no
effort was made to construct a detector for photons or electrons. However
this changed when it was realized in about 2002 that the already operational
WASA detector would soon become available.

WASA was installed at the CELSIUS (Uppsala) storage ring and it was
originally designed for the study of rare $\pi^0$ decays but its remit was
extended to look for the more interesting $\eta$ decays. The operation of the
detector at CELSIUS is described in Ref.~\cite{BAR2008}. Following the
closure of the CELSIUS ring in 2004, the detector was transferred to
J\"{u}lich and installed in the COSY ring. As envisaged in the WASA
proposal~\cite{ADA2004}, the much higher maximum proton energy available at
COSY (2.9~GeV) compared to CELSIUS (1.4~GeV) meant that major upgrades were
needed, especially in the forward detector. but also in the readout system,
to allow heavier mesons, such as the $\eta^{\prime}$, to be studied. An
up-to-date description of experiments using this detector is to be found in
Ref.~\cite{ADL2015}.

\begin{figure}[htb]
\begin{center}
\includegraphics[width=0.95\columnwidth]{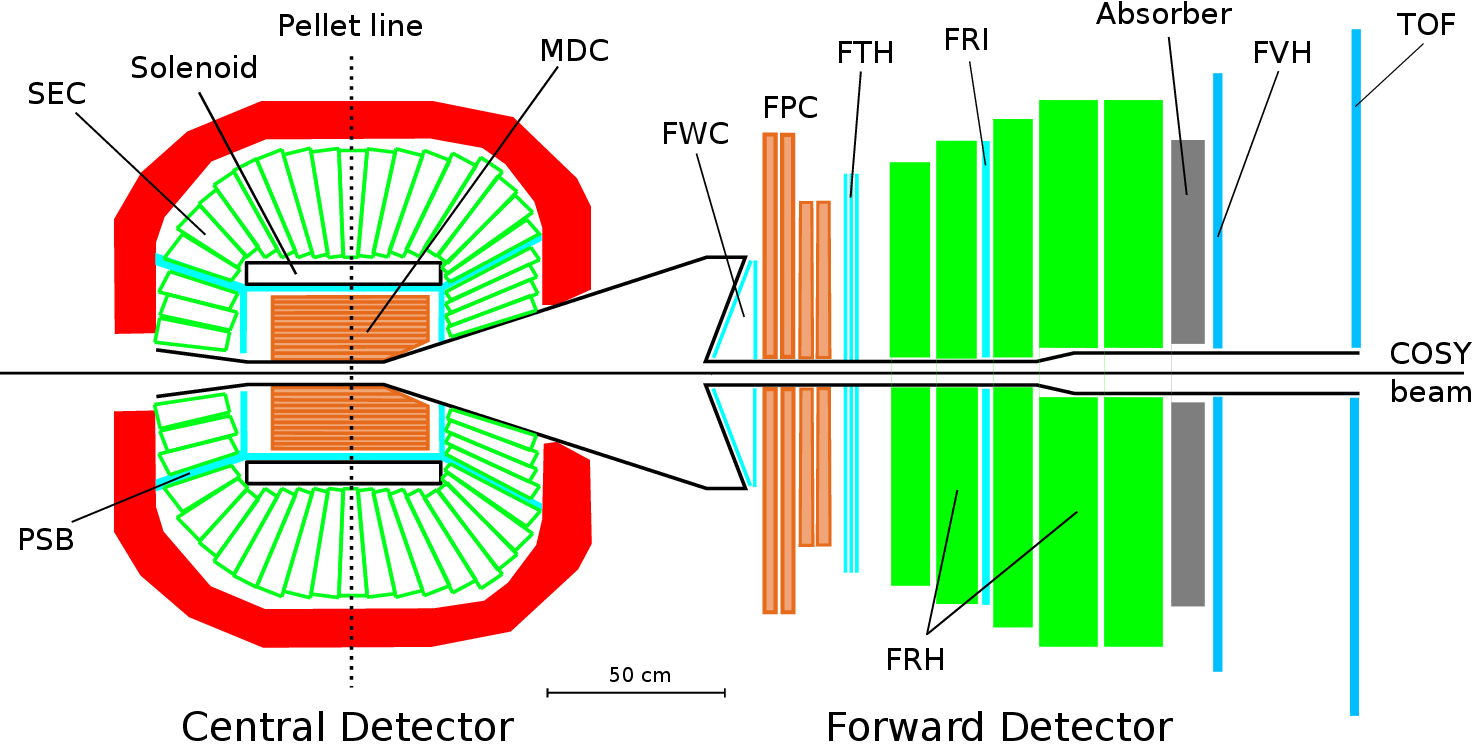}
\caption{\label{fig:wasadet} A cross-sectional diagram of the WASA
detector~\cite{ADA2004}, with the beam coming from the left. Hadronic
ejectiles are measured with the forward detector on the right while meson
decay products are measured with the central detector on the left. The
individual components are described in the text. }
\end{center}
\end{figure}

The forward detector of the WASA spectrometer is designed to measure hadronic
ejectiles and the central detector to measure light mesons or their decay
products. A cross-sectional view of the apparatus is shown in
Fig.~\ref{fig:wasadet}. The forward detector, which registers particles
emitted with polar angles from about $3^{\circ}$ to $18^{\circ}$, consists of
an arrangement of thin and thick plastic scintillators and drift chambers
covering the full azimuthal angle.  Thick scintillators in the forward range
hodoscope (FRH) are designed to measure energy loss via ionization. Thin
scintillator layers in the forward window counter (FWC) and forward trigger
hodoscope (FTH) provide precise timing information. The kinetic energy and
the particle type can be determined from the pattern of energy deposits in
the thin and thick scintillator layers.  A proportional chamber system (FPC)
consists of eight layers, each with 260 aluminized Mylar straws. Layers of
the forward detector beyond the first layer of the FRH included the Forward
Range Interleaving Hodoscope (FRI) detector and the Forward Veto Hodoscope
(FVH). The kinetic energy $T$ of a proton from say a $pp\to pp\eta$ reaction
can be reconstructed with a resolution of $\sigma(T)/T \sim 1.5-3\%$ for $T <
400$~MeV.

The central detector, which is designed to measure photons, electrons, and
charged pions, is surrounded by a CsI(Na) electromagnetic calorimeter with
1012 elements (SEC). Contained within the calorimeter is a superconducting
solenoid providing a uniform 1~T magnetic field in the space directly
surrounding the interaction region. Charged particle tracking is provided by
the mini drift chamber (MDC), which is surrounded by an 8~mm thick plastic
scintillator barrel (PSB) that provides precise timing and particle
identification. The MDC consists of 4, 6, and 8 mm diameter straw tubes
arranged in 17 layers that are alternately axial or skewed by $+3^\circ$ or
$-3^\circ$ relative to the beam axis in order to provide three-dimensional
tracking. An iron return yoke, shown in red in Fig.~\ref{fig:wasadet},
surrounds the central detector and shields the photomultplier tubes of the
SEC from the magnetic field. As suggested in the figure, the polar angle
range is from $20^{\circ}$ to $169^{\circ}$, which represents about 96\% of
the geometrical acceptance. The energy resolution of the calorimeter is
$\sigma(E)/E \approx 5\%/\sqrt{E{\rm [GeV]}}$.

The WASA facility is equipped with an internal target, where frozen pellets
of hydrogen or deuterium are injected at rates of several thousand per
second~\cite{EKS1996} perpendicular to the COSY beam, as indicated in
Fig.~\ref{fig:wasadet}. The pellets have typical diameters of the order of
30~$\mu$m, which provide a target density on the order of
$10^{15}$~atoms/cm$^2$. Smaller pellets might be desirable but there is then
the danger of blocking the nozzle producing the pellets. Though vacuum pumps
are positioned as closely as possible to the interaction region, a certain
amount of residual gas is present in the region around the target due to the
evaporation of pellets. In tests with a deuterium target, $pd\rightarrow
{^{3}\textrm{He}}\,\pi^{+}\pi^{-}$ events were selected and the vertex
determined from the pion tracks~\cite{ADL2015}. Over 90\% of these events
originated within one centimetre of the centre of the interaction region.

In a reaction such as $pp\to pp\eta$, the two recoil protons are measured in
the forward detector and this allows the $\eta$ to be selected via the
missing mass in the reaction. The $\eta$ decay products are then measured in
the central detector. Although this is the only facility at COSY that is
capable of such measurements, it must be realized that, in the absence of a
magnetic spectrometer, the missing-mass resolution is not optimal and that
this might lead to extra background, depending upon the particular
experiment.

\subsubsection{The Time-of-Flight detector} 
\label{TOF}

Most of the spectrometers used at COSY rely on the analysis of trajectories
in a magnetic field. In contrast, the COSY-TOF spectrometer is based on the
measurements of the velocity vectors of all the charged products by combining
the hit positions in various detectors together with very careful
time-of-flight determinations~\cite{BOH2000}. The identification of the final
particles is then achieved through the study of what reaction could produce
such a velocity distribution at that particular energy and then optimizing
through a kinematic fitting procedure.

Some of the most important advances achieved through the use of TOF have been
in the field of strangeness production, where the delayed decays of neutral
particles can lead to very characteristic patterns. Thus, in the reaction
$pp\to K^+p\Lambda$, there may initially be only two tracks, corresponding to
the charged particles $K^+$ and $p$ but, after the decay $\Lambda \to
p\pi^-$, four charged tracks can be seen in TOF. The design of COSY-TOF was
certainly influenced by the experience gained with the PS185 spectrometer
used at the CERN Low Energy Antiproton Ring (LEAR)~\cite{BAR1991}. In
particular the technique to detect a $\Lambda$ hyperon through its delayed
decay was developed here.

The requirements that were set for the COSY-TOF spectrometer
were~\cite{HAU2014}:
\begin{itemize}
\item Full geometrical reconstruction of all charged particles of a
    reaction,%
\item Reconstruction of the primary and secondary vertices,%
\item Reconstruction of the momenta through Time of Flight for additional
    kinematic information,%
\item High background rejection. \end{itemize}

\begin{figure}[htb]
\begin{center}
\includegraphics[width=0.95\columnwidth]{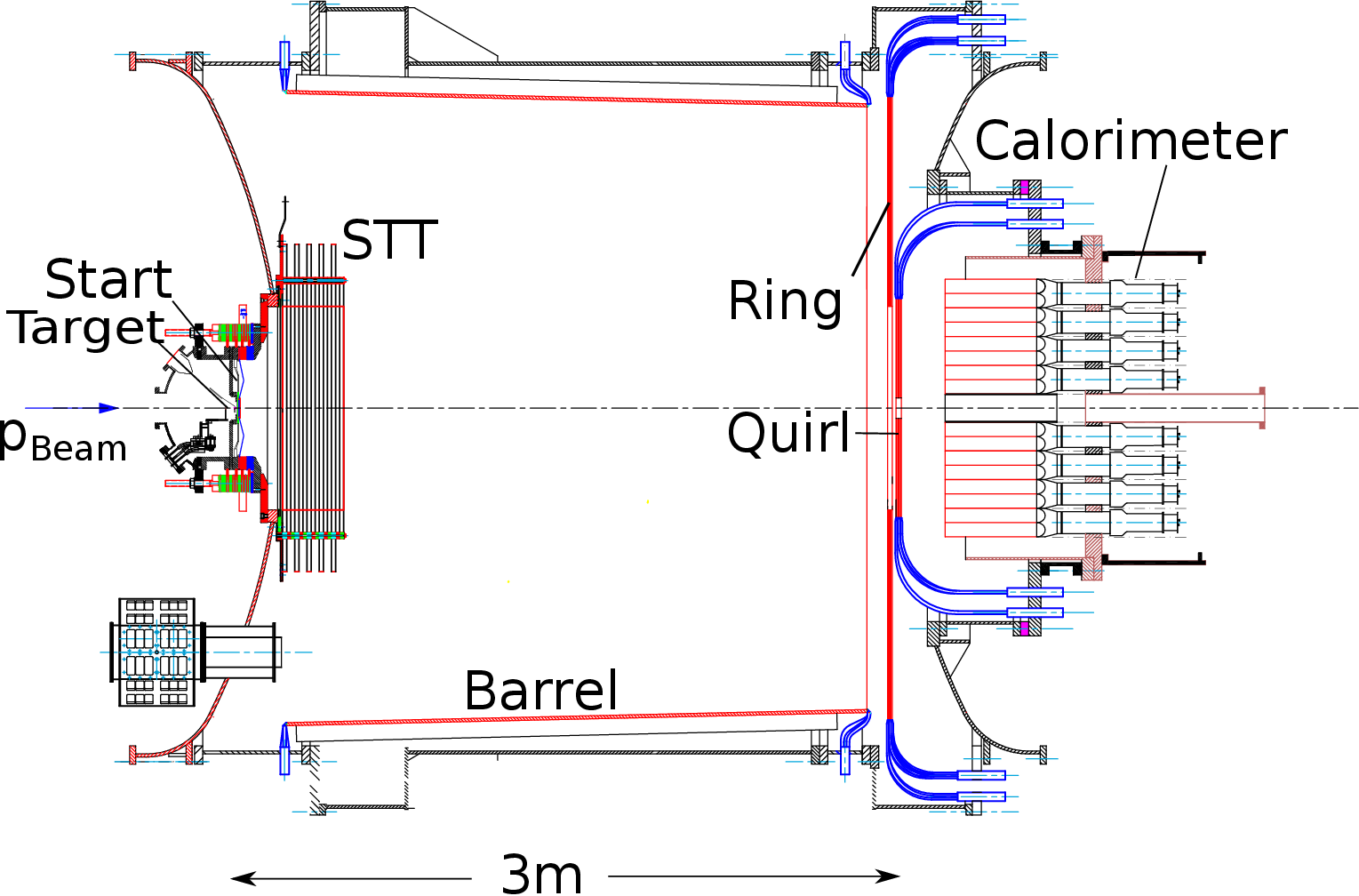}
\caption{\label{fig:TOF} Schematic drawing of the COSY-TOF detector,
Following the direction of the beam, after the target there are the start
counter (Start), the straw tube tracker (STT), the barrel scintillators, the
inner ring (Quirl), the outer ring (Ring), and the Calorimeter. All detectors
and the liquid target are located inside the vacuum
vessel~\cite{JOW2016}.}
\end{center}
\end{figure}

To fulfill all these requirements, TOF was built in a modular way such that
it is possible to change the detector length and the position of different
sub-detectors. A typical setting with the so-called long barrel is shown in
Fig.~\ref{fig:TOF}~\cite{JOW2016}; the calorimeter that gives energy
information is generally only used in conjunction with the shorter barrel. It
is seen from the figure that the diameter of the stainless steel barrel is
also around 3~m, which makes it physically the largest detector installed at
COSY.

Despite the large size, all the detectors as well as the liquid hydrogen or
deuterium target are positioned inside the vacuum tank, where the residual
pressure of less than $7\times 10^{-4}$~mbar minimizes the effects of
secondary interactions and multiple scattering.

The charged particles produced in interactions in the target are registered
in three different groups of detectors. The first consists of a set of
plastic scintillators, providing a trigger and the start signal for the TOF
measurement, and a silicon quirl detector for precise track information near
the primary vertex. The latter is segmented into 128 Archimedian spirals,
each of which covers an azimuthal angular range of 180$^{\circ}$.

All the recent experiments at TOF were carried out with the next crucial
element, the Straw Tube Tracker (STT), that is situated 25~cm downstream from
the target. This consists of 2704 straw tubes, each of which is a cylindrical
minidrift chamber. The STT actually gives the most precise information for
track and vertex reconstruction. For example, the achieved resolution in
momentum and invariant mass of the $pK^+\Lambda$ final state is significantly
better than that obtained purely from the time-of-flight measurement.

The end detector region consists of scintillators covering the full
cylindrical inner surface of the vacuum tank and the end cap (Quirl and Ring
detector) and the calorimeter. The Quirl detector consists of three layers of
scintillators, the first of which is structured in 48 wedge-shaped slices,
with the other two being in the form of 24 Archimedian spirals, oriented in
opposite directions. The structure of the Ring detector is similar to the
Quirl but with twice the number of elements per layer,

Though, as seen clearly in Fig.~\ref{fig:TOF}, the TOF detector covers only a
$2\pi$ solid angle in the laboratory system, very few particles go backwards
in this frame and so the geometric coverage is almost complete. The loss of
events due to the  hole that allows the primary beam to pass through the
start and silicon quirl detectors can also be minimized. Charged particles
can be triggered by the stop detector and evaluated by the straw detector
starting from $2^{\circ}$ but, for these low angles, the start timing and
start trigger has to be provided by a second charged particle, for example
the proton if the track of the kaon in a $pK^+\Lambda$ final state lies very
close to the forward direction.

Of course, without a magnetic field it is not possible to directly determine
the sign of the charge of a meson and so it is then not possible to study the
full structure in, for example, $pp\to ppK^+K^-$, which is discussed in
sect.~\ref{ppKK}.

%
%

\subsubsection{The Big Karl spectrometer} 
\label{BK}

It was mentioned in the introduction that the SATURNE accelerator was
equipped with a series of magnetic spectrometers placed on external beam
lines. The only similar facility at COSY was the magnetic spectrometer Big
Karl~\cite{MAR1983}. This was actually designed as a QQDDQ facility for
measurements at the JULIC cyclotron which, as discussed in sect.~\ref{COSY},
is now used as the injector for COSY providing, for example, 45~MeV protons.
Big Karl was used from about 1979 for studies of nuclear levels production
in, for example, $(p,p^{\prime})$ or $(p,d)$ reactions.

\begin{figure}[htb]
\begin{center}
\includegraphics[width=0.8\columnwidth]{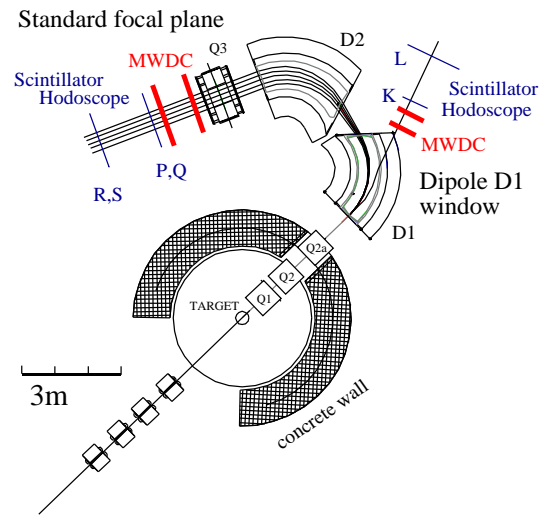}
\caption{\label{fig:BK} Top view of the Big Karl magnetic
spectrometer~\cite{BOJ2002}. In the standard 3Q2DQ mode of operation,
particles exiting from the target cell pass through the quadrupoles Q1, Q2
and Q2a and are bent to the focal plane by the dipoles D1 and D2. In the 3QD
mode, high rigidity particles produced in the target are also registered
emerging from the side hole of the first dipole D1. }
\end{center}
\end{figure}

The spectrometer's design was subject to an initial modification in order to
carry out experiments at the higher energies available at
COSY~\cite{DRO1998}. In particular, the two entrance quadrupole magnets were
replaced by three quadru\-pole magnets having larger geometrical acceptances
and higher maximum magnetic field strengths. This resulted in a version of
Big Karl that was a high resolution 3Q2DQ spectrometer, though the final
quadrupole was often found not to be needed. The lay-out of the spectrometer
is illustrated schematically in Fig.~\ref{fig:BK}. It is equipped with two
sets of multi-wire drift chambers (MWDC) for position measurement and two
layers of scintillating hodoscopes for time-of-flight and energy loss
information, used for particle identification.

The most important parameters of the 3Q2DQ version of Big Karl are summarized
in Table~\ref{BKtab}~\cite{DRO1998}. One feature that is clearly relevant for
several of the experiments carried out at this facility is the momentum
acceptance. Thus, when measuring, for example, inclusive $\pi^+$ or $K^+$
production in proton-proton collisions more than one setting of the central
momentum is required in order to span the physics region of interest. In such
cases it is important to ensure significant overlap between the settings.

\begin{table}[hbt]
\centering
\begin{tabular}{|l|l|} \hline
Magnet structure& 3Q2DQ \\
Central radius& 1.94~m\\
Magnetic rigidity& $0 \leq B\rho \leq 3.6$~Tm\\
Max.\ proton momentum& 1080~MeV/$c$\\
Momentum resolution& $\Delta p/p \approx 10^{-4}$\\
Dispersion& 6.47~cm/\%\\
Momentum acceptance& $\pm 4.5\%$\\
Max.\ horizontal angular acceptance& $\pm28$~mrad\\
Max.\ vertical angular acceptance& $\pm 110$~mrad \\
\hline
\end{tabular}
\caption{Selected properties of the 3Q2DQ version of the Big Karl magnetic
spectrometer~\cite{DRO1998}.} \label{BKtab}
\end{table}

However, the initial modifications described in Ref.~\cite{DRO1998} were not
sufficient to exploit the full possibilities offered by the increased energy
available at COSY. In particular, the maximum momentum per charge of
1080~MeV/$c$ was too low, for example, to measure the fast tritons from the
$pd\to{}^3$H$\,\pi^+$ reaction over much of the COSY energy range. For such a
two-body reaction, where the identification of one particle is sufficient to
isolate the reaction and momentum resolution is less critical, a
supplementary mode of Big Karl was installed~\cite{BOJ2002}. As illustrated
in Fig.~\ref{fig:BK}, high rigidity particles could be measured by putting
detectors after the exit of the first dipole so that for these particles Big
Karl works as a 3QD spectrometer, where the momentum range is extended up to
3240~MeV/$c$. This would allow the measurement of the $pd\to{}^3$H$\,\pi^+$
and $pd\to{}^3$He$\,\pi^0$ cross sections under similar conditions to check
charge independence, as discussed in sect.~\ref{pd3hepi}.

Additional detectors were often used in combination with Big Karl in order to
register extra particles. The GEM detector of sect.~\ref{GEM} could also be
used in a stand-alone mode but the MOMO detector of sect.~\ref{MOMO} and
the ENSTAR detector of sect.~\ref{ENSTAR} were designed to be used in
conjunction with Big Karl.

\subsubsection{The GEM detector} 
\label{GEM}

\begin{figure}[htb]
\begin{center}
\includegraphics[width=0.8\columnwidth]{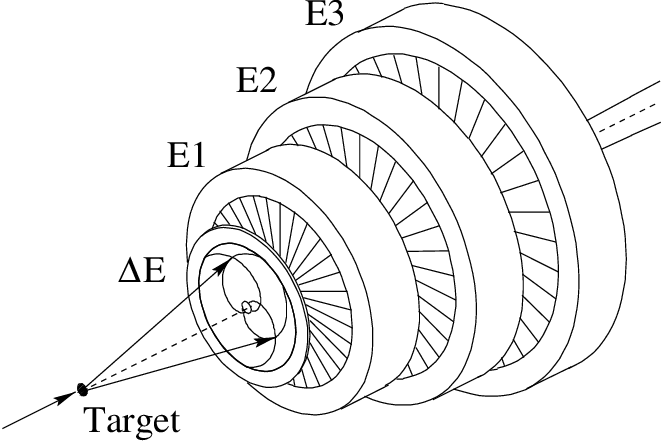}
\caption{\label{fig:GEN} Sketch of the GEM detector~\cite{BET1999a}.}
\end{center}
\end{figure}

Though the Big Karl spectrometer has excellent resolution, its limited
angular acceptance meant that it has to be moved several times in order to
produce an angular distribution. For two-body reactions, such as
$pd\to{}^{3}$He$\,\eta$, a detector with lower resolution but larger angular
coverage might be preferable. A sketch of the GEM detector, designed for this
purpose, is shown in Fig.~\ref{fig:GEN}.

The first element in the GEM detector~\cite{BET1999a} (the so-called Quirl)
measures the position and energy loss of penetrating particles. The active
area of this diode is divided on both sides by 200 grooves. Each groove is
shaped as an Archimedian spiral covering an angular range of $2 \pi$, turning
in opposite directions on the front and rear of the detector. This is
followed by three high purity germanium detectors with radial symmetry with
respect to the beam axis, as shown in the figure. These are mainly used for
measuring the energy loss of penetrating particles or the total kinetic
energy of stopped particles. These detectors are divided into 32 wedges to
reduce the counting rate per division and this leads to a higher maximum
total counting rate of the total detector.

As used in experiments, GEM subtended an opening angle of about
16.5$^{\circ}$ at the target and this limited in particular the excess
energies up to which it could be operated. In addition, in the centre of each
detector there was a hole of angular size $1.6^{\circ}$ that allowed the
passage of the primary beam. Though GEM was designed primarily as a
stand-alone facility, the presence of the central hole permitted it to be
used in combination with Big Karl, which then worked as a zero-degree
spectrometer. Thus the GEM collaboration worked with either the GEM detector
or Big Karl, or with both.

\subsubsection{The MOMO detector} 
\label{MOMO}

\begin{figure}[hbt]
\begin{center}
\includegraphics[width=0.7\columnwidth]{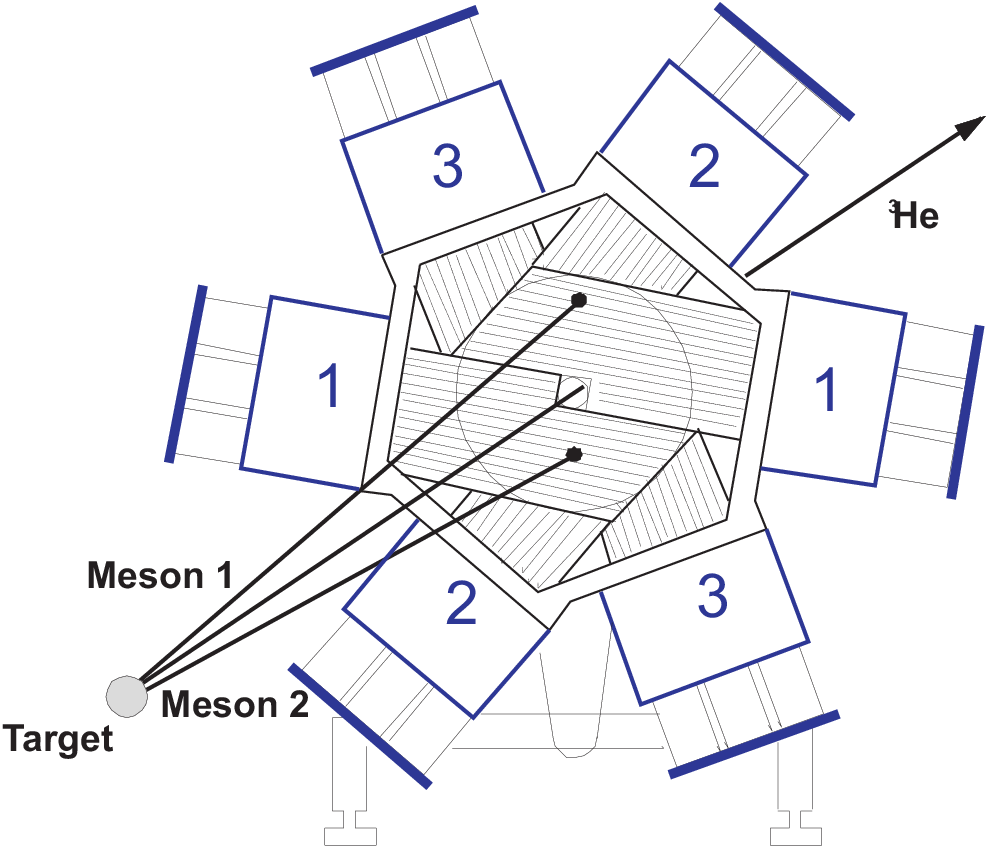}
\caption{Front view of the MOMO vertex detector~\cite{BEL1999,BEL2007}, with
the indication of a typical event. Both the primary beam and the recoil
$^3$He detected in Big Karl pass through the central hole. The numbers denote
the different layers and the three boxes at the end of each read--out
symbolize the phototubes.} \label{fig:MOMO}
\end{center}
\end{figure}

The MOMO (Monitor-of-Mesonic-Observables) vertex detector was specifically
designed for the measurement of the charged mesons $X^{\pm}$ from the
$pd(dp)\to{}^{3}\mbox{\rm He}\,X^+X^-$ reaction~\cite{BEL1999,BEL2007}, where
the $^3$He would be analyzed in Big Karl. A schematic view of the detector is
shown in Fig.~\ref{fig:MOMO}.

MOMO consists of 672 scintillating fibres, arranged in three planes, denoted
by (1,2,3) in Fig.~\ref{fig:MOMO}. The fibres are individually read out by
16-anode multichannel photomultipliers. The fibres in the three planes are
rotated by 60$^{\circ}$ with respect to each other and hits in three layers
are required in order to avoid combinatorial ambiguities. It is important to
note that the sign of the charge on each of the mesons $X^{\pm}$ is not
determined and this automatically leads to the symmetrization of some of the
differential distributions.

In its applications, the MOMO detector was placed perpendicular to the beam
direction 20~cm downstream of the target, outside a vacuum chamber, the end
wall of which was a 5~mm thick aluminum plate. The central hole, which
subtended an angle of 6$^{\circ}$ at the target, allowed the passage of the
primary beam and also the $^3$He that were detected in Big Karl. The maximum
angle of 45$^{\circ}$ was set by the physical dimensions of MOMO.

Each scintillating fibre is 2.5~mm thick but, when operating with a deuteron
beam, these were too thin to provide reliable energy information. The MOMO
wall was therefore complemented by a hodoscope consisting of 16 wedge-shaped
2~cm thick scintillators. This hodoscope was already used in the study of the
$pd\to {^3\text{He}\,K^+K^-}$ reaction~\cite{BEL2007}. High above threshold
the acceptance for a $pd \to{}^3\textrm{He}\,X^+X^-$ event is low, even for
small $^3$He angles, because one of the mesons $X$ would miss the MOMO
detector. On the other hand, very close to threshold there are significant
losses of events from mesons escaping through the central hole. Nevertheless,
due to the forward boost, the acceptance of MOMO for $\pi^+\pi^-$ production
is much higher for a deuteron beam than for a proton beam at the same c.m.\
energy.

\subsubsection{The ENSTAR detector} 
\label{ENSTAR}

\begin{figure}[ht]
\begin{center}
\includegraphics[width=0.7\columnwidth]{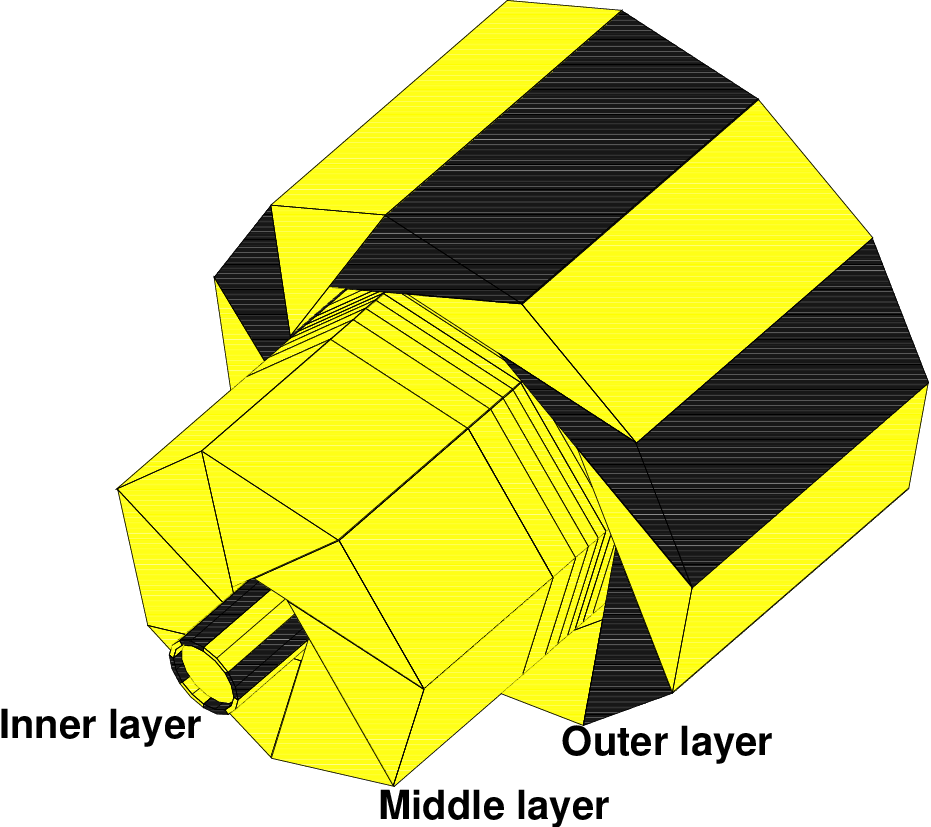}
\caption{One half of the ENSTAR detector~\cite{BET2007} surrounding the
target. It consists of wedges made of scintillating material and the read-out
is ensured by scintillating fibres collecting the light in grooves milled in
the wedges and transporting it to phototubes. For clarity, some elements of
the middle and outer layers have been moved along the beam direction to
provide an inner view.} \label{fig:ENSTAR}
\end{center}
\end{figure}

The ENSTAR detector~\cite{BET2007} was designed to detect a pair of
relatively low energy particles emerging from a target, with a fast particle
being measured in the Big Karl spectrometer. A typical example, discussed in
sect.~\ref{dd4heeta}, is where a $^3$He is measured in Big Karl and a
$\pi^-p$ pair, emitted from the target almost back to back, is registered in
ENSTAR~\cite{BUD2009b}. The basic design of this detector~\cite{BET2007} is
illustrated in Fig.~\ref{fig:ENSTAR}.

ENSTAR is cylindrically shaped, with three layers of plastic scintillators
that are used to generate $\Delta E- E$ spectra for particle identification
and to obtain total energy information for the stopped particles. Each layer
is divided into a number of pieces to obtain angular information. The
detector consists of two identical half cylinders that are placed
symmetrically on either side of the target. There is effectively full
azimuthal angular coverage, but the modest resolution in $\phi$ of about
$45^{\circ}$ is sufficient for the envisaged $\eta$-mesic nucleus
search~\cite{BET2007}. The corresponding limits for the polar angle are
$15^{\circ} < \theta_{\rm lab} < 165^{\circ}$.

Though, like MOMO, there is no magnetic field to help with the particle
identification, the background could be suppressed by demanding strict timing
coincidences between ENSTAR and Big Karl~\cite{BET2007,MAC2015}.

\subsubsection{The EDDA detector} 
\label{EDDA}

It could be argued that EDDA has been the most successful detector employed
at COSY because it was specifically optimized for one series of experiments.
It relied on the stability and reproducibility of the circulating proton beam
because data were taken at a continuum of energies during an acceleration or
deceleration mode in COSY and the necessary statistics were acquired through
the addition of data from many such cycles. Although designed for the
measurement of proton-proton elastic scattering, the principles could be
extended to other two-body reactions, such as $pp\to d\pi^+$, where two final
particles are detected in coincidence and the resulting geometrical
constraints eliminate much of the background.

\begin{figure}[htb]
\begin{center}
\includegraphics[width=0.95\columnwidth]{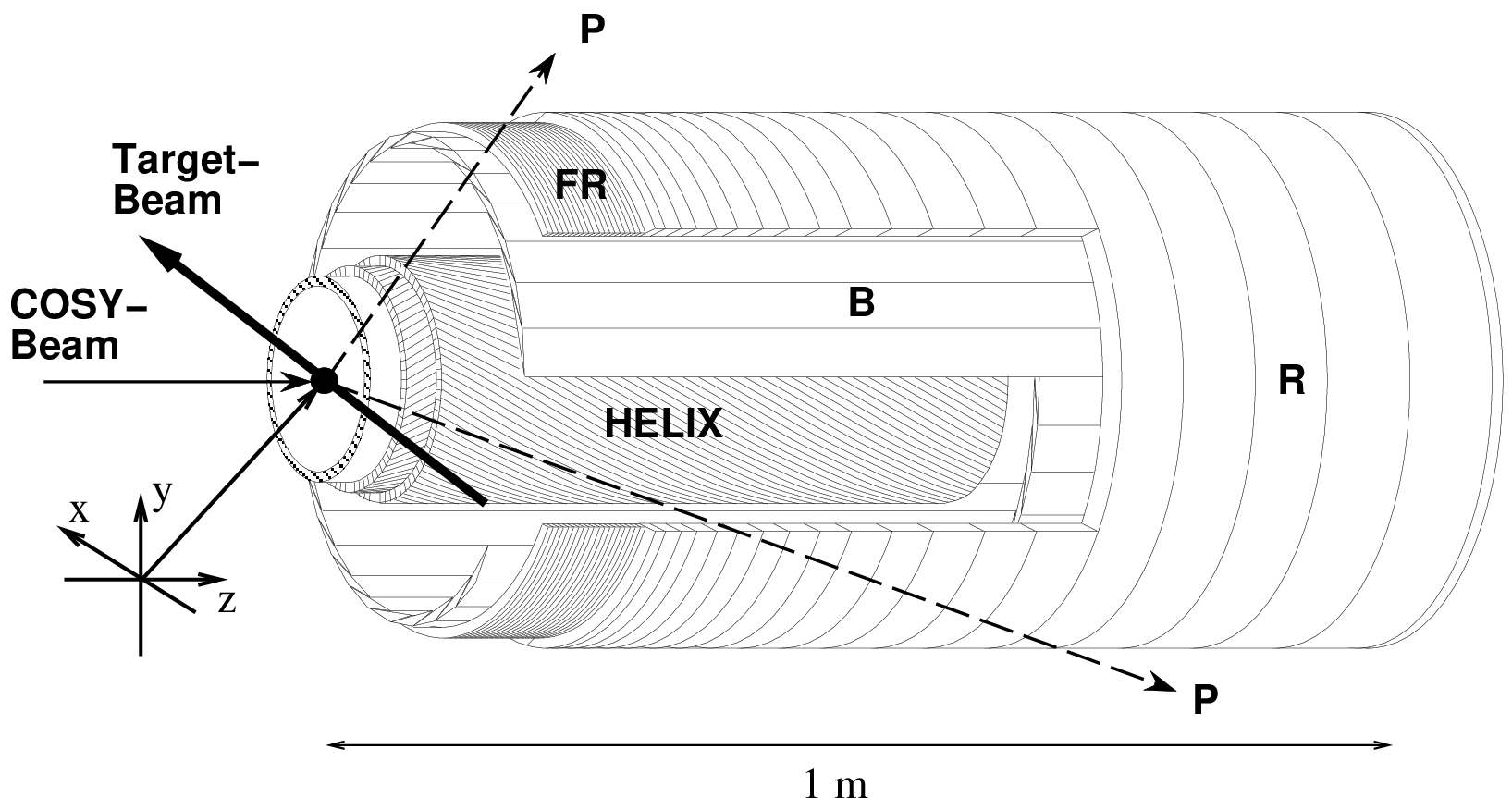}
\caption{\label{fig:Albers0} Schematic drawing of the EDDA detector as used
for the polarized target experiments~\cite{ALT2005}. The outer hodoscope
consists of scintillator bars $B$, scintillator semi-rings $R$ and semi-rings
made of scintillating fibres \textit{FR}. The inner hodoscope HELIX consists
of four layers of scintillating fibres, helically wound in opposite
directions.}
\end{center}
\end{figure}

The azimuthal symmetry of EDDA, which is so important when studying
spin-dependent observables, is well illustrated by the drawing in
Fig.~\ref{fig:Albers0}. This shows the form of the detector used for
measuring analyzing powers and spin correlations in $pp$ elastic
scattering~\cite{ALT2005}. Less redundancy was required in the measurement of
the unpolarized cross section and an even simpler version could be used for
the final EDDA incarnation as the default polarimeter for experiments at
COSY. Since the detector was used only for the measurement of $pp\to pp$,
most of the details of its application there are to be found in
sect.~\ref{pp_elastic} and we here concentrate on describing the relevant
hardware.

The need to detect both final particles in a reaction puts strong constraints
on the angular acceptance. Thus for very small angle $pp$ scattering the
recoil proton emerges almost perpendicular to the beam with low momentum and
would not be detected. In practice, therefore, the EDDA $pp$ elastic
scattering measurements were restricted to c.m.\ angles $30^{\circ} \lesssim
\theta_{cm} \lesssim 150^{\circ}$.

EDDA consists of two cylindrical detector shells, though only the outer shell
was needed for the unpolarized cross section measurements. This shell
consists of 32 scintillator bars ($B$) which are mounted parallel to the beam
axis. They are surrounded by scintillator semi-rings ($R$) and semi-rings
made of scintillating fibres (\textit{FR}). The resulting polar and azimuthal
angular resolutions are about $1^{\circ}$ and $1.9^{\circ}$ FWHM,
respectively.

In experiments involving polarized hydrogen, the target is far from
point-like. There can be a non-negligible background associated with
scattering events from the unpolarized hydrogen atoms surrounding the
polarized hydrogen beam. This can be reduced significantly if the reaction
vertex is well identified, which is achieved through the implementation of
the inner detector shell, called HELIX in the figure. HELIX is a cylindrical
hodoscope consisting of four layers of 640 plastic scintillating fibres of
2.5~mm diameter, helically wound in opposite directions so that a coincidence
of hits in the two spirals determines the point where the ejectile traversed
the hodoscope. Combined with the spatial resolution of the outer detector
shell, the helix fibre detector allows vertex reconstruction with a FWHM
resolution of 1.3~mm in the transverse directions and 0.9~mm in the COSY beam
direction. Using a fit of the vertex and scattering angles with constraints
imposed by $pp$ elastic scattering kinematics the resulting polar and
azimuthal angular resolutions are. respectively, about 0.3$^{\circ}$ and
1.3$^{\circ}$ FWHM.

After the completion of the EDDA physics programme, a stripped-down version
of the detector has been used extensively at COSY as an on-line beam
polarimeter. For this purpose the central helix shown in
Fig.~\ref{fig:Albers0} was removed. Only carbon fibre targets were used and
these could be moved remotely into the beam in order to measure the
polarization. The resulting polarimeter consists of 29 pairs of half-rings
placed to the left and right of the beam to detect coincidences from
quasi-free scattering from the carbon. The asymmetry is determined
individually for each pair of half-rings and the weighted average evaluated.
This is converted into a value of the beam polarization using dedicated
C/CH$_2$ measurements and the EDDA values of the elastic $pp$ analyzing
powers at that particular energy~\cite{ALT2005}. The systematic uncertainty
in the polarizations is estimated to be $\approx \pm3\%$ at each
energy~\cite{WEI2000}. It should be noted that, unless great care is
taken~\cite{GUI2016}, the interaction of the beam with the target makes the
residual beam unusable for precision experiments, so that the EDDA
polarimeter is only employed at the end of a COSY cycle.

\subsection{Targets and equipment} 
\label{targets}

\subsubsection{The Atomic Beam Source} 
\label{ABS}

For experiments with an external beam, the particles pass only once through
the target so that, in order to obtain meaningful counting rates, the targets
have to be ``thick''.  This causes particular problems for polarized targets
because it has not been possible to produce a polarized target of pure
hydrogen or deuterium on a macroscopic scale. For example, the alcohol
pentanol, which has often been used for a polarized target, has a hydrogen
content of less than 16\% and this clearly reduces the figure of merit for
any experiment~\cite{LEH2015}.

The situation is very different for experiments carried out inside a storage
ring such as COSY because the beam traverses the target a myriad of times and
so much thinner targets must be used. Polarized hydrogen and deuterium ions
are routinely produced using an Atomic Beam Source (ABS) and, although this
may not lead to targets that are sufficiently thick, this can be compensated
by using the ABS in combination with a gas cell, as described in
sect.~\ref{cells}.

Several ABS systems have been used at COSY, including those at
EDDA~\cite{ALT2005}, ANKE~\cite{MIK2013}, and PAX~\cite{WEI2015}. These are
technologically very complex devices and the interested reader is directed to
these references. Only the basic principles of an ABS, which are completely
analogous to the devices used for producing polarized beams, are outlined
here. In a static magnetic field the energy levels of an atom split into four
(hydrogen) or six (deuterium) distinct lines though, as summarized by the
Breit-Rabi diagrams~\cite{HAE1967}, the relative separations between the
lines change significantly with the field strength.

Transitions between the hyperfine states are induced by the magnetic
component of an \emph{rf} field, and this leads to changes in the populations
of the states, and hence to a possible polarization. However, very different
effects can be achieved in weak fields, where the total atomic spin is a good
quantum number, from medium and strong fields. For hydrogen it is in
principle possible to produce spin-``up'' protons with transitions in a
medium strength field whereas spin-``down'' requires supplementary
transitions in a weak field. It should therefore not come as a surprise if
the polarizations `up'' and ``down'' differed in magnitude.

The situation with deuterium is much more complicated because, in addition to
the vector polarization $p_z =(N_+-N_-)/(N_+ + N_0 +N_-)$, there is also the
tensor polarization $p_{zz} =(N_++N_--2N_0)/(N_+ + N_0 +N_-)$ to be
considered. Here the $N_i$ are the populations of the three magnetic
sub-states for a spin-one particle in the quantization direction of the
source. In order to get different mixtures of vector and tensor
polarizations, various combinations of hyperfine transitions in weak, medium,
and strong fields must be introduced.

In the ideal case, the settings on the transition units would specify the
polarizations of the source but, due to imperfections, this does not happen
in practice and the resultant polarizations have to be measured
independently. This can be done quite precisely with the help of a Lamb Shift
Polarimeter (LSP)~\cite{ENG2005}, which can also be used to optimize the
polarization of the atomic hydrogen and deuterium beams delivered by the ABS.

\begin{table}[hbt]
\centering
\begin{tabular}{|l|c|c|c|} \hline
Beam & $(p_z,p_{zz})_{\rm ideal}$ & $p_z$ & $p_{zz}$\\
\hline
Hydrogen & $(+1,-)$ & $+0.89\pm0.01$ & ---\\
         & $(-1,-)$ & $-0.96\pm0,01$& ---\\
\hline
Deuterium & $(+1,+1)$ & $+0.88\pm0.01$ & $+0.88\pm0.03$\\
          & $(-1,+1)$ & $-0.91\pm0.01$ & $+0.85\pm0.03$\\
          & $(\phantom{-}0,+1)$ & $+0.005\pm0.003$ & $+0.90\pm0.01$\\
          & $(\phantom{-}0,-2)$ & $+0.005\pm0.003$ & $-1.71\pm0.03$\\
\hline
\end{tabular}
\caption{Values of the polarizations achieved for hydrogen and deuterium in a
test experiment with the ABS used at ANKE~\cite{MIK2013}.} \label{ABStab}
\end{table}

In tests carried out on the ABS used at the ANKE target station, the values
of the polarizations achieved for hydrogen and deuterium are reported in
Table~\ref{ABStab}~\cite{MIK2013}. However, these are just typical examples
and the choices of hyperfine transitions, and hence polarizations, can be
tailored to the needs of a particular experiment. Note also that there may be
polarization losses when a beam from an ABS is used inside COSY.

\subsubsection{Polarized gas cell targets} 
\label{cells}

Though an ABS can produce high quality polarized beams of hydrogen and
deuterium, they are generally too weak to provide acceptable luminosities in
a storage ring such as COSY. The general solution to this dilemma is to use
the ABS beam to supply a storage cell that holds the polarized atoms in the
vicinity of the passage of the high energy circulating beam. In this way the
target density can be increased by up to two orders of magnitude compared to
the direct ABS beam~\cite{STEF2003}. Such a storage cell was routinely used
by the HERMES collaboration working at the DESY electron storage
ring~\cite{AIR2005} and at COSY they have formed parts of the EDDA, ANKE, and
PAX programmes.

\begin{figure}[htb]
\begin{center}
\includegraphics[width=0.9\columnwidth,trim={0 0 6.4cm 0},clip]{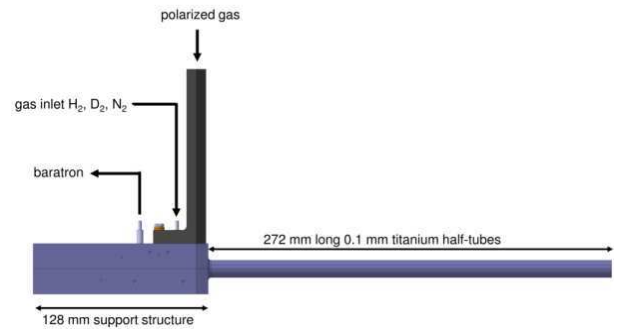}
\caption{\label{fig:cell} Design for the openable storage cell for use at
ANKE~\cite{DYM2014} showing the typical T-shape layout, with the vertical
feeding tube that connects to the ABS and the storage cell itself that lies
along the COSY beam. Unpolarized hydrogen and deuterium gas can be introduced
through the smaller tube to the left, as can the nitrogen used for background
studies. There is also an outlet to the baratron pressure monitor. The cell
can be opened (or closed) vertically along its length with the use of a
precision piezoelectric drive. }
\end{center}
\end{figure}

The basic design of the T-shaped system illustrated on the left of
Fig.~\ref{fig:cell} is fairly general. There is a vertical feeding tube that
catches the gas flow from the ABS to guide it into the horizontal tube of the
storage cell that lies along the circulating beam of the accelerator. The
minimum feeding tube diameter, which is of the order of 10~mm, is determined
by the extension of the focused gas beam from the ABS. The areal density of
the  target increases roughly like $L^2/d^3$, where $d$ is the diameter of
the storage cell and $L$ its length~\cite{STEF2003}. The highest density is
therefore achieved with a long target cell with the smallest possible
diameter.

The minimum diameter of the storage cell is defined by the beam extension
and, depending on the beam energy and the cooling capabilities of the
machine, diameters $d$ from 10 to 12~mm are reasonable. With a storage cell
of length $L=390$~mm, this would lead to target areal densities from 3 to
$6\times10^{13}$~cm$^{-2}$ compared to the $1\times10^{12}$~cm$^{-2}$
obtained directly from the ABS.

A storage cell with a diameter of 10-12~mm would, however, pose a serious
obstacle for the COSY beam at injection and during electron cooling. This
would restrict the beam intensity and hence the luminosity. One way to
overcome this limitation is to design a cell that is open at the start of a
COSY cycle and only closes once the beam has been well prepared and is
stable. Such a cell is illustrated on the right of Fig.~\ref{fig:cell}, which
shows the open two halves of the cell. These close and open around the beam
at each cycle of the accelerator and in this way the storage cell is not the
limiting factor in the COSY beam intensity.

The choice of material for the walls of the storage cell is critical.
Aluminium is suitable for a rigid cell because the target polarization is not
destroyed and all but the lowest energy ejectiles can pass through the 0.2~mm
walls. However, for an openable cell, such walls are too flexible to provide
firm closure over the 272~mm length shown in Fig.~\ref{fig:cell}. Stainless
steel has the necessary rigidity but, even after coating with PTFE, there is
serious loss of polarization due to recombination effects on the walls. The
successful cell was made from titanium, which was as thin as 0.1~mm in the
region of interest, coated with a 0.005~mm layer of PTFE. The rigid support
structures allow the precise and reproducible positioning of the half-tubes
of the cell~\cite{RAL2016}.

The cell polarizations were measured for both $\pol{\rm H}$ and $\pol{\rm D}$
targets with the 580~MeV proton beam, using the free or quasi-free $pp\to
d\pi^+$ reaction for the polarimetry, with both the $d$ and $\pi^+$ being
measured in ANKE, with identification made on the basis of time of flight.
The value of the polarization for hydrogen in the cell was $86\pm 5$\%, which
is only slightly lower than the highest ABS jet polarization observed in
laboratory tests. As expected, the polarization of the rest-gas around the
cell was very low. The vector polarization of the deuterium target was $61\pm
10$\%, which is also consistent with there being little polarization loss on
the titanium walls~\cite{DYM2014}.

It therefore seems that the openable storage cell technology represents a
major advance for the use of polarized targets in storage rings.

\subsubsection{Silicon Tracking Telescopes} 
\label{STT}

The original motivation for the design of the Silicon Tracking Telescope
(STT) was the detection of low energy protons emerging from a deuterium
target~\cite{SCH2003}. In a hard process, such as $pd\to p_{\rm sp}d\eta$, a
recoiling proton with a momentum less than say 150~MeV/$c$ might be
considered to be a \emph{spectator}, $p_{\rm sp}$, which only influences the
reaction through the kinematic changes that it induces. In this case the
reaction can be interpreted in terms of quasi-free $pn\to d\eta$.

Spectator detection in internal measurements at storage rings is made easier
because the low energy protons are not lost in a liquid target and an initial
trial of the method was carried out at the CELSIUS ring~\cite{BIL2001a}.
However, more dedicated equipment has been constructed at COSY~\cite{SCH2003}
and the STT have found other uses, such as the measurement of recoil protons
from $pp$ elastic scattering or facilitating the vertex location when using a
long polarized cell target.

The COSY STT have been developed to trigger, identify, and track low energy
protons and deuterons. Three layers of silicon-strip detectors act as a range
telescope, combining particle trigger and time-of-flight information with
particle tracking and energy-loss determination over a wide dynamic range.
Stopped particles are unambiguously identified by the $\Delta E/E$ method and
their four-momentum determined. With the STT acting as modular building
blocks, an extended vertex detector covering a large acceptance can be setup
depending on the needs of an individual experiment.

A single STT is made up of three layers of 70, 300 and 5000~$\mu$m thick
double-sided structured silicon-strip detectors to guarantee particle
triggering and tracking over the full energy-loss range of 0.05--50~MeV. Each
detector has an active area of $64\times 64$~mm$^2$. The 70 and 300~$\mu$m
thick detectors have 128 strips (0.5 mm pitch) per side whereas the
5000~$\mu$m thick detector has 64 segments (1~mm pitch) on each side. The
segmentation and geometry have been chosen taking into account the
limitations due to small angle scattering within the detector planes. The
electronics that provide information for each individual strip are placed
behind the detectors so as not to disturb the particle detection. There are
independent cooling branches so that the electronics can be kept at room
temperature whereas the detectors can be cooled down to $-20^{\circ}$C.

\begin{figure}[htb]
\begin{center}
\includegraphics[width=0.8\columnwidth]{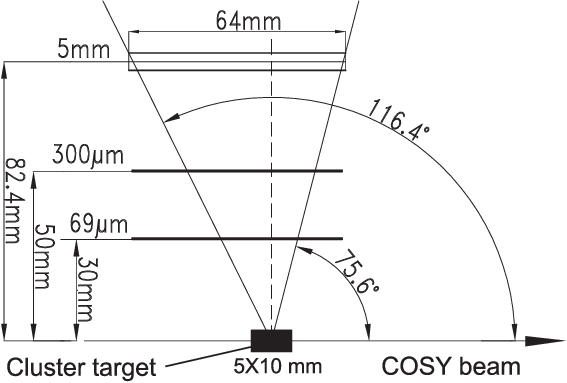}
\caption{\label{fig:Ralf1} Schematic diagram of one possible layout of the
three STT layers inside the ANKE target chamber when used in conjunction with
a cluster-jet target~\cite{SCH2003}.}
\end{center}
\end{figure}

The layout of the STT inside the high vacuum of the ANKE target chamber is
shown schematically in Fig.~\ref{fig:Ralf1}. There is some flexibility in the
location of the detector but the first silicon layer could be placed as close
as 1~cm from the cluster-jet target so that the angular acceptance would then
be even larger than that indicated in the figure. In order to pass through
the three layers, the protons must have  kinetic energies of at least
2.5~MeV, 6~MeV, and 30~MeV, respectively. The first of these criteria is the
most severe because about half the spectator protons from a deuterium target
have energies below this and so such events cannot be reconstructed.

\begin{figure}[htb]
\begin{center}
\includegraphics[width=0.8\columnwidth]{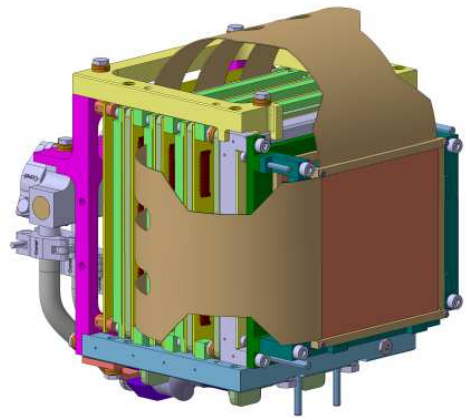}
\caption{\label{fig:Ralf2} A fully assembled STT detector~\cite{SCH2003}.}
\end{center}
\end{figure}

For stopping protons with energies below 30~MeV the particle identification
is unambiguous and greater precision in the angle of the recoiling proton is
achieved by deducing it from the energy measured in the telescope rather than
from a direct angular measurement. However, by studying the energy deposited
principally in the third layer, it is also possible to deduce the energy of
punch-through protons up to 90~MeV, thus expanding considerably the angular
coverage of the telescope. The fully assembled STT detector illustrated in
Fig.~\ref{fig:Ralf2} is compact and transportable.

The demand that the STT identifies and determines the track of a charged
particle means that this must pass through the first layer and hit (and
possibly stop) in the second layer of the detector. This allows the STT to be
used with a long target, such as the cell filled with polarized gas. The
downside is the fact that a proton must have a minimum momentum of about
70~MeV/$c$ to pass through the first layer. In the case of elastic
proton-proton scattering it can only be used for momentum transfers
$|t|>0.005~(\textrm{GeV}/c)^2$. To access smaller momentum transfers requires
changed criteria and a different design, an example of which is the KOALA
detector.

\subsubsection{The KOALA detector} 
\label{KOALA}

The prime motivation for the development of the KOALA (Key experiment fOr
pAnda Luminosity determinAtion) detector by the PANDA collaboration is the
study of anti\-proton-proton scattering at small momentum transfers at HESR.
Since the evaluation of the pure Coulomb differential cross section, which
proportional to $1/t^2$, is unambiguous, a measurement in the region of
Coulomb dominance would determine the $\bar{p}p$ luminosity in an independent
way and allow parameters of the $\bar{p}p$ interaction to be extracted. For
this to be feasible the device must allow smaller values of momentum
transfers to be studied than is possible with the STT. This in turn requires
that the particle be registered on the front layer of a detector and that the
track be determined by demanding the beam-target interaction to be
point-like. As described in sect.~\ref{pp_elastic}, such a detector could
also be used to investigate proton-proton elastic scattering at COSY.

\begin{figure}[htb]
\begin{center}
\includegraphics[width=0.7\columnwidth]{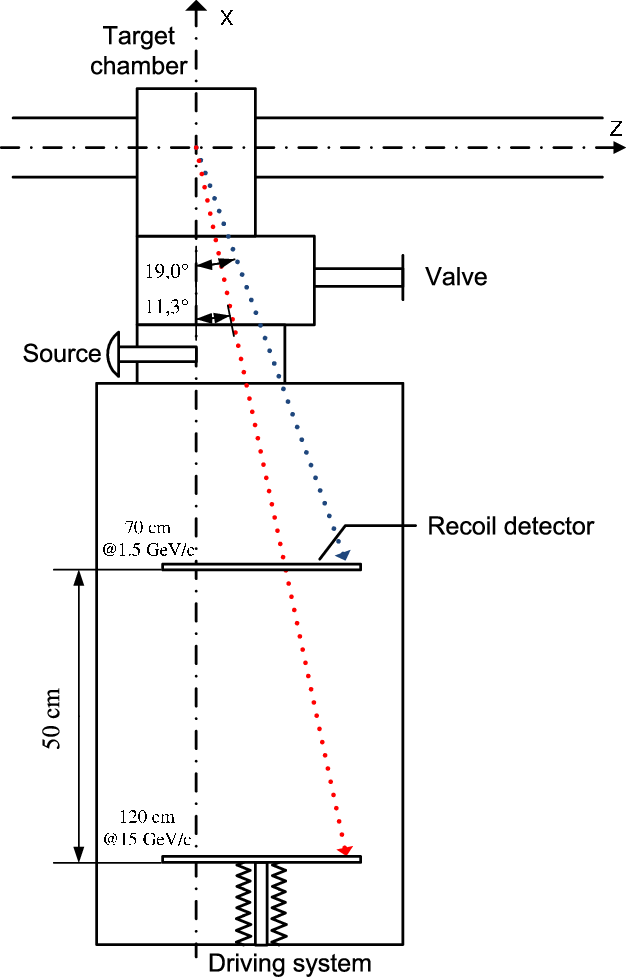}
\caption{\label{fig:KOALA1} Schematic view of the KOALA recoil
detector~\cite{HUX2014}, showing the complete setup with a movable detector
plane in order to cover the desired range of recoil angles depending upon the
chosen beam momentum.}
\end{center}
\end{figure}

The general layout of the KOALA detector is shown in
Fig.~\ref{fig:KOALA1}~\cite{HUX2014}. In order to optimize the settings for
different beam momenta, it is possible to adjust remotely the distance of the
detector plane from the interaction point. The recoil detector will measure
both the kinetic energy and the polar angle of the recoil protons which will
provide two determinations of the momentum transfer $t$.

As used at COSY, the KOALA detector plane, containing two 76.8~mm $\times$
50~mm $\times$ 1~mm silicon strip sensors, was positioned about 1~m from the
target. Each silicon detector has 64 strips with 1.2~mm pitch. In order to
measure higher energy protons, two germanium strip detectors with 5 and 11~mm
thickness were also added. These each have have 67 readout strips and a strip
pitch of 1.2~mm.

\subsection{Technical experiments} 
\label{technical}

Two of the more difficult challenges that must be faced when carrying out
experiments inside a storage ring are the evaluation of the beam-target
luminosity $\mathcal{L}$ and the precise determination of the beam momentum.
At an external target position the beam is generally much smaller than the
area of the target and, by taking a target of uniform thickness,
$\mathcal{L}$ will not depend on fine details of the beam properties. If the
fluxes of the incident and scattered particles are measured, the absolute
cross section of a reaction can be determined. Even here there may be
complications due, for example, to the bulging of the windows of a liquid
target.

In an internal experiment the beam-target overlap and the target thickness
are very hard to estimate from macroscopic measurements and so another method
to determine the luminosity must be sought. To avoid the associated
normalization uncertainties, many experiments at COSY have derived cross
sections by comparing production rates with those for processes with known
differential cross sections, often elastic or quasi-elastic scattering.

A much more ingenious method was implemented by the EDDA collaboration in
their measurement of proton-proton scattering. As described in
sect.~\ref{pp_elastic}, this involved evaluating the numbers of electrons
knocked out of the target by the proton beam. Since this is an
electromagnetic process, its cross section can be calculated quite reliably.
A different approach to the problem, which is also based upon electromagnetic
processes, viz.\ the energy losses of energetic charged particles as they
pass through matter, is discussed in sect.~\ref{Schottky}. The energy loss of
the stored beam, which is proportional to the target thickness, builds up
steadily in time and leads to a shift in the revolution frequency of the
machine, which can be determined by measuring the Schottky spectra. If the
characteristics of the machine are known, the effective target thickness can
be deduced~\cite{STE2008}.

Regarding the second challenge, when performing a precision measurement, such
as determining the mass of the $\eta$ meson discussed in
sect.~\ref{pd3heeta}, it is important to know the COSY beam momentum to a
fraction of a MeV/$c$. Though the circulation frequency is known to better
than $10^{-5}$, the same cannot be said for the length of the COSY orbit
because there may be small but significant and uncontrolled deviations from
the ideal path. An alternative approach, such as that described in
sect.~\ref{Paul1}, is necessary if great accuracy is required.

The momentum of a stored polarized proton or deuteron beam in COSY can be
determined by sweeping an \emph{rf} magnetic dipole or solenoid field over a
spin resonance. This perturbation induces a beam depolarization that is
maximal at the resonance frequency. Taken together, the resonance and beam
revolution frequencies completely determine the beam's Lorentz $\gamma$
factor. This allows the corresponding beam momentum to be determined at least
one order of magnitude more precisely than with macroscopic
methods~\cite{GOS2010}.

There were other \emph{technical} experiments, especially several carried out
by the SPIN@COSY collaboration~\cite{KRI2007,MOR2009}, but these had less
direct influence on the hadron physics programme and will not be described
here. The one that is described explicitly is the PAX programme to study the
production of polarized protons by spin filtering. Though the PAX experiments
could have no direct influence on the hadronic physics programme at COSY, it
has been suggested that such a technique could be used to produce polarized
antiproton beams at FAIR.

\subsubsection{Determination of beam-target luminosity} 
\label{Schottky}

When the particles in a closed orbit in COSY lose energy in passing through
the target, the fractional change in the momentum $p$ is proportional to the
fractional change in the frequency $f$ of the machine:
\begin{equation}
\frac{1}{p}\frac{\dd p}{\dd t} = \frac{1}{\eta} \frac{1}{f}\frac{\dd f}{\dd t},
\label{eq:5}
\end{equation}
where $\eta$ is the so-called \emph{frequ\-ency-slip parameter}. Once this
constant of proportionality is known, the rate of change of frequency
determines the effective target thickness $n_T$ through~\cite{STE2008}:
\begin{equation}
n_T = \left(\frac{1+\gamma}{\gamma}\right)\frac{1}{\eta}
\frac{1}{(\dd E/\dd x) m}\,\frac{T}{f^2}\,\frac{\dd f}{\dd t}\,,
\label{eq:4}
\end{equation}
where $T$ is the kinetic energy of the beam particles of mass $m$, $\gamma$
the Lorentz factor, and $\dd E/\dd x$ the stopping power of the target
material.

\begin{figure}[htb]
\begin{center}
\includegraphics[width=0.49\columnwidth]{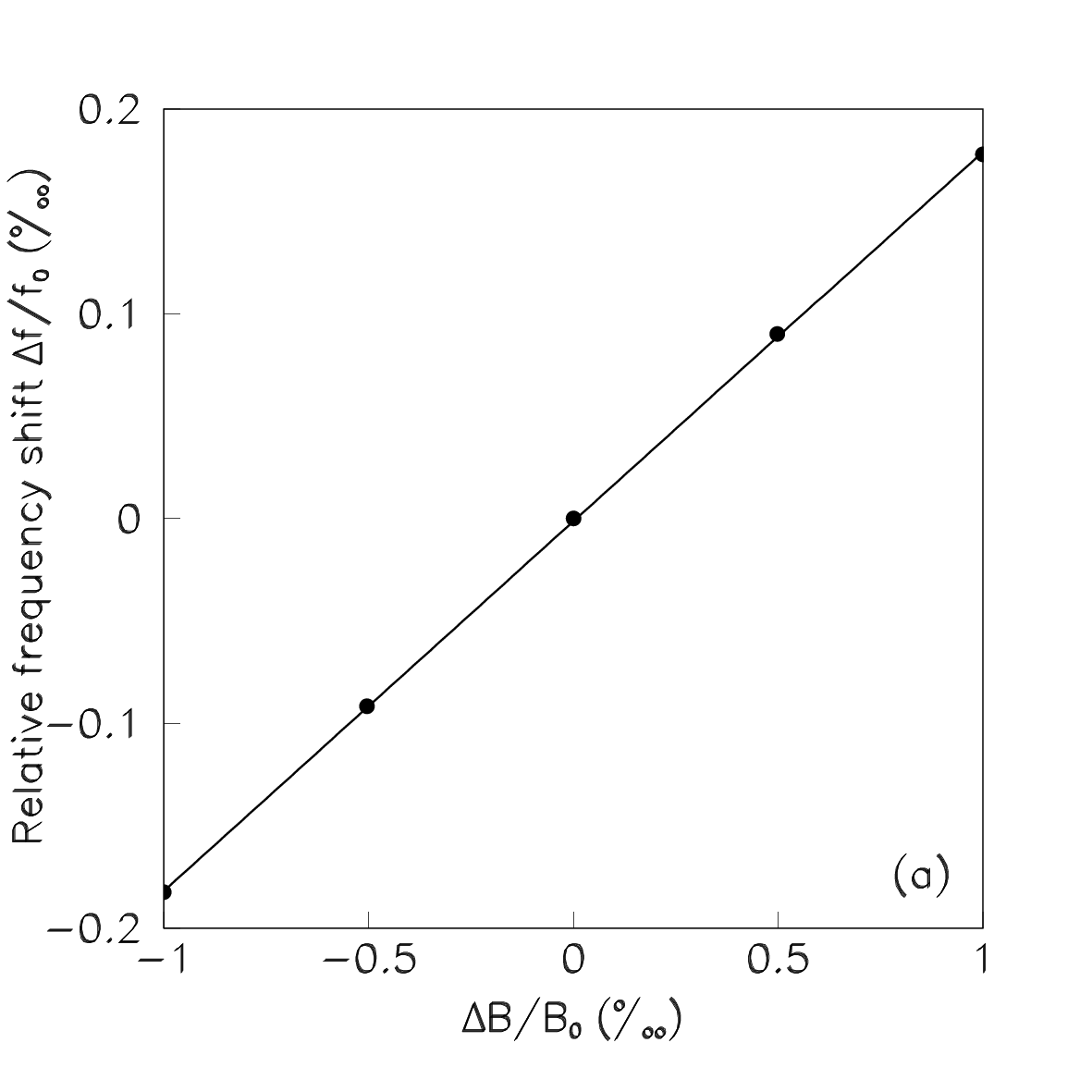}
\includegraphics[width=0.49\columnwidth]{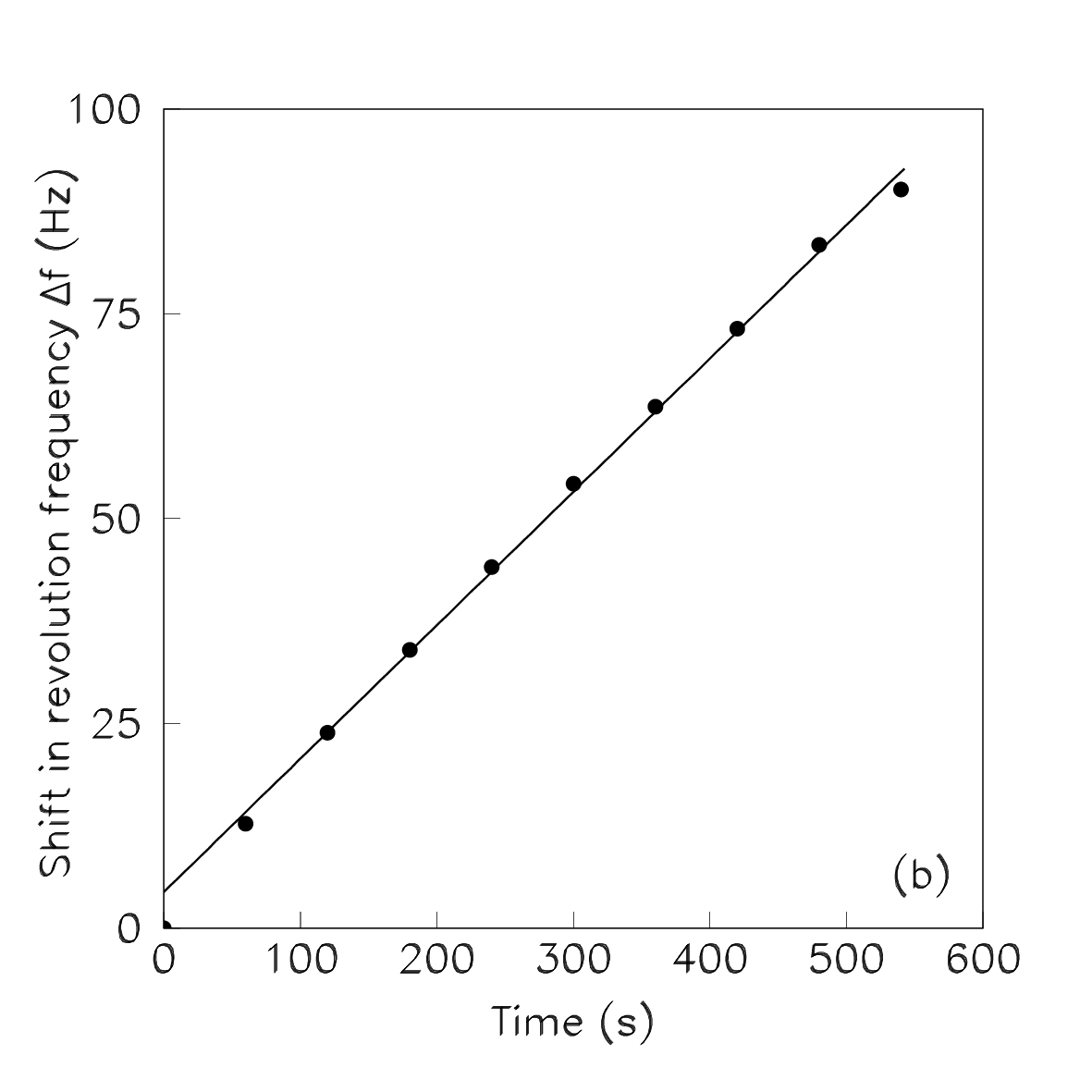}
\caption{\label{fig:Stein} (a) Variation of the revolution frequency with
field strength in the bending magnets, both in parts per thousand. (b)
Typical mean frequency shifts derived from the Schottky power spectra at ten
equally spaced intervals of time. Both measurements were conducted for
settings corresponding to protons with momenta 3.463~GeV/$c$~\cite{STE2008}.}
\end{center}
\end{figure}

The $\eta$-parameter, which reflects a competition between the slowing down
due to the energy loss and an apparent speeding up following an orbit
adjustment, is a property of COSY that is quite independent of the particular
target. Though it can be estimated from the general machine parameters, it is
best measured by varying the field strength in the bending magnets by a few
parts per thousand. As seen from Fig.~\ref{fig:Stein}a, obtained for a proton
beam with a momentum of about 3.463~GeV/$c$, the resulting frequency change
is quite linear and the slope $\alpha$ leads to a value of
$\eta=1/\gamma^2-\alpha = -0.115\pm0.003$~\cite{STE2008}.

The frequency shift is measured by analyzing the Schottky noise power
spectrum of the coasting proton beam at a sequence of times. The origin of
the Schottky noise is the statistical distribution of the particles in the
beam, which gives rise to current fluctuations that induce a voltage signal
at a beam pick-up. The centroids of these power spectra are shown in
Fig.~\ref{fig:Stein}b at ten equally spaced intervals in time. After
neglecting the first and last points, the slope of these data gives $\dd
f/\dd t = (0.163\pm 0.003)\,$Hz/s for $pp$ interactions at
3.463~GeV/$c$~\cite{STE2008}. By inserting the values of $\dd f/\dd t$ and
$\eta$ into Eq.~(\ref{eq:4}), a value of the effective target thickness can
be deduced which, without any loss of precision, can be converted into one of
luminosity by multiplying by the number of beam particles measured in the
same cycle with a high precision beam current transformer.

The measurement relies on the particles in the beam passing through the
target more or less the same number of times so that they build up similar
energy shifts. This is confirmed by the Schottky spectrum at the end of a
cycle being similar in shape to that at the beginning. Several corrections,
described in detail in Ref.~\cite{STE2008}, have to be applied before a
reliable value of the luminosity can be obtained. The biggest of these is to
account for the energy loss caused by the interactions with the residual gas
in the ring. After making such corrections, the uncertainty in the luminosity
in this initial experiment was estimated to be on the 5\%
level~\cite{STE2008}.

Even greater precision was achieved when the Schottky technique was used in
the normalization of the ANKE measurements of the proton-proton differential
cross section described in sect.~\ref{pp_elastic}. However, it is important
to note that, with the actual settings in COSY, $\eta$ changes sign for a
proton kinetic energy around 1.3~GeV. Due to the resulting large error bars,
the Schottky technique is of little value for energies in this neighbourhood.

\subsubsection[Precision determination of beam\\ momentum]{Precision determination of beam momentum} 
\label{Paul1}

The determination of the momentum of a polarized electron beam through the
study of induced depolarizing resonances was used at the VEPP accelerator to
measure the masses of a variety of neutral mesons from the $\phi$ to the
$\Upsilon$~\cite{BUK1978}. A similar technique has also been used at COSY to
measure the momentum of a vector polarized deuteron beam~\cite{GOS2010}. If
$f_{\rm res}$ is the frequency of the depolarizing \emph{rf} field and $f_0$
the revolution frequency of the beam in COSY then the total energy $E_d$ of a
deuteron in the beam is given by
\begin{equation}
\label{eq:3} E_d = \frac{m_dc^2}{|G_d|}\left(k-\frac{f_{\rm res}}{f_0}\right)\!,
\end{equation}
where $k$ is an integer. Under the actual conditions of COSY, $k=1$. Thus, by
measuring the two frequencies it is possible to determine $E_d$ in terms of the
deuteron mass $m_d$ and its gyromagnetic anomaly $G_d = -0.1429873$.

\begin{figure}[htb]
\begin{center}
\includegraphics[width=0.9\columnwidth]{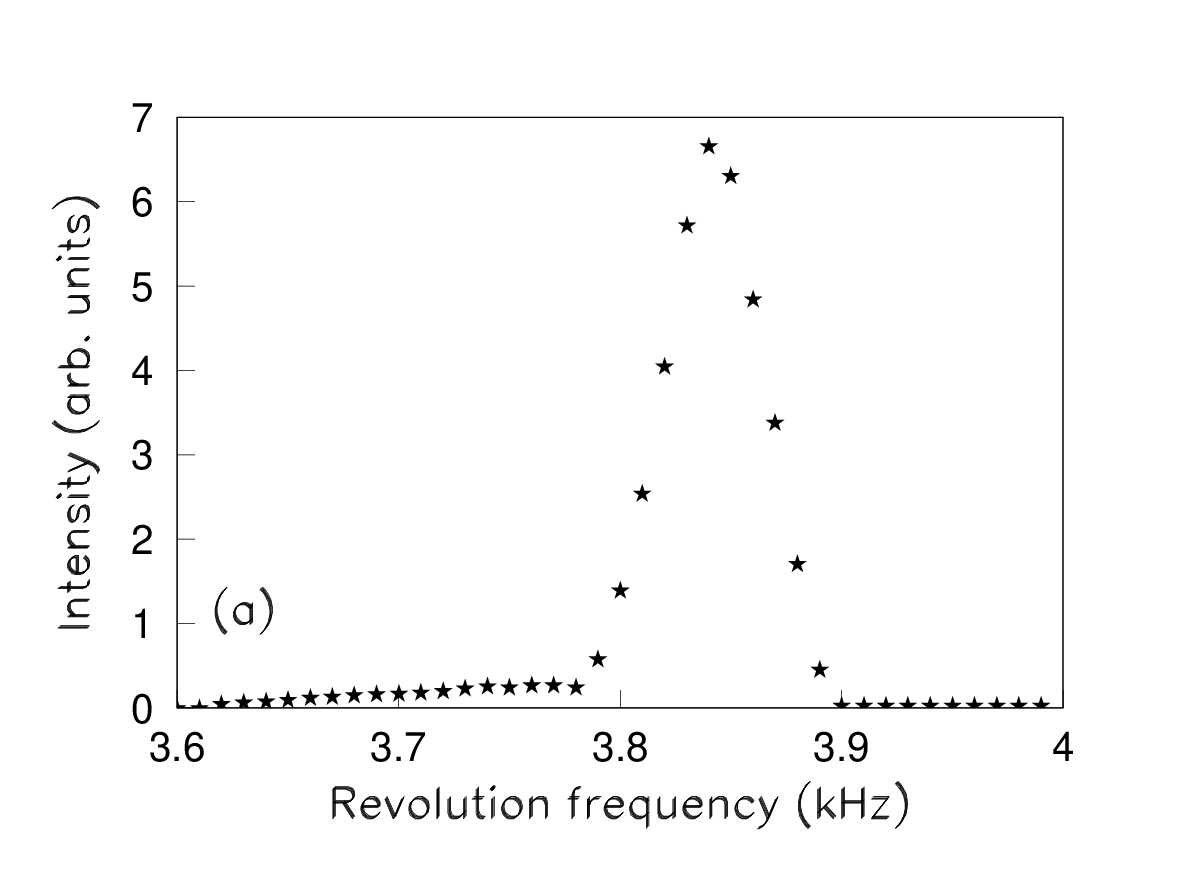}
\includegraphics[width=0.9\columnwidth]{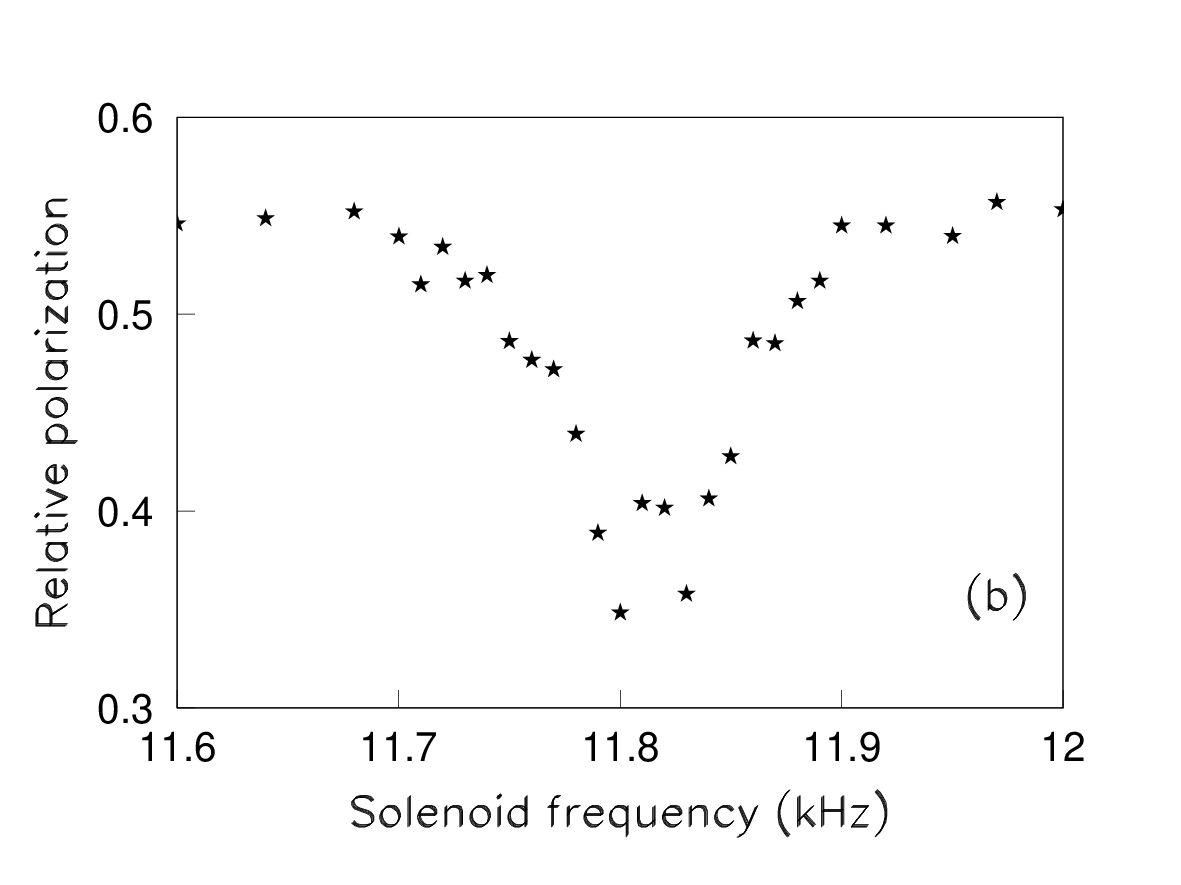}
\caption{\label{fig:Paul} (a) Mean Schottky power spectrum extracted from
measurements over five days at a single energy. (b) Spin-resonance
measurements at a single energy. The frequencies shown were displaced by (a)
1.4~MHz and (b) 1.0~MHz. The measurements were carried out with a deuteron
beam of momentum $\approx 3.1$~GeV/$c$~\cite{GOS2010}.}
\end{center}
\end{figure}

The revolution frequency $f_0$ was measured by using once again the Schottky
noise of the beam. From all the spectra taken over five days that were
measured under the same conditions at a particular energy, one mean spectrum
was constructed, an example of which is presented in Fig.~\ref{fig:Paul}. The
small tail at lower frequencies is well understood. The FWHM of the peak is
below 50~Hz and an average revolution frequency of $\overline{f_0} =
1403831.75 \pm 0.12$~Hz was deduced~\cite{GOS2010}. The tiny statistical
error here is dwarfed by the systematic uncertainty of $\Delta f_0 \approx
6$~Hz that arose from the limited preparation of the Schottky analyzer used
in the experiment. As a consequence the value of $\overline{f_0}$ was only
determined with a relative precision $\approx 4\times 10^{-6}$.

At COSY a horizontal \emph{rf} field from a solenoid was used to induce a
depolarizing resonance in the deuteron beam~\cite{GOS2010}. The polarization
was measured with the EDDA polarimeter discussed in sect.~\ref{pp_elastic}
using deuteron-proton elastic scattering. The absolute calibration of this
polarimeter is unimportant because it is only the frequency dependence of the
signal that is relevant. Figure~\ref{fig:Paul} shows the measured deuteron
beam polarization as a function of the \emph{rf} solenoid frequency $f_{\rm
res}$ with statistical errors. The major contribution to the width of the
signal comes from the momentum spread of the deuteron beam inside COSY, which
is about $\delta p/p\approx 2\times 10^{-4}$. This value agrees with that
deduced from the study of the kinematics of the $dp\to\,^{3}$He$\,\eta$
reaction described in sect.~\ref{pd3heeta}.

The structure in the middle of the resonance frequency scan of
Fig.~\ref{fig:Paul} is due to the interaction with a broad-band
barrier-bucket (stabilizing) cavity in COSY and does not affect the mean
position of the resonance frequency, which was determined with a precision of
$\approx 1.5\times 10^{-5}$. It is this uncertainty that dominates the error
in the extraction of the deuteron energy, and hence its momentum, on the
basis of Eq.~(\ref{eq:3}). This resulted in a limit of $\Delta p_d/p_d
\leqslant 6\times 10^{-5}$ for $p_d\approx 3.1$~GeV/$c$ ~\cite{GOS2010}. This
is over an order of magnitude better than ever reached before for a standard
experiment in the COSY ring and is quite sufficient for measuring the $\eta$
mass, as discussed in sect.~\ref{pd3heeta}.

The method described here is far more general than the one applied on an
external proton beam using the Big Karl spectrometer~\cite{BET1999}. This
technique was specific for the momentum 1930.477~MeV/$c$, where the
forward-going pion from the $pp\to d\pi^+$ reaction has the same momentum as
the backward-going deuteron. An experiment based upon this principle resulted
in a precision of about $5\times 10^{-5}$, though small corrections had to be
made to account for the energy losses in the target and its windows.
%
%
\subsubsection{Spin-filtering experiments} 
\label{PAXsection}

The ultimate aim of the PAX (Polarized Antiproton eXperiment) collaboration
is to produce beams of polarized antiprotons that can be used to study a
variety of double-polarized $\bar{p}p$ interactions~\cite{PAX2006,LEN2013}.
Of particular interest here is the transversity distribution that can be
measured in the Drell-Yan production of lepton pairs in double-polarized
$\bar{p}p$ collisions. The basic principle of the PAX approach is quite
straightforward but the methodology should first be refined for proton beams.

For beam and target polarized transversally in the $y$ direction, the
proton-proton total cross section has the spin structure
\begin{equation}
\sigma_{\rm tot} = \sigma_0 + \sigma_1PQ,
\end{equation}
where $P$ is the beam polarization and $Q$ that of the target. Thus, if
$\sigma_1$ is non-zero, the $pp$ total cross section will depend on the
relative orientations of the beam and target spin orientations. In this case,
when an unpolarized beam passes through a polarized target, one of the beam
spin orientations is preferentially absorbed and the residual beam thus
acquires a polarization. The obvious disadvantage of this spin-filtering
method is that, unless $\sigma_1/\sigma_0$ is very large, the beam intensity
would be much reduced before a significant polarization could be built up.
There are some similarities here with the measurements of the beam lifetime
discussed in connection with the TRIC experiment, the difference being that
in the study of $A_{y,y}$ the beam was initially polarized~\cite{EVE2009}.

A proof-of-principle experiment was carried out at the Test Storage Ring at
Heidelberg using 23~MeV protons passing through a cell filled with
transversally polarized hydrogen supplied by an atomic beam source. Beam
polarizations (both ``up'' and ``down'') of over 1\% were achieved after
about 80~min of spin filtering~\cite{RAT1993}.

A similar experiment was carried out by the PAX collaboration at COSY close
to the injection energy of 45~MeV \cite{AUG2012}. The polarized hydrogen
target cell was installed in one of the straight sections of COSY, in the
position previously taken by the PISA detector shown in Fig.~\ref{fig:COSY}.
As at Heidelberg, this was fed from an atomic beam source but the target
polarization was controlled much more precisely through the inclusion of a
Breit-Rabi polarimeter. The electron cooler was used to compensate for
multiple small-angle Coulomb scattering and the energy loss in the target and
the residual gas in the machine~\cite{WEI2015}. The beam polarization was
measured in the other straight section of COSY using the ANKE facility, which
does not seriously affect the beam quality. The setup for this was very
similar to that used for the measurement of the analyzing power in $pp$
scattering~\cite{BAG2014}, except that the target was in the form of a
deuterium cluster jet. The asymmetry of elastically scattered deuterons was
measured in a pair of STT placed on either side of the target.

The rate of polarization build-up through spin filtering was much slower in
the COSY experiment compared to that of Heidelberg, due mainly to the
differences in the machine frequencies and the relative sizes of the
spin-dependent cross sections. Thus about 0.8\% was achieved after 270~min.
It is important to realize that a beam proton has to scatter through some
minimum angle in order to leave the ring acceptance and this has to be taken
into account when making estimates of $\sigma_1$ from phase-shift
amplitudes~\cite{OEL2009}. After making this correction, there is good
agreement with the COSY data at 49.3~MeV~\cite{AUG2012}.

Faster polarization build-up would be possible with an increased target
thickness and this would be allowed with an openable cell of the type
discussed in sect.~\ref{cells}. Also, in order to be independent of the ANKE
facility, a dedicated polarimeter, with better azimuthal coverage is under
construction. Although the COSY experiment has shown that spin filtering
works and is well understood, the test only applies to transverse
polarizations. Any test of longitudinal spin filtering would involve a
rotation of the beam polarization and this will require the use of the new
Siberian snake.

%
%
\section{Nucleon-nucleon elastic scattering} 
\label{NN_elastic}\setcounter{equation}{0}%

A good understanding of the nucleon-nucleon ($NN$) interaction is one of the
principal goals of nuclear and hadron physics. Apart from their intrinsic
importance for the study of nuclear forces, $NN$ elastic scattering data are
also necessary ingredients in the modeling of meson production and other
nuclear reactions at intermediate energies. It is therefore clear that all
facilities must try to fill in any remaining gaps in our knowledge in the
area. In this respect COSY has certainly taken its responsibilities seriously
because the measurements of proton-proton elastic scattering carried out at
this machine have completely revolutionized the $pp$ database above 1~GeV,
where previously there had been relatively few systematic
experiments~\cite{ARN2000}. However, as shown in sect.~\ref{np_elastic},
significant advances in the measurements of neutron-proton elastic scattering
have also been made at COSY.

\subsection{Proton-proton elastic scattering} 
\label{pp_elastic}

It is important at the outset to realize that the beam intensities at COSY
are not sufficient to make the study of the polarizations of recoil particles
through double scattering experiments a very attractive proposition. On the
other hand, the possibility of using a thin windowless target at an internal
target station of COSY offers significant advantages over the methods used in
standard external experiments. By exploiting the repeated passage of the
recirculating polarized or unpolarized proton beam through such a polarized
or unpolarized target in COSY it was possible to measure the differential
cross section~\cite{ALB1997,ALB2004}, the proton analyzing
power~\cite{ALT2000,ALT2005}, and spin-correlations~\cite{BAU2003,BAU2005}
with the EDDA detector described in Section~\ref{main} and illustrated in
Fig.~\ref{fig:Albers0}. This approach allowed these observables to be studied
over effectively a continuum in energy though, for presentational reasons,
the final results were necessarily published in finite energy bins. Though a
similar technique was used earlier at the SATURNE synchrotron, the
measurements there were restricted to the unpolarized cross section and only
at centre-of-mass angles close to $\theta_{cm}=90^{\circ}$~\cite{GAR1985}.

Proton-proton elastic scattering can be cleanly identified by geometry
if the directions of the two recoiling protons are measured. In this case
the laboratory polar angles of fully coplanar events must satisfy
\begin{equation}
\label{coincidence}
\cot\theta^1_{\rm lab}\cot\theta^2_{\rm lab} = 1 + T_p/2m_p,
\end{equation}
where $T_p$ is the laboratory kinetic energy of the beam and $m_p$ the proton
mass. Imposing this condition suppresses significantly the contributions of
multibody final states, such as meson production. In contrast to the
single-arm measurements that are discussed later in this section, a detailed
study of the recoil energies is not required.

In all measurements of a differential cross section in an internal experiment
at an accelerator, a crucial element is the determination of the absolute
normalization, i.e., of the beam-target luminosity $\mathcal{L}$. The
technique used at EDDA involved the evaluation of the numbers of electrons
kicked out of the target through purely electromagnetic proton-electron
scattering. The requirements of this procedure had a significant influence on
the target design, to which we now turn.

The EDDA targets for the cross section measurements were made of strips of
polypropylene (CH$_2$) of dimension $4\times5~\mu$m$^2$ coated with a very
thin layer of aluminium. These were then strung horizontally between the
prongs of a metal fork such that they could be moved remotely into the path
of the COSY beam. The carbon background under the $pp$ elastic scattering
peak was already much reduced by the coplanarity cut and the correlation
requirement of Eq.~\eqref{coincidence}. In addition, the shape of the
residual background could be determined with high accuracy by using similar
carbon fibres as targets.

\begin{figure}[htb]
\begin{center}
\includegraphics[width=0.9\columnwidth]{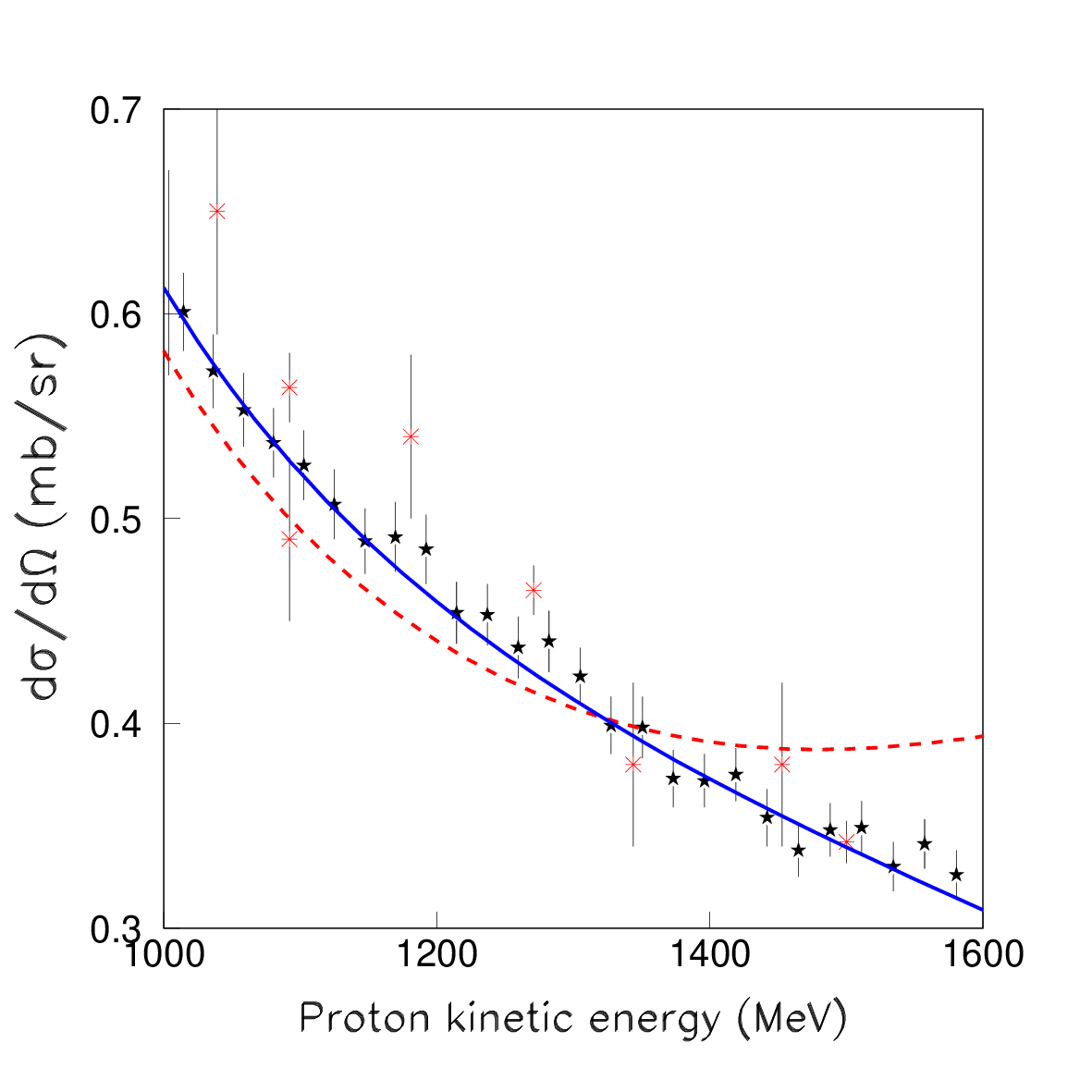}
\caption{\label{fig:Albers1} Differential cross section for elastic
proton-proton scattering at a centre-of-mass angle $\theta_{cm}=89^{\circ}$.
The (black) stars are EDDA points~\cite{ALB2004} whereas the (red)
crosses are a selection of earlier results taken from the SAID
database~\cite{ARN2000}. The (red) dashed curve represents SAID solution SM94
before taking the EDDA data into account whereas the (blue) solid line
is that of SP07, where all the EDDA data are included~\cite{ARN2000}.}
\end{center}
\end{figure}

A key feature of the EDDA cross section measurement is that it possessed two
independent methods to determine the electron loss from the target fibres.
Though these studied electrons in widely different kinematic regions, on
average the deviations between the two methods was below 1.5\%. The secondary
electron monitor (SEM) measured the electric current through the metal
supporting fork of the target. This current of electrons replaced the
low-energy secondary electrons emanating from the target surface due to the
interaction with the proton beam. The alternative technique involved
detecting higher energy $\delta$ electrons kicked out of the target in two
PIN-diodes placed downstream of the target. It must be stressed that, due to
uncertainties in the geometry etc., neither of these could give reliable
absolute normalizations. However, since the proton-electron cross section and
its energy dependence are calculable, both methods provide excellent relative
luminosity determinations.

Since the target used for the differential cross section experiment was
relatively narrow, the inner shell of the EDDA detector, called the helix in
Fig.~\ref{fig:Albers0}, was not needed for the vertex reconstruction and so
only the outer shell was left in position. There are, of course, many
detailed refinements in the very careful analysis of the EDDA
data~\cite{ALB2004} but, by normalizing the results on an angular integral of
the precise LAMPF measurements at $T_p=793$~MeV~\cite{SIM1993}, values of the
$pp$ elastic differential cross section could be obtained from 230 to
2590~MeV with a 1\% overall systematic uncertainty. In total about
$4\times10^7$ good $pp$ elastic scattering events were registered and so the
statistical uncertainties were also very low.

\begin{figure}[htb]
\begin{center}
\includegraphics[width=0.9\columnwidth]{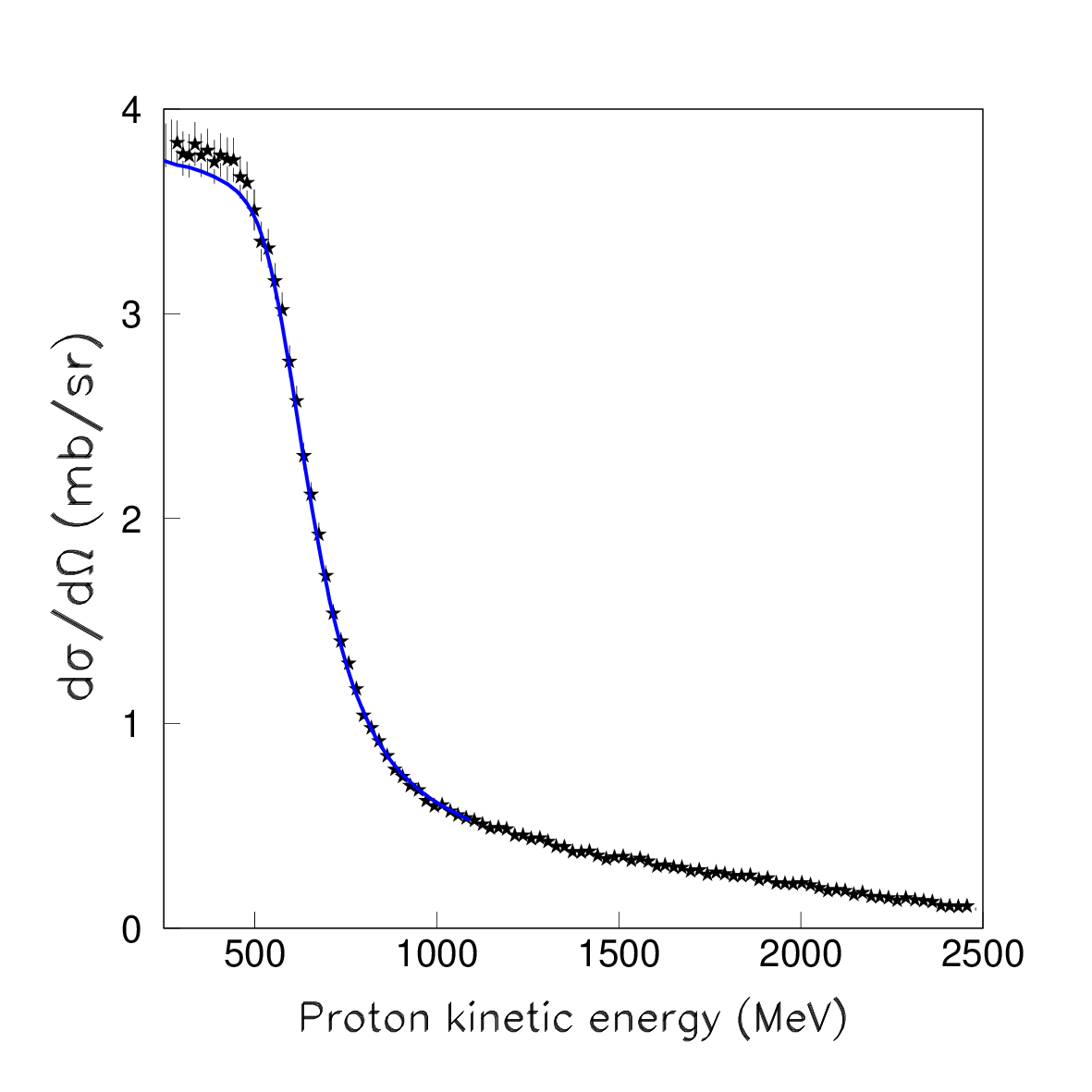}
\caption{\label{fig:Albers2} Differential cross section for elastic
proton-proton scattering at $\theta_{cm}=89^{\circ}$. The EDDA
data~\cite{ALB2004} are compared to the SP07 SAID partial wave
solution~\cite{ARN2000}.}
\end{center}
\end{figure}

To illustrate the influence of the (combined) EDDA results on the phase shift
analysis, we show in Fig.~\ref{fig:Albers1} these data at
$\theta_{cm}=89^{\circ}$ compared to SAID solutions before and after the EDDA
results were available~\cite{ARN2000}. It should be noted that the earlier
SAID solution was only valid up to 1.6~GeV. Also shown are a selection of
results from previous experiments. The improvements in the data and their
representation are manifest. The complete EDDA $89^{\circ}$  data set is
shown in Fig.~\ref{fig:Albers2}, where deviations from the SAID SP07 are only
evident at very low energies, where many results from other experiments are
available to constrain the solution.

Due to the necessity to detect both protons in the detector, the major
drawback in all the EDDA data sets is the lack of acceptance near the forward
or backward directions. The lowest c.m.\ angle for the differential cross
sections was typically $\theta_{cm} \approx 35^{\circ}$, depending upon the
beam energy, though this was reduced to $\approx 32.5^{\circ}$ for the
spin-dependent observables.

\begin{figure}[htb]
\begin{center}
\includegraphics[width=0.9\columnwidth]{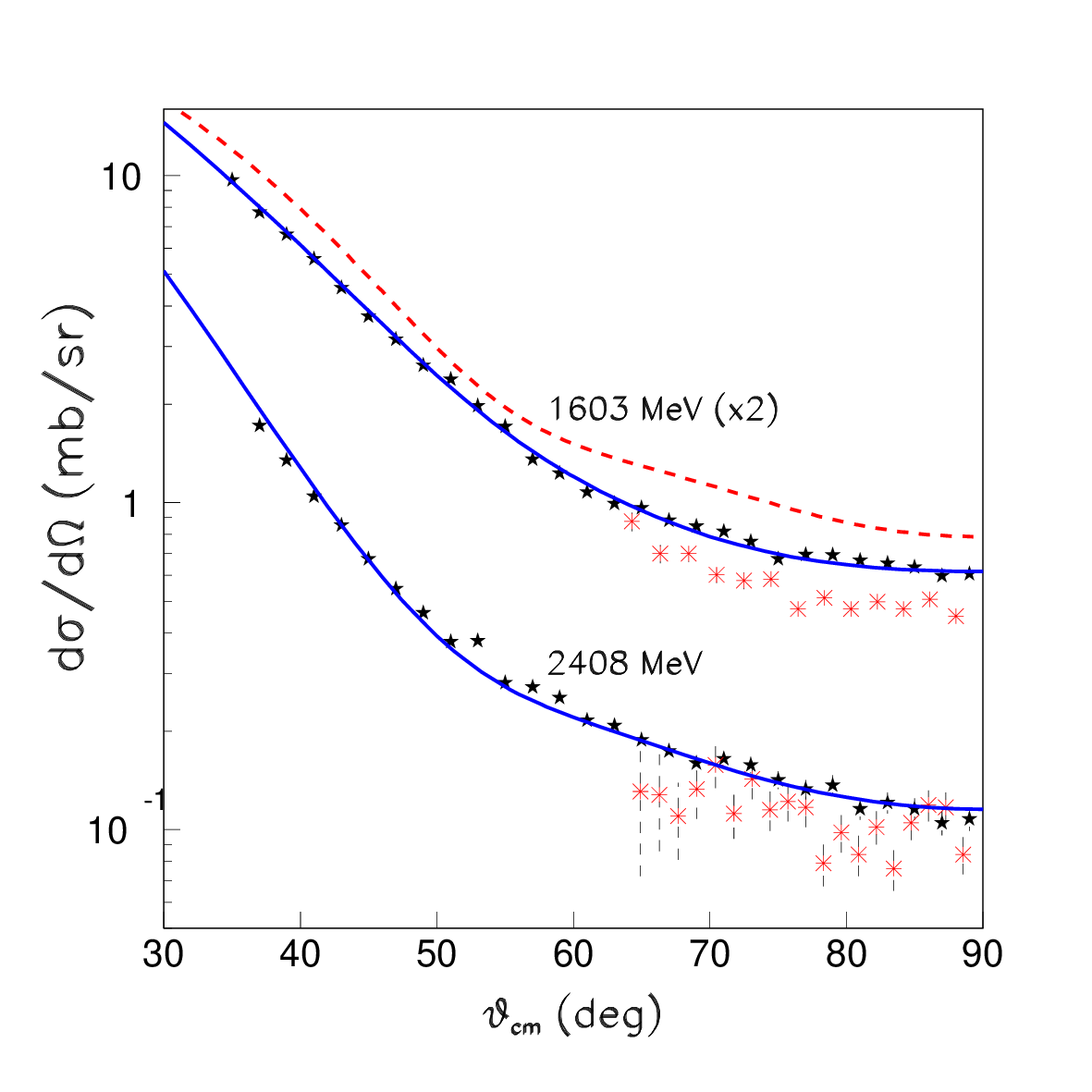}
\caption{\label{fig:Albers3} Differential cross sections for elastic
proton-proton scattering. The EDDA data (stars)~\cite{ALB2004} at the two
marked beam energies are compared with the SP07 SAID solution and,
at the lower energy, with the SM94 solution (dashed curve)~\cite{ARN2000}.
Early data from Ref.~\cite{JEN1980} are also shown as (red) crosses.}
\end{center}
\end{figure}

The proton-proton differential cross sections measured with the EDDA detector
at two beam energies are shown in Fig.~\ref{fig:Albers3}, where they are
compared with the SP07 and, at the lower energy, with a pre-EDDA partial wave
solution~\cite{ARN2000}. It must be stressed that these are just two out of
the very many EDDA measurements, as already indicated by the $89^{\circ}$
data set shown in Fig.~\ref{fig:Albers2}. The change between the SM94 and
SP07 solutions is clear and the fact that the SP07 SAID solution passes
through almost all of the EDDA points proves that these data completely
dominate the SAID database above about 1~GeV. One reason for this may, of
course, be the limitations of the other data shown in Fig.~\ref{fig:Albers3}.

A full partial wave analysis clearly needs information from spin-dependent
measurements and so we now turn to the contributions in this field made by
the EDDA collaboration. The EDDA proton analyzing power, $A_y$, measurements
were carried out using the unpolarized COSY proton beam incident on a
polarized hydrogen target~\cite{ALT2000,ALT2005}. This approach avoids the
difficulties associated with the depolarizing resonances in COSY so that all
the EDDA $A_y$ results are inversely proportional to the same factor, namely
the polarization of the target. Its value can be most reliably deduced by
comparing the EDDA data at 730~MeV with the results of a precise external
target experiment in the angular range $45\lesssim \theta_{cm} \lesssim
70^{\circ}$~\cite{MCN1990}.

The target used in the EDDA experiments was a polarized hydrogen gas jet fed
from the atomic beam source (ABS). The typical width of the jet in the region
of intersection with the COSY beam was about 12~mm. The resulting
luminosities of almost $10^{28}$~cm$^{-2}$s$^{-1}$ were adequate for the
programme because of the comparatively large cross section for proton-proton
elastic scattering. The effective polarization seen by the COSY beam was
diluted by the background of unpolarized hydrogen gas in the COSY ring. This
resulted in a variation of the effective polarization across the jet width
and its effects in the longitudinal direction were especially important at
small angles and had to be taken into consideration.

Because of the symmetry of the detector in the  azimuthal angle $\phi$ shown
in Fig.~\ref{fig:Albers0}, EDDA is ideally suited for measuring asymmetries.
The polarized differential cross section may be written
\begin{equation}
\frac{\dd\sigma}{\dd\Omega}=\left(\frac{\dd\sigma}{\dd\Omega}\right)_{\!0}
\left[1+A_y(Q_y\cos\phi + Q_x\sin\phi)\right],
\label{Alt1}
\end{equation}
where $({\dd\sigma}/{\dd\Omega})_0$ is the unpolarized cross section and
$Q_y$ and $Q_x$ are the target polarizations in the (transverse) $y$ and $x$
directions, respectively, which were cycled around the $+x$, $-x$, $+y$, and
$-y$ directions. Since, using the $\phi$ dependence of Eq.~\eqref{Alt1}, each
of these polarizations would be sufficient for the extraction of $A_y$, the
extra redundancy allowed the authors to eliminate the false asymmetries that
would still be observed with no target polarization.

\begin{figure}[htb]
\begin{center}
\includegraphics[width=0.9\columnwidth]{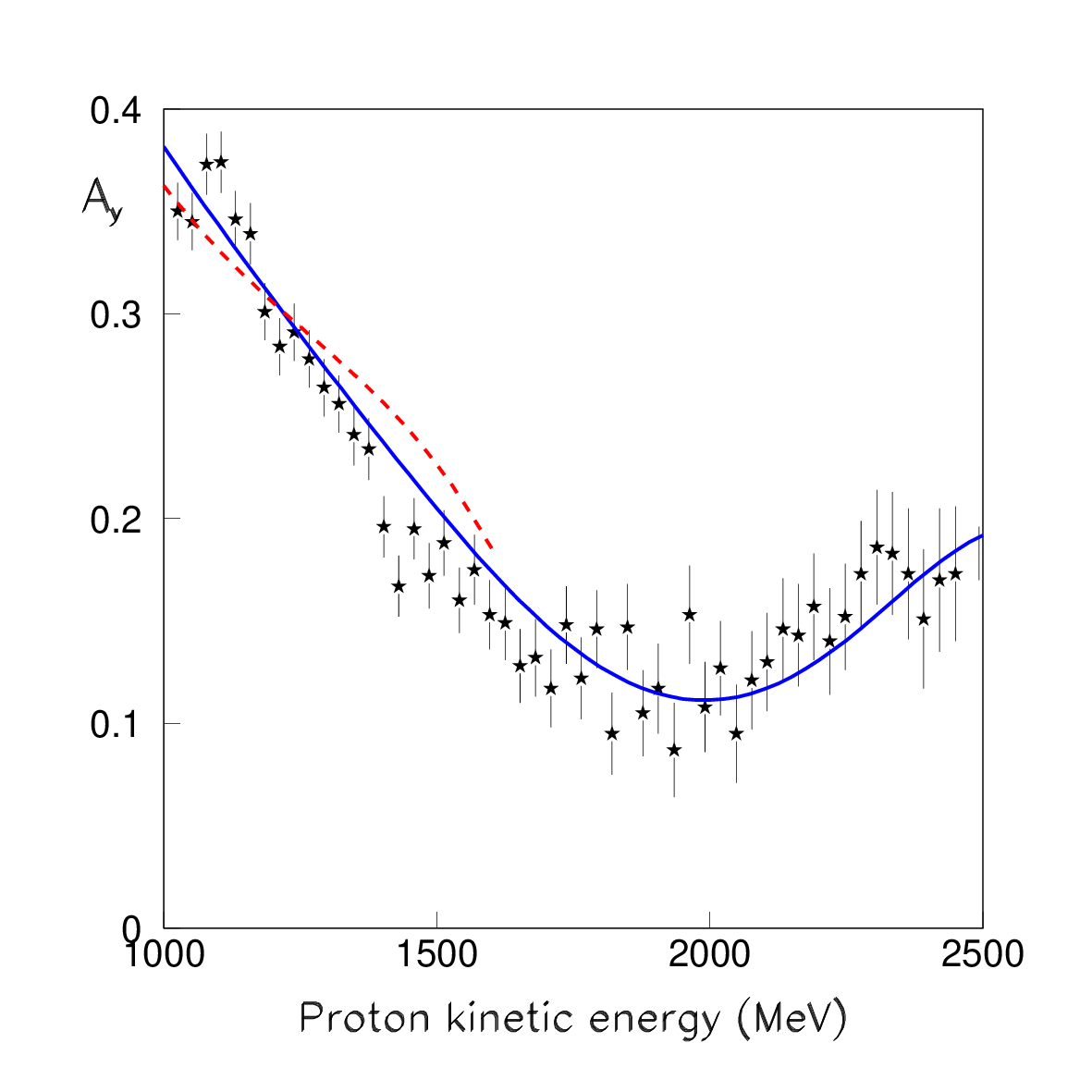}
\caption{\label{fig:Altmeier1} Proton analyzing power in $pp$ elastic
scattering measured by the EDDA collaboration at $\theta_{cm}=56^{\circ}$ as
a function of the proton beam energy~\cite{ALT2005}. The curves correspond to
the SAID SP07 (blue solid line) and SM94 (red dashed line) solutions, though
the latter is only valid up to 1.6~GeV.}
\end{center}
\end{figure}

A typical excitation function at $\theta_{cm}=56^{\circ}$ of the EDDA $pp$
elastic analyzing power above 1~GeV is shown in Fig.~\ref{fig:Altmeier1},
where it is compared to pre- and post-EDDA solutions. This is just a small
fraction of the total EDDA data set and the statistical fluctuations would be
reduced if wider energy bins were used. However, such a wider binning would
not be of any real benefit for the partial wave fits. As an extra consistency
check, data were also taken during the deceleration of the COSY proton beam
as well as in the acceleration mode.

\begin{figure}[htb]
\begin{center}
\includegraphics[width=0.9\columnwidth]{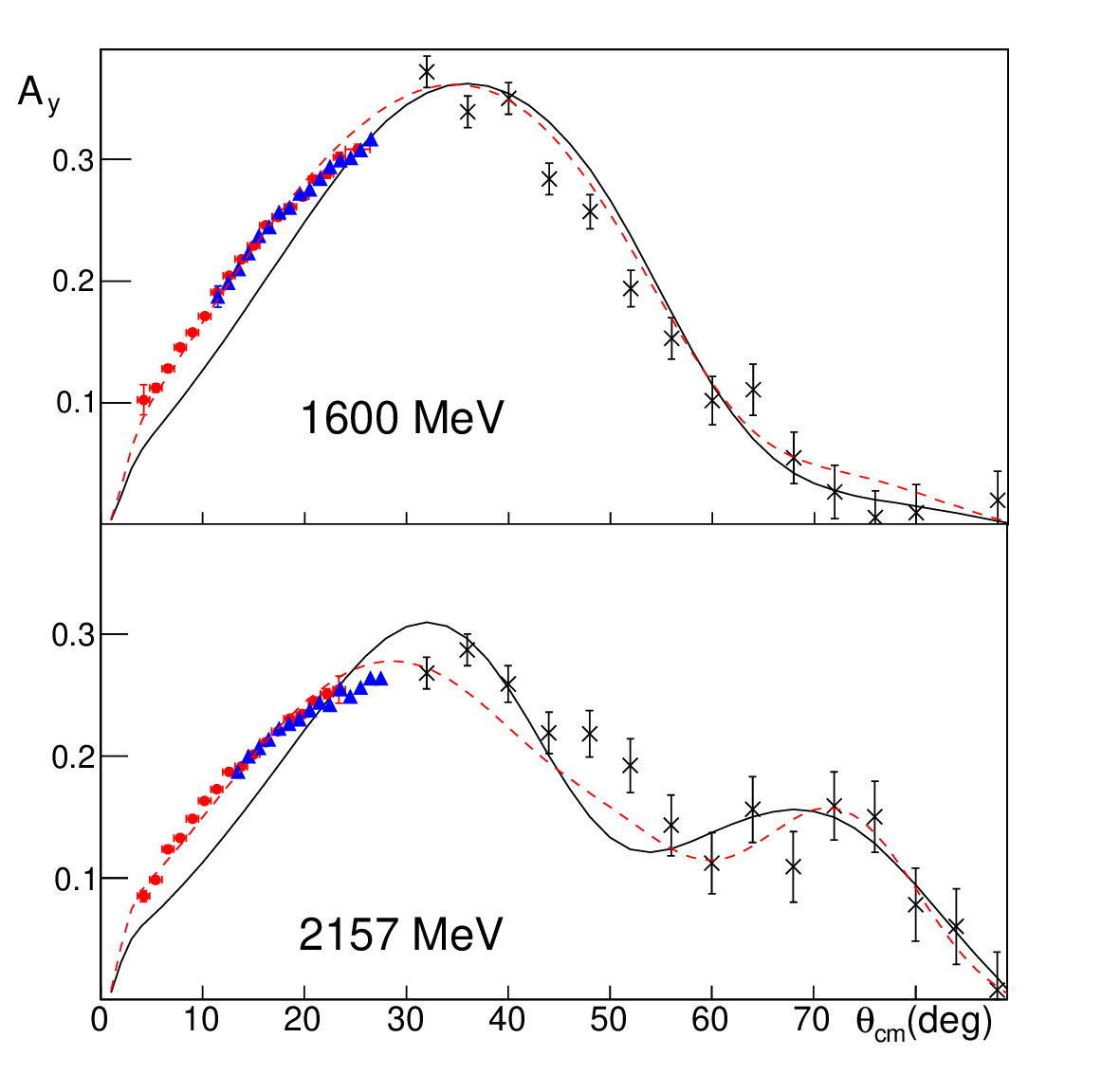}
\caption{\label{fig:Zara1} Comparison of the ANKE measurements of the proton
analyzing power in $pp$ elastic scattering at two (out of six) energies using
the STT (red filled circles) and FD (blue filled triangles)
systems~\cite{BAG2014} with results from EDDA (black crosses)~\cite{ALT2005}
for energies that differed by no more than 7~MeV. If continuity in energy
were imposed, many of the EDDA statistical fluctuations would be
significantly diminished. The curves correspond to the SAID SP07 (solid black
line) solution~\cite{ARN2000} and a revised one (dashed red)~\cite{WOR2016a}.}
\end{center}
\end{figure}

Angular distributions of $A_y$ are shown at two energies in
Fig.~\ref{fig:Zara1}, where the EDDA results~\cite{ALT2005} are compared with
small-angle data obtained at ANKE~\cite{BAG2014} and partial wave
solutions~\cite{ARN2000,WOR2016a}, which will be discussed later in this
section.

\begin{figure}[htb]
\begin{center}
\includegraphics[width=0.9\columnwidth]{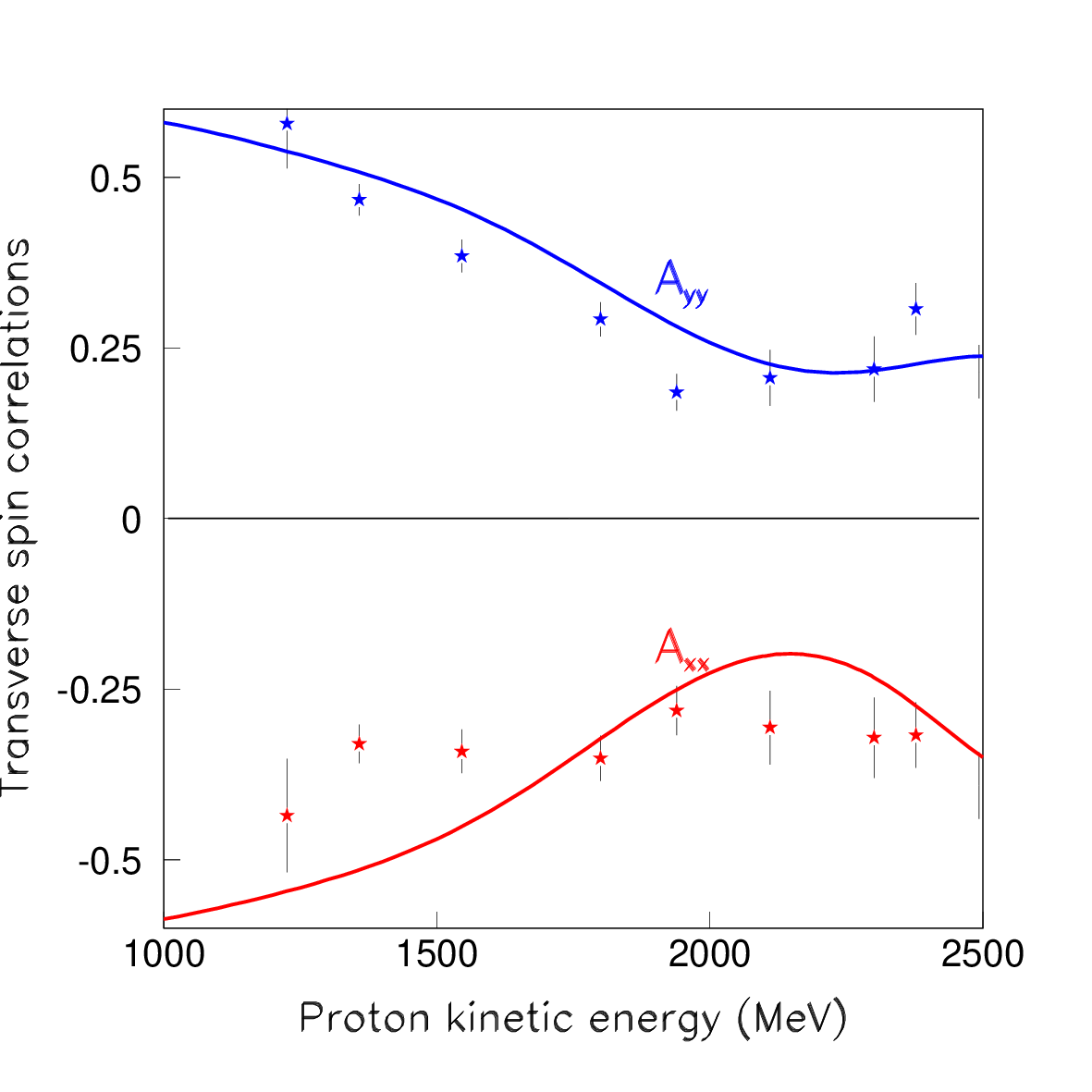}
\caption{\label{fig:Bauer1} Transverse spin correlations measured by the EDDA
collaboration at nine flat tops above 1~GeV at the fixed angle of
$\theta_{cm}=57.5^{\circ}$~\cite{BAU2005}. These are compared to the results
of SAID solution SP07~\cite{ARN2000}.}
\end{center}
\end{figure}

The culmination of the innovative EDDA campaign was provided by measurements
of the spin correlations $A_{yy}$, $A_{xx}$, and $A_{xz}$ in $pp$ elastic
scattering~\cite{BAU2003,BAU2005}. For this purpose the operation of the
polarized gas jet target described for the $A_y$ measurements was extended so
that the polarization cycle included also $\pm z$ modes as well as $\pm y$
and $\pm x$. When using a polarized proton beam in conjunction with such a
target, the dependence on the azimuthal angle $\phi$ is more complicated than
that of Eq.~\eqref{Alt1}~\cite{BAU2003}. However, by studying the $\phi$
variation for different target polarizations it was possible to extract the
value of the asymmetry due to the beam in terms of that of the target. Since
the target polarization, or equivalently the target proton analyzing power,
had been precisely measured in the $A_y$ studies with an unpolarized
beam~\cite{ALT2005}, this led to accurate values of the beam polarizations.

The spin-correlation measurements were carried out using two modes, either as
a study of excitation functions with a quasi-continuous beam energy or as a
series of ten fixed energies, the so-called flat tops. One critical problem
when accelerating polarized protons in a circular machine is handling the
depolarizing resonances which, in the most serious cases, is achieved by
flipping the proton polarization. Thus, in the vicinity of the
$p_p=2443$~MeV/$c$ ($T_p=1678$~MeV) resonance the statistics in the
excitation functions are rather poor, due to the low beam polarization. The
data for $A_{yy}$ and $A_{xx}$ shown in Fig.~\ref{fig:Bauer1} were those
obtained at the nine flat tops above 1~GeV at the fixed c.m.\ angle of
$\theta_{cm}=57.5^{\circ}$~\cite{BAU2005} compared to the SAID SP07
solution~\cite{ARN2000}.

\begin{figure}[h!]
\begin{center}
\includegraphics[width=0.9\columnwidth]{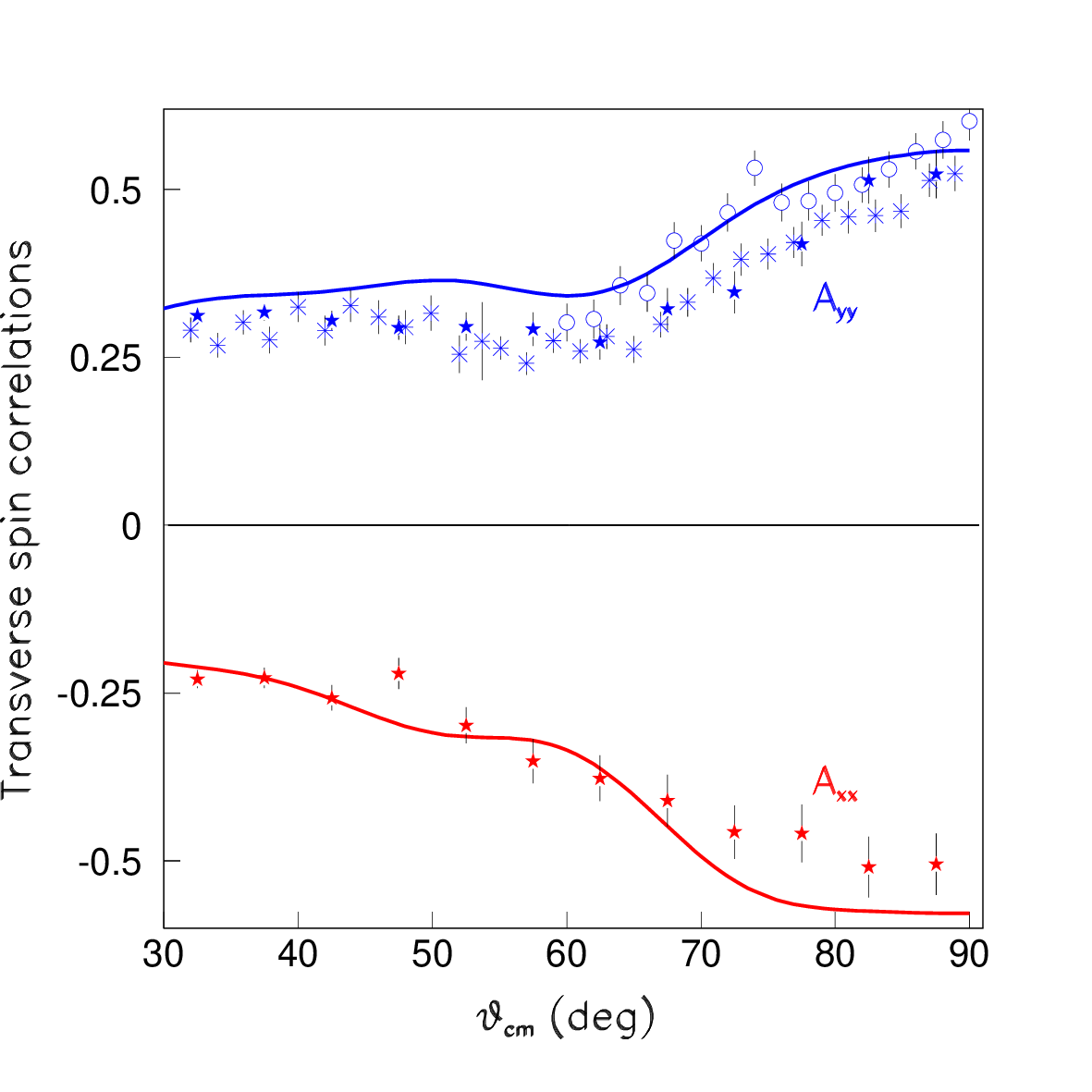}
\caption{\label{fig:Bauer2} Transverse spin correlations measured by the EDDA
collaboration at 1.8~GeV as functions of the scattering angle $\theta_{cm}$
(stars)~\cite{BAU2005}. The crosses~\cite{LEH1987} and open
circles~\cite{ALL2000} represent $A_{yy}$ data taken in external target
experiments at SATURNE. The results are compared to the SAID solution
SP07~\cite{ARN2000}.}
\end{center}
\end{figure}

The angular distributions of $A_{yy}$ and $A_{xx}$ at 1.8~GeV are shown in
Fig.~\ref{fig:Bauer2} as functions of the scattering angle. Note that both
these observables are symmetric about $90^{\circ}$. The EDDA $A_{yy}$ results
are in reasonable agreement with the earlier SATURNE
experiment~\cite{LEH1987} but far less so with the later one~\cite{ALL2000}.
There are no other data to which one could compare the $A_{xx}$ results. It
is seen from both Figs.~\ref{fig:Bauer1} and \ref{fig:Bauer2} that the SAID
SP07 solution still needs further refinement if it is to accommodate the EDDA
transverse spin correlations. Though considerable effort was put into the
measurement of $A_{xz}$, all the EDDA results for this observable are
consistent with zero and this is in accord with the SAID partial wave
analysis~\cite{ARN2000}.

It is hard to overstate the impact that the EDDA data have had on the partial
wave analysis of proton-proton elastic scattering above 1~GeV. However, the
basic design of EDDA means that measurements could not be carried out with
this apparatus for c.m.\ angles below about $30^{\circ}$. To fill some of the
void left by EDDA in this region, measurements of the unpolarized
differential cross section~\cite{MCH2016} and proton analyzing
power~\cite{BAG2014} have been carried out at small angles at several
discrete energies at the ANKE facility.

The ANKE measurements differed from those carried out with EDDA in several
important respects. At ANKE a polarized proton beam was incident on a
hydrogen cluster-jet target so that no C/CH$_2$ subtraction was required.
Furthermore, the ANKE experiments involved single-arm measurements, where the
energy of one of the final protons as well as its direction were measured.
This allowed the second proton to be identified through the peak in the
missing-mass distribution.

Fast protons were measured in the forward detector, which covered
$10^{\circ}-30^{\circ}$ in c.m.\ polar angles for $pp$ elastic scattering and
$\pm 30^{\circ}$ in azimuth, though the polar angle range was cut in order to
minimize acceptance uncertainties. In the analyzing power experiment the
slower recoil protons were detected in one of the Silicon Tracking Telescopes
(STT), described in sect.~\ref{targets}. These were placed inside the vacuum
chamber near the target, symmetrically to the left and right of the beam.
Although there was a large overlap in acceptance angle between the FD and STT
data, the latter allowed measurements in c.m.\ angle down to $\approx
5^{\circ}$. For protons stopping in the third layer of the STT, greater
precision in the angle of the recoiling proton was achieved by deducing it
from the energy measured in the telescope rather than from a direct angular
measurement.

Although there were events where one proton was measured in the FD and the
other in the STT, unlike the EDDA experiment, such a coincidence requirement
was not placed on the trigger. However, for events where both of the protons
were simultaneously measured in the two detectors it was possible to make two
determinations of the scattering angle and typically $\theta_{cm}({\rm STT})
- \theta_{cm}({\rm FD}) \approx 0.3^{\circ}$. This offset is fortunately much
smaller than the bin width used to present the data.

Just as for the EDDA spin-correlation experiments~\cite{BAU2003,BAU2005}, the
use of a (vertically) polarized proton beam necessitated overcoming the
depolarizing resonances in COSY. The analyzing power measurements were
carried out at 796~MeV and five other fixed beam energies between 1.6 and
2.4~GeV that were well away from the resonances. The values of the six beam
polarizations were determined at the end of each COSY cycle using the EDDA
polarimeter that was discussed in sect.~\ref{EDDA}~\cite{WEI2000}.

Though not possessing the same azimuthal acceptance as the EDDA detector, the
symmetric positioning of the STT did allow the left-right asymmetry to be
robustly evaluated. On the other hand, the ANKE forward detector only covered
part of one hemisphere and an asymmetry could only be deduced if the relative
luminosities for polarizations ``up'' and ``down'' could be determined to
high precision. This was achieved by comparing the rates of charged particle
production in angular regions where the beam polarization could have no
influence. As is seen in Fig.~\ref{fig:Zara1}, the two very different methods
gave remarkably consistent results in the overlap region, differences being
typically on the 1\% level. This agreement suggests that most of the
systematic errors in the asymmetry determinations are under control and that
the dominant uncertainty arises from the $\pm 3\%$ of the EDDA
polarimeter~\cite{WEI2000}.

Figure~\ref{fig:Zara1} also compares the small-angle ANKE results at two beam
energies with larger angle taken with EDDA at neighbouring energies. Such a
presentation does not reflect fairly the EDDA statistics because their data
were obtained at many closely spaced energies. Although there is no overlap
in angle between the EDDA and ANKE data, there is no obvious discrepancy
between the two experiments. The solid lines represent the predictions of the
SAID solution SP07~\cite{ARN2000}, which was heavily influenced by the
combined EDDA data set. Since these curves do not reproduce the rapid
variation of the ANKE measurements at small angles, partial wave solutions
were sought that included these results along with the EDDA and other data in
the fitting process. This resulted in the broken curves in
Fig.~\ref{fig:Zara1}, which reproduce far better the \emph{shapes} defined by
the ANKE points, though even here there seem to be systematic differences at
the larger ANKE angles~\cite{WOR2016a}.

\begin{figure}[htb]
\centering
\includegraphics[width=0.8\columnwidth, angle=0]{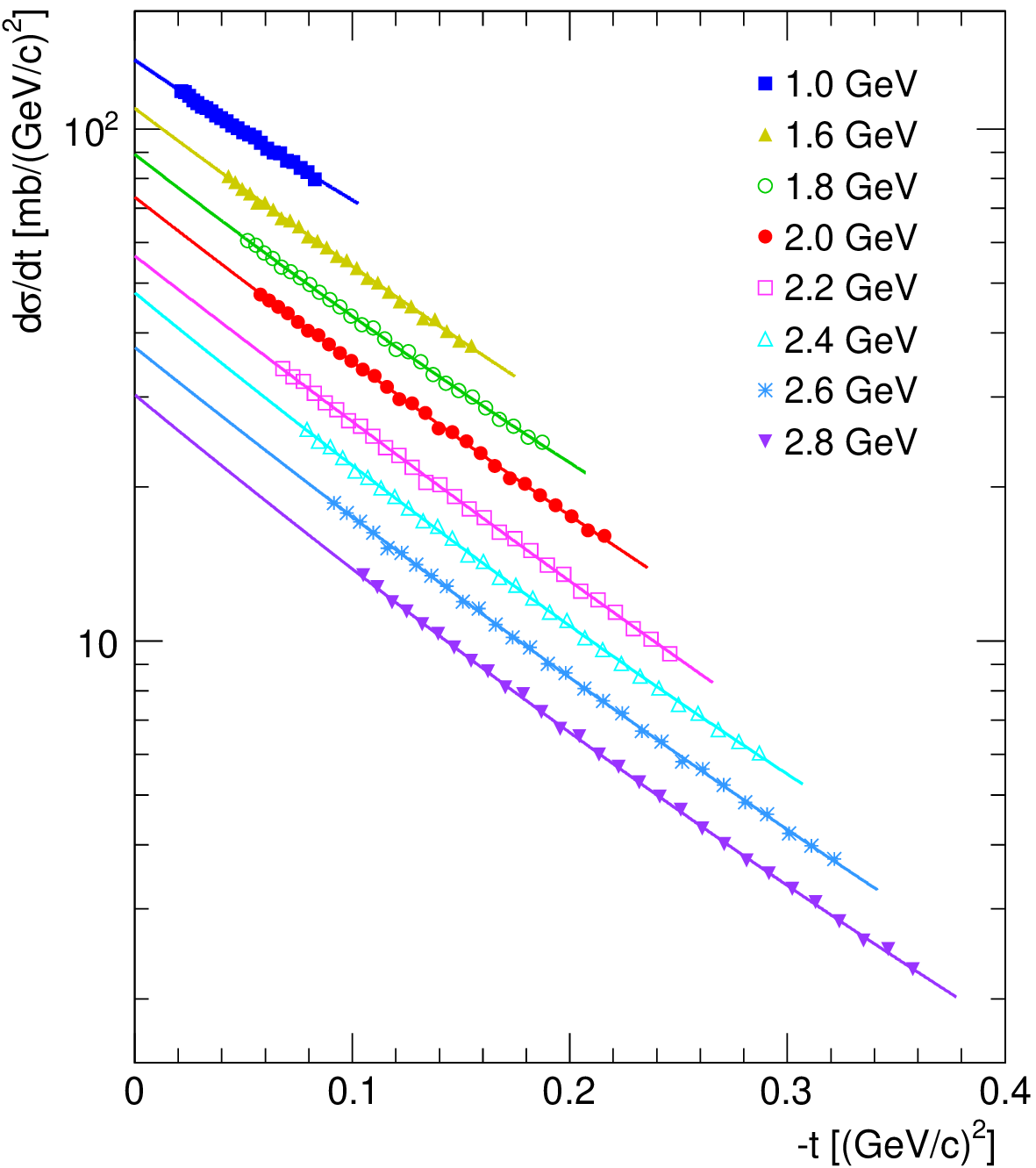}
\caption{ \label{fig:dsdt} Combined ANKE data set at eight beam energies of
$pp$ elastic differential cross sections with respect to the four-momentum
transfer $t$~\cite{MCH2016} compared to fits made on the basis of
Eq.~(\ref{pars}). The correct values are shown at 1.0~GeV but, for clarity of
presentation, the data are scaled down sequentially in energy by factors of
1.2. The typical systematic uncertainty is of the order of 3\%. }
\end{figure}

The luminosity in the ANKE measurement of the differential cross section for
elastic $pp$ scattering was evaluated using the so-called Schottky technique
discussed in sect.~\ref{Schottky}. It was shown here that this was even more
accurate than the $\pm 4\%$ found in the initial experiment~\cite{STE2008}.
Though the cross section can be deduced from the count rates in either the FD
or STT, the acceptance can be more reliably estimated for the forward
detector and the results shown in Fig.~\ref{fig:dsdt} were obtained using
this system.

The energy dependence of the ANKE measurements of the $pp$ elastic
differential cross sections~\cite{MCH2016} can be seen most clearly in terms
of the four-momentum transfer $t$ and the results at the eight energies are
shown in Fig.~\ref{fig:dsdt}. The data in the measured region vary very
smoothly on this logarithmic plot and can be well represented by
\begin{equation}
\label{pars}
\frac{\dd\sigma}{\dd t} = A\,\exp\left(-B|t|+C|t|^2\right).
\end{equation}
Good fits could be obtained at low energies with $C=0$.

The perfect agreement with the ANL data at 2.2 and 2.83~GeV~\cite{AMB1974}
may be fortuitous because these measurements have a quoted normalization
uncertainty of $\pm 4\%$. The other data that overlap the ANKE results were
obtained at Gatchina at several energies up to 992~MeV using the IKAR recoil
detector~\cite{DOB1983}. These 992~MeV points are about 8\% lower in
normalization than the ANKE 1000~MeV values.

\begin{figure}[htb]
\centering
\includegraphics[width=0.8\columnwidth]{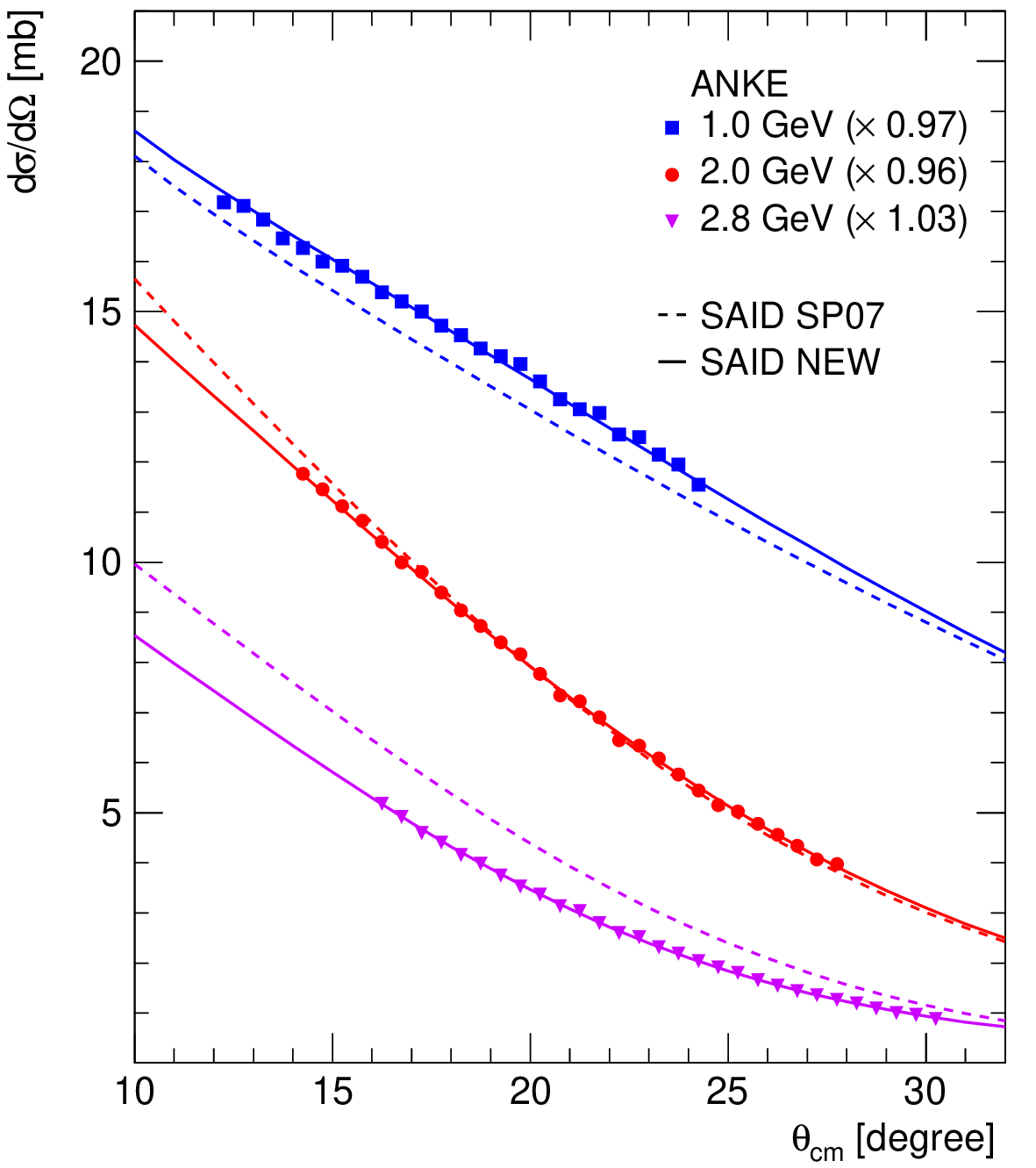}
\caption{Scaled ANKE proton-proton elastic differential cross sections at
1.0, 2.0, and 2.8~GeV with statistical errors~\cite{MCH2016} compared to the
SAID SP07 solution~\cite{ARN2000} and a ``new'' partial wave solution where
the ANKE data have been taken into account. For presentational reasons the
2.0 and 2.8~GeV data and curves have been reduced by factors of 0.5 and 0.25,
respectively. The best agreement with the new partial wave data was achieved
by scaling the ANKE data with factors 0.97, 0.96, and 1.03 at the three
energies. The deviations from unity are consistent with the overall
systematic uncertainties.} \label{fig:RON}
\end{figure}

Although the ANKE results are not inconsistent with the EDDA
data~\cite{ALB1997,ALB2004}, the gap in angle between the two data sets means
that one cannot use this as direct evidence in favour of the ANKE
normalization. On the other hand, the modified (``new'') SAID partial wave
solution of Fig.~\ref{fig:RON} can describe both data sets provided that the
ANKE results are scaled by factors that are consistent with the overall
systematic uncertainties~\cite{WOR2016a}.

\begin{figure}[h!]
\centering
\includegraphics[width=0.9\columnwidth, angle=0]{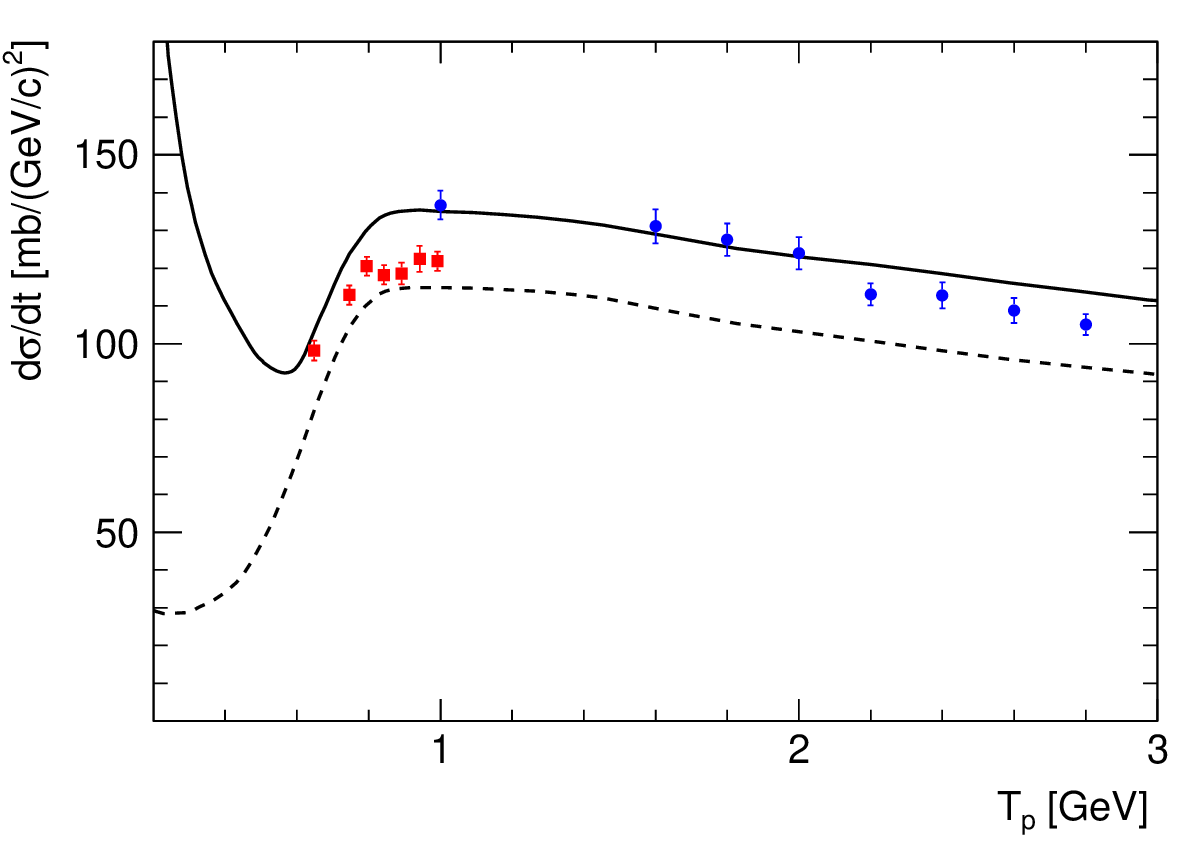}
\caption{ \label{fig:GK} The predictions of Grein and Kroll~\cite{GRE1982}
for the values of the forward $pp$ elastic differential cross section (solid
line), the corresponding lower limit provided by the spin-independent optical
theorem being indicated by the broken line. The
extrapolated ANKE data are shown with their quoted errors by the (blue)
circles~\cite{MCH2016}, whereas the (red) squares are the published Gatchina
values obtained with the IKAR recoil detector~\cite{DOB1983}. }
\end{figure}

In the forward direction the number of proton-proton elastic scattering
amplitudes reduces from five to three and the imaginary parts of these
amplitudes are determined completely by the spin-averaged and spin-dependent
total cross sections through the generalized optical theorem. The
corresponding real parts have been estimated from forward dispersion
relations, where these total cross sections provide the necessary
input~\cite{GRE1982}. All the terms contribute positively to the value of $A$
and, using the optical theorem, the lower bound, $ A\ge (\sigma_{\rm
tot})^2/16\pi, $ is obtained by taking the $pp$ spin-averaged total cross
section $\sigma_{\rm tot}$. This lower bound and the full Grein and Kroll
estimates for $A$~\cite{GRE1982} are both shown in Fig.~\ref{fig:GK}.

Before extrapolating to $t=0$ the data have to be corrected for Coulomb
effects. In the Gatchina case this was done by adding a spin-independent
Coulomb amplitude~\cite{DOB1983} whereas the ``new'' SAID fit with and
without Coulomb provided a means to correct the ANKE data~\cite{MCH2016}.
Both sets of extrapolated values are shown in Fig.~\ref{fig:GK} and the
agreement of the ANKE results with the theoretical curve is very encouraging,
It is, however, unfortunate that similar data were not taken below 1~GeV.

\begin{figure}[hbt]
\centering
\includegraphics[width=1.0\columnwidth, angle=0]{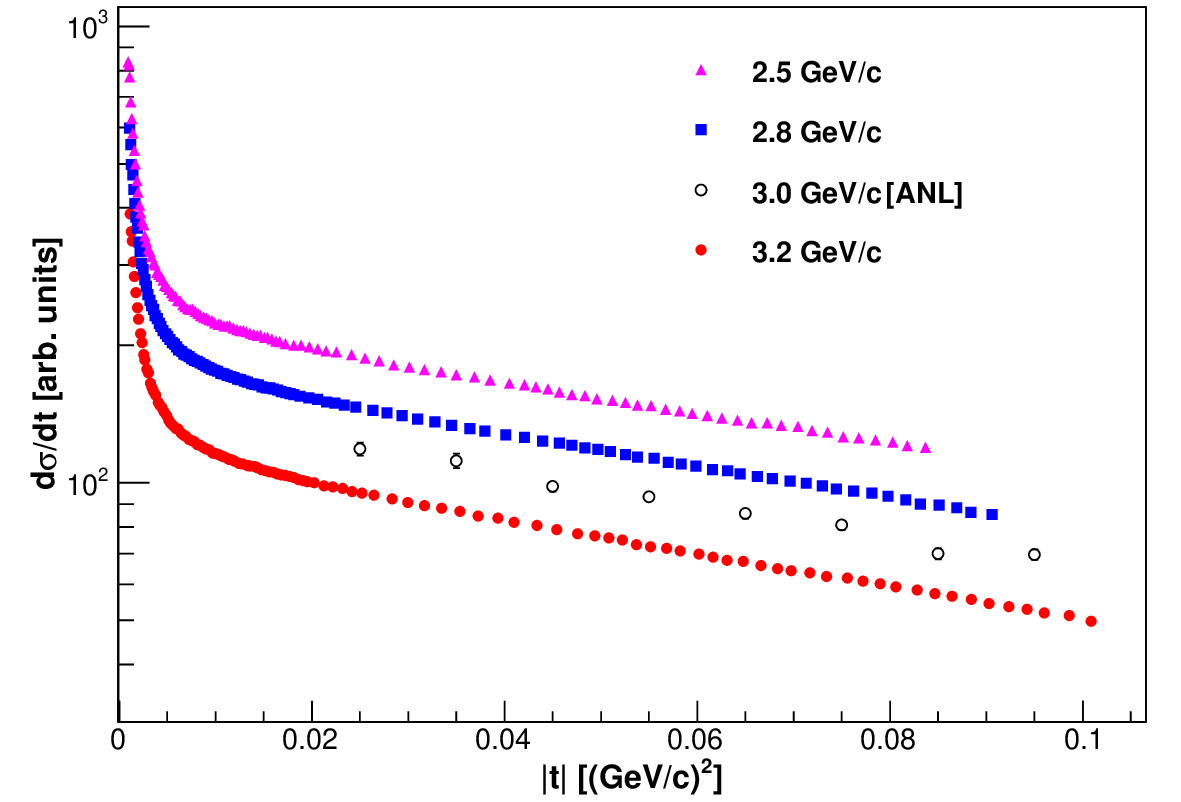}
\caption{ \label{fig:KOALA} Elastic proton-proton differential cross sections
at small momentum transfers $t$. The preliminary values measured by the PANDA
collaboration with the KOALA detector at 2.5, 2.8, and
3.2~GeV/$c$~\cite{HUX2016} are compared to the angular dependence measured at
ANL at 3.0~GeV/$c$~\cite{AMB1974}. The normalizations of the KOALA data at
the three momenta are still under study~\cite{HUX2016a} and all data are
given in arbitrary units.}
\end{figure}

The KOALA detector described in sect.~\ref{KOALA} is capable of studying the
proton recoils from $pp$ or $\bar{p}p$ elastic scattering in the momentum
transfer range $10^{-3} \lesssim |t| \lesssim
10^{-1}$~(GeV/$c$)$^2$~\cite{HUX2014} so that it covers different regions
where the Coulomb, the Coulomb-nuclear interference, and the nuclear are all
significant. The detector was tested at COSY by investigating proton-proton
elastic scattering at 2.5, 2.8, and 3.2~GeV/$c$~\cite{HUX2016}. The
preliminary values obtained for the differential cross sections are shown in
Fig.~\ref{fig:KOALA}, where they are compared with the results from ANL at
2.2~GeV~\cite{AMB1974}. The overall normalizations of the KOALA data have not
yet been determined because the completely pure Coulomb region was not fully
accessed and the analysis is still proceeding~\cite{HUX2016a}.

It is intriguing to note that, if two KOALA arms had been used at small
momentum transfers in the ANKE proton-proton analyzing power
experiment~\cite{BAG2014}, the resulting data could have been sensitive to
magnetic moment effects and this would have provided extra constraints on the
amplitude analysis.

%
%
%

\subsection{Neutron-proton elastic scattering} 
\label{np_elastic}

Since COSY was not designed for the production of external neutron beams, the
only contributions that could be made in neutron-proton elastic scattering
have been through quasi-free scattering with a deuterium target or a deuteron
beam and several measurements of interest have been performed in this way.

As part of their programme to study the properties of the possible dibaryon
found in two-pion production described in sect.~\ref{nppipi}, the
WASA-at-COSY collaboration measured the analyzing power in $np$ quasi-elastic
scattering in the vicinity of the dibaryon mass of
2.38~GeV/$c^2$~\cite{ADL2014,ADL2014a}. For this they used a polarized
deuteron beam of the maximum energy available at COSY, viz.\ 2.27~GeV, and
detected the recoil proton in coincidence with the \textit{spectator} proton.
The values of the vector and tensor polarizations of the beam were determined
using deuteron-proton elastic scattering that had been measured at
ANKE~\cite{MCH2013a}. The methodology was also checked by studying in
parallel the analyzing power in $pp$ quasi-elastic scattering.

\begin{figure}[htb]
\begin{center}
\includegraphics[width=0.9\columnwidth]{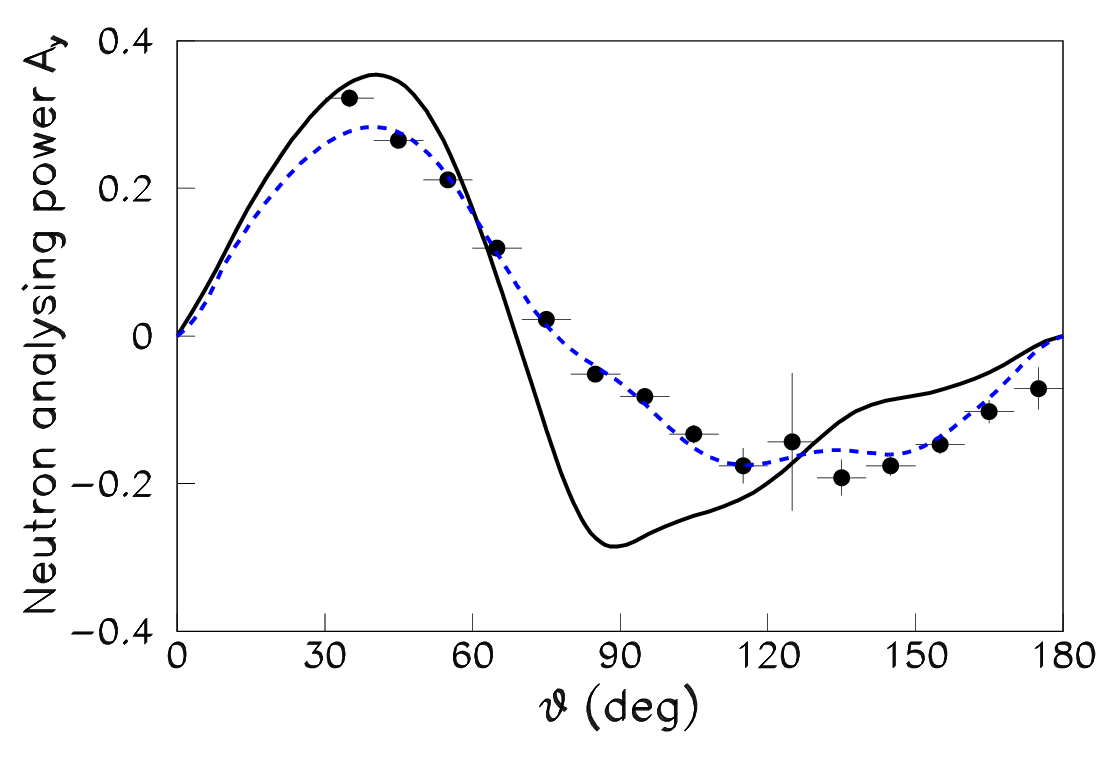}
\caption{\label{fig:Heinz1}  Angular distribution of the neutron analyzing
power measured in $np$ quasi-elastic scattering in the $\sqrt{s}=2.377$~GeV
region by the WASA-at-COSY collaboration~\cite{ADL2014}. Also shown by the
(black) solid line are the SP07 predictions~\cite{ARN2000}, averaged over the
deuteron momentum distribution and the WASA acceptance. An updated SAID
solution, similarly averaged, is shown by the (blue) dashed line.}
\end{center}
\end{figure}

The design of the WASA detector, with its azimuthal symmetry, makes it very
suitable for measuring analyzing powers and the full set of results is shown
in Fig.~\ref{fig:Heinz1}. Although the SAID $np$ analysis has been used up to
a beam energy of 1.3~GeV~\cite{ARN2000}, the data upon which it is based are
rather sparse above 1.0~GeV and this leads to large uncertainties in its
application at the higher energies.

By detecting the spectator proton it is also possible to separate the data
into intervals in the total c.m.\ energy $\sqrt{s}$, as for the dibaryon
studies in sect.~\ref{nppipi}. However, this was not done for the results
shown in Fig.~\ref{fig:Heinz1} and, instead, the SP07
predictions~\cite{ARN2000} have been averaged over the c.m.\ energies
produced by the deuteron Fermi momenta, as moderated by the WASA acceptance.
It is immediately clear that the SP07 partial wave solution does not
reproduce satisfactorily the new WASA measurements. A modified solution that
does take these data into account is shown by the dashed
curve~\cite{WOR2016}. The major change compared to SP07 is in the $^{3\!}D_3$
wave, where a pole has been generated. This structure therefore gives extra
evidence in support of the dibaryon hypothesis discussed in
sect.~\ref{nppipi}.

In summary, the WASA results show that the isoscalar part of the SP07
solution is in conflict with data in the 1.135~GeV region. Given the limited
number of other measurements available above 1~GeV, it was possible to modify
this solution to yield the satisfactory agreement shown in
Fig.~\ref{fig:Heinz1} though, it must be stressed, there may still be
ambiguities in this revised solution. The partial wave solutions are
considered in Ref.~\cite{WOR2016}, where double-spin experiments that might
clarify the dibaryon hypothesis are discussed. There are, in fact, other
measurements at COSY that have questioned the validity of the SAID $np$ SP07
partial wave solution at 1.135~GeV, and we now turn to these.

It has been known for many years that the charge exchange of deuterons on
hydrogen, $\pol{d}p\to \{pp\}_{\!s}n$, is very sensitive to the deuteron
tensor polarization provided that the excitation energy $E_{pp}$ in the
recoiling proton pair is low~\cite{BUG1987}. In this case the diproton is
dominantly in an $S$-wave and the Pauli principle then demands that the
proton spins are antiparallel in the \Szero\ configuration. There is
therefore a spin-isospin flip from the $(S,I)=(1,0)$ of the deuteron to
$(0,1)$ of the diproton. At small momentum transfers between the deuteron and
diproton the transition amplitudes are well described in impulse
approximation in terms of the three spin-spin small angle neutron-proton
charge-exchange amplitudes, i.e., the three spin-spin large angle
neutron-proton elastic amplitudes~\cite{BUG1987,CAR1991}.

There has been an extensive programme at ANKE to study the charge exchange of
polarized deuterons on hydrogen and, by using a polarized target, this was
also extended to include measurements of deuteron-proton spin correlations.
The polarized deuterium ion source at COSY is capable of producing beams with
a variety of vector ($p_z$) and tensor ($p_{zz}$) polarizations. The
$z$-direction indicated here is the quantization axis in the polarized source
system and this is relabeled as the $y$-direction in the COSY frame. This is
perpendicular to the COSY ring, i.e., along the direction of the holding
fields, and it is only in this direction that the polarization is not
modified by the spin precession.

Although some information on the beam polarizations was available from the
low energy and EDDA polarimeters at 75.8~MeV and 270~MeV, respectively, the
ANKE collaboration wished to measure the polarizations at the energy of the
primary experiment. This was achieved by comparing the results for various
nuclear reaction, viz.\ $\pol{d}p\to {}^{3}$He$\,\pi^0$, $\pol{n}p\to
d\,\pi^0$, and $\pol{d}p$ elastic scattering, with values obtained in
external target experiments~\cite{CHI2006}. Following this procedure, no
evidence was found for deuterons being depolarized during acceleration and
this is completely consistent with the absence of depolarizing resonances for
deuterons over the entire COSY energy range. Using these values of the beam
polarization, together with neutron-proton elastic scattering amplitudes
taken from the SAID PWA~\cite{ARN2000}, the impulse approximation described
well the $dp\to \{pp\}_{\!s}n$ differential cross section and tensor
analyzing powers at 1.17~GeV~\cite{CHI2009}. Since this reaction is easily
identified, and has a large figure of merit, the results at the neighbouring
energy of 1.2~GeV were used in all the subsequent ANKE experiments to
determine the tensor polarization of the beam. This is very much in the
spirit of the original work, which proposed that the reaction could be used
as the basis of a tensor polarimeter for deuterons~\cite{BUG1987}.

\begin{figure}[htb]
\begin{center}
\includegraphics[width=0.8\columnwidth]{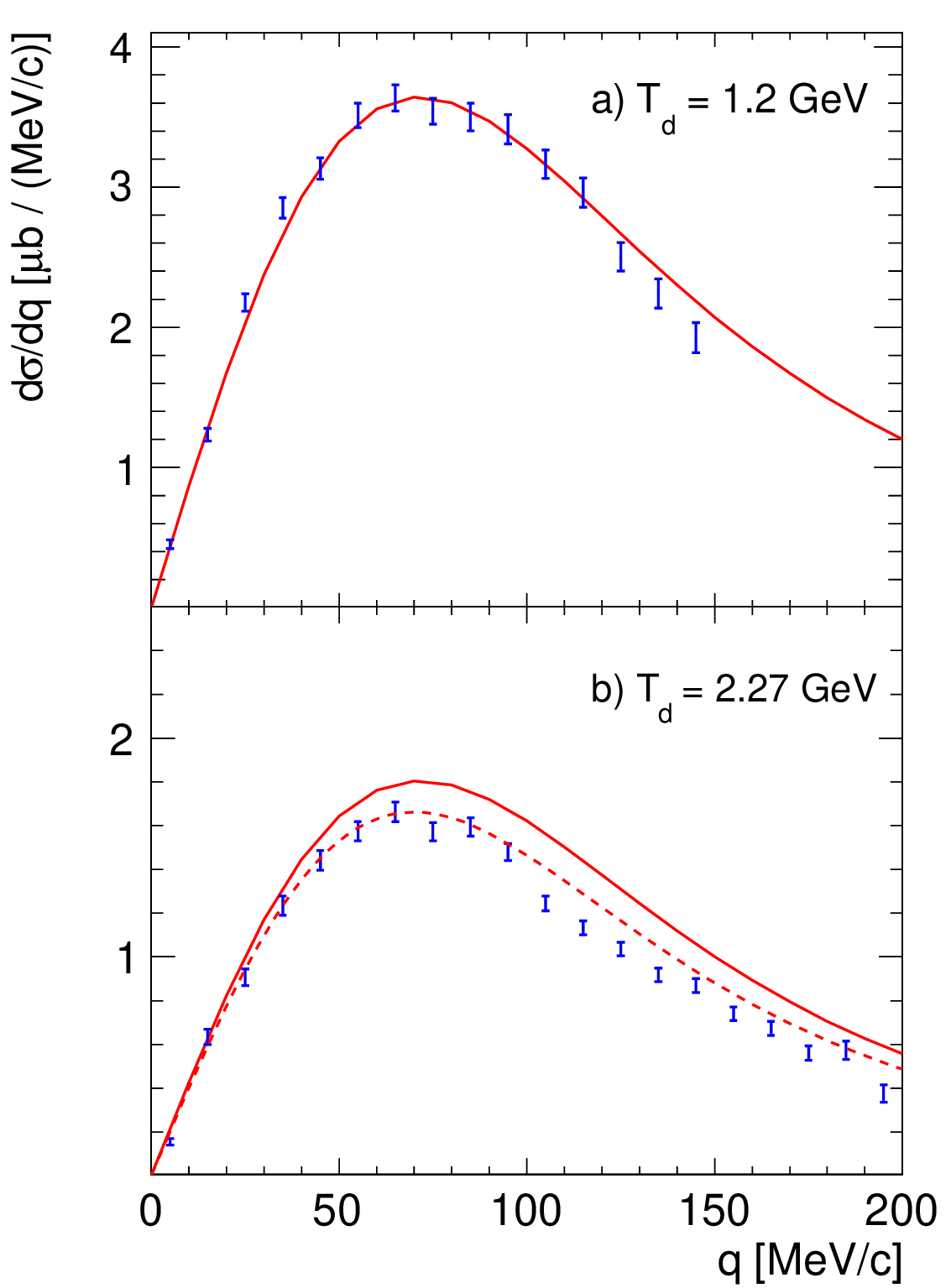}
\caption{\label{fig:Mch1} Differential cross sections for the $dp\to
\{pp\}_{\!s}n$ reaction at 1.2 and 2.27~GeV~\cite{MCH2013} compared with
impulse approximation predictions based upon the SAID SP07
solution~\cite{ARN2000}. The data are integrated over $E_{pp}<3$~MeV. Only
statistical errors are shown; The systematic uncertainty of $\approx 5\%$ is
particularly large at 2.27~GeV. The dashed curve at this energy corresponds
to the longitudinal $np$ spin-spin amplitude being reduced by $25\%$.}
\end{center}
\end{figure}

\begin{figure}[h!]
\begin{center}
\includegraphics[width=0.8\columnwidth]{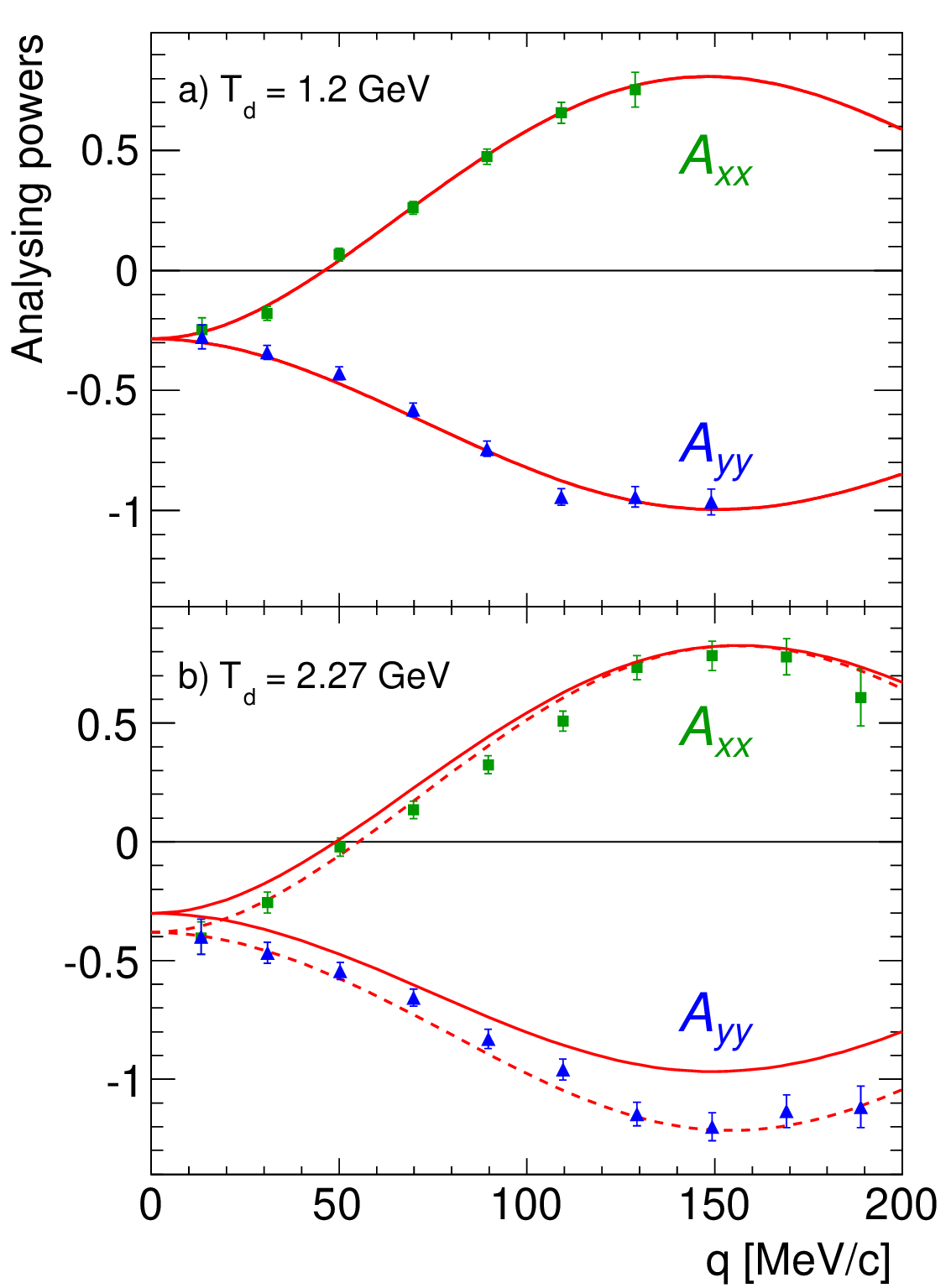}
\caption{\label{fig:Mch2} Tensor analyzing powers $A_{xx}$ (squares) and
$A_{yy}$ (triangles) of the $\pol{d}p\to \{pp\}_{\!s}n$ reaction at 1.2 and
2.27~GeV for $E_{pp}<3$~MeV~\cite{MCH2013} compared to impulse approximation
predictions based upon the SAID SP07 solution~\cite{ARN2000}. The dashed
curves at $2.27$~GeV correspond to a uniform reduction of the
spin-longitudinal amplitude by $25\%$. In addition to the error bars shown,
there could be an overall uncertainty of up to 4\% in the beam polarization
at 2.27~GeV.}
\end{center}
\end{figure}

Measurements of the differential cross section and tensor analyzing powers of
the $\pol{d}p\to \{pp\}_{\!s}n$ reaction were made at 1.2, 1.6, 1.8, and
2.27~GeV at ANKE using an unpolarized hydrogen cluster-jet
target~\cite{MCH2013,MCH2013b}. The results at the lowest and highest energy
are shown in Figs.~\ref{fig:Mch1} and \ref{fig:Mch2}, where they are compared
to impulse approximation predictions based upon the SAID SP07
solution~\cite{ARN2000}. Whereas the data are well reproduced at 1.2~GeV, as
they are also at 1.6~GeV and 1.8~GeV, serious discrepancies are evident at
2.27~GeV, despite the fact that the impulse approximation approach should get
better as the energy is increased. In order to see whether these could be
explained as arising from the $np$ input, the spin-longitudinal input
amplitude at 1.135~GeV was reduced uniformly by an \emph{ad hoc} factor of
$0.75$. This results in much better agreement for the analyzing powers and
reduces significantly the discrepancy for the cross section at 2.27~GeV,
though it should be noted that the systematic uncertainty in the luminosity
(5\%) is particularly large at this energy.

The spin-correlation experiments were carried out at 1.2 and 2.27~GeV, with a
deuteron beam with a limited set of polarization modes, incident on a
teflon-coated (closed) aluminum storage cell target, fed with a jet of
polarized atomic hydrogen~\cite{MCH2013,MCH2013b}. The polarization of the
target was flipped between ``up'' and ``down'' every five seconds. Both the
polarization of the hydrogen target and the vector polarization of the
deuteron beam were determined from measurements of the analyzing power of the
$\pol{n}p\to d\pi^0$ reaction.

By using a polarized gas target, it was possible to measure also the proton
analyzing power in the \mbox{$d\pol{p}\to \{pp\}_{\!s}n$} reaction%
. The impulse approximation model reproduces the $A_y^p$ data very well at
1.2~GeV but fails completely at 2.27~GeV, which is a strong indication of
problems with the spin-orbit amplitude in the isospin-zero part of the SP07
solution~\cite{ARN2000}.

\begin{figure}[h!]
\begin{center}
\includegraphics[width=0.8\columnwidth]{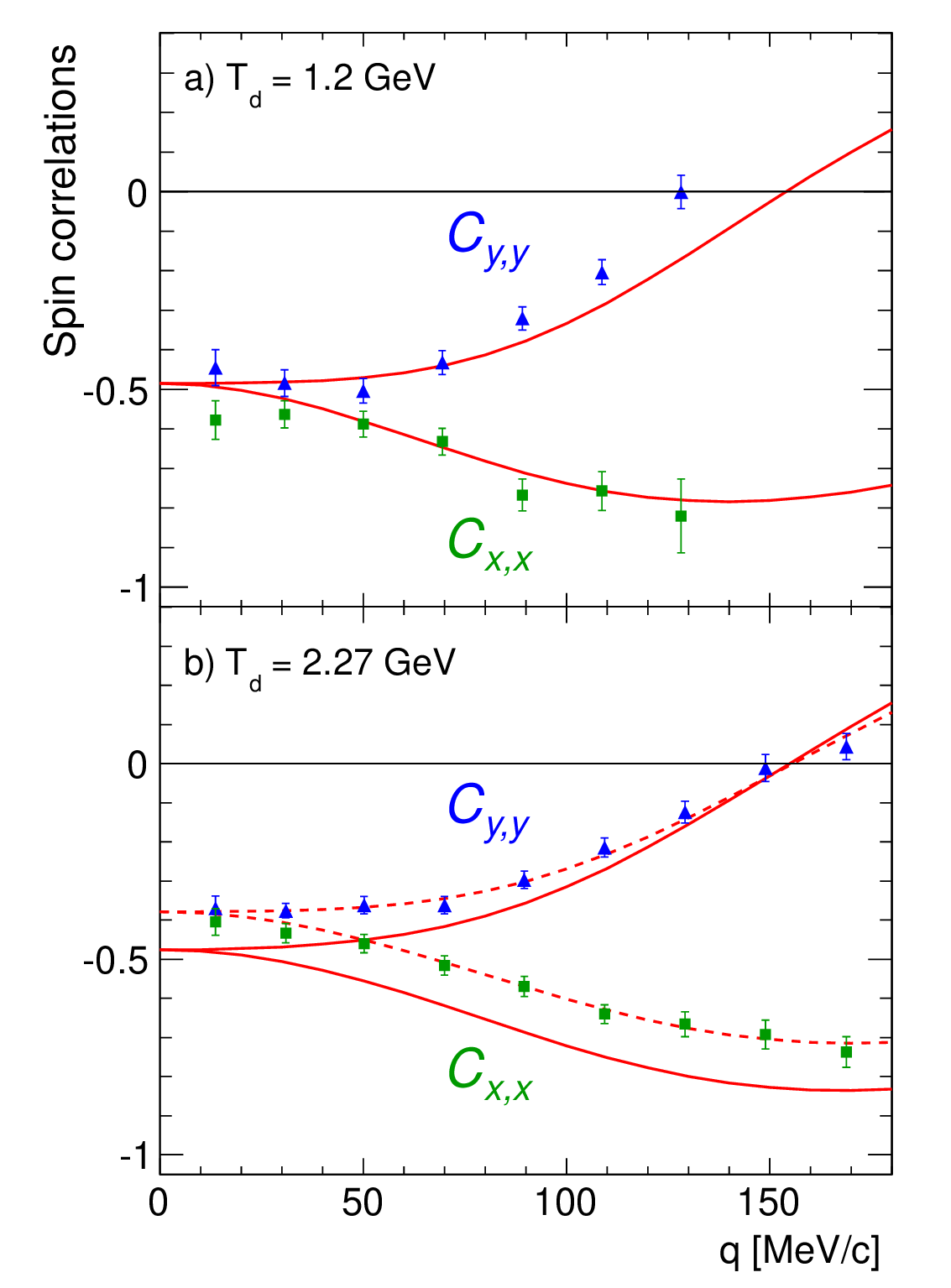}
\caption{\label{fig:Mch4} Spin-correlations $C_{x,x}$ and $C_{y,y}$ for the
$dp\to \{pp\}_{\!s}n$ reaction at $T_d = 1.2$~GeV and $2.27$~GeV for
$E_{pp}<3$~MeV~\cite{MCH2013} compared to impulse approximation
predictions~\cite{CAR1991}. The dashed curves at $2.27$~GeV correspond to the
longitudinal $np$ spin-spin amplitude being reduced by $25\%$.}
\end{center}
\end{figure}

The spin-correlation results are shown in Fig.~\ref{fig:Mch4} and the message
here is very similar to that given by the other ANKE data. Though there is
broad agreement with the model at 1.2~GeV, at 2.27~GeV the theoretical
description is improved significantly if the longitudinal $np$ spin-spin
input is reduced by $25\%$.

It is immediately obvious from looking at the four figures that the ANKE data
are reasonably well described by the model at 600~MeV per nucleon but the
agreement is less satisfactory at 1.135~GeV, where the model should be more
reliable. The deviations in the differential cross section are on the 10-15\%
level, as they are also for the deuteron tensor analyzing powers and the spin
correlations. These discrepancies can be largely eliminated if the $np$
longitudinal spin-spin amplitude is reduced by a factor of $0.75$. This
should not be taken as proof of the validity of such a reduction; rather it
indicates that a revised $np$ partial wave solution could give a much better
description of the ANKE data. The situation is even more extreme for the
proton analyzing power 
since this shows that the spin-orbit term is very badly described in the SP07
solution.

To put some of these discrepancies into context, and to link up to the WASA
neutron analyzing power experiment~\cite{ADL2014a}, one sees from
Figs.~\ref{fig:Mch2} and \ref{fig:Mch4} that the measured values of $A_{xx}$
and $C_{x,x}$, extrapolated to $q=0$, are $-0.38\pm0.03$ and $-0.39\pm 0.05$,
respectively, where uncertainties in the beam and target polarizations have
been included. These are to be compared to the SP07 predictions shown in the
figures of $-0.30$ and $-0.48$. On the other hand, the revised $np$ partial
wave solution discussed in connection with the WASA data~\cite{ADL2014a}
yields rather $-0.42$ and $-0.31$, respectively. Given that the new solution
is far from being unambiguous with respect to observables in this region, the
fact that the changes from SP07 are both in the right direction and are of
the right order of magnitude, is very promising. Thus both the WASA $np$
analyzing power measurement and the ANKE deuteron charge-exchange data
indicate in similar ways that the SAID SP07 solution~\cite{ARN2000} requires
modification at 1.135~GeV.

In order to go to higher energies in quasi-free interactions on the deuteron
at COSY one must use a deuterium target rather than a deuteron beam. A
feasibility test was carried out at ANKE with a 600~MeV unpolarized proton
beam incident on the cell target that had been fed with polarized deuterium
atoms from the ABS~\cite{GOU2015}. Protons from a charge exchange then had
very low energies and these were detected in STT placed to the left and right
of the target. The values of $A_{yy}$ extracted from these data at low
$E_{pp}$ showed good continuity in three-momentum transfer $q$ with those
obtained with a deuteron beam~\cite{MCH2013}. The major drawback of the
deuterium target approach is that it is not possible with the current STT to
investigate the region where $E_{pp}$ and $q$ are simultaneously small.

Since there were successful measurements at ANKE of the analyzing power in
elastic $pp$ scattering at small angles that were shown in
Fig.~\ref{fig:Zara1}~\cite{BAG2014}, it is natural to wonder whether similar
data could be obtained in quasi-free $\pol{p}n$ elastic scattering. For this
purpose the collision of a polarized proton beam with a deuterium cluster-jet
target was studied at 796~MeV and five other beam energies from 1.6 to
2.4~GeV and these data are still under analysis~\cite{BAR2016}. The fast
proton was measured in the forward detector and the supposed \emph{spectator}
proton in one of the two STT that were placed symmetrically inside the vacuum
chamber to the left and right of the beam. The $\pol{p}d\to ppn$ reaction
could then be identified through the missing-mass peak corresponding to the
undetected neutron. Just as in the $pp$ case, the proton beam polarization
was established on the basis of measurements with the EDDA polarimeter.
Although the data are broadly similar to the measured free $np$ elastic
analyzing power at 796~MeV, the paucity of the database above 1.3~GeV has
limited the influence of the ANKE results.

\begin{figure}[htb]
\begin{center}
\includegraphics[width=0.8\columnwidth]{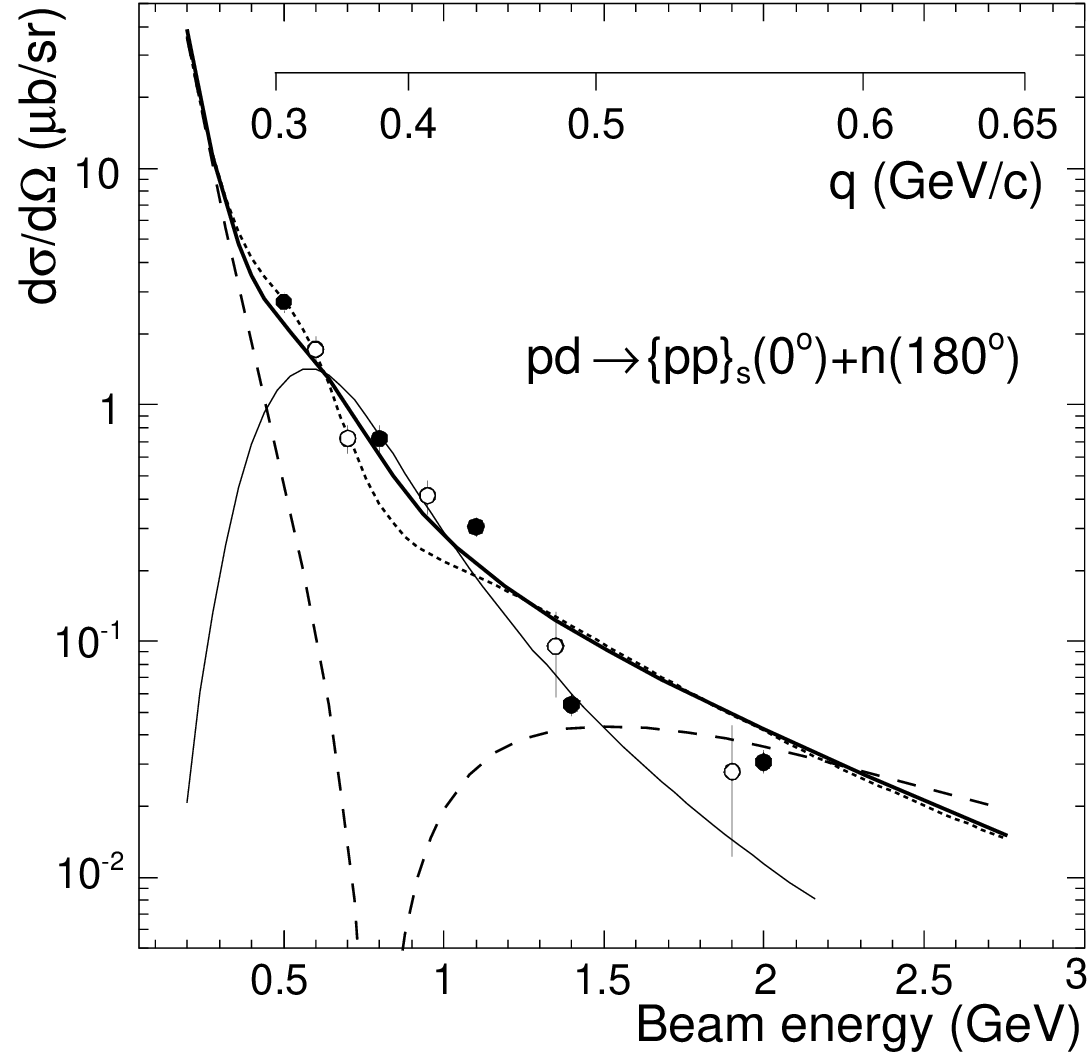}
\caption{\label{fig:Uzikov} Differential cross section in the c.m.\ frame for
the $pd\to \{pp\}_{\!s}n$ reaction averaged over the angular interval
$172^{\circ}-180^{\circ}$ versus the beam energy. The experimental points are
taken from the ANKE experiments of Ref.~\cite{KOM2003} (open circles) and
Ref.~\cite{DYM2010a} (closed circles). The predictions of one nucleon
exchange are shown by the dashed line and those of the $\Delta$ mechanism by
the thin solid line. The total predictions of the model, including small
single-scattering contributions, are shown by the thick solid
line~\cite{DYM2010a}. The upper scale shows the internal momentum $q$ of the
nucleons in the deuteron (or diproton) for the one nucleon exchange.}
\end{center}
\end{figure}

The deuteron charge exchange reactions discussed in this section so far are
\emph{soft} processes, where the momentum transfer between the initial
deuteron and the final diproton is very low. These are not to be confused
with the large momentum transfer $pd\to \{pp\}_{\!s}n$ reactions where, in
the c.m.\ frame, the diproton emerges in the backward direction with respect
to the incident deuteron. The kinematics are then very similar to those of
proton-deuteron backward elastic scattering. The unpolarized differential
cross section for \emph{hard} proton-deuteron breakup was measured at ANKE at
ten proton beam energies from 0.5~GeV to 2.0~GeV with the standard
$E_{pp}<3$~MeV cut ~\cite{KOM2003,DYM2010a} and the results are shown in
Fig.~\ref{fig:Uzikov}.

It should be noted that, throughout the energy range shown in
Fig.~\ref{fig:Uzikov}, the differential cross section for the backward $pd\to
\{pp\}_{\!s}n$ reaction is about two orders of magnitude less than that for
$pd \to dp$. However, it was suggested many years ago that backward
proton-deuteron elastic scattering at high energies may be driven by the
virtual excitation of the $\Delta(1232)$ isobar and estimates of its effects
were derived in a phenomenological model using a $\pi^+d\to pp$
input~\cite{CRA1969,KOL1973} and a similar approach was initiated for
diproton production~\cite{UZI2007}.

At low energies one would expect the reaction to be dominated by purely
nucleonic degrees of freedom, so that the main driving term is that of one
proton exchange, which depends on the wave functions of the deuteron and
diproton. In the calculations of Ref.~\cite{DYM2010a}, this predicts a node
for $T_p\approx 0.8$~GeV and in this region the $\Delta$ provides the main
contribution. Taking into account also small effects from impulse
approximation terms, the overall theoretical description given in
Fig.~\ref{fig:Uzikov} is reasonable and it reinforces the suggestion that
$\Delta$ degrees of freedom cannot be ignored in high momentum transfer
reactions even if they only involve initial and final nucleons.

In a further investigation, the proton analyzing power $A_y$ in the
$\pol{p}d\to \{pp\}_{\!s}n$ reaction was also measured near the backward
direction at 0.5~GeV and 0.8~GeV~\cite{YAS2005}. It is interesting to note
that at 0.8~GeV $A_y$ remains small over the measured angular range from
$167^{\circ}$ to $180^{\circ}$ whereas at 0.5~GeV it becomes very large below
$170^{\circ}$. This different behaviour might be linked to the dominance of
different driving terms seen in Fig.~\ref{fig:Uzikov}. However, since $A_y$
represents an interference between amplitudes, it is hard to draw firm
conclusions from such data.

%
%
\section{Single non-strange meson production in nucleon-nucleon collisions} 
\label{NNX}\setcounter{equation}{0}

The energy range of COSY is such that it is possible to produce non-strange
mesons in nucleon-nucleon collisions with masses up to that of the $\phi$.
This section will detail the COSY efforts in this field, though the $\phi$
itself will be considered in sect.~\ref{NNphi} since its detection is
intimately connected with kaon pair production.

\begin{figure}[htb]
\begin{center}
\includegraphics[width=1.0\columnwidth]{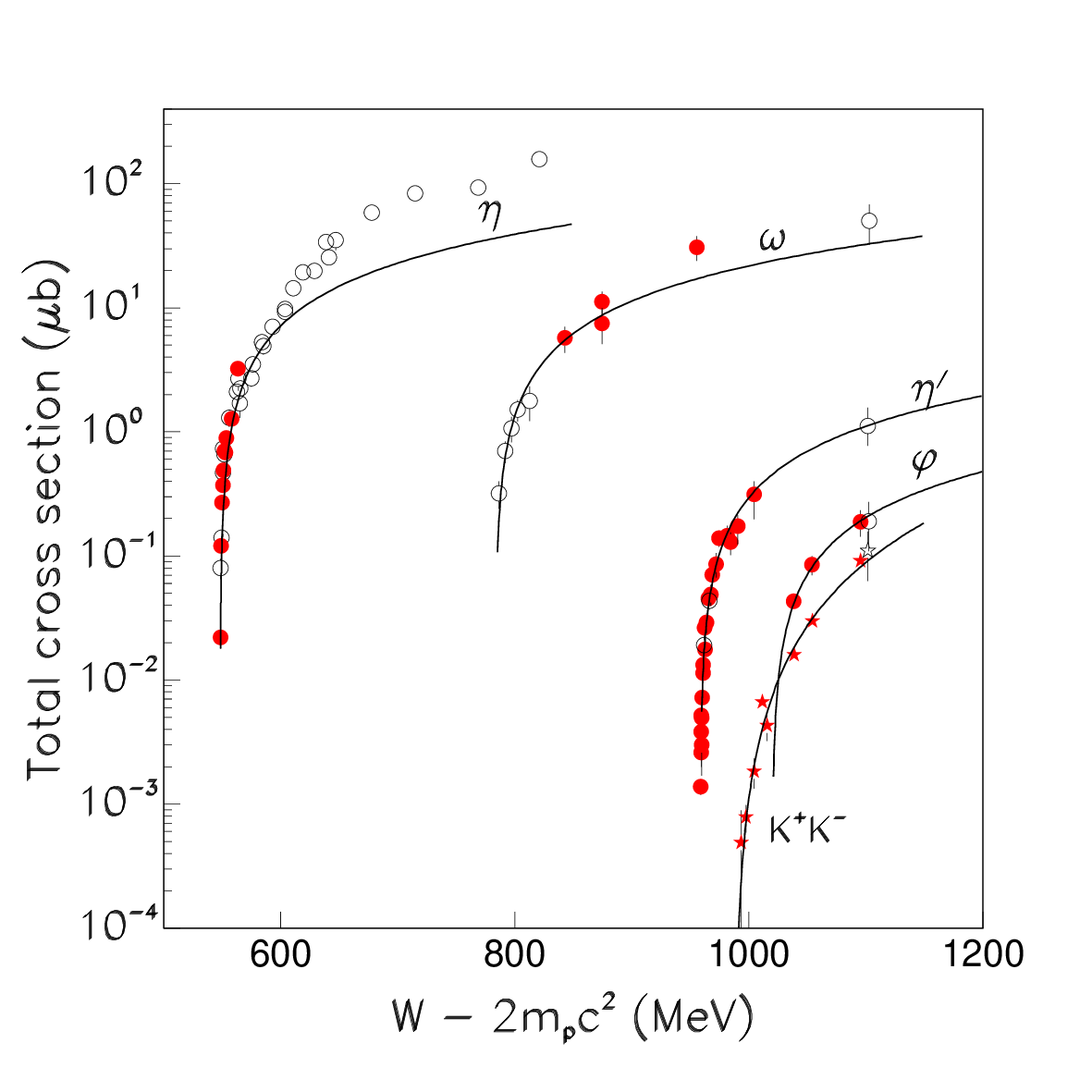}
\caption{\label{fig:Overview} Total cross sections near threshold for $pp\to
pp\eta$, $pp\to pp\omega$, $pp\to pp\eta^{\prime}$, $pp\to pp\phi$, and
$pp\to ppK^+K^-$ as functions of the total c.m.\ energy available for meson
production. The (red) circles and stars represent COSY measurements while the
open (black) symbols were obtained at other laboratories. The
phenomenological curves are discussed with the specific reactions.}
\end{center}
\end{figure}

However, before we go into details of specific reactions, we show in
Fig.~\ref{fig:Overview} the total cross sections for meson production near
threshold in proton-proton scattering in terms of the total c.m.\ energy $W$
minus twice the proton mass. This figure, which is an update of that
presented in a 2002 review~\cite{MOS2002}, shows the overwhelming
contribution that COSY has made in the field from the $\eta$ to kaon pair
production. It is extended even further in sect.~\ref{hyperon} to encompass
hyperon production, where the COSY data are completely dominant.

\subsection{Phenomenological description} 
\label{Phenomenology}

In a reaction such as $pn\to pn\eta^{\prime}$ near threshold, the interaction
of the $\eta^{\prime}$ with either of the two recoiling nucleons is quite
weak. In contrast there is a very strong Final State Interaction (FSI)
between the neutron and proton which might even lead to the formation of a
deuteron. Although this could be taken into account by constructing an $np$
final-state wave function, it is nevertheless useful to have simple
closed-form expressions that describe semi-quantitatively the main effects
observed in the production of both non-strange and strange mesons. This has
been achieved in a series of papers~\cite{FAL1996,BOU1996,FAL1997,FAL1997a}
and the principal results of the approach are outlined here before the
individual experiments are discussed.

The starting point is the observation that for a real local potential the
bound state wave function $u(r)$ and scattering state wave function $v(k,r)$
are intimately linked. If these are chosen to satisfy real boundary
conditions, then in the $S$-wave
\begin{equation}
\label{3}
v(k,r) \approx -\frac{1}{\sqrt{2\alpha(\alpha^2+k^2)}}\,u(r)\:.
\end{equation}
Here $k$ is the relative momentum in the scattering state and
$\alpha^2=m_{\rm red}B$, where $B$ is the binding energy of two particles
with reduced mass $m_{\rm red}$. The result is exact when extrapolated to the
pole at $k^2=-\alpha^2$, but it is generally a good first approximation at
small $r$ and $k$.

If the meson production operator is of short range, Eq.~(\ref{3}) shows that
the final $S$-wave triplet contribution to the differential cross section for
$pp \rightarrow np \pi^{+}$ should be related to that for $pp \rightarrow d
\pi^{+}$ through
\begin{eqnarray}
\nonumber
\frac{\dd^{2} \sigma}{\dd \Omega\, \dd x} (pp \rightarrow np \pi^{+})
\approx \frac{q(x)}{q(-1)} \frac{\sqrt{x}}{2 \pi (x+1)}
\frac{\dd \sigma}{\dd \Omega} (pp \rightarrow d \pi^{+})\:.\\
\label{5}
\end{eqnarray}%
The dimensionless variable $x$ is defined as $x = Q_{pn}/B_{I=0} =k^2/2m_{\rm
red}B_{I=0}$, where $q(x)$ and $q(-1)$ are the momenta of the pion in the
three- and two-body reactions, respectively. The excess energy $Q_{pn}$ is
the total c.m.\ energy in the $pn$ system minus the rest masses. In some of
the literature it is denoted by $\varepsilon$.

It is important to note that Eq.~(\ref{5}) only predicts the $S$-wave
spin-triplet $np$ production in a truly model-independent way as $x\to -1$.
Thus it breaks down at large $x$ when $P$ and higher waves become important
and the shape dependence of the $S$-wave is significant. It also assumes that
the distortion introduced by pion-nucleon scattering is similar for the two-
and three-body final states. More subtly, it ignores the coupling between the
$S$ and $D$ waves through the $np$ tensor force. These drawbacks will come to
the fore when discussing the Big Karl experiment in sect.~\ref{comparison}.

Equation~(\ref{5}) is useful in the description of the high energy tail of
inclusive pion production, i.e., small $x$, even away from threshold but it
can also be integrated analytically to predict the energy dependence of the
total cross section near threshold. If the deuteron production cross section
varies like phase space, $\sigma_d\approx D\sqrt{Q}$, then, using
non-relativistic integration, the three-body total cross section behaves as
\begin{equation}
\label{simple}
\sigma_{I=0}(Q) \approx \fmn{1}{4}D\sqrt{Q}\left(\frac{Q}{B_{I=0}}\right)^{\!\!3/2}\left(1+\sqrt{1+Q/B_{I=0}}\,\right)^{\!-2},
\end{equation}
where $Q$ is the excess energy for either the two- or three-body final state.
It is, however, important to repeat that this formula only corresponds to
$S$-wave spin-triplet $np$ final states. The formula can be extended to take
into account the more complex energy dependence seen in $pp\to d\pi^+$ near
threshold~\cite{FAL1997a}.

For the spin-singlet final states, which are relevant in the $pp\to pp\pi^0$
reaction, there is no bound state to which one can normalize the cross
section, i.e., there is no equivalent of Eq.~(\ref{simple}). Nevertheless
there is a pole corresponding to a virtual state very close to threshold with
a ``binding'' energy $B_{I=1}\approx 0.5$~MeV. We would then expect this
feature to dominate the $pp\to pp\pi^0$ reaction close to threshold to give
an energy dependence of the form
\begin{equation}
\label{simple2}
\sigma_{I=1}(Q) \approx C\left(\frac{Q}{B_{I=1}}\right)^{\!2}\left(1+\sqrt{1+Q/B_{I=1}}\,\right)^{\!-2},
\end{equation}
Equation~(\ref{simple2}) is often used to describe reactions such as $pp\to
K^+\Lambda p$ or $pp\to pp\eta^{\prime}$ but in the latter case one must
realize that there is some ambiguity in the value chosen for $B_{I=1}$
because the formula neglects the Coulomb repulsion between the two protons.

The approach can easily be extended to the case where the final state
interaction is described by the more complete Jost function
$(k-i\alpha)/(k+i\beta)$. The results of Eq.~(\ref{5}) should be multiplied
by $(\beta^2+k^2)/(\beta^2-\alpha^2)$ and an analytic formula has been
derived for the near-threshold energy dependence of the corresponding total
production cross section~\cite{SIB2006}.

%
%

\subsection[Hard bremsstrahlung in proton-proton scattering]{Hard bremsstrahlung in proton-proton scattering} 
\label{gamma}%

The photon is clearly not a meson, but it is nevertheless convenient to
consider its production here because the principle of measuring the $pp\to
pp\gamma$ reaction by detecting the two protons and reconstructing a
missing-mass peak is identical to that used extensively at COSY for several
mesons. Indeed the cross section for hard bremsstrahlung production is often
obtained as a by-product of a study of the $pp\to pp\pi^0$ reaction.

The threshold for $\pi^0$ production in $pp$ collisions is at about 280~MeV
and so, when the first $pp\to pp\gamma$ experiment was carried out at
COSY-TOF at 293~MeV~\cite{BIL1998a}, pion production was much reduced by the
proximity to threshold. Since no large-acceptance photon detector was then
available at COSY, the reaction was studied by detecting the two protons in
the most basic version of the COSY-TOF spectrometer. Peaks in the $pp$
missing-mass distribution were seen that corresponded to the production of
the $\pi^0$ and the $\gamma$. The latter suffered from a large random
background, whose shape was determined from empty-target measurements.

In most of the earlier hard bremsstrahlung experiments carried out at other
laboratories, the two protons were measured in pairs of counters placed on
either side of the beam line and, as a consequence, they had little or no
acceptance at small $pp$ excitation energy. A total cross section was
estimated by summing data taken at different angles. In contrast, a large
fraction of the $pp\to pp\gamma$ phase space was covered in a single setting
at COSY-TOF and this allowed a Dalitz plot to be constructed. For low $pp$
invariant masses, i.e., energetic photons, there was a coverage of over 95\%.
The coplanar photon angular distributions that could be extracted from the
COSY-TOF data were shown to be in good agreement with earlier results, though
there was an overall 20\% normalization uncertainty.

A notable feature of these data is that there was no obvious evidence for the
production of the \Szero\ enhancement of the $pp$ final state, which was seen
in the TOF data on pion production. This could of course be a statistical
fluctuation arising from the small number ($<1500$) of events, of which only
a tiny fraction would fall in the FSI region. Alternatively, it was argued
that the effect might be caused by the electromagnetic transition operator
coupling only weakly to a spin-singlet $pp$ state~\cite{BIL1998a}. This
latter possibility was excluded by a later experiment at CELSIUS, where a
large FSI enhancement was seen in the 58,000 $pp\to pp\gamma$ events measured
using a similar missing-mass technique~\cite{JOH2009,JOH2011}. The beam
energy of 310~MeV was only slightly higher than that used at COSY-TOF and the
windowless target in the CELSIUS experiment meant that the random background
was almost non-existent. However, it must be noted that the CELSIUS data at
small $pp$ excitation energy are strongly forward peaked~\cite{JOH2009} and
this might affect the measured COSY-TOF statistics~\cite{BIL1998a}.

\begin{figure}[h!]
\begin{center}
\vspace{5mm}
\includegraphics[width=0.7\columnwidth]{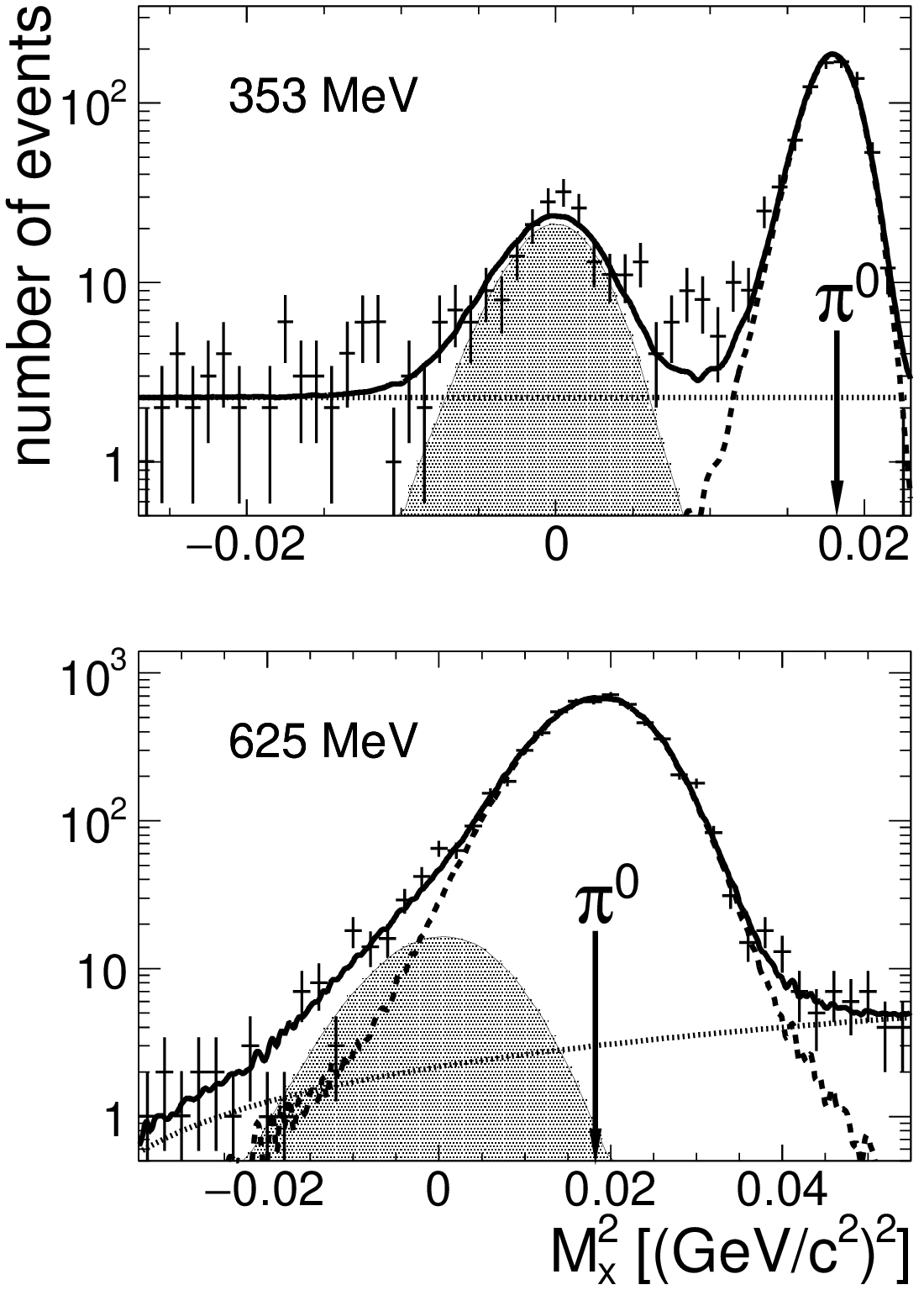}
\caption{\label{fig:vik} Distributions in the square of the missing mass in
the $pp\to \{pp\}_{\!s}X$ reaction at beam energies of 353 and 625~MeV for
proton pairs with $\theta_{pp} < 20^{\circ}$~\cite{KOM2008,TSI2010}. The
expected $\pi^0$ position is indicated by the arrow. In the fits, the shaded
area corresponds to the $\gamma$ peak, the dashed line to the $\pi^0$ peak,
the dotted to the linear accidental background, and the solid to the sum of
these three contributions. }
\end{center}
\end{figure}

In complete contrast to the large geometric acceptance of the COSY-TOF
detector, the ANKE facility can only measure fast protons emerging at small
angles with respect to the beam direction. This means that the acceptance for
two protons from a $pp\to ppX$ reaction is maximal when these have similar
momentum vectors, i.e., the excitation energy $E_{pp}$ in the final $pp$ rest
frame is small. In such cases the Pauli principle requires the diproton to be
in the \Szero\ configuration. Taking a cut with $E_{pp}<3$~MeV, the group
made small angle measurements of $pp\to \{pp\}_{\!s}\gamma$ at six energies
between 353~MeV and 800~MeV~\cite{KOM2008,TSI2010}. Such small angle studies
were possible at ANKE because of the absence of a beam-pipe hole.

Missing-mass distributions from the ANKE experiment are shown for two beam
energies in Fig.~\ref{fig:vik}~\cite{KOM2008,TSI2010}. Although there is no
real difficulty in identifying the $\gamma$ peak at 353~MeV, the same is not
true at 625~MeV where, for kinematic reasons, the $\pi^0$ peak is
considerably wider. The bremsstrahlung reaction could then only be isolated
through a careful fitting process.

\begin{figure}[h!]
\begin{center}
\includegraphics[width=0.9\columnwidth]{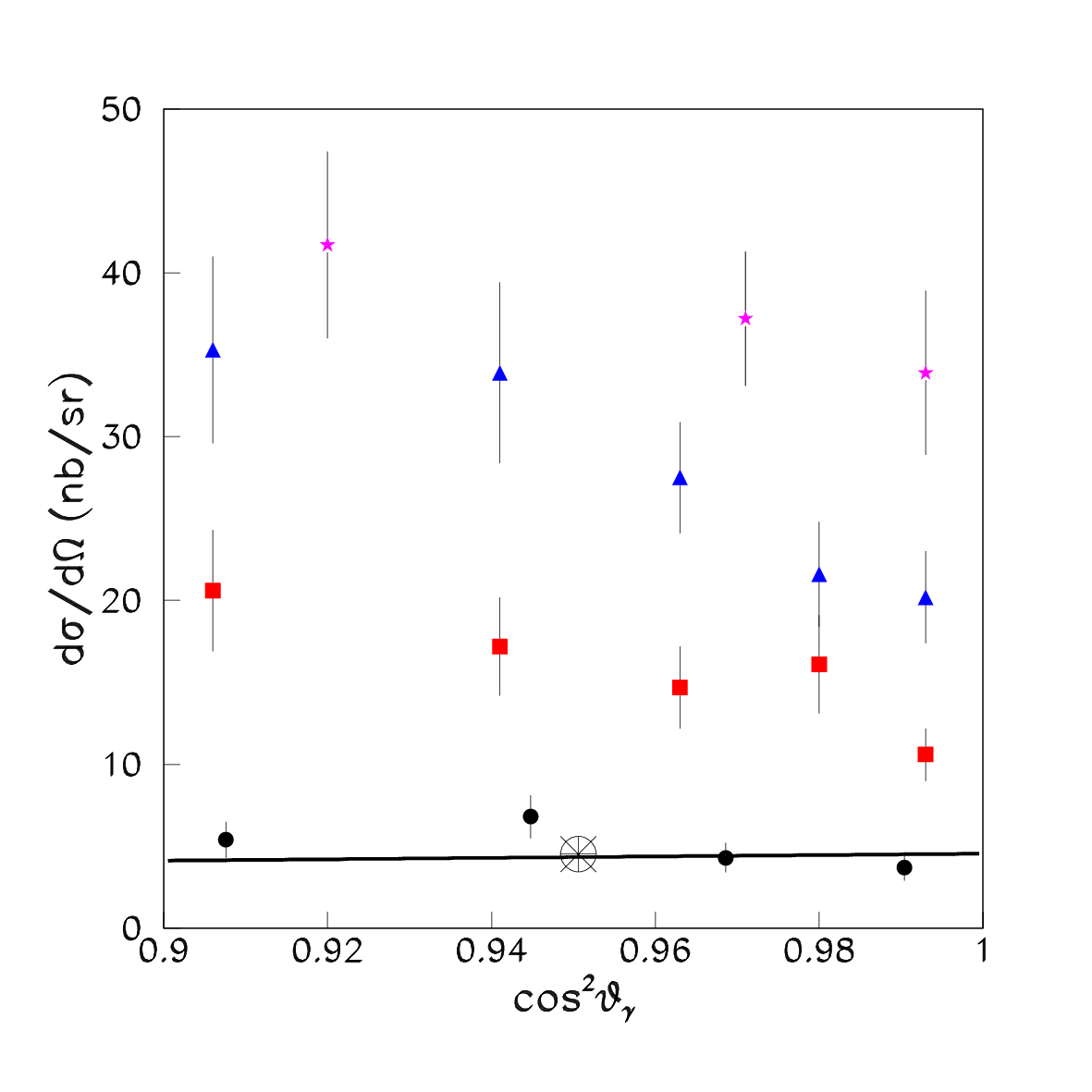}
\caption{\label{fig:vik2} Angular dependence of the differential cross
section for the $pp\to\{pp\}_{\!s}\gamma$ reaction at four beam energies
measured at ANKE~\cite{KOM2008,TSI2010}. Experimental results are shown at
353~MeV by (black) circles, at 500~MeV by (red) squares, at 550~MeV by (blue)
triangles, and at 700~MeV by (magenta) stars. Also shown by the (black)
circled cross is the one point in this angular domain measured at CELSIUS at
310~MeV~\cite{JOH2009}; the straight line is the fit to all the 35 CELSIUS
points .}
\end{center}
\end{figure}

The luminosity $\mathcal{L}$ in the ANKE experiments was determined from the
number of elastically scattered protons detected in parallel and these led to
normalization uncertainties that varied between about 3\% and 5\%, depending
upon the energy. In the c.m.\ frame the $\gamma$ polar angle
$\theta_{\gamma}$ is $180^{\circ}$ minus the diproton angle and in
Fig.~\ref{fig:vik2} the ANKE data at four energies are shown in terms of
$\cos^2\theta_{\gamma}$. The CELSIUS 310~MeV data have also been evaluated
for the $E_{pp}<3$~MeV cut~\cite{JOH2009} and the one point that falls within
the ANKE domain is also shown. At 310~MeV it might be reasonable to keep just
the lowest multipoles, viz.\ $E1$ and $M2$, and these would lead to a linear
dependence of the cross section on $\cos^2\theta_{\gamma}$. The fit to the
CELSIUS data, $0.27+4.27\cos^2\theta_{\gamma}$, which is also plotted,
corresponds to a forward enhancement, whereas the ANKE higher energy data
show evidence for some suppression for small $\theta_{\gamma}$.

The integral of the ANKE cross section for $0^{\circ} < \theta_{\gamma} <
20^{\circ}$ is maximal for a beam energy at around 650~MeV and the
authors~\cite{TSI2010} argue that this might be associated with a
$\Delta(1232)N$ intermediate state in a relative $P$-wave, the $S$-wave being
forbidden by selection rules.

\begin{figure}[h!]
\begin{center}
\includegraphics[width=0.9\columnwidth]{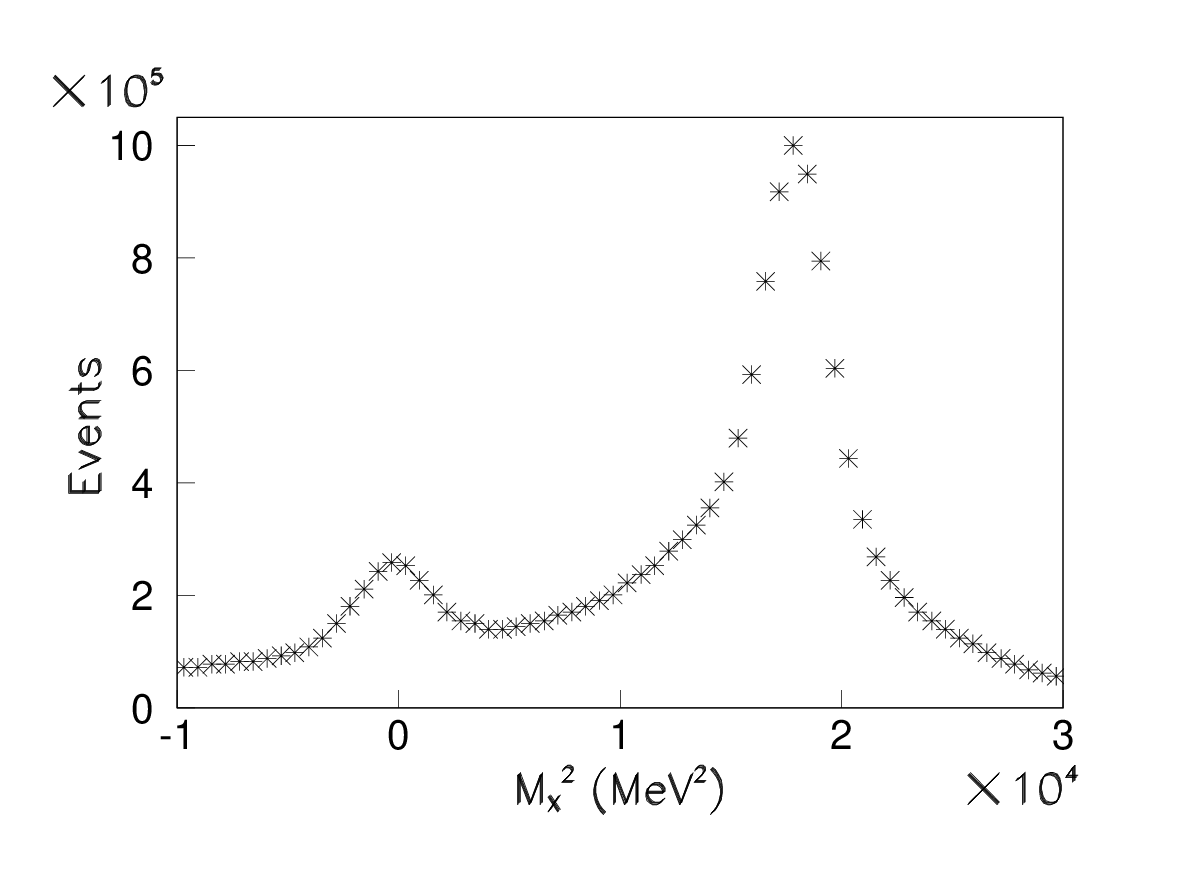}
\caption{\label{fig:Jozef2} Missing-mass-squared distribution for the two
protons from the $pp\to pp\gamma$ measured with the COSY-WASA detector at
550~MeV~\cite{ZLO2013}. These preliminary results were obtained by measuring
the photon in coincidence and putting a cut on its c.m.\ energy relative to
that of the two protons.}
\end{center}
\end{figure}

It is clear from looking at Fig.~\ref{fig:vik} that the identification of the
$pp\to pp\gamma$ reaction through the missing-mass peak becomes more
difficult as the beam energy is raised. This problem can be circumvented by
measuring directly the photon as well as the two protons in the WASA
detector. In the preliminary missing-mass spectrum taken at a beam energy of
550~MeV and shown in Fig.~\ref{fig:Jozef2} there is a peak containing about
$1.3\times 10^6$ events at zero missing mass sitting on a smooth background
associated mainly with photons that come from $\pi^0$ decay~\cite{ZLO2013}.
Here both protons were detected in the WASA Forward Detector and about the
same statistics are available where one of the protons was measured in the
Central Detector, though it must be realized that the resolution on the
momentum of this proton is poorer than that of the one entering the Forward.
The only cut applied to these data is that the missing energy of the two
protons in the c.m.\ frame should be at least 100~MeV bigger than the photon
energy. This criterion, which does not eliminate good $pp\to pp\gamma$
events, is responsible for producing the clear bremsstrahlung peak in
Fig.~\ref{fig:Jozef2} at 550~MeV despite there being little sign of it in
Fig.~\ref{fig:vik} at 625~MeV.

Though the energy cut is useful way of obtaining a reliable estimate on the
number of bremsstrahlung events, a much more robust way of eliminating the
background comes from noting that, in the c.m.\ frame, a single photon should
move in the opposite direction to the proton pair. Thus, in a two-dimensional
plot of $\theta_{pp}+\theta_{\gamma}$ versus $|\phi_{pp}-\phi_{\gamma}|$
there is a clear island of $pp\to pp\gamma$ events centred at
$180^{\circ}\times 180^{\circ}$ which extends only a few degrees in either
direction.

This data set probably represents the largest collection of clean $pp\to
pp\gamma$ events ever obtained above the pion production threshold but
results are not yet available on the differential cross
sections~\cite{ZLO2013}.

\subsection[Single pion production in nucleon-nucleon collisions]{Single pion production in nucleon-nucleon collisions} 
\label{pion}

At the start up of most new accelerators some of the first experiments that
are performed involve pion production from nuclear targets. The general aim
of such tests is to find optimal conditions for the creation of pion beams
but, since at COSY there were no plans to use secondary pion beams, all
research in this area was focussed on the understanding of the underlying
reaction mechanisms. A useful description of meson production more generally
is to be found in Ref.~\cite{HAN2004}.

It has already been stressed that COSY-TOF has the big advantage of a large
geometric acceptance which is so important when measuring reactions with
three or more particles in the final state far away from threshold. The only
other spectrometer at COSY with large acceptance is WASA, but this has a hole
in the detector to allow the unscattered beam to emerge. This subtends an
angle of about $3^{\circ}$ in the laboratory and, if either proton from say a
$pp\to ppX$ reaction is produced within this cone, the event is lost. This
becomes more problematic at low energies where, as threshold is approached,
more and more events fall into the beam pipe trap. Since there is a strong
proton-proton final state interaction, it has been argued~\cite{KUH2003} that
results of the near-threshold missing-mass measurements carried out at
IUCF~\cite{MEY1992} and the PROMICE-WASA facility at CELSIUS~\cite{BIL2001}
have significant model dependence caused, in part, by uncertainties in the
Monte Carlo estimation of the acceptance.

A $pp\to pp\pi^0$ experiment was carried out at three energies very close to
threshold by detecting the two final protons at COSY-TOF~\cite{KUH2003}. The
luminosity was established by measuring in parallel elastic proton-proton
scattering for which the uncertainty in the differential cross section is
less than 5\%. Since the cross section for pion production varies very fast
with energy in the near-threshold region, it was equally important to
establish this energy to high precision. This was achieved by measuring both
final particles in the two-body $pp\to d\pi^+$ reaction, which determined the
proton beam energy to 0.3~MeV.

The measured total cross sections for excess energies between about 6 and
9~MeV seem to be about 50\% higher than those found at IUCF~\cite{MEY1992}
and CELSIUS~\cite{BIL2001} and the COSY-TOF authors speculated that this
might be associated with the beam-pipe problem in these two experiments. This
could have significant implications for the value of the $s$-wave $\pi^0$
production amplitudes but a greater energy range would be needed to confirm
this. Such a programme was indeed carried out for $\eta$ and $\eta^{\prime}$
production at COSY11, and this will be discussed in sect.~\ref{eta}.

Pion production in unpolarized proton-proton collisions was also studied well
away from threshold at a beam energy of 397~MeV~\cite{SAM2006a,SAM2009}.
Since the momenta of all three particles in the $pp\to pp\pi^0$ reaction were
measured or reconstructed over a very large fraction of phase space, an
intrinsic problem was choosing which variables to use in the presentation of
the data. One that is of great interest to other experiments at COSY and
CELSIUS is that of the angular distribution of the pion in bins of the
excitation energy $E_{pp}$ in the final two-proton system.

If the data in the c.m.\ frame are parameterized in the form
$\dd\sigma/\dd\Omega \sim 1+b\cos^2\theta_{\pi}$, the COSY-TOF data yield
$b=-1.00\pm0.02$ for $E_{pp}<3$~MeV and $b=-0.17\pm0.01$ for the whole data
sample. The corresponding numbers obtained in the PROMICE-WASA (CELSIUS)
experiment at 400~MeV~\cite{BIL2001} are $b=-0.58\pm0.03$ and
$b=+0.19\pm0.01$, respectively. There are therefore unresolved systematic
differences that are much larger than the statistical uncertainties. As
discussed in sect.~\ref{PWA_pi}, the ANKE data at 353~MeV~\cite{TSI2012} seem
to be consistent with the PROMICE-WASA results at 360~MeV~\cite{BIL2001}.

Since the two incident protons are identical, the pion c.m.\ angular
distribution must be symmetric about $90^{\circ}$ and this is one test of
whether systematic effects are under control. Although this was successfully
passed for the $\pi^0$ at COSY-TOF~\cite{SAM2006a}, the group only measured
$\pi^+$ data in one hemisphere. There were also clearly problems in one
hemisphere of the CELSIUS experiment when the two photons from the $\pi^0$
decay were detected in coincidence with the two protons~\cite{THO2007}.

\begin{figure}[h!]
\begin{center}
\includegraphics[width=1.0\columnwidth]{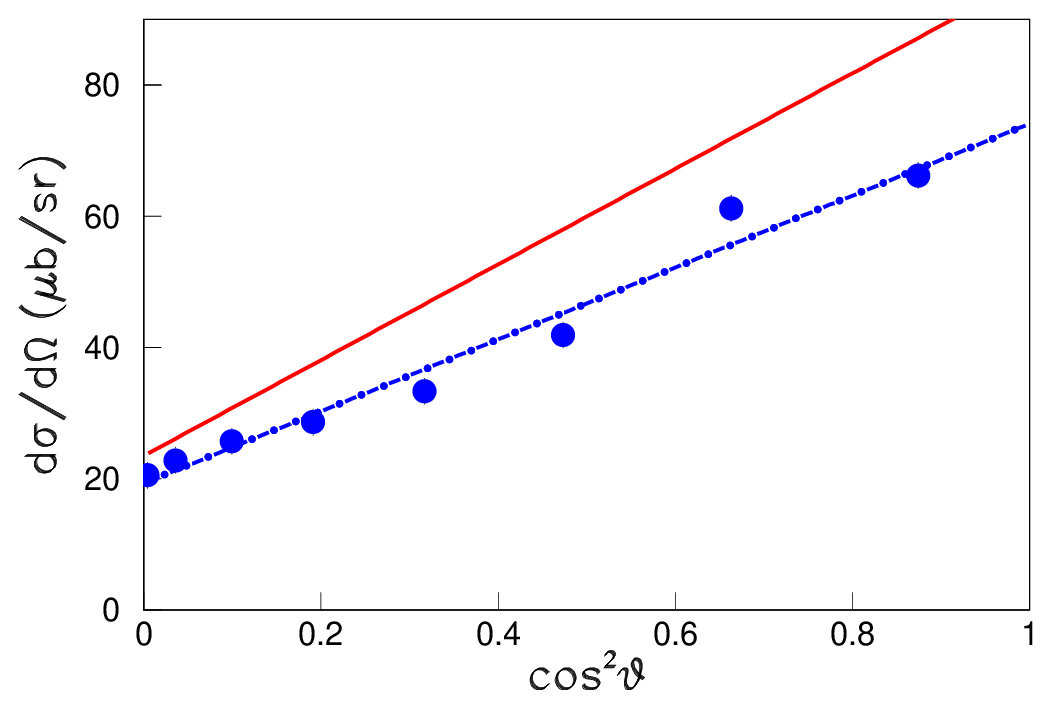}
\caption{\label{fig:hc1} Differential cross section for the $pp\to pn\pi^+$
reaction at 397~MeV in terms of the pion angle $\theta$~\cite{SAM2009}. The
best straight line fit (dashed blue) gives an integrated cross section of
$(470\pm 20)~\mu$b whereas the FSI model of Ref.~\cite{FAL1997a} (solid red)
predicts about 600~$\mu$b.}
\end{center}
\end{figure}

Though the experimental procedures were very similar for the $pp\to pn\pi^+$
reaction measured in parallel at 397~MeV~\cite{SAM2009}, the results were
significantly different because the data were dominated by the onset of the
$J^P=2^+$ $\Delta(1232)N$ intermediate state. A cut had to be imposed to stop
leakage from the $pp\to d\pi^+$ reaction, where the deuteron broke up in a
secondary reaction. The strong angular dependence, where the pion is produced
preferentially along the beam direction, which is illustrated in
Fig.~\ref{fig:hc1}, is very similar to that of $pp\to d\pi^+$. The measured
slope parameter is $b=2.8\pm0.1$ compared to $b \approx 3.6$ for the $pp\to
d\pi^+$ reaction~\cite{ARN1993}.

Of course, in the case of the three-body final state there are also $S$-wave
spin-singlet $pn$ contributions and these have been estimated from the $pp\to
pp\pi^0$ data in the phenomenological model presented in
sect.~\ref{Phenomenology}~\cite{FAL1997a}, whose results are also shown in
Fig.~\ref{fig:hc1}. In this approach it is assumed that the $\pi^+d$ final
state is produced by a fusion of the spin-triplet proton and neutron pair in
the three-body $\pi^+pn$ channel in a final state interaction. Starting from
experimental data on $pp\to d\pi^+$, an evaluation of the final state
interaction model at 397~MeV gives $b\approx 3.1$, which is very close to the
COSY-TOF value~\cite{SAM2009}.

The FSI model also predicts~\cite{FAL1997a} a $pp\to pn\pi^+$ total cross
section of about 600~$\mu$b at 397~MeV compared to an experimental value of
$(470\pm 20)~\mu$b~\cite{SAM2009}, where the error bar is the authors'
estimate of the systematic uncertainty. Given the uncertainties in both the
measurement and the calculation, the experimental result could be consistent
with the theoretical estimate. This reinforces the COSY-TOF conclusion that
it is the same intermediate state that governs both the $d\pi^+$ and
$pn\pi^+$ channels. While accepting this, it may be a step too far to agree
with the authors' suggestion that both production are associated with an
isovector $2^+$ dibaryon~\cite{SAM2009}!

To obtain information on pion production in the isospin-zero channel requires
data from neutron-proton collisions and, in the absence of neutron beams at
COSY, this necessarily involves the use of a deuteron beam or target. The
former is clearly more suitable for COSY-TOF because the \emph{spectator}
proton is then fast and all four charged particles from $dp\to ppp\pi^-$ can
be measured. In an initial experiment~\cite{ABD2006} the group showed that it
was possible to isolate the spectator proton ($p_{\rm sp}$) so that the
resulting data could be interpreted in terms of quasi-free $np\to pp\pi^-$
production.

Due to the Fermi motion of the neutron inside the deuteron, a measurement of
$dp\to p_{\rm sp}pp\pi^-$ provides a scan of $np\to pp\pi^-$ over a range of
excess energies. This was exploited in the second COSY-TOF
experiment~\cite{ABD2008a}, which was carried out at a beam energy of
759~MeV. The value of the excess energy $Q$ depended primarily on the
momentum vector of the spectator proton and the overall energy resolution was
estimated to be about 8~MeV. The data were therefore put into six bins in $Q$
between 0 and 90~MeV.

\begin{figure}[h!]
\begin{center}
\includegraphics[width=1.0\columnwidth]{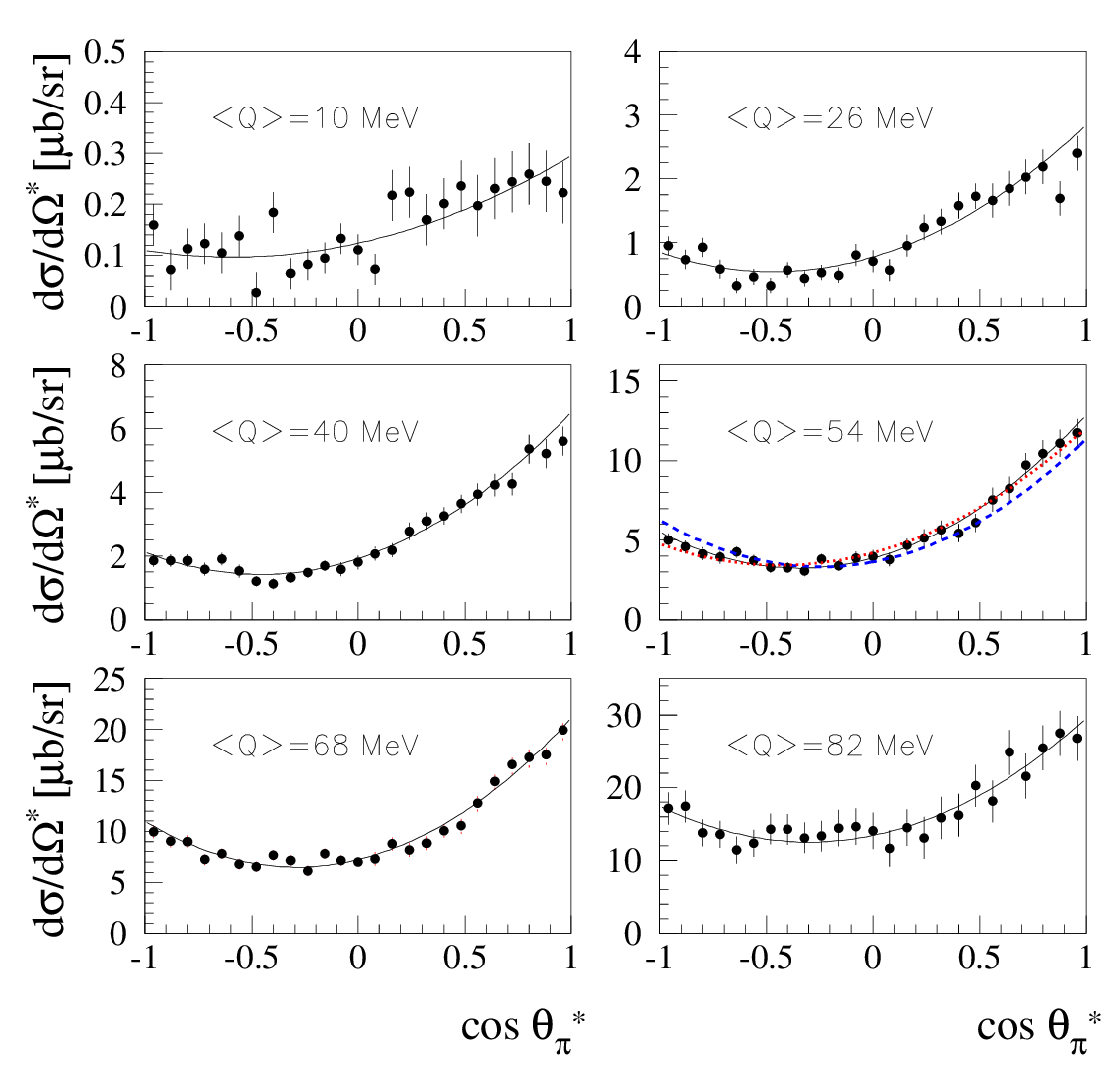}
\caption{\label{fig:Eberhard7} Angular distributions of the $\pi^-$ from the
quasi-free $np\to pp\pi^-$ reaction with respect to the neutron direction for
selected mean excess energies $\langle Q \rangle$. The COSY-TOF
data~\cite{ABD2008a} are shown together with quadratic fits in
$\cos\theta_{\pi}$. The data are normalized at $\langle Q \rangle =54$~MeV to
the PSI results~\cite{DAU2002}, shown by the dashed (blue) curve. The earlier
results of Handler~\cite{HAN1965} at this energy are parameterized by the
dotted (red) curve.}
\end{center}
\end{figure}

The pion angular distributions for the six mean values of $Q$ are shown in
Fig.~\ref{fig:Eberhard7}. It should be noted that the angle is here defined
with respect to the  direction of the incident (virtual) neutron and not the
proton that was used for the ANKE data~\cite{DYM2013} shown in
Fig.~\ref{fig:DYM}, where only events with the final $pp$ excitation energy
$E_{pp}<3$~MeV were considered. After taking this angle definition into
account, one sees some similarities between the COSY-TOF and ANKE data.

It is seen from Fig.~\ref{fig:Eberhard7} that, even at the highest excess
energy, the data can be represented by a quadratic in $\cos\theta_{\pi}$,
which is a reflection of the dominance of low partial waves in this reaction.
No attempt was made to determine the absolute luminosity and the data were
normalized to PSI measurements with free neutrons~\cite{DAU2002} that are
shown in the figure at 54~MeV.

Though the COSY-TOF data show evidence for a $pp$ FSI, the data do not allow
the angular distributions to be displayed in bins of $pp$ excess energy. A
simple estimate of the total cross section for $E_{pp}<3$~MeV from the $pp$
effective mass distributions of Ref.~\cite{ABD2008a} would suggest a somewhat
lower value than the ANKE result~\cite{DYM2013} discussed in
sect.~\ref{PWA_pi}, but such an evaluation is very crude.

\subsubsection{The $pp\to d\pi^+$ reaction} 
\label{dpiplus} 

Over the last sixty years there have been countless measurements of the
$pp\to d\pi^+$ or the inverse $\pi^+d\to pp$ reaction. One therefore has to
wonder if it is possible for COSY to add useful information in this field.
Nevertheless, there are two experiments that are worthy of note.

The GEM collaboration measured the $pp\to d\pi^+$ differential cross section
with the Big Karl spectrometer at five c.m.\ energies up to
3.6~MeV~\cite{DRO1996,DRO1998}. They checked isospin invariance by comparing
the integrated Coulomb-corrected cross sections with those obtained for
$np\to d\pi^0$. This is clearly not without problems because, unless one has
a good reaction model, Coulomb corrections are somewhat ambiguous and pion
mass differences are significant. Furthermore, absolute cross sections always
present a challenge with neutron beams.

\begin{figure}[h!]
\begin{center}
\includegraphics[width=0.8\columnwidth]{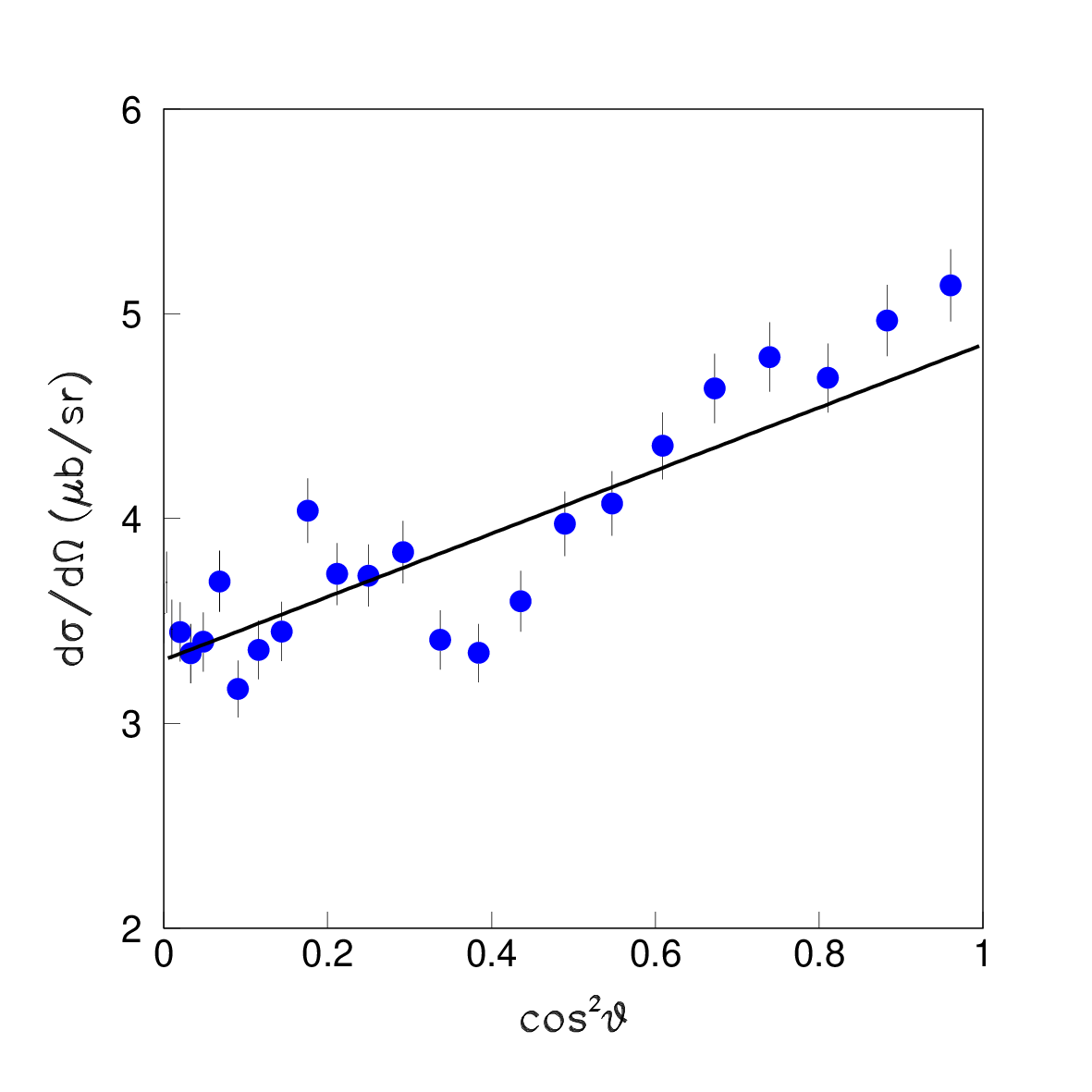}
\caption{\label{fig:GEM1} The differential cross section for the $pp\to
d\pi^+$ reaction 3.6~MeV above threshold in the c.m.\
frame~\cite{DRO1996,DRO1998}. The curve is the group's straight line fit to
the data in terms of $\cos^2\theta$.}
\end{center}
\end{figure}

Far less contentious were the group's measurements of the angular
distributions, an example of which is shown in Fig.~\ref{fig:GEM1}. Due to
having identical protons in the initial channel, the cross section is
symmetric about $90^{\circ}$ in the c.m.\ frame, so that it is a function of
$\cos^2\theta$. It also means that there can be no interference between even
and odd pion waves so that the first deviations from isotropy must arise from
either $s$-$d$ interference or the squares of $p$-wave amplitudes. Since
$s$-wave pion production is expected to be generally weak, the second of
these options would seem to be the more likely.

The data in Fig.~\ref{fig:GEM1} are consistent with a linear behaviour,
$\dd\sigma/\dd\Omega = a_0+a_2\cos^2\theta$ and the same is true for the GEM
data at lower energies. At 3.6~MeV the parameters are
$a_0=(3.31\pm0.59)~\mu$b/sr and $a_2=(1.54\pm0.33)~\mu$b/sr so that, even
very close to threshold, the differential cross section displays significant
anisotropy. The near-threshold GEM data are consistent with $a_2/a_0 \approx
11\,\eta^2$, where $\eta$ is the pion c.m.\ momentum in units of the pion
mass. The ratio is, of course, independent of the uncertainties in the
absolute normalization, and may provide a more robust method to check charge
independence. Some evidence of isospin breaking in the pion $p$-waves was
shown many years ago in measurements of $\pi^{\pm}p$ and $\pi^{\pm}d$ total
cross sections~\cite{PED1978} and this might be relevant for the value of
$a_2$.

\begin{figure}[h!]
\begin{center}
\includegraphics[width=0.9\columnwidth]{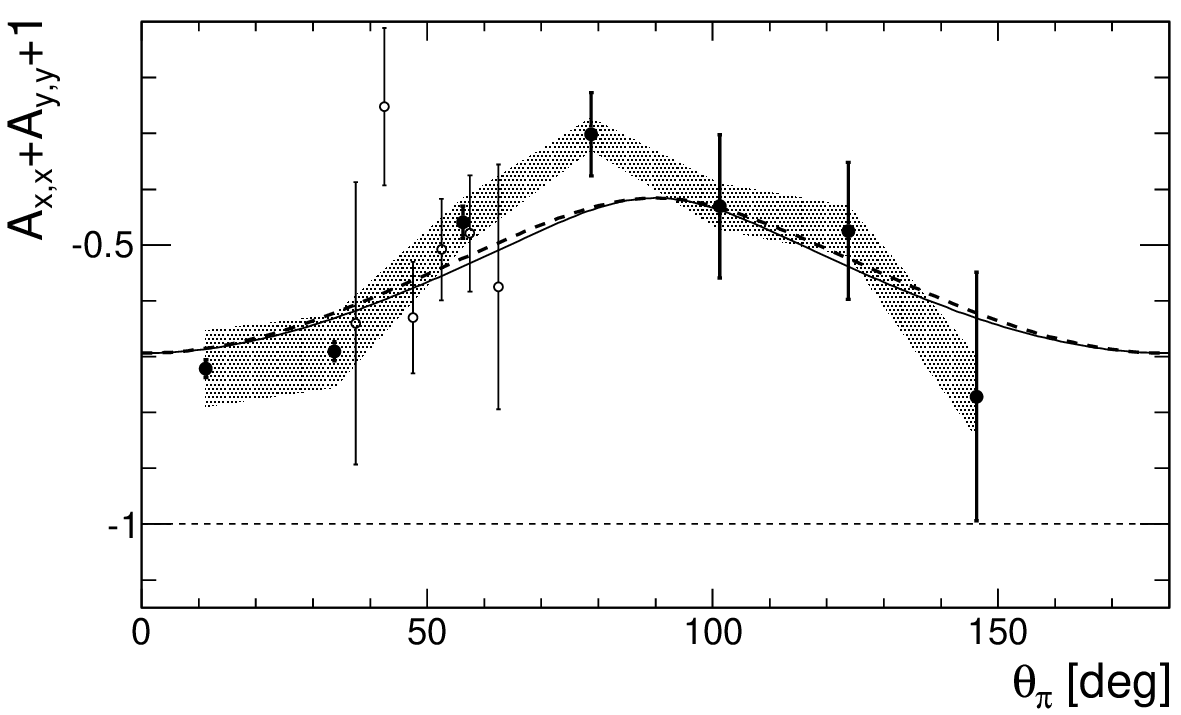}
\caption{\label{fig:Azz353} The combination $1+A_{x,x}+A_{y,y}$ measured in
the $\pol{n}\pol{p}\to d\pi^0$ reaction at 353~MeV~\cite{SHM2013} as a
function of the pion c.m.\ angle $\theta_{\pi}$ (closed circles) compared with the
SAID $\pol{p}\pol{p}\to d\pi^+$ predictions (dashed curve)~\cite{ARN1993}.
Statistical uncertainties are shown with error bars; systematic uncertainties
are illustrated with shaded bands.  Also presented are IUCF data for
$A_{z,z}$ taken at 350~MeV for the $\pol{p}\pol{p}\to d\pi^+$ reaction (open
circles)~\cite{PRZ2000} and the SAID prediction for this observable (solid
curve).}
\end{center}
\end{figure}

Since there is only one isospin amplitude, all the spin observables in the
$pp\to d\pi^+$ and $np\to d\pi^0$ reactions should be identical. As
by-products of other studies, there were measurements at ANKE of the spin
correlations $A_{x,x}$ and $A_{y,y}$ in the quasi-free $np\to d\pi^0$
reaction at 353 and 600~MeV~\cite{SHM2013}. Figure~\ref{fig:Azz353} shows the
ANKE measurements of the combination $1+A_{x,x}+A_{y,y}$ at 353~MeV, together
with the SAID prediction~\cite{ARN1993} of this observable. If one neglects
pion $d$-waves, which the SAID analysis suggests is a very good approximation
at 353~MeV, this observable is identical to the longitudinal spin-correlation
parameter $A_{z,z}$, for which there are some measurements from
IUCF~\cite{PRZ2000}. Though these have large error bars, they are not
incompatible with the ANKE results. All the ANKE data~\cite{SHM2013} are
consistent with the current SAID $pp\to d\pi^+$ solution and no sign was
found for any breaking of isospin invariance.

\subsubsection{Partial wave analysis of the $NN\to \{pp\}_{\!s}\pi$ reaction} 
\label{PWA_pi} 

There were pioneering measurements of the $pp\to \{pp\}_{\!s}\pi^0$
differential cross section carried out at the CELSIUS storage ring at a
series of beam energies from close to threshold up to 425~MeV~\cite{BIL2001}.
Here the $\{pp\}_{\!s}$ denotes a proton-proton pair with low excitation
energy $E_{pp}$ such that the Pauli principle forces them to be in the
$^{1\!}S_0$ state. The CELSIUS group chose to impose the cut $E_{pp}<3$~MeV
and, even if this is a little arbitrary, it became the standard for most
subsequent experiments carried out at ANKE.

In some ways the $pp\to \{pp\}_{\!s}\pi^0$ reaction is an ideal one to study
at COSY because the $^{1\!}S_0$ diproton configuration means that there are
no spin degrees of freedom in the final state that have to be determined
through a double-scattering measurement. The differential cross section was
measured at 800~MeV at ANKE~\cite{DYM2006} by detecting both protons in the
Forward Detector, with the $\pi^0$ being recognized from the peak in the
missing-mass spectrum. However, this restricted the angular coverage to
diproton c.m.\ angles less than about $15.4^{\circ}$. The measurements were
later extended to cover the energy range from about 500~MeV to 2.0~GeV, but
always in a similar small angle region~\cite{KUR2008}.

The quasi-two-body $pp\to \{pp\}_{\!s}\pi^0$ reaction is kinematically very
similar to that of $pp\to d\pi^+$, though it must be noted that an
intermediate $S$-wave $\Delta(1232)N$ state, which is so important for
$\pi^+$ production, is not allowed in the $\pi^0$ case. As a consequence, it
is expected that the maximum cross section would occur at a somewhat higher
energy, and that was precisely what was seen in the ANKE data~\cite{KUR2008}.
Furthermore, the ratio of $\pi^0$ production to that of $\pi^+$ through these
(quasi) two-body processes increases with energy, albeit from a very low
base. However, the only \emph{ab initio} theoretical estimation of $\pi^0$
production with a low $E_{pp}$ cut~\cite{NIS2006} fails completely to
describe the experimental data. It should be noted though that there can be
delicate cancelations between different partial waves. This was shown more
clearly in an amplitude analysis that took into account new ANKE measurements
of the analyzing power. This provided evidence for resonance structure in
both $s$- and $d$-wave $\pi^0$ production~\cite{KOM2016}.

The most ambitious programme of measurements was carried out at 353~MeV per
nucleon, where the aim was to perform a full spin-isospin analysis of pion
production leading to the $^{1\!}S_0$ diproton state. It was then hoped that
this would lead to the isolation of a term that was relevant for chiral
perturbation theory~\cite{BAR2014}. Such experiments are only possible at
ANKE close to threshold where the spectrometer acceptance covers a large
fraction of the solid angle. The programme involved the measurement of the
differential cross section and analyzing power in $pp\to
\{pp\}_{\!s}\pi^0$~\cite{TSI2012}, the differential cross section and
analyzing power in quasi-free $\pol{p}n\to \{pp\}_{\!s}\pi^-$ with a
deuterium target~\cite{DYM2012}, and the analyzing powers and
spin-correlations in quasi-free $\pol{n}\pol{p}\to \{pp\}_{\!s}\pi^-$ with a
polarized deuteron beam and a polarized hydrogen cell target~\cite{DYM2013}.

\begin{figure}[h!]
\begin{center}
\includegraphics[width=0.8\columnwidth]{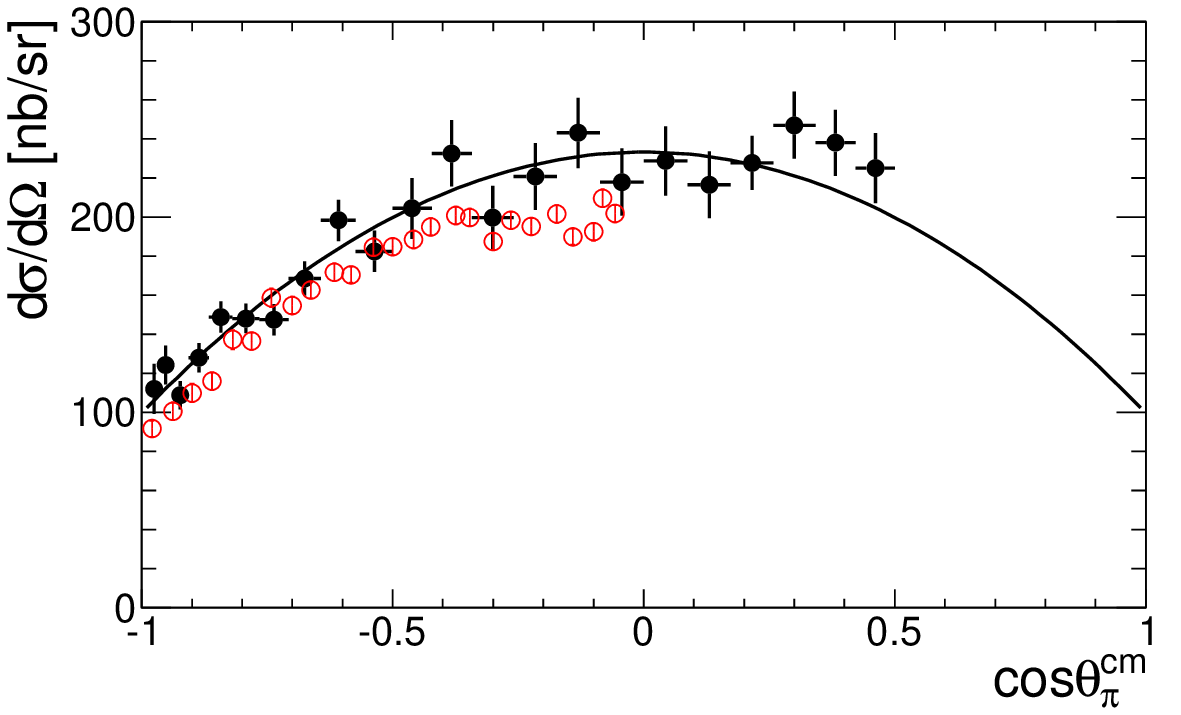}
\includegraphics[width=0.8\columnwidth]{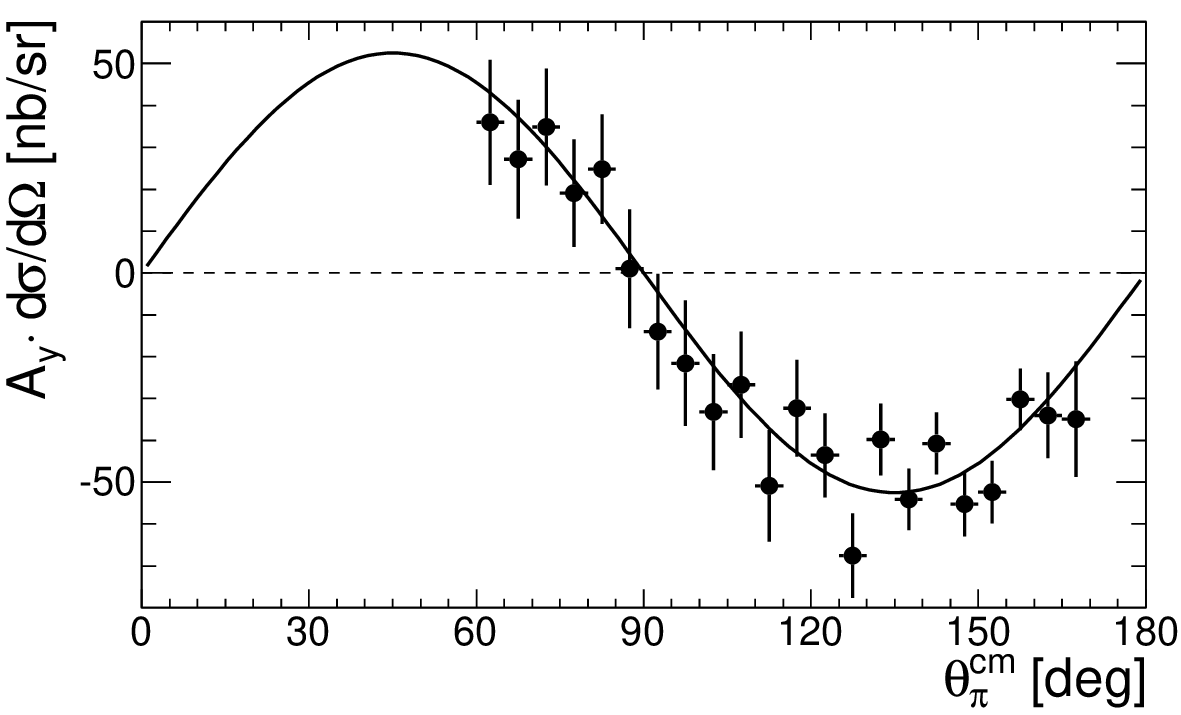}
\caption{\label{fig:TSI} Upper panel: Differential cross section for the
$pp\to \{pp\}_{\!s}\pi^0$ reaction at 353~MeV as a function of the cosine of
the pion c.m.\ angle. The solid (black) circles represent the ANKE
measurements~\cite{TSI2012} whereas the open (red) circles are CELSIUS data
obtained at 360~MeV~\cite{BIL2001}. The curve is a linear fit in
$\cos^2\theta_{\pi}$ to the ANKE data. Lower panel: The corresponding product
of the measured analyzing power and differential cross section for the
$\pol{p}p\to \{pp\}_{\!s}\pi^0$ reaction~\cite{TSI2012}. The curve is of the
form $b\sin2\theta$, where the parameter $b$ is fitted to the data.}
\end{center}
\end{figure}

The $pp\to \{pp\}_{\!s}\pi^0$ study was carried out in a similar manner to
the earlier $\pi^0$ experiments at ANKE~\cite{DYM2006,KUR2008}, but it is
important to realize that the differential cross section is symmetric about
$90^{\circ}$ and the analyzing power is antisymmetric. This feature provides
a useful extension to the limited ANKE acceptance. So close to threshold one
would expect only low partial waves to contribute and the ANKE data shown in
Fig.~\ref{fig:TSI} are consistent with the behaviour $\dd\sigma/\dd\Omega =
a_0+a_2\cos^2\theta$ and $A_y\dd\sigma/\dd\Omega =b_1\sin2\theta$. Even if
one only considers terms up to pion $d$-waves, there are three complex
amplitudes that can contribute and the information from the three real
parameters must be supplemented by other constraints in order to achieve a
full amplitude decomposition.

Studies of the $pn\to \{pp\}_{\!s}\pi^-$ reaction were carried out in two
stages, the first with a polarized proton beam incident on a deuterium pellet
target~\cite{DYM2012}. The full kinematics were determined by detecting
either the spectator proton in an STT or the produced $\pi^-$ in the negative
detector, in coincidence with the diproton pair. In the subsequent
experiment, a polarized deuteron beam collided with a polarized hydrogen
cell~\cite{DYM2013}. In this case the fast spectator proton was measured in
the forward detector of ANKE and the diproton pair in the positive detector.
In both cases the momentum of the spectator in the deuteron rest frame was
restricted such that the spread in effective energies was not as wide as in
the COSY-TOF experiment~\cite{ABD2008a}.

\begin{figure}[h!]
\begin{center}
\includegraphics[width=1.0\columnwidth]{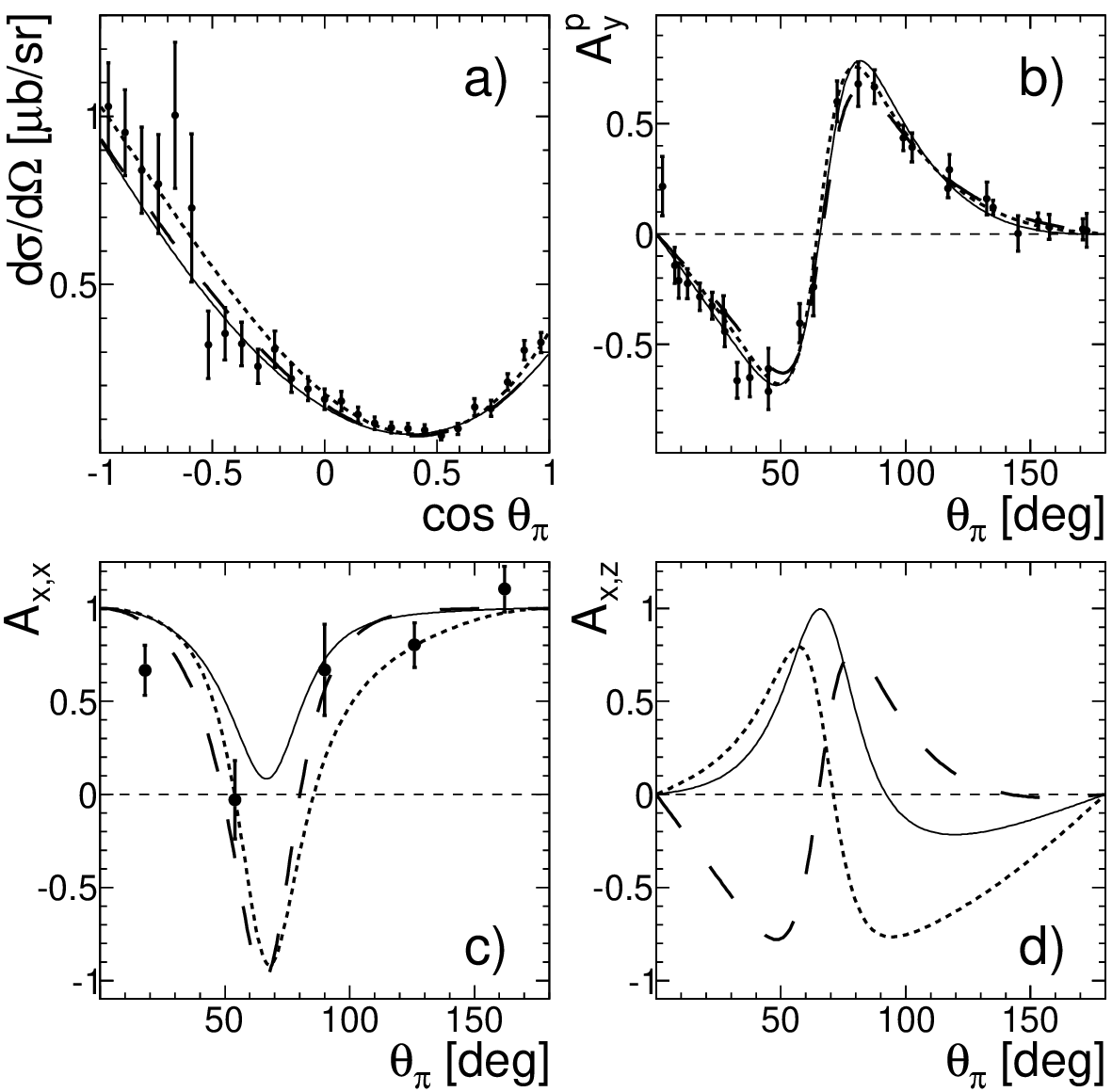}
\caption{\label{fig:DYM} Predictions of the partial wave analysis for the
polarized $pn\to \{pp\}_{\!s}\pi^-$ reaction at 353~MeV with the
${E}_{pp}<3$~MeV cut. The full, long-dashed, and short-dashed lines
correspond to solutions 1, 2, and 3, respectively. The ANKE experimental data
with statistical errors correspond to a) the differential
cross-section~\cite{DYM2012}, b) $A_y^p$~\cite{DYM2013}, and c)
$A_{x,x}$~\cite{DYM2013}. There are no experimental data to compare with the
predictions for $A_{x,z}$ shown in panel d).}
\end{center}
\end{figure}

The most important information obtained from the $np$ experiments is
summarized in Fig.~\ref{fig:DYM}. This shows the measurements of the
differential cross section, the proton analyzing power $A_y^p$, and the
transverse spin correlation $A_{x,x}$, where $x$ lies in the horizontal COSY
plane, perpendicular to the beam direction $z$. Also shown are the
predictions of three possible amplitude analyzes.

In addition to the three isospin-one amplitudes needed to describe the $pp\to
\{pp\}_{\!s}\pi^0$ observables up to pion $d$ waves, two more isospin-zero
amplitudes are required for the $pn\to \{pp\}_{\!s}\pi^-$ data. Since the
$^{3\!}P_0$, $^{3\!}P_2$, and $^{3\!}F_2$ $pp$ waves are either uncoupled or
weakly coupled, the authors~\cite{TSI2012} assumed the Watson theorem and
took the phases of the pion production amplitudes to be the same as those of
elastic proton-proton scattering. Taken together with the data shown in
Fig.~\ref{fig:TSI}, this fixes completely the values of the three complex
$I=1$ amplitudes given in Table~\ref{tab:fit}. Particularly striking is the
fact that the production from the initial $^{3\!}F_2$ state is consistent
with zero.

On the other hand the $^{3\!}S_1$ and $^{3\!}D_1$ waves are strongly coupled
by the tensor force and so it is much harder to justify invoking the Watson
theorem in this case. Fortunately, the extra $np$ information contained in
Fig.~\ref{fig:DYM} allows a complete amplitude reconstruction, apart from
some discrete ambiguities. The three possible solutions are listed in
Table~\ref{tab:fit} and the resulting predictions for the observables are
shown in Fig.~\ref{fig:DYM}.

\begin{table}
\centering
\begin{tabular}{|c|r|r|} \hline
Amplitude  &  Real\phantom{xxll} & Imaginary\phantom{l}\\ \hline
\multicolumn{3}{|c|}{Solution 1: $\chi^2/\textit{ndf}=101/82$ } \\ \hline
$^{3\!}P_0\to\, ^{1\!}S_0s$ & $  53.4 \pm    1.0$ & $ -14.1 \pm    0.3$ \\
$^{3\!}P_2\to\, ^{1\!}S_0d$ & $ -25.9 \pm    1.4$ & $  -8.4 \pm    0.4$ \\
$^{3\!}F_2\to\, ^{1\!}S_0d$ & $  -1.5 \pm    2.3$ & $   0.0 \pm    0.0$\\
$^{3\!}S_1\to\, ^{1\!}S_0p$ & $ -37.5 \pm    1.7$ & $  16.5 \pm    1.9$\\
$^{3\!}D_1\to\, ^{1\!}S_0p$ & $ -93.1 \pm    6.5$ & $ 122.7 \pm    4.4$\\ \hline
         \multicolumn{3}{|c|}{Solution 2: $\chi^2/\textit{ndf}=103/82$} \\ \hline
$^{3\!}P_0\to\, ^{1\!}S_0s$ & $  52.7 \pm    1.0$ & $ -13.9 \pm    0.3$ \\
$^{3\!}P_2\to\, ^{1\!}S_0d$ & $ -28.9 \pm    1.6$ & $  -9.4 \pm    0.5$ \\
$^{3\!}F_2\to\, ^{1\!}S_0d$ & $   3.4 \pm    2.6$ & $   0.0 \pm    0.0$ \\
$^{3\!}S_1\to\, ^{1\!}S_0p$ & $ -63.7 \pm    2.5$ & $  -1.3 \pm    1.6$ \\
$^{3\!}D_1\to\, ^{1\!}S_0p$ & $-109.9 \pm    4.2$ & $  52.9 \pm    3.2$ \\ \hline
         \multicolumn{3}{|c|}{Solution 3: $\chi^2/\textit{ndf}=106/82$} \\ \hline
$^{3\!}P_0\to\, ^{1\!}S_0s$ & $  50.9 \pm    1.1$ & $ -13.4 \pm    0.3$ \\
$^{3\!}P_2\to\, ^{1\!}S_0d$ & $ -26.3 \pm    1.5$ & $  -8.5 \pm    0.5$ \\
$^{3\!}F_2\to\, ^{1\!}S_0d$ & $   2.0 \pm    2.5$ & $   0.0 \pm    0.0$ \\
$^{3\!}S_1\to\, ^{1\!}S_0p$ & $ -25.4 \pm    1.9$ & $  -7.3 \pm    1.5$ \\
$^{3\!}D_1\to\, ^{1\!}S_0p$ & $-172.2 \pm    5.6$ & $  92.0 \pm    6.2$ \\ \hline
\end{tabular}
\caption{Values of the real and imaginary parts of the amplitudes for the
five lowest partial waves deduced from fits to the ANKE measurements at
353~MeV. } \label{tab:fit}
\end{table}

The phase assumptions in the $I=1$ case mean that there are no discrete
ambiguities in the $pp\to \{pp\}_{\!s}\pi^0$ analysis and the three solutions
presented in Table~\ref{tab:fit} lead to indistinguishable curves in the two
panels of Fig.~\ref{fig:TSI}. In contrast, in the $np$ case there are three
solutions that are statistically very similar. As shown in
Fig.~\ref{fig:DYM}d, a measurement of the spin correlation parameter
$A_{x,z}$ in the $pn\to \{pp\}_{\!s}\pi^-$ reaction would allow one to
resolve these discrete ambiguities. However, such a measurement could only be
carried out with a polarized deuterium cell and would require a Siberian
snake to rotate the spin of the incident proton into the beam direction. The
latter facility was not available before the termination of the hadron
physics programme at COSY.

It is nevertheless interesting to study the phases of the two pion $p$-wave
amplitudes from the three solutions given in Table~\ref{tab:fit}, which are
$(\textrm{Im}(^{3\!}S_1\to\, ^{1\!}S_0p)/\textrm{Re}(^{3\!}S_1\to\,
^{1\!}S_0p)$,\ $\textrm{Im}(^{3\!}D_1\to\,
^{1\!}S_0p)/\textrm{Re}(^{3\!}D_1\to\, ^{1\!}S_0p)) = (-0.44,-1.32)$,
$(0.02,-0.48)$, and $(0.29,-0.53)$ for solutions 1, 2, and 3,
respectively\footnote{We are here using the standard notation of $L\ell$,
where $L$ is the angular momentum in the $NN$ system and $\ell$ is the
angular momentum of the meson with respect to the $NN$.}. These can be
compared with the nucleon-nucleon phase-shift analysis values of
$(\tan\delta_{^{3\!}S_1},\tan\delta_{^{3\!}D_1})=(0.03,-0.46)$~\cite{ARN2000},
which are well within the error bars of Solution 2. If Solution 2 were
indeed the correct one, then it would suggest that the concerns over the use
of the Watson theorem for coupled channels might be less serious than feared
and this can have important consequences for the modeling of meson
production. However, if the \emph{truth} corresponded to one of the other
solutions, one would have to explain why the phases had suffered such severe
modifications.

\subsubsection{Comparison of $pp\to pn\pi^+$ and $pp\to d\pi^+$} 
\label{comparison} 

The FSI approach discussed in sect.~\ref{Phenomenology} predicts the
differential cross sections for $S$-wave spin-triplet $np$ production in the
$pp\to\pi^+pn$ reaction in terms of the cross section for
$pp\to\pi^+d$~\cite{BOU1996}, as shown in Eq.~(\ref{5}). The theorem is
derived on the assumption that the $np$ potential is local and that the
coupling induced by the tensor force can be neglected. Though, under these
conditions, the theorem is exact when extrapolated to the bound state pole,
deviations in the physical region are minimized if the pion production
operator is of short range. The relation given in Eq.~(\ref{5}) was tested in
several experiments at COSY.

The ANKE $pp\to\pi^+np$ experiment at 492~MeV~\cite{ABA2001} used a CH$_2$
target but isolated the production on the proton by measuring the final
$p\pi^+$ pairs in coincidence. The normalization was assured by measuring
also pions coming from the $\pi^+d$ final state, though greater care had to
be taken here with the carbon background and this was one contributor to the
overall systematic uncertainties of approximately 8\%. The width of any
spin-singlet contribution was determined by the angular integration in the
ANKE data rather than the intrinsic energy resolution of the apparatus.

The ANKE results were successfully discussed in the framework of the FSI
theorem of Eq.~(\ref{5}). The shape of the $np$ spectrum for an excitation
energy $E_{np}<3$~MeV can be explained in terms of pure spin-triplet
production and an upper limit of any spin-singlet production of about 10\%
was found. This limit is not really competitive with the direct measurements
of spin-singlet production in the $pp\to pp\pi^0$ channel discussed earlier
in this section. To improve on this limit requires better effective
resolution and also the simultaneous measurement of the $\pi^+d$ and
$\pi^+np$ final states under identical conditions in order to eliminate any
normalization ambiguities. A pure hydrogen target is clearly preferable for
this purpose. These criteria were met in experiments carried out by the GEM
collaboration using the Big Karl spectrometer~\cite{ABD2005a,BUD2009a}.

\begin{figure}[hbt]
\begin{center}
\includegraphics[width=0.8\columnwidth]{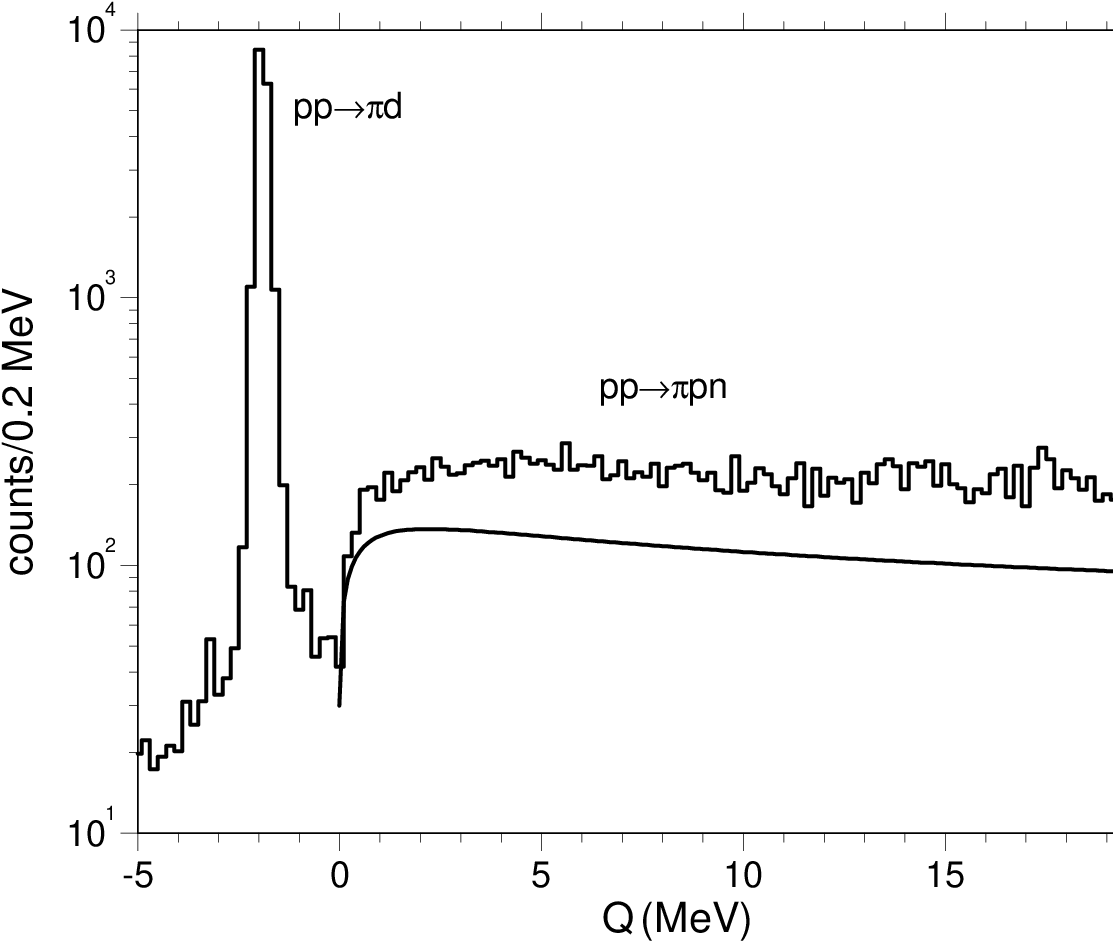}
\end{center}
\caption{The results from the first Big Karl experiment
(histogram)~\cite{ABD2005a} compared with the prediction (curve) of the
$S$-wave final state interaction theorem of
Eq.~(\ref{5})~\cite{BOU1996} for pure spin-triplet $pn$ production.}
\label{FW_model}
\end{figure}

The Big Karl experiments were carried out at 401, 601, and 951~MeV. Only the
$\pi^+$ was detected in the spectrometer, set close to the forward direction.
In order to optimize the resolution, a liquid hydrogen target of only 2~mm
thickness was used, with windows made of 1~$\mu$m Mylar. As can be seen from
the 951~MeV data~\cite{ABD2005a} shown in Fig.~\ref{FW_model}, a resolution
of $\sigma = 97$~keV was achieved on the deuteron peak and this was
sufficient to put stringent bounds on the production of the very narrow peak
that would correspond to the $np$ spin-singlet final state.

Though corrections for acceptance, etc., were included, these varied
relatively little for $Q<20$~MeV so that the data provide a robust
measurement of the ratio of the production of the $\pi^+pn$ and $\pi^+d$
final states. Also shown in the figure is the prediction of Eq.~(\ref{5}),
where the normalization was taken from the area of the deuteron peak. Though
the shape is similar to that of the experimental spectrum, it is too low by a
factor of $N = 2.2\pm 0.1$. This factor is reduced to about 1.8 if the
model-dependent term $(\beta^2+k^2)/(\beta^2-\alpha^2)$ is included with the
value of $\beta=0.927$~fm$^{-1}$, taken from the spin-triplet scattering
length and effective range.

Deviations from Eq.~(\ref{5}) were also studied at the two lower energies
measured, where normalization factors of $N=0.51\pm 0.06$ and $1.06\pm 0.04$
were required at 401~MeV and 601~MeV, respectively~\cite{BUD2009a}. The
authors could not explain the energy dependence of $N$ in terms of the $np$
tensor force and the resulting deuteron $D$-state. Instead they suggested
that the culprit might be the long-range part of the pion production operator
associated with on-shell intermediate pions. It should be noted in this
context that $N$ changes from below to above unity at an energy that
corresponds to the $\Delta$ threshold.

\subsection[$\eta$ production in proton-proton collisions]{$\boldsymbol{\eta}$  production in proton-proton collisions} 
\label{eta}

\begin{figure}[hbt]
\begin{center}
\includegraphics[width=1.0\columnwidth]{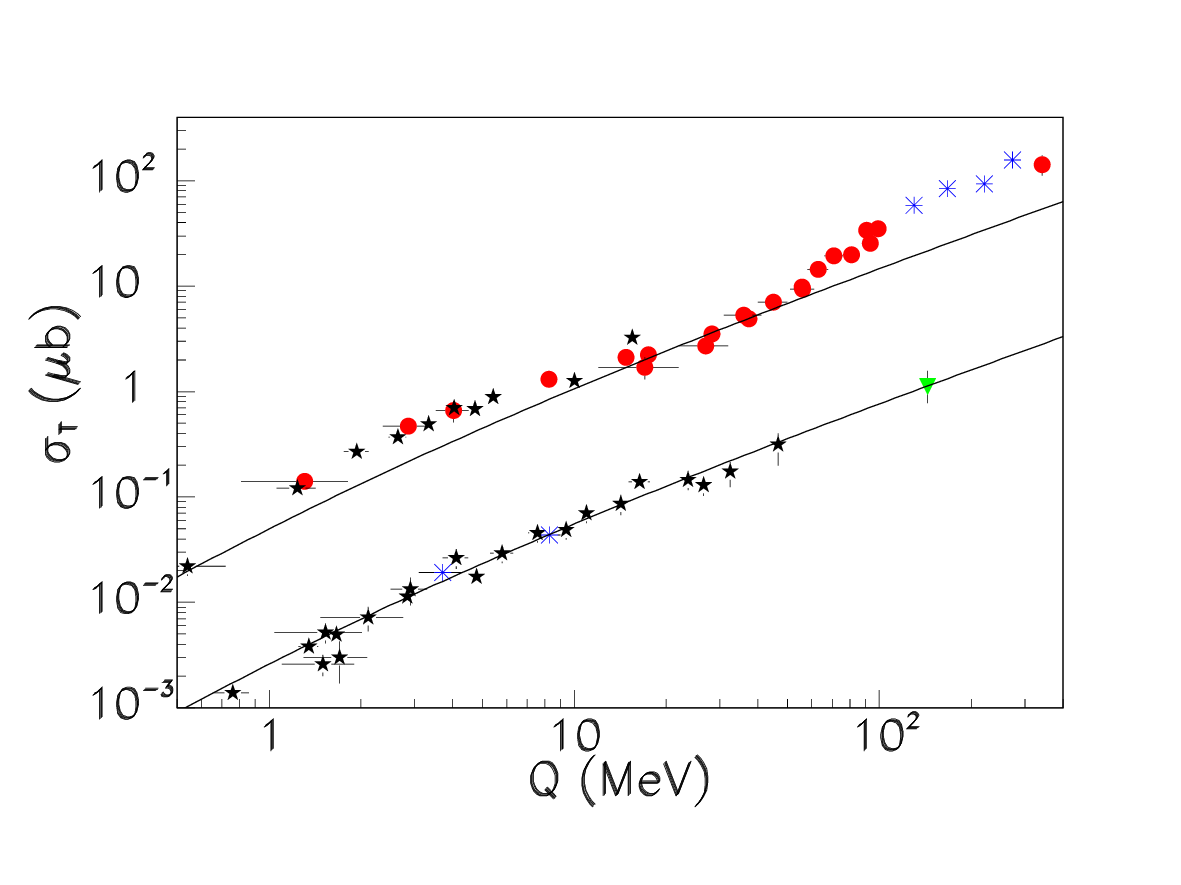}
\caption{\label{fig:etaetaprime} Total cross sections for $pp\to pp\eta$
(upper points) and $pp\to pp\eta^{\prime}$ (lower points). The $\eta$ data
are taken from Refs.~\cite{CHI1994a,CAL1996,BER1993,HIB1998} (closed red
circles), COSY-11~\cite{SMY2000,MOS2004,MOS2010} (closed black stars),
\cite{MAR2001} (blue crosses), and \cite{AGA2012} (green star). The
$\eta^{\prime}$ data are from Ref.~\cite{BER1993,HIB1998} (blue crosses),
\cite{BAL2000} (green star), and
COSY-11~\cite{MOS1998,MOS2000,KHO2004,CZE2014} (closed black stars). The
solid curves are arbitrarily scaled $pp$ FSI predictions of
Eq.~(\ref{simple2}). }
\end{center}
\end{figure}

By far the most extensive series of measurements of the $pp\to pp\eta$
reaction near threshold was undertaken by the COSY-11
collaboration~\cite{SMY2000,MOS2004,MOS2010} and this has led to the bulk of
the low energy points shown in Fig.~\ref{fig:etaetaprime}. The two emerging
protons were identified and their momenta measured and the reaction isolated
by finding the missing-mass peak corresponding to the production of the
$\eta$ meson. Very close to threshold this peak stands out clearly from the
multipion background, though more care has to be taken at the higher excess
energies.

Also shown in Fig.~\ref{fig:etaetaprime} is a curve corresponding to the
energy dependence expected according to the $pp$ FSI model of
Eq.~(\ref{simple2}). This assumes that the data are dominated by $S$-wave
$pp$ final states but this is in conflict with the differential data at
$Q=72$~MeV, where the valley along the diagonal of the Dalitz plots shows
strong evidence for the production of $Pp$ or higher waves~\cite{PET2010}.
This suggests that the deviations from the curve in
Fig.~\ref{fig:etaetaprime} for $Q\gtrsim 40$~MeV might be associated with the
excitation of higher partial waves. This cannot be the explanation of the
relatively high cross sections at low $Q$, which are probably driven by the
strong $\eta p$ interaction, which is already well known in the
$pd\to{}^3$He$\,\eta$ reaction to be discussed in sect.~\ref{pd3heeta}.

\begin{figure}[hbt]
\begin{center}
\includegraphics[width=0.8\columnwidth]{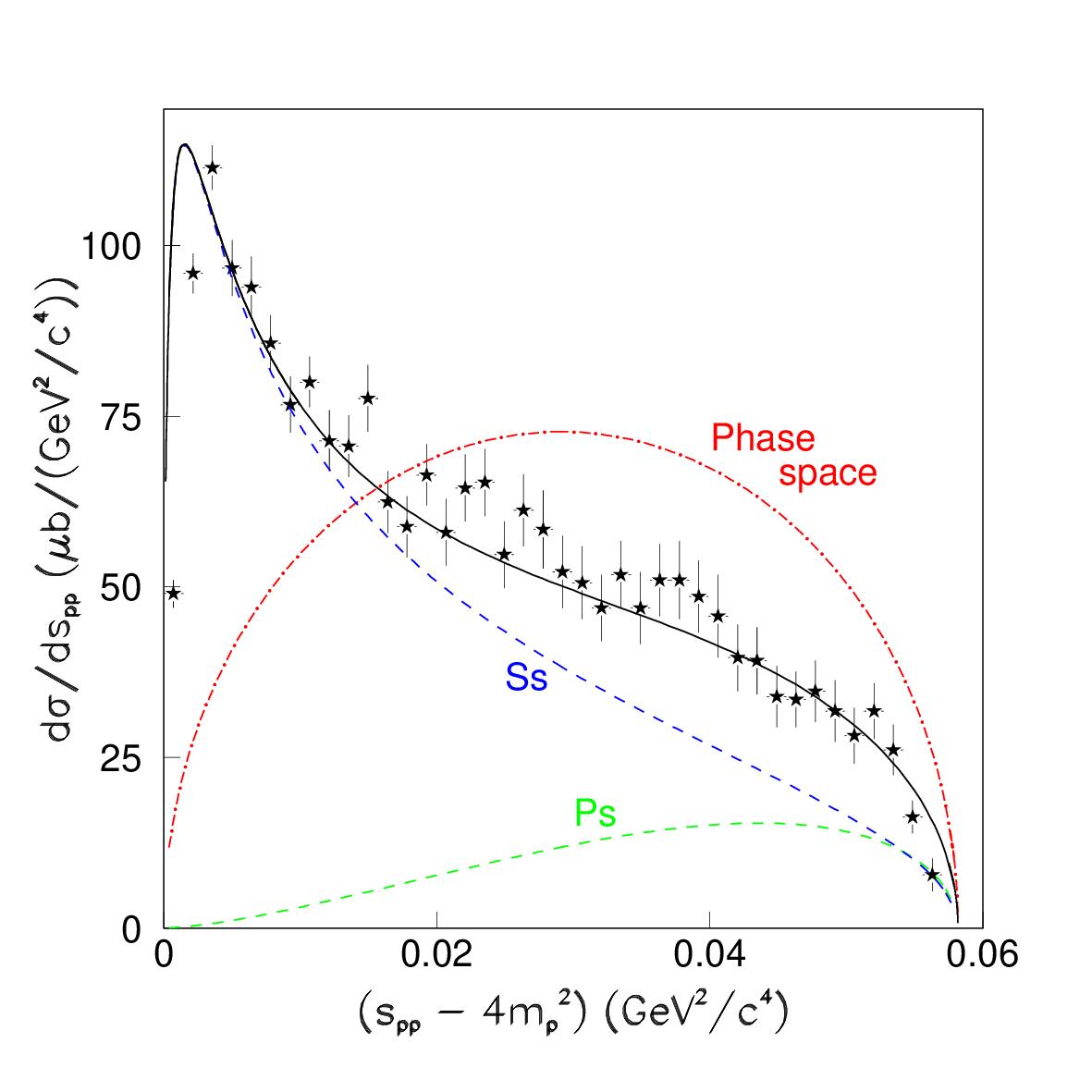}
\caption{\label{fig:pp15} One-dimensional distribution measured in the $pp\to
pp\eta$ reaction by the COSY-11 collaboration at $Q=15.5$~MeV~\cite{MOS2004};
to a very good approximation the abscissa represents $4m_pE_{pp}$. The (red)
chain curve corresponds to a phase space distribution and weighting this
arbitrarily with the $pp$ $S$-wave FSI or a $P$-wave factor gives the (blue
or green) dashed curves. The sum of $Ss$ and $Ps$ contributions (solid black
curve) describes well the shape of the data.}
\end{center}
\end{figure}

More detailed information can be obtained from looking at differential
distributions and the spectrum of the excitation energy $E_{pp}$ in the $pp$
system is shown in Fig.~\ref{fig:pp15} at an excess energy of $Q\approx
15.5$~MeV~\cite{MOS2004}. What is immediately striking here is the sharp
peaking of the experimental data at very low $E_{pp}$ that is due to the
dominance of the $Ss$ wave and the very strong final state interaction
between the two protons. There are minor differences in the literature on how
the FSI is modeled and the curve shown in Fig.~\ref{fig:pp15} does not
include Coulomb repulsion or experimental resolution, Nevertheless, it is
clear that the model falls well below the data at large $E_{pp}$. A natural
assumption is that at large $E_{pp}$ there are contributions associated with
$Ps$ final waves and the combination of $Ss$ and $Ps$ final waves describes
the COSY-11 data very well. Fully reconstructed $pp\to pp\eta$ events were
also obtained in a COSY-TOF experiment at 15~MeV and 41~MeV~\cite{ABD2003a}
and at the higher energy these seem to show an even larger fraction of events
at large $pp$ invariant masses.

\begin{figure}[h!]
\begin{center}
\includegraphics[width=0.8\columnwidth]{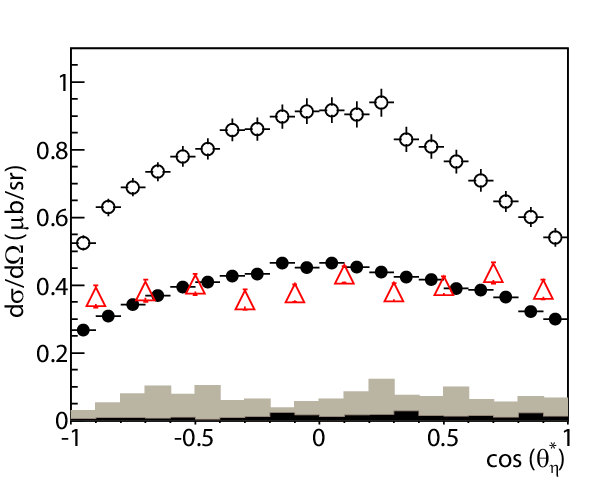}
\caption{\label{fig:PET3} The $pp\to pp\eta$ cross section as a function of
the cosine of the $\eta$ polar angle in the overall CM frame. The data from
the CELSIUS-WASA experiment~\cite{PET2010} at $Q=72$~MeV (open circles) and
$Q=40$~MeV (filled circles) are compared with those of
COSY-TOF~\cite{ABD2003a} at $Q=41$~MeV (triangles). The $Q=72$~MeV points
have been shifted down by a factor $0.5$ for ease of presentation. The
systematic uncertainties in the CELSIUS data are shown by the grey and black
histograms, where the black corresponds to the $Q=40$~MeV data.}
\end{center}
\end{figure}

The COSY-TOF experiment also produced angular distributions of both the
$\eta$ and the $pp$ relative momentum~\cite{ABD2003a} and the $\eta$
differential cross section at 41~MeV is shown in Fig.~\ref{fig:PET3}, where
it is compared with CELSIUS-WASA measurements at 40~MeV and
72~MeV~\cite{PET2010}. The shape of the distribution seems to be better
defined by the CELSIUS experiment.

Data were also obtained at ANKE in a missing-mass experiment at $Q=55$~MeV
and 270~MeV, where a cut of $E_{pp}<3$~MeV was placed on the excess energy of
the outgoing $pp$ pair~\cite{DYM2009}. At such high excess energies only data
at small $\eta$ angles were accessible and, putting a further cut of
$\cos\theta_{\eta}>0.95$, a preliminary cross section of $(4.3\pm0.8)$~nb at
55~MeV could be extracted, where only the statistical error is quoted. The
CELSIUS-WASA experiment~\cite{PET2010} was not very sensitive to this
kinematic region but the ANKE value does raise questions regarding the
restricted amplitude analysis used at CELSIUS to extrapolate into this
region.

Although the WASA programme on $pp\to pp\eta$ was primarily directed towards
the study of the rarer $\eta$ decays described in sect.~\ref{eta_decays}, the
azimuthal symmetry of the detector makes it an ideal instrument with which to
measure analyzing powers. In order to avoid unwelcome rotations of the proton
spin and the consequent loss of beam polarization, the field in the solenoid
was switched off~\cite{WASA2015}. This was possible because at WASA the
$\eta$ could be detected through the $\eta\to 2\gamma$ and $\eta\to 3\pi^0$
decays as well as a missing-mass peak. The beam polarizations were deduced
from measurements of elastic $pp$ scattering, with one proton being
registered in the forward detector and the other in the central detector.
These asymmetries were converted into polarizations using the analyzing
powers measured by the EDDA collaboration~\cite{ALT2005}. The experiments,
which were carried out at $Q=40$~MeV and 72~MeV, complemented the earlier
COSY-11 measurements at 10~MeV and 36~MeV, which showed no significant
signal~\cite{WIN2002,CZY2007}.

\begin{figure}[hbt]
\begin{center}
\includegraphics[width=1.0\columnwidth]{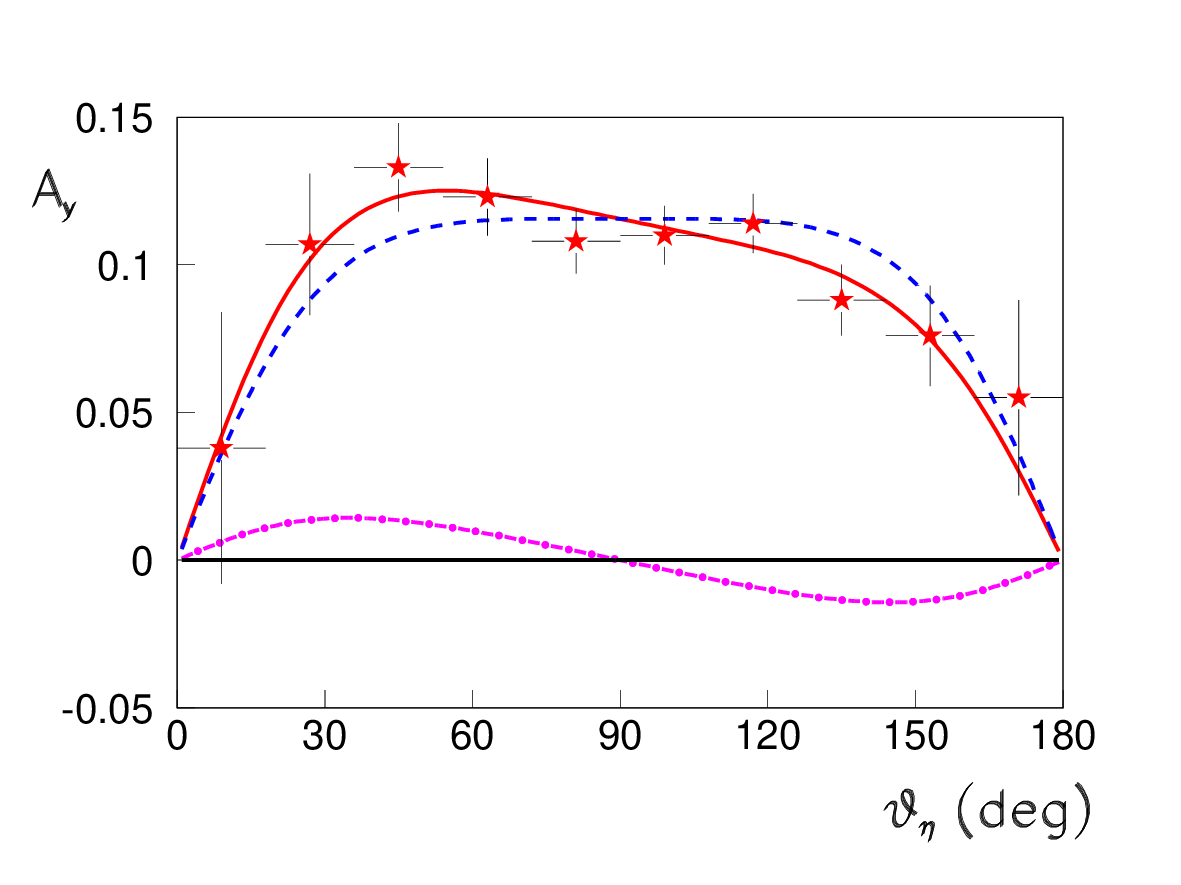}
\caption{\label{fig:Pawel}
Measurement of the proton analyzing power in the $\pol{p}p\to pp\eta$
reaction with respect to the $\eta$ polar angle in the c.m.\ frame at
$Q=72$~MeV~\cite{WASA2015}. Only statistical errors are shown. The fits on
the basis of Eq.~(\ref{PM_fit}) used differential cross sections measured at
CELSIUS~\cite{PET2010}. The dashed (blue) curve represents the
$\sin\theta_{\eta}$ component, the dot-dashed (magenta) curve corresponds to
the $\sin2\theta_{\eta}$ term, whereas the solid (red) curve is their sum. It
appears from these data that the analyzing power is driven mainly by $Ps:Pp$
interference.}
\end{center}
\end{figure}

Figure~\ref{fig:Pawel} shows the WASA data on the analyzing power of the
$\pol{p}p\to pp\eta$ reaction at an excess energy of $Q=72$~MeV as a function
of the c.m.\ polar angle of the $\eta$~\cite{WASA2015}. Near threshold we can
expect $A_y\dd\sigma/\dd\Omega$ to be a linear combination of
$\sin\theta_{\eta}$ and $\sin2\theta_{\eta}$. Taking the unpolarized angular
distribution at this energy from the earlier WASA measurement at
CELSIUS~\cite{PET2010} shown in Fig.~\ref{fig:PET3}, the $A_y$ data are well
described by
\begin{equation}
\label{PM_fit}
A_y=(C_1\sin\theta_{\eta}+C_2\sin2\theta_{\eta})/(0.88+0.92\sin^2\theta_{\eta}),
\end{equation}
where $C_1=0.208\pm0.008$ and $C_2=0.018\pm0.009$. The error bars are
statistical but there may be systematic effects in $C_2$ arising from
slightly different acceptances in the two hemispheres. The fit is shown in
Fig.~\ref{fig:Pawel} along with separate curves corresponding to the
$\sin\theta_{\eta}$ and $\sin2\theta_{\eta}$ components. It is clear from
this that the former, which arises principally from $Ps$ interfering with
$Pp$ waves, is much bigger than the latter, which is probably driven mainly
by $Ss:Sd$ interference. This is not a total surprise since the CELSIUS
data~\cite{PET2010} have shown that the $Pp$ contribution is very strong at
72~MeV and there is evidence in Fig.~\ref{fig:pp15} for a $Ps$ contribution
already at 15.5~MeV. In contrast, the group found no significant analyzing
power at $Q=15$~MeV~\cite{WASA2015} but this is also not unexpected so close
to threshold.

\subsection[$\eta$ production in proton-neutron collisions]{$\boldsymbol{\eta}$  production in proton-neutron collisions} 
\label{deta}

Although there were measurements of the cross section for the quasi-free
$pn\to pn\eta$ reaction at CELSIUS~\cite{CAL1998}, these were not carried out
in the immediate vicinity of the threshold. What is remarkable in these data
is the large ratio of $\eta$ production in $pn$ compared to $pp$ collisions,
with cross section ratios being typically $R_{pn/pp}\approx 7$. The $pd\to
p_{\rm sp}pn\eta$ reaction was measured closer to threshold by adding the
neutral particle detector to the standard COSY-11 facility~\cite{MOS2009}.
The outgoing neutron was then measured in this detector, which delivered
information about the position and time at which the registered neutron
induced a hadronic reaction. By using a single beam energy (1.34~GeV) and
exploiting the spread in excitation energies $Q$ caused by the deuteron Fermi
motion, the COSY-11 authors extracted values of the $pn\to pn\eta$ total
cross section in three bins in $Q$. Although the value of $R_{pn/pp}$ was
consistent with the CELSIUS results for $10<Q<15$~MeV, the COSY-11 results
closer to threshold were much lower than the CELSIUS factor of seven.

As the authors pointed out~\cite{MOS2009}, much of this variation could arise
from the different final state interactions in the nucleon-nucleon $I=1$ and
$I=0$ channels. Following the arguments given in sect.~\ref{Phenomenology},
if this factor alone is included, one might expect that near threshold the
ratio should vary as~\cite{FAL1996}
\begin{equation}
R_{pn/pp}=0.5+C\,\frac{B_{I=1}}{B_{I=0}}\left(\frac{1+\sqrt{1+Q/B_{I=1}}}{1+\sqrt{1+Q/B_{I=0}}}\right)^{\!2},
\end{equation}
where $B_{I=0}=2.23$~MeV and the authors assumed $B_{I=1}\approx 0.68$~MeV.
The 0.5 corresponds to the $I=1$ component in the initial $pn$ wave and the
parameter $C$ reflects the basic production mechanism that is outside the
remit of the FSI approach. By taking $C\approx 7$, which is consistent with
high energy data, much of the near-threshold decrease in $R_{pn/pp}$ could be
explained, though more refined data would be required to isolate clearly this
effect~\cite{MOS2009}.

\begin{figure}[hbt]
\begin{center}
\includegraphics[width=1.0\columnwidth]{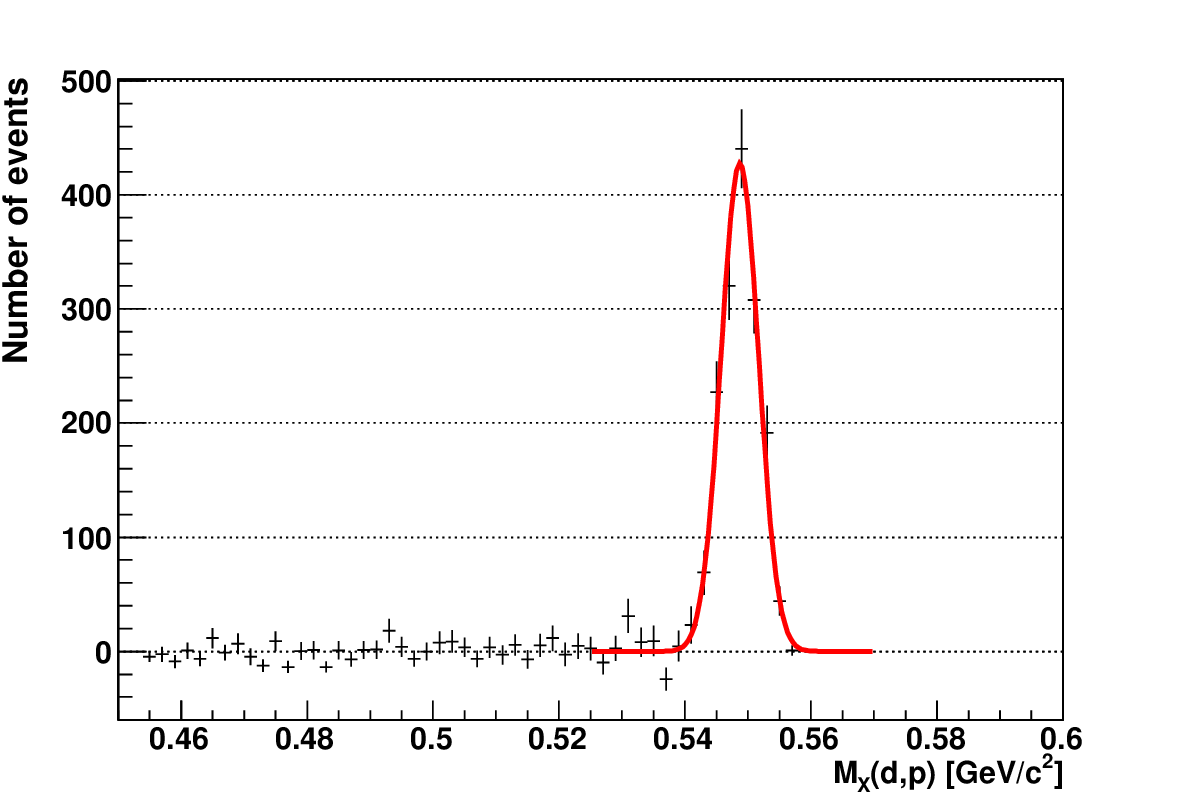}
\end{center}
\caption{ Missing-mass spectrum after background subtraction for
the $dp\to d p X$ reaction at 2.27~GeV from preliminary ANKE data for
$5<Q<10$~MeV~\cite{DYM2010}.}
\label{SD_dp}
\end{figure}

An attempt was made at ANKE to measure the cross section for $np\to d\eta$ by
using a deuteron beam with the maximum COSY energy of 2.27~GeV. Since the
central neutron energy is below the $d\eta$ threshold, which is at about
1.26~GeV, only the upper part of the Fermi momentum contributed to $\eta$
production. As can be judged from the missing-mass distribution shown in
Fig.~\ref{SD_dp}, it is straightforward to identify the reaction but, being
so far below threshold, it has not been possible to separate quasi-free
production cleanly from more complicated three-nucleon effects.

To go higher in energy at COSY, the experiment must use a deuterium target
and isolate the $pd\to p_{\rm sp}d\eta$ by detecting explicitly the spectator
proton. Such an experiment was undertaken at ANKE, where the spectator proton
was detected in one of the two  STT~\cite{KHO2012}. The data, which covered
the range $0 < Q < 100$~MeV in excess energy, are still under analysis but
preliminary results in the range below 20~MeV indicate an $\eta d$ scattering
length of magnitude $|a_{\eta d}|\approx 1.2$~fm~\cite{SCH2016}.

\subsection[$\omega$ production in proton-proton scattering]{$\boldsymbol{\omega}$ production in proton-proton scattering} 
\label{omega}

With the possible exception of the $\eta^{\prime}$, the natural widths of the
other mesons discussed in this section are much less than the resolution of
the detectors available at COSY. As a consequence, any improvement in the
resolution automatically improves the signal-to-background ratio. This is no
longer true for $\omega$ production because in the COSY experiments the
missing-mass peak is dominated by the natural meson width of $\Gamma_{\omega}
= 8.49\pm0.08$~MeV/$c^2$~\cite{OLI2014}. There is therefore a large
background of mainly two- and three-pion production that has to be mastered.

In some experiments only the two final protons were
measured~\cite{HIB1999,BAR2007} but in more refined approaches the $\pi^+$
and $\pi^-$ from the three-pion decay of the $\omega$ were detected in
coincidence~\cite{BAL1998a,BAL2001,SAM2001,ABD2007a,ABD2010a}. However, in
all cases one is still faced with the problem of separating the $\omega$
signal from the background. Although this can be done by fitting smooth
curves on either side of the $\omega$ peak, there are two other procedures
that deserve closer attention.

\begin{figure}[h!]
\begin{center}
\includegraphics[width=0.9\columnwidth]{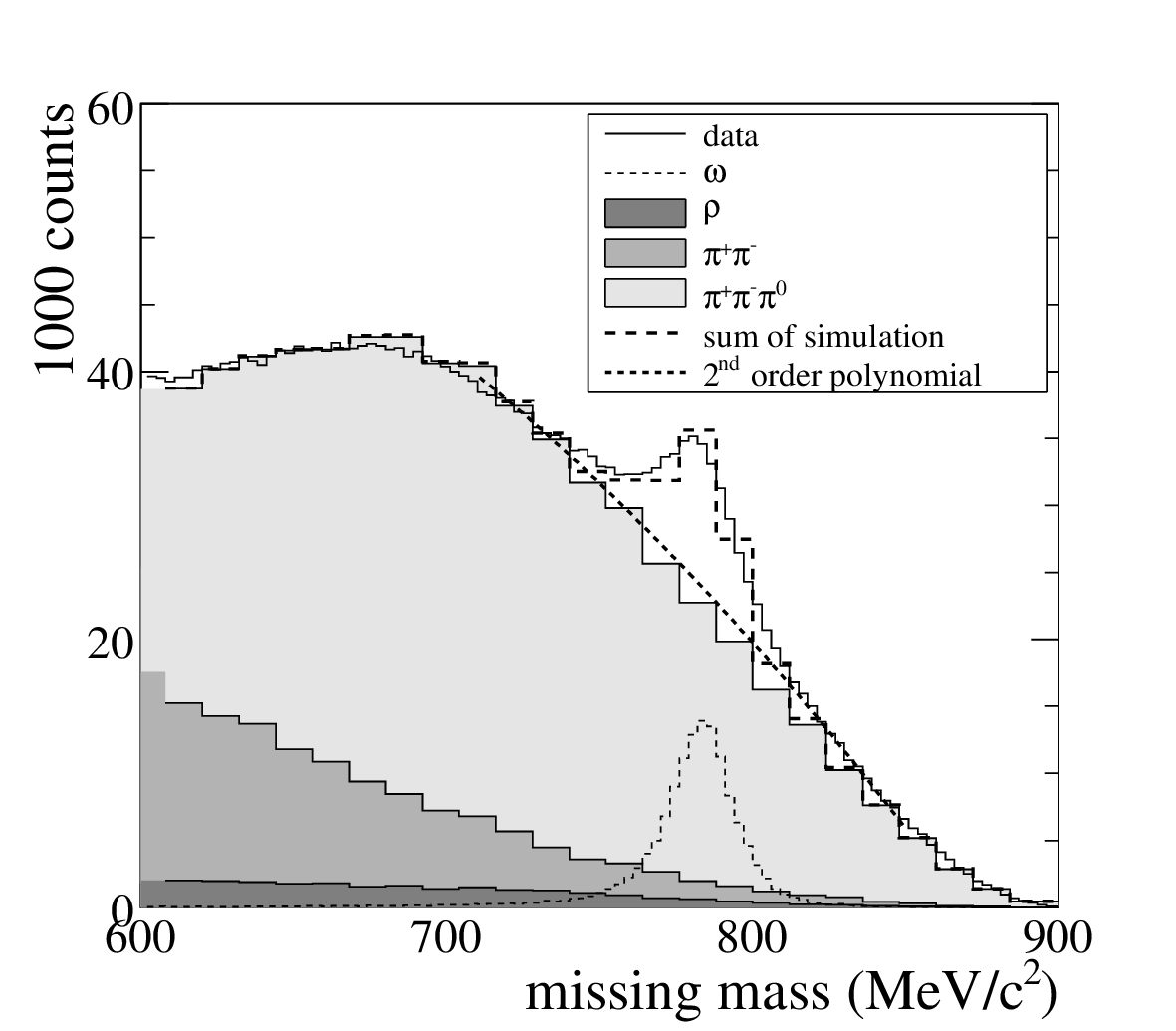}
\caption{\label{fig:HF5} Proton-proton missing-mass spectra at an excess
energy of $Q = 128$~MeV with respect to the nominal $\omega$
threshold~\cite{ABD2010a}. Also shown are normalized Monte Carlo
distributions obtained for resonant $pp\to pp\,(\rho\to\pi^+\pi^-)$ and
plane-wave two- and three-pion production. The sum describes the non-$\omega$
background very well.}
\end{center}
\end{figure}

In the COSY-TOF missing-mass spectrum of Fig.~\ref{fig:HF5}, the background
was explicitly modeled in terms of $\rho$ production and decay plus larger
contributions coming from direct two- and three-pion production. By allowing
each of these contributions to be included with fitted weights, a good
overall description of the multipion background could be achieved. However,
it was assumed in this analysis that these non-$\rho$ pion productions
followed phase space although in reality the spectra would be distorted in
some way by $\Delta$ or $N^*$ isobar production.

In an alternative, more empirical approach, it was assumed that in
experiments with magnetic spectrometers the shape of the multipion background
was determined mainly by the limited spectrometer acceptance, so that the
shape of the background was taken from sub-threshold measurements. Rather
than simply shifting the data from negative $Q$ to positive $Q$, in the
Saclay experiment the sub-threshold data were analyzed as if they had been
taken at the energy of the $\omega$ signal~\cite{HIB1999}. The ANKE data were
taken at only two energies, well above threshold, but a similar procedure was
adopted, taking the background away from the $\omega$ peak at one energy and
using it under the $\omega$ peak at the other~\cite{BAR2007}. The practice of
using sub-threshold data to determine the shape of multipion background is
quite common at COSY, for example in the measurement of $dp\to{}^3$He$\,\eta$
at COSY-11~\cite{SMY2007} and ANKE~\cite{MER2007}.

\begin{figure}[h!]
\begin{center}
\includegraphics[width=0.9\columnwidth]{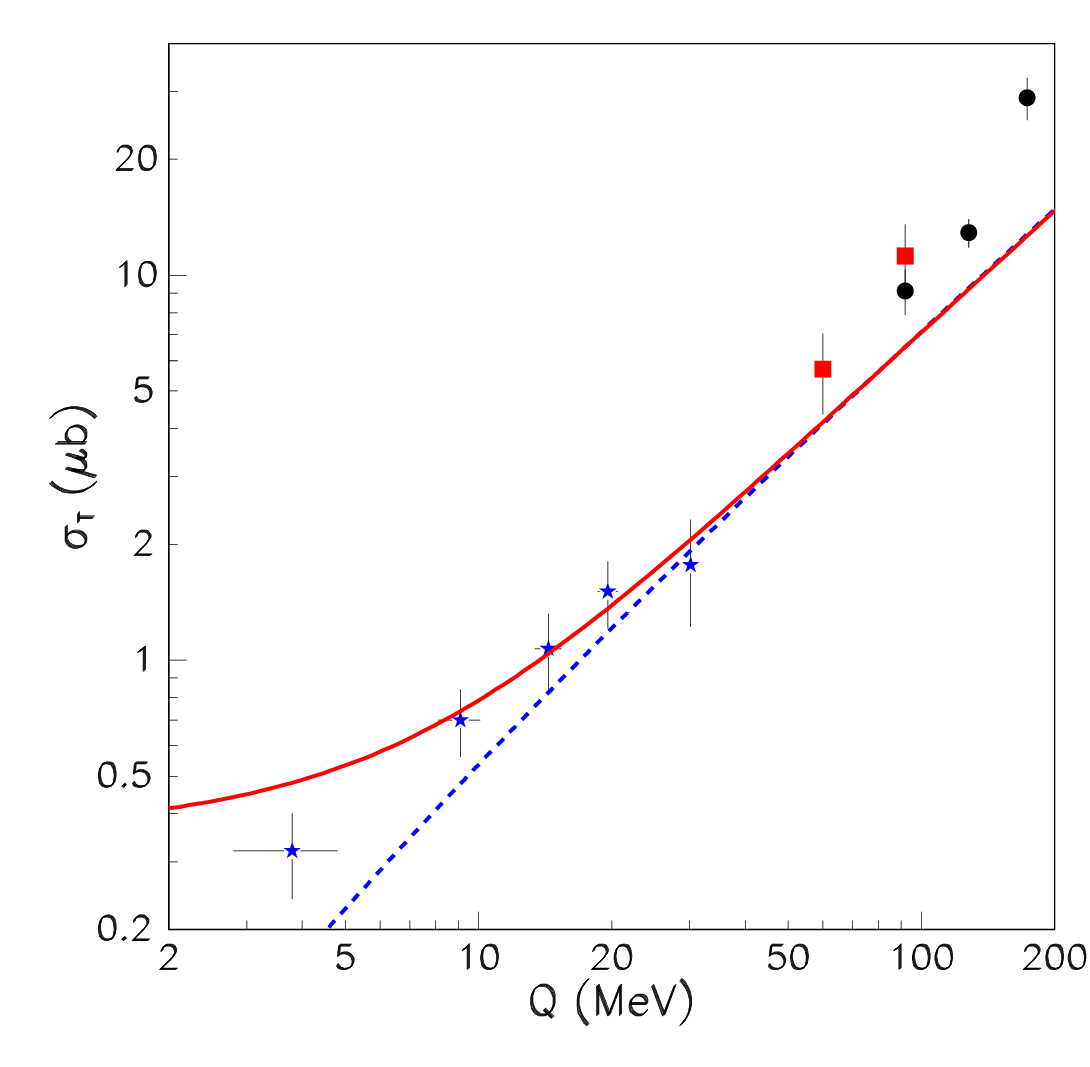}
\caption{\label{fig:omega} Total cross sections for $pp\to pp\,\omega$ in terms
of the nominal value of $Q$, i.e., neglecting the $\omega$ width. The data
are taken from Refs.~\cite{HIB1999} (blue stars), \cite{BAR2007} (red
squares), and \cite{ABD2010a} (black circles). The (blue) dashed curve is an
arbitrarily scaled $pp$ FSI prediction of Eq.~(\ref{simple2}), whereas the
(red) solid one has been smeared over the $\omega$ width. }
\end{center}
\end{figure}

The values of the low energy $pp\to pp\,\omega$ total cross sections
extracted from the COSY and earlier Saclay experiments are shown in
Fig.~\ref{fig:omega} along with the arbitrarily normalized pure $pp$ FSI
prediction of Eq.~(\ref{simple2}). This assumes that the final $pp$ pair is
in the \Szero\ state and the deviations at large $Q$ are, like the $\eta$
production data of Fig.~\ref{fig:etaetaprime}, due to $P$ and higher waves in
the $pp$ system. The deviations close to threshold probably arise from the
finite width of the $\omega$ and smearing the $pp$ FSI predictions over this
width gives the modified curve that is also shown in the figure.

Apart from reducing somewhat the multipion background, detecting the decay
products of the $\omega$ allows one to determine the tensor polarization (the
alignment) $\rho_{00}$ of this meson. At threshold the only transition
allowed is $^{3\!}P_1\to\!{}^{1\!}S_0s$ and this leads to $\omega$ spin
projections of $m_{\omega}=\pm 1$ along the beam axis and hence
$\rho_{00}=0$. The COSY-TOF collaboration has measured $\rho_{00}$ at excess
energies of $Q=92$, 128, and 173~MeV and seen a significant departure from
the threshold value towards the unpolarized value of
$\rho_{00}=1/3$~\cite{ABD2010a}. The results at the three energies could be
parameterized as $\rho_{00}=Q/3(Q+A)$, where $A=90\pm 35$~MeV.

In principle the various angular distributions in the $pp\to pp\,\omega$
reaction could be investigated by just measuring the two final protons but
the large data sample of relatively clean $\omega$ events in the COSY-TOF
experiments, especially at 128~MeV, has allowed angular distributions to be
extracted in the c.m., the helicity, and the Jackson frame~\cite{ABD2010a}.
At this energy the proton angular distribution is relatively flat in all
three frames though the $\omega$ c.m.\ dependence is quite strong, varying
like $1+(0.97\pm0.21)\cos^2\theta_{\omega}$. These two facts suggest that the
other important final wave is $^{1\!}S_0p$, though this would give an angular
dependence to $\rho_{00}$.

The other valuable piece of evidence that is relevant for the presence of
higher partial waves comes from the measurements of the proton analyzing
power of the $\pol{p}p\to pp\,\omega$, where a value consistent with zero was
found at $Q=129$~MeV~\cite{ABD2008}. This would follow if the extra partial
waves were spin-triplet since there would then be no interference with the
threshold spin-singlet for the analyzing power. However, this would also be
the case for the differential cross section, so that this may not be the
origin of the observed non-isotropy in $\theta_{\omega}$. Of course, the
vanishing of the analyzing power could be the result of an accident in the
phases of the production amplitudes. The rich COSY-TOF $pp\to pp\,\omega$
data set~\cite{ABD2010a,ABD2008} will certainly provide a challenge for
modelers.

There was one measurement of quasi-free $pn\to d\,\omega$ production, where
the spectator proton in the $pd \to p_{\rm sp}d\,\omega$ was measured in two
energy regions in an early version of a Silicon Tracking
Telescope~\cite{BAR2004}. Though the total cross sections could be a little
larger than those for $pp\to pp\,\omega$, the error bars are large, due to
the limited statistics and the very significant multipion background.

\subsection[$\eta^{\prime}$ production in proton-proton scattering]{$\boldsymbol{\eta^{\prime}}$ production in proton-proton scattering}  
\label{etaprime}

\begin{figure}[h!]
\begin{center}
\includegraphics[width=0.9\columnwidth]{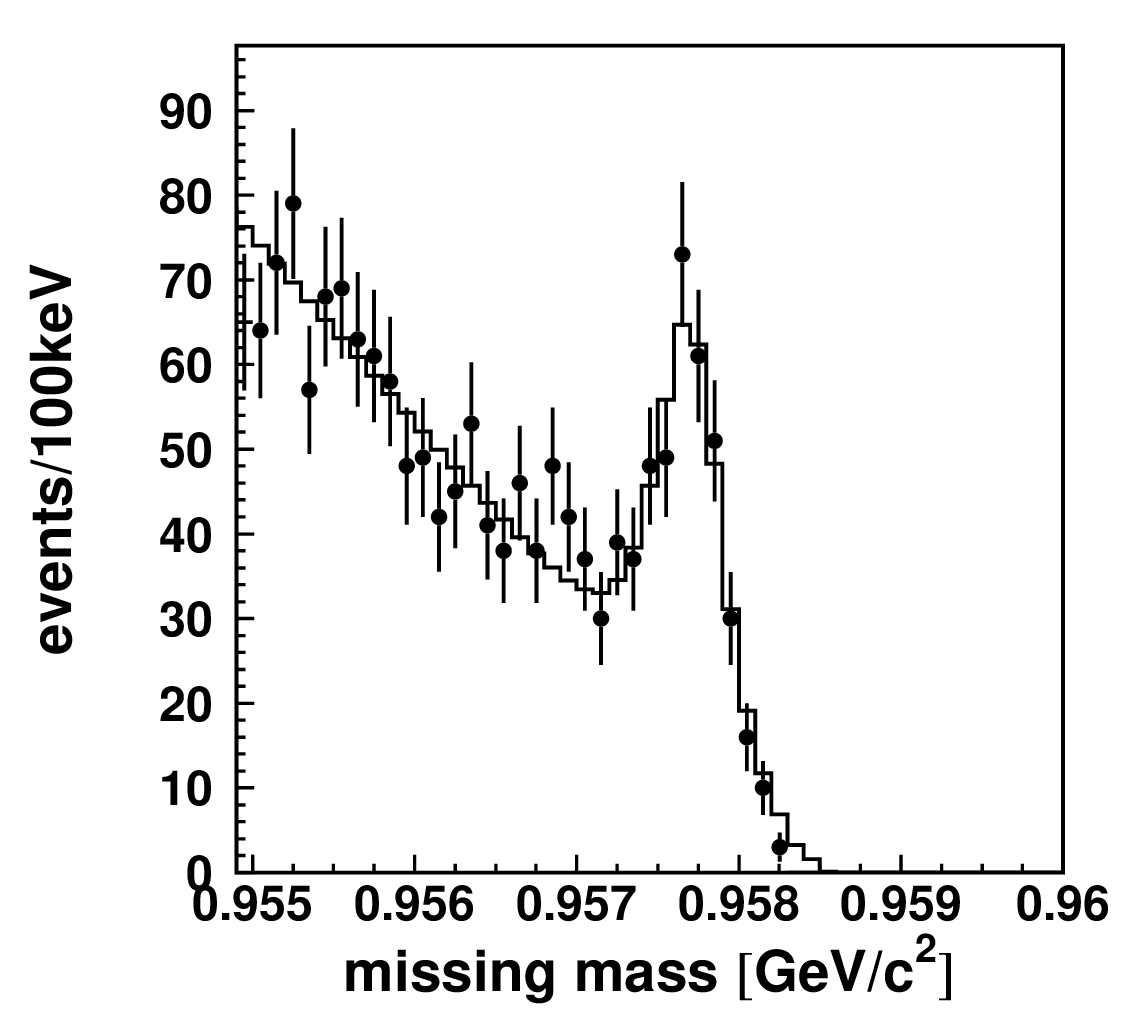}
\caption{\label{fig:Eryk} The missing-mass spectrum for the $pp\to ppX$
reaction at an excess energy of $Q=0.8$~MeV with respect to the
$pp\eta^{\prime}$ threshold~\cite{CZE2010}. The experimental data are
presented as points while the line corresponds to the sum of the Monte
Carlo generated signal for the $pp\to pp\eta^{\prime}$ reaction with
$\Gamma_{\eta^{\prime}}=0.226$~MeV/c$^2$ and the background obtained from
another energy. Having a reliable parametrization of the background, it is
straightforward to extract the numbers of $\eta^{\prime}$ mesons produced
from the missing-mass peak. }
\end{center}
\end{figure}

A particularly interesting case of near-threshold meson production is the
COSY-11 study of the $pp\to pp\eta^{\prime}$ reaction. This is because the
meson has a natural width of some hundreds of keV/$c^2$, which is comparable
to the resolution of the spectrometer. The COSY-11 measurement of the $pp\to
ppX$ missing-mass distribution at an excess energy of $Q=0.8$~MeV with
respect to the $\eta^{\prime}$ threshold is shown in
Fig.~\ref{fig:Eryk}~\cite{CZE2010}. Since the COSY beam momentum was not
known with sufficient precision from macroscopic measurements, it could be
fixed to $\pm 0.2$~MeV/$c$ by using the standard value of the $\eta^{\prime}$
mass, as given in the PDG tables~\cite{OLI2014}.

The acceptance of the COSY-11 or ANKE spectrometer for such reactions
increases as threshold is approached from above because the two final protons
are then squeezed into a smaller and smaller forward cone. This advantage no
longer holds for, e.g., the MOMO and WASA detectors where particles are lost
down the beam pipe. The missing-mass resolution also improves near threshold
because, just as for the $pd\to{}^3$He$\,\eta$ reaction discussed in
sect.~\ref{pd3heeta}, the value of $M_X$ is stationary at threshold. As a
consequence, the missing-mass peak in Fig.~\ref{fig:Eryk} stands out very
clearly.

The background in the figure is mainly due to multipion production which
changes very smoothly with beam energy. The shape of the background, which is
fixed dominantly by the characteristics of the spectrometer, can be
determined by fitting data taken at a different energy and then shifting the
spectrum so that the kinematic limits coincide. This method gave a very good
description of the background at all the energies studied and it allowed the
$\eta^{\prime}$ peak to be isolated for the different values of $Q$. Even at
the highest COSY-11 energy of $Q=46.6$~MeV, the rather flat angular
distribution was consistent with pure $s$-wave production~\cite{KHO2004}.

The series of measurements at COSY-11~\cite{CZE2014,CZE2014a,CZE2010} yielded
two important Physics results. The 21 COSY-11 points completely dominate the
energy dependence of the $pp\to pp\eta^{\prime}$ total cross section
displayed in Fig.~\ref{fig:etaetaprime}, where the values are typically a
factor of 30 or more below those for $\eta$ production. It is also
immediately apparent from this figure that any enhancement of the cross
section at low $Q$ is much less than that for $\eta$ production. It would
therefore seem that the magnitude of the $\eta^{\prime}p$ scattering length,
$a_{\eta^{\prime}p}$, must be significantly smaller than that of $\eta\,p$.
Numerical estimates of $a_{\eta^{\prime}p}$ were given in
Ref.~\cite{CZE2014}, though there is some model dependence in the imaginary
part. The relatively weak interaction of the $\eta^{\prime}$ with the proton
has significant consequences for the chances of this meson binding to nuclei.

The Particle Data Group obtained a value of the natural width of the
$\eta^{\prime}$ by making fits to 51 measurements of partial widths,
integrated cross sections, and branching ratios. The existing direct
measurements of the line width had very large uncertainties and a more
accurate one was clearly highly desirable. This was achieved with the COSY-11
data~\cite{CZE2010}. In the $\eta^{\prime}$ threshold region the circulating
proton beam has a momentum spread of $\textrm{FWHM}= 2.5$~MeV/$c$. However,
due to the position of the COSY-11 target in a dispersive region of COSY, the
momentum spread seen at the target could be reduced down to a mere $\pm
0.06$~MeV/$c$. This therefore gave a negligible contribution to the total
experimental missing-mass resolution of $\textrm{FWHM}\approx
0.33$~MeV/$c^2$.

By analyzing simultaneously data taken at five excess energies, from 0.8 to
4.8~MeV, the width was determined to be $\Gamma_{\eta^{\prime}} = (0.226 \pm
0.017(\textrm{stat}) \pm
0.014(\textrm{syst}))~\textrm{MeV}/c^2$~\cite{CZE2010}. This direct
measurement is to be compared with the PDG fit value of
$(0.198\pm0.009)$~MeV/$c^2$, which was obtained by summing the partial cross
sections and normalizing on the $\gamma\gamma \to\eta^{\prime}$ production
rate. The agreement is very reassuring.

%
%
\section{Two-pion production in nucleon-nucleon collisions} 
\label{2pi}\setcounter{equation}{0}%

\subsection{Two-pion production in proton-proton collisions} 
\label{pppipi}

\begin{figure}[hbt]
\begin{center}
\includegraphics[width=0.9\columnwidth]{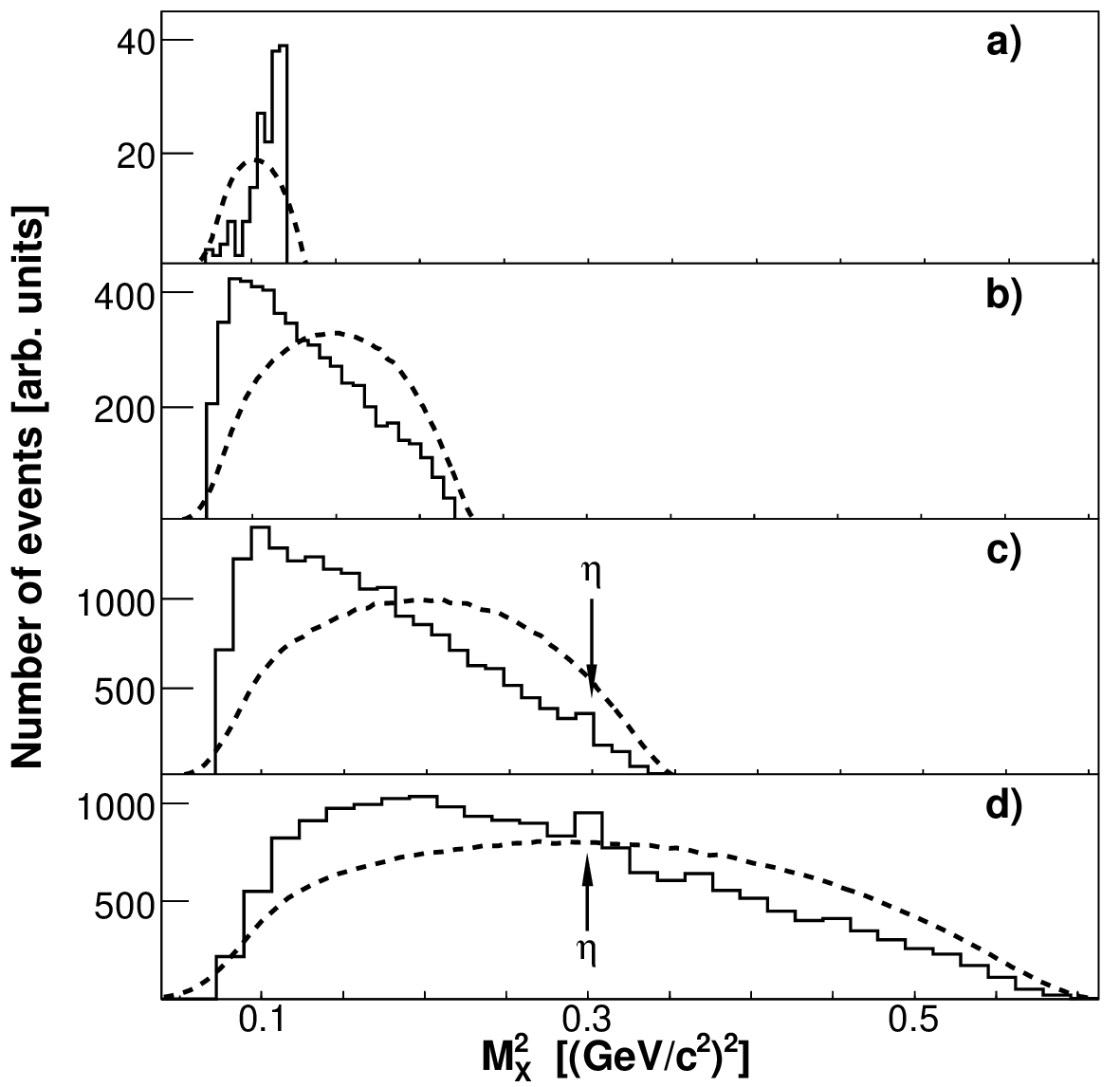}
\caption{\label{fig:ppABC} Distribution in missing-mass squared for the
$pp\to \{pp\}_{\!s}X$ reaction for $E_{pp}<3$~MeV and
$\cos\theta_{pp}>0.95$ at a) 0.8, b) 1.1, c) 1.4, and d)
2.0~GeV~\cite{DYM2009}. The $\eta$ signal seen at the the two higher energies
was already mentioned in sect.~\ref{eta}. The curves represent normalized
simulations within a phase-space model. }
\end{center}
\end{figure}

Although much of the excitement in recent years has been connected with
two-pion production in neutron-proton collisions, COSY has also made some
useful contributions in proton-proton collisions. The simplest of these to
discuss is the ANKE experiment where, in an inclusive measurement, only two
protons from the $pp\to ppX$ reaction were detected at $pp$ excitation
energies $E_{pp}<3$~MeV~\cite{DYM2009}. This was already mentioned in
connection with $\eta$ production in sect.~\ref{eta}, where it was stressed
that, under these conditions, the diproton acts kinematically like a single
particle. The acceptance at ANKE is restricted to very small angles and the
raw data shown in Fig.~\ref{fig:ppABC} were obtained for the cosine of the
c.m.\ diproton angle bigger than 0.95.

There is a region of missing masses $270 \lesssim M_X \lesssim 420$~MeV/$c^2$
where the data in Fig.~\ref{fig:ppABC} must correspond to two-pion production
and it was argued~\cite{DYM2009} that even at higher $M_X$ two-pion
production probably dominates. Even if this were true, one has no way of
knowing the relative weights of $\pi^+\pi^-$ and $\pi^0\pi^0$ in the final
state. Nevertheless, there is one intriguing feature to note in the data. At
1.1~GeV and possibly also at 1.4~GeV there is a strong enhancement compared
to phase space at low dipion masses. This is the so-called ABC
effect~\cite{ABA1963} that will be described in some detail in
sect.~\ref{pd3hepipi}. On the other hand, at 0.8~GeV, i.e., $Q\approx
80$~MeV, there is a kind of anti-ABC effect where the enhancement comes at
the largest dipion masses. Exactly the same behaviour is observed in the
$pd\to{}^{3}\mbox{\rm He}\,\pi^+\pi^-$ reaction by the MOMO
collaboration~\cite{BEL1999}, as discussed in sect.~\ref{pd3hepipi}. This
striking behaviour should be reproduced in any modeling of the ABC
phenomenon.

Whereas the ANKE measurements~\cite{DYM2009} covered only a tiny region of
phase space, much more global studies were undertaken in experiments carried
out at COSY-WASA and COSY-TOF. The TOF measurements of $pp\to pp\pi^+\pi^-$
were carried out at 747 and 793~MeV using a polarized proton
beam~\cite{BAR2008a}. Although both protons fell within the geometric
acceptance of TOF, the same was not true for the pions. However, since the
detection of two protons and one pion was sufficient to reconstruct the event
in TOF, most of the reaction phase space was covered, especially for the
unpolarized cross section, which is symmetric in the c.m.\ frame.

\begin{figure}[hbt]
\begin{center}
\includegraphics[width=0.49\columnwidth]{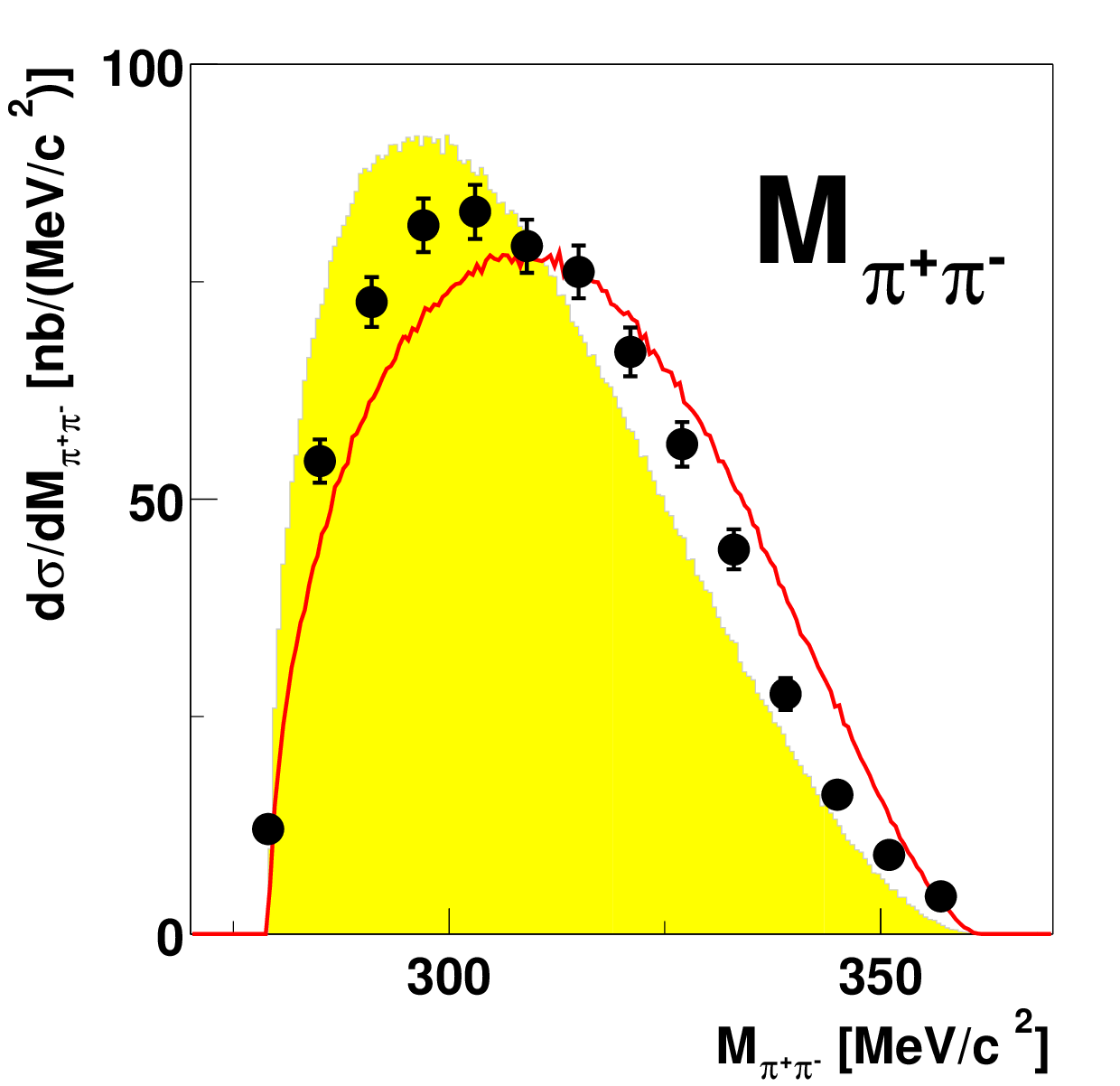}
\includegraphics[width=0.49\columnwidth]{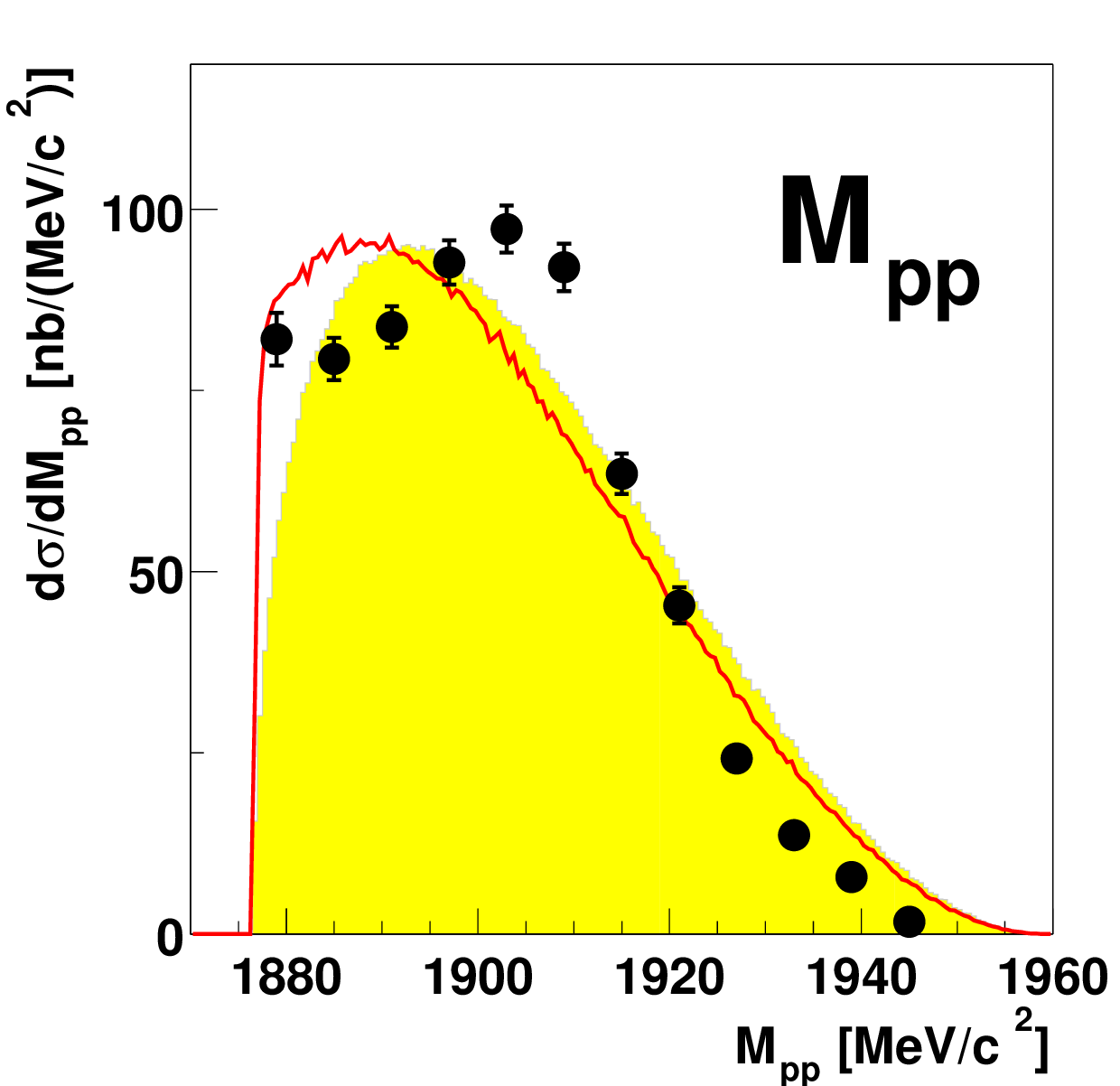}
\caption{COSY-TOF measurements of the differential distributions of the
invariant masses $M_{\pi\pi}$ and $M_{pp}$ in the $pp\to pp\pi^+\pi^-$
reaction at $T_p = 793$~MeV~\cite{BAR2008a}. They are compared to phase-space
distributions (shaded areas) as well as to an $N^*$-Roper-inspired model
(solid lines).\label{fig:toy}}
\end{center}
\end{figure}

The fully reconstructed events allowed the authors to extract a wide variety
of of one-dimensional distributions in invariant masses and proton and pion
angles and in Fig.~\ref{fig:toy} we show the differential cross section at
793~MeV in terms of the $\pi^+\pi^-$ and $pp$ masses. The ANKE
data~\cite{DYM2009} would correspond to just the first point in the $pp$
distribution and, if we make simple assumptions on the angular distributions,
it is clear that the normalizations of the ANKE and TOF data are at least
broadly consistent at 800~MeV. On the other hand, the shapes of the $\pi\pi$
distributions in Figs.~\ref{fig:ppABC} and \ref{fig:toy} at this energy look
very different and so any anti-ABC behaviour is only apparent for very low
values of $E_{pp}$.

\begin{figure}[hbt]
\begin{center}
\includegraphics[width=0.9\columnwidth]{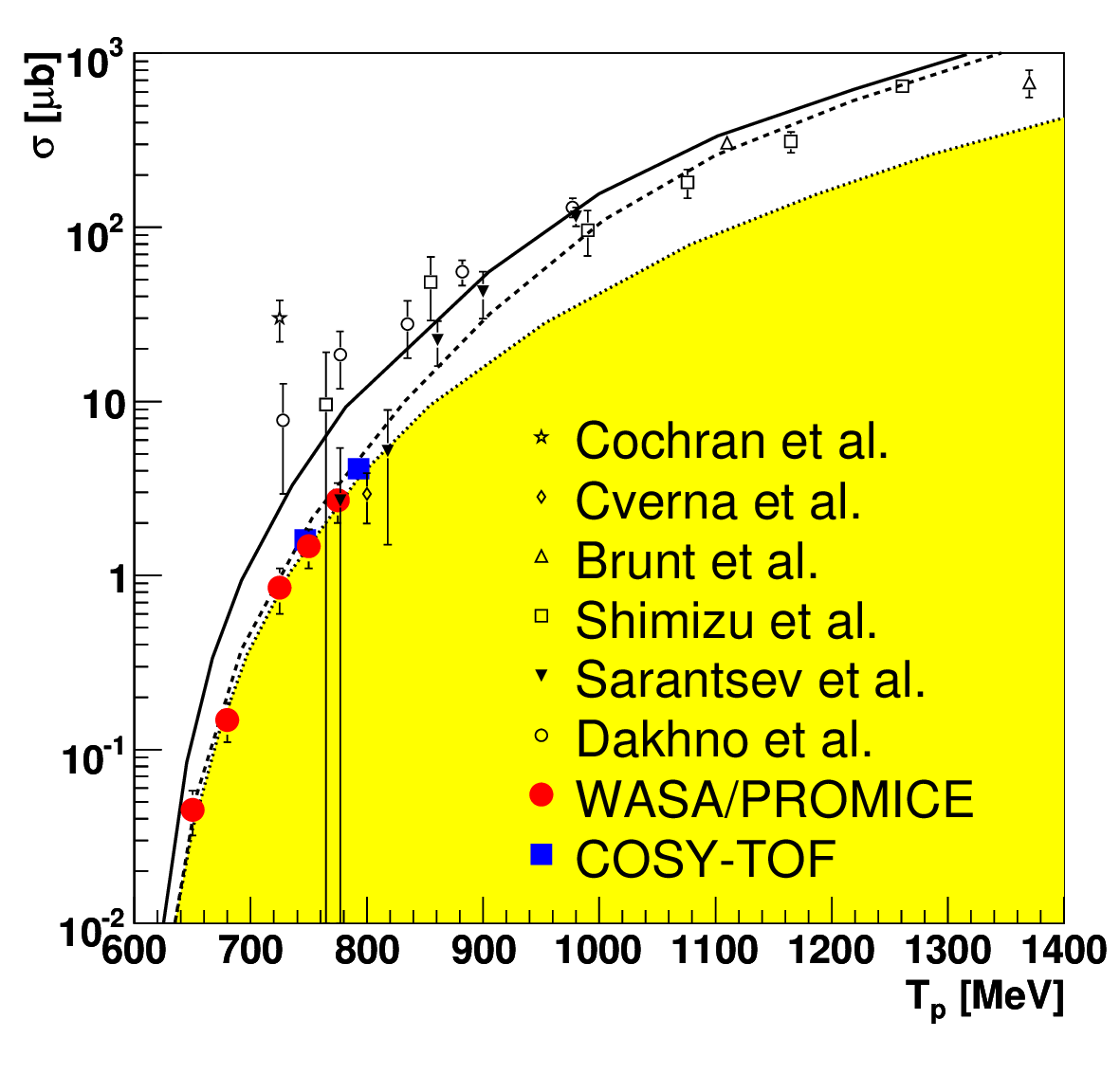}
\caption{Total cross section for the $pp\to pp\pi^+\pi^-$ reaction as a
function of the proton beam energy. Closed (red) circles are WASA data taken
at CELSIUS~\cite{PAT2003} whereas the (blue) squares were obtained by the
collaboration at COSY. A collection of older data is also
shown~\cite{OLDLOTT}. Solid and dashed curves correspond to theoretical
calculations of Ref.~\cite{ALV1998} with and without the $pp$ FSI. The shaded
area represents the phase space dependence adjusted arbitrarily  to the value
at $T_p = 750$~MeV. \label{fig:sigpipi} }
\end{center}
\end{figure}

Having a good absolute normalization is, of course, critical when one is
looking at the energy dependence of the total cross section measured at
different facilities. Shown in Fig.~\ref{fig:sigpipi} are the points obtained
by the COSY-TOF collaboration~\cite{BAR2008a} and values from an early
version of WASA at CELSIUS~\cite{PAT2003}. These data seem to behave more or
less like phase space, whose dependence is shown by the shaded area. On the
other hand, these points are low compared to other data in the
literature~\cite{OLDLOTT}, which are also shown. Two calculations by the
Valencia group~\cite{ALV1998}, one with and one without the $pp$ final state
interaction, are also illustrated. Since the $pp$ FSI must exist, it is clear
that more work is required on the theoretical modeling of this reaction.

For a four-body final state the analyzing power $A_y$ can be measured with
respect to several different planes. Thus the values of $A_y$ for the
final pion and dipion directions in the c.m.\ frame showed some small non-zero
signals at 750~MeV and, since the associated unpolarized cross sections were
fairly isotropic, $A_y$ could be fitted directly in terms of $\sin\theta$ and
$\sin 2\theta$~\cite{BAR2008a}.

On the basis of all the differential distributions, it was claimed that the
dominant mechanism involved the excitation of the Roper resonance, which
decayed through the emission of an $s$-wave dipion. This would not, of
course, describe the $pp\to\pi^+\pi^+nn$ reaction, which was searched for at
800~MeV by the COSY-TOF collaboration~\cite{SAM2009a}. Only an upper limit
was found and this was over an order of magnitude less than the cross
sections for producing other two-pion channels. The result did not therefore
invalidate the $N^*$(Roper) hypothesis for $\pi^+\pi^-$ production at low
energies.

However the situation for $pp\to pp\pi^0\pi^0$ was clarified significantly by
a subsequent measurement by the COSY-WASA collaboration at 1.4~GeV, where
about $5\times 10^5$ events were analyzed~\cite{TOL2012}. The $\pi^0p$
invariant mass distributions showed that at this energy the reaction was
driven mainly by intermediate $\Delta(1232)\Delta(1232)$ states. It would
therefore seem that there could be two competing mechanisms involved, with
the Roper $N^*$ being dominant only close to threshold.

%
%

\subsection{Two-pion production in neutron-proton collisions} 
\label{nppipi}

One of the most exciting measurements in medium energy nuclear physics in
recent decades was that of the differential cross section for the $pn\to
d\pi^0\pi^0$ reaction carried out by the WASA collaboration, first at
CELSIUS~\cite{BAS2009} and then more extensively at
COSY~\cite{PAD2011,PAD2013}. In quasi-free production on a deuterium target,
the centre-of-mass energy $W$ in the $pn$ system has to be reconstructed from
the measurements in WASA of the deuteron and the photons arising from the
decays of the two $\pi^0$. On the other hand, the deuterium target allowed
the measurement to be carried out at a range of values of $W$ while keeping
the proton beam energy fixed. The results were later confirmed at COSY by
using a (polarized) deuteron beam, where the fast spectator proton could be
measured explicitly~\cite{PAD2016}.

\begin{figure}[hbt]
\begin{center}
\includegraphics[width=1.0\columnwidth]{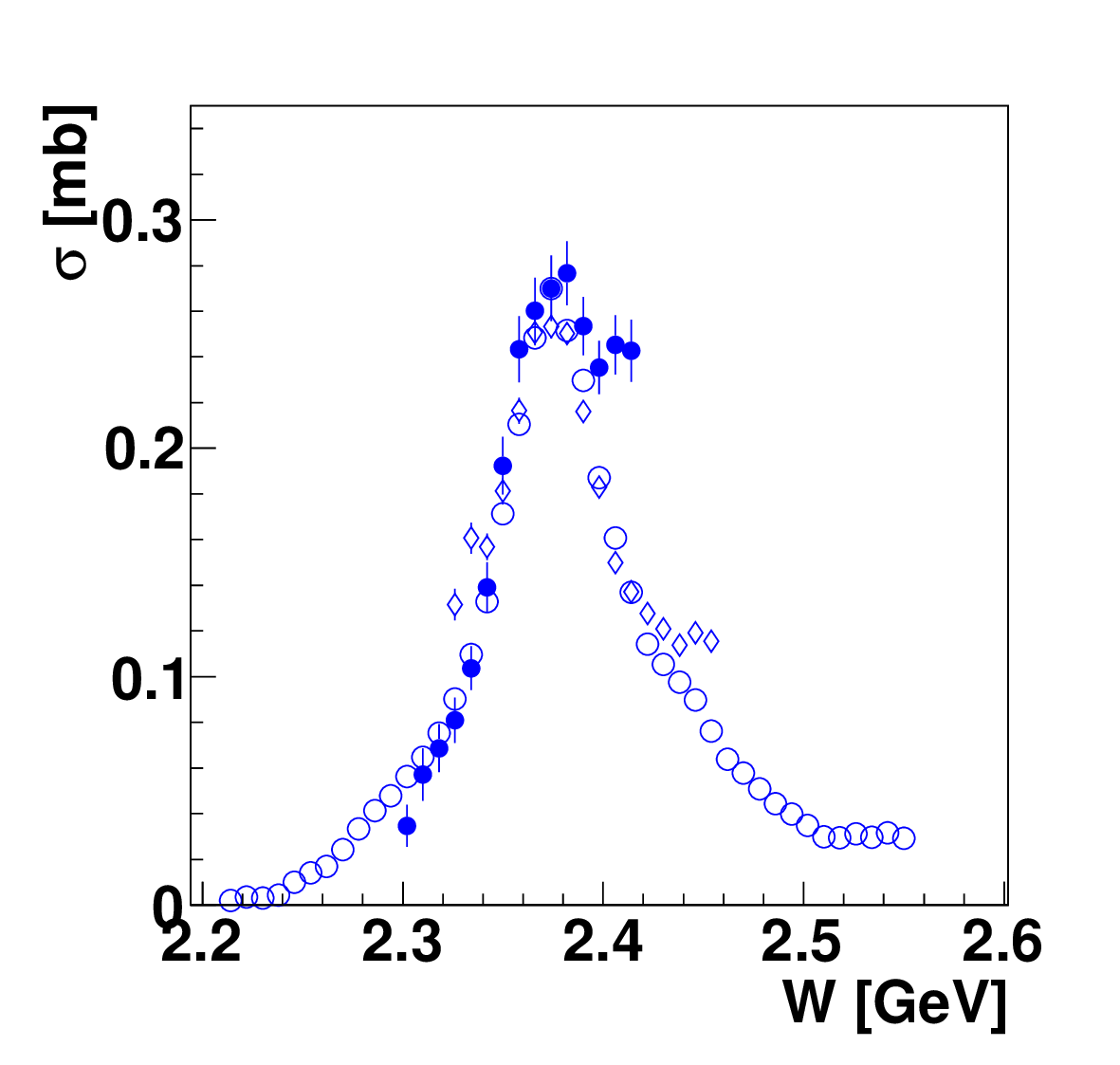}
\caption{Total cross section for the quasi-free $pn\to d\pi^0\pi^0$ reaction
as a function of the total energy $W$ in the c.m.\ frame. The data with a
deuterium target are taken from Refs.~\cite{PAD2011} (open circles) and
\cite{PAD2013} (open diamonds). Results obtained with a deuteron beam come
from Ref.~\cite{PAD2016} (closed circles). The data from the two other sets
were normalized to those of Ref.~\cite{PAD2013} at the maximum of the
peak.\label{fig:peak}. }
\end{center}
\end{figure}

Figure~\ref{fig:peak} shows the WASA measurements of the total cross section
for the quasi-free $pn\to d\pi^0\pi^0$ reaction~\cite{PAD2011,PAD2013} and
the $np\to d\pi^0\pi^0$ reaction~\cite{PAD2016}. All these data are
consistent and indicate a sharp peak at a mass of 2.38~GeV/$c^2$ and a width
of about 70~MeV/$c^2$. It should be noted at this point that, in order to
avoid using events corresponding to very large Fermi momenta in the deuteron,
the full width of the peak in Fig.~\ref{fig:peak} was scanned by using more
than one setting of the proton beam momentum.

It was suggested by the WASA authors that this peak corresponded to a
resonance with baryon number equal to two, i.e., that it was a dibaryon,
which they denoted by $d^*(2380)$. The other interesting piece of
experimental information is that the two-pion spectrum associated with this
peak seems to display the ABC effect, which shows up as an enhancement in the
$\pi^0\pi^0$ mass distribution that is strongest around
310~MeV/$c^2$~\cite{ABA1963}. At such a low mass the pions must be dominantly
in a relative $s$-wave so that the dipion has then quantum
numbers\footnote{Isospin-two pion pairs are ruled out for the $d\pi^0\pi^0$
final state by overall isospin conservation.} $(J^P,I)=(0^+,0)$, which means
that the peak in Fig.~\ref{fig:peak} must also be in the isospin $I=0$
channel. If, as seems likely, the dynamics of two-pion production are driven
by an intermediate $\Delta(1232)\Delta(1232)$ state then the Pauli principle
requires them to be antisymmetric so that the peak must correspond to either
the $J^P=1^+$ or $3^+$ wave. It was argued that the angular distributions
strongly favoured the $J^P=3^+$ assignment~\cite{PAD2011}. A similar
conclusion was also reached using very different reasoning, to which we now
turn.

In sect.~\ref{Phenomenology} a method was presented to make simple estimates
of the cross section for the production of $S$-wave isoscalar $np$ states in
the reaction $pp\to \{pn\}\pi^+$ in terms of that for $pp\to d\pi^+$. This
can be extended to estimate the rate for $pn\to\{pn\}\pi\pi$ in terms of that
for $pn\to d\pi\pi$~\cite{FAL2011,ALB2013}. Due to the different kinematic
factors, these estimates have to be made separately for the two spin
hypotheses but, when this is done for the $J^P=1^+$ case, it is seen that the
sums of the $d\pi\pi$ and $\{pn\}\pi\pi$ productions significantly exceeds
the total inelastic cross section in the SAID SP07 solution~\cite{ARN2000} in
the combined $^{3\!}D_1+^{3\!}S_1$ states. This argument is, of course, not
watertight because the neutron-proton input to the SAID solution is rather
incomplete above 1~GeV. Nevertheless, it does suggest that, if there is a
dibaryon resonance, then it is more likely to be in the $3^+$ wave, where the
inelasticity constraints are much less severe.

The use of the deuteron beam at COSY allowed the WASA group to measure the
quasi-free $dp\to p_{\rm sp}pn\pi^0\pi^0$ cross section in the vicinity of
the $d^*(2380)$~\cite{PAD2015}. Only six values close to the resonance peak
were obtained and the maximum in the $pn\to pn\pi^0\pi^0$ total cross section
was found to be $295\pm14\pm29~\mu$b compared to the 275~$\mu$b for $np\to
d\pi^0\pi^0$ shown in Fig.~\ref{fig:peak}. Of course the $pn\to pn\pi^0\pi^0$
reaction also has contributions associated with isovector $np$ pairs. Since
there is also production of isospin-two $\pi^0\pi^0$ pairs in the $pn\to
pn\pi^0\pi^0$ reaction, it is not possible to make model-independent
estimates of the extra $I=1$ $np$ contributions.

The WASA authors~\cite{PAD2015} presented results based upon a modified
Valencia model that had been tuned to fit the $pp\to pp\pi^0\pi^0$
data~\cite{ALV1998}. This predicts about 100~$\mu$b for the $I=1$ $np\to
np\pi^0\pi^0$ cross section at the $d^*(2380)$ peak but is is impossible to
quantify the associated theoretical uncertainty. Nevertheless, if we accept
this value, it means that $\sigma(np\to\{np\}_{I=0}\pi^0\pi^0)/\sigma(np\to
d\pi^0\pi^0) \approx 0.7$ compared to the 0.8--0.9 predicted in the simplest
FSI model~\cite{FAL2011}. Though this does not prove the dibaryon assertion,
it clearly does not invalidate it.

In the search for extra data to test the $d^*(2380)$ hypothesis, the WASA
collaboration also extracted analyzing powers $A_y$ of the quasi-free
$\pol{n}p\to d\pi^0\pi^0$ from the polarized deuteron beam
data~\cite{PAD2016}. It is expected that a non-zero value of $A_y$ would
arise from an interference of the $d^{*}$ with the non-resonant background.
However, the data are hard to interpret, in part due to the limited range of
masses covered. Below resonance the analyzing powers with respect to the
final deuteron direction are small. They do increase with $W$, but it is
difficult to see in these data the rapid phase variation associated with the
$d^*(2380)$ pole.

\begin{figure}[hbt]
\begin{center}
\includegraphics[width=1.0\columnwidth]{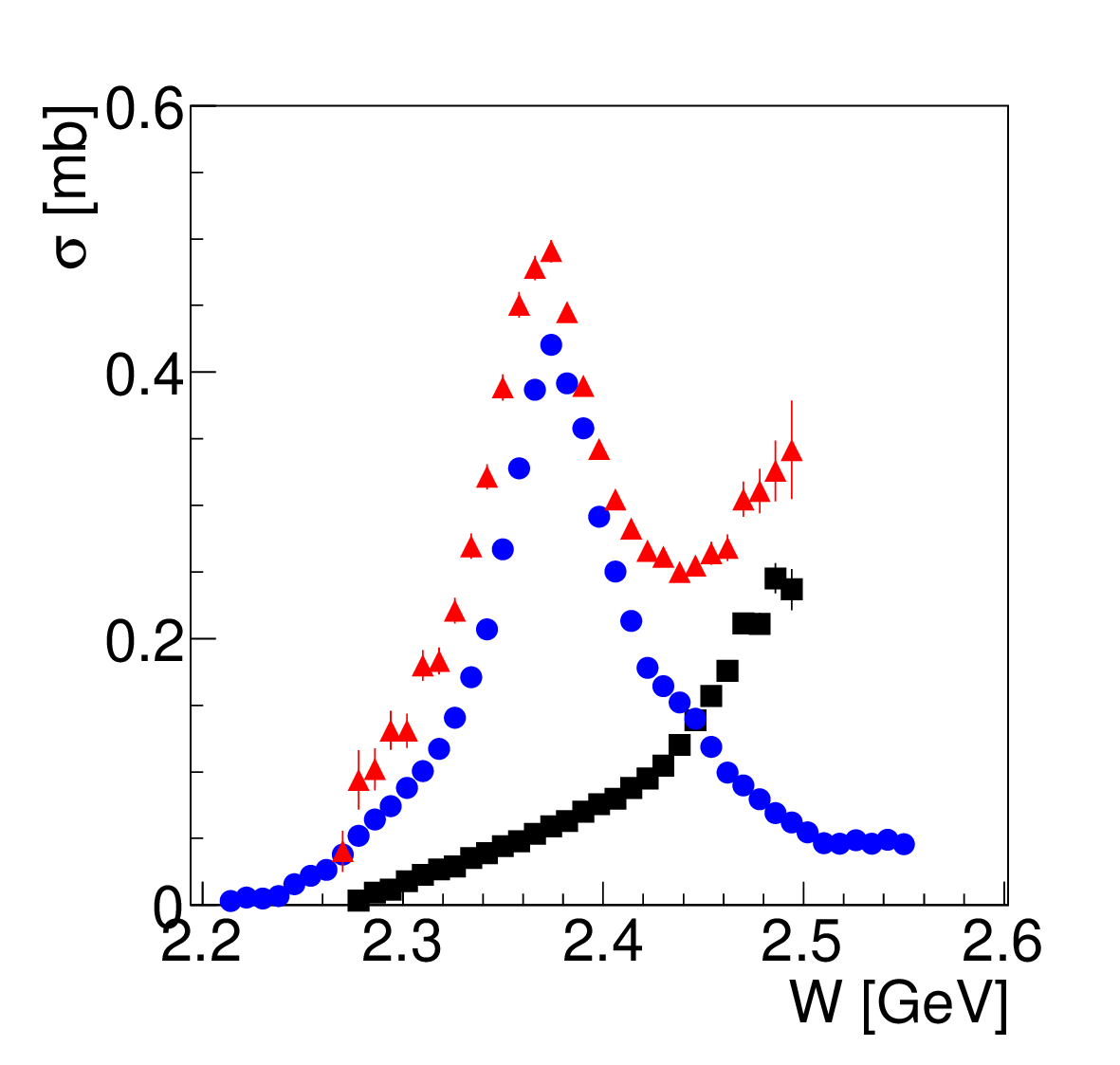}
\caption{Dependence of the total cross section for the $pn\to d\pi^+\pi^-$
reaction (red triangles) and its isospin decomposition into an isoscalar part
that should be compared to $2\sigma(pn\to d\pi^0\pi^0)$ (blue circles) and an
isovector part corresponding to $\frac{1}{2}\sigma(pp\to d\pi^+\pi^0)$ (black
squares) as functions of the centre-of-mass energy
$W$~\cite{PAD2013}.\label{fig:BAS8} }
\end{center}
\end{figure}

The group also made measurements at COSY of the closely related $pn\to
d\pi^+\pi^-$ and $pp\to d\pi^+\pi^0$ reactions in the $d^*(2380)$
region~\cite{PAD2013}. In the first reaction the $pn$, and hence the
$\pi^+\pi^-$ system, is a mixture of isospin $I=0$ and $I=1$, whereas the
$\pi^+\pi^0$ system must be purely $I=1$. Since these amplitudes do not
interfere in the expression for the total cross section, they can be
subtracted to give the pure $I=0$ cross section and this has been done in
Fig.~\ref{fig:BAS8}. Within experimental uncertainties the directly measured
$I=0$ total cross section shown in Fig.~\ref{fig:peak} is consistent with the
one measured indirectly and presented in Fig.~\ref{fig:BAS8}. Both data sets
show the very strong peaking for $W\approx 2380$~MeV.

The $d^*(2380)$ interpretation has been questioned~\cite{ALB2013} on the
basis of the comparison of $pn\to d\pi^+\pi^-$ and $pn\to pn\pi^+\pi^-$ data
but, as discussed in the context of a higher energy
experiment~\cite{AGA2015}, any apparent discrepancy might be connected with
the limited $pn\to pn\pi^+\pi^-$ database in the vicinity of the resonance
peak.

Although the original idea~\cite{RIS1973} that the $\Delta\Delta$ channel
might serve as the entrance channel for the $np\to d\pi^0\pi^0$ reaction, and
hence for the ABC effect, might be valid, no realistic calculations have yet
reproduced the striking behaviour seen in Fig.~\ref{fig:peak}. One obvious
problem is that the width of the structure is of the order of 70~MeV compared
to the 120~MeV that one normally associates with the $\Delta(1232)$. In part
this difference may be connected with the reduction in the average $\Delta$
mass to 1190~MeV/$c^2$, which would certainly have an influence on the
$p$-wave decays. Pauli blocking may also have some effect but this might be
compensated by the extra width coming from the $\Delta\Delta \to np$ decay.

Even if the dibaryon exists, the interesting question is, of course, whether
the relevant degrees of freedom are those of six quarks or those of
$\Delta\Delta$, i.e., pions and nucleons. The nucleon-nucleon force that
gives rise to the only stable dibaryon, the deuteron, is conventionally
described in terms of nucleons and mesons with the binding depending
critically on many refinements, such as the $S$--$D$ coupling driven by the
tensor force of the one-pion-exchange. The coupled $NN$:$\Delta\Delta$ force
is likely to be even more complicated. Even if such a force could generate a
$d^*(2380)$ pole it will not necessarily describe quantitatively the $pn\to
d\pi^0\pi^0$ reaction; all the many angular and Dalitz plot distributions
extracted by the WASA collaboration must also be explored.

The $d^*(2380)$ peak in the $pn\to d\pi^0\pi^0$ reaction seems to be
associated with the ABC effect in the $\pi^0\pi^0$ system. This is not
unexpected if both are driven by the $\Delta\Delta$ intermediate state. The
ABC effect is seen also in $pd\to{}^{3}$He$\,\pi^0\pi^0$ and, most
spectacularly, in $dd\to{}^{4}$He$\,\pi^0\pi^0$~\cite{KEL2009,PAD2012}.
However, in the $dd$ case the cross section and deuteron tensor analyzing
power have been described using conventional physics without the need for the
$d^*(2380)$~\cite{GAR1999}. It may therefore be that there is no one single
mechanism that is responsible for generating the ABC effect for all
reactions. More evidence is certainly required before it is safe to assume
that the existence of an ABC effect must be a signal for the importance of
the $d^*(2380)$ in a particular reaction.

In summary, the WASA collaboration have found a very striking peak in the
$pn\to d\pi^0\pi^0$ reaction and all the data, including the energy
dependence of the neutron analyzing power in $np$ elastic scattering, seem to
be consistent with the $d^*(2380)$ dibaryon hypothesis. Even if it is later
shown that the dibaryon assumption is untenable, the group will still have
made a remarkable and most unexpected discovery in the domain of hadron
physics.

%
%
\section[Inclusive strangeness production]{Inclusive strangeness production} 
\label{InclusiveK}\setcounter{equation}{0}%

\subsection[The $pp\to K^+X^+$ reaction]{The $\boldsymbol{pp\to K^+X^+}$ reaction} 
\label{HIRES}

If the momentum of a well-identified $K^+$ emerging from proton-proton
collisions is measured then this will provide information on the recoiling
system, $X^+$, which has baryon number $+2$ and strangeness $-1$. For missing
masses below the $\Sigma N$ threshold, $m_X < m_{\Sigma} + m_N$, the only
strong interaction channel that is allowed is $pp\to K^+ \{\Lambda p\}$ so
that these measurements provide a simple way to investigate the $\Lambda p$
interaction. The situation is far more complicated at higher $m_X$ because,
in addition to real $\Sigma$ production, there is strong channel coupling
between $\Lambda p$ and $\Sigma N$ final states. This problem will be
discussed in connection with the cusp phenomenon in sect.~\ref{cusp} and
$\Sigma^+$ production in sect.~\ref{Sigmap}.

The first high-resolution measurement of the $pp\to K^+X^+$ reaction was
carried out at SATURNE II with proton beam energies of 2.3 and 2.7~GeV at
four fixed laboratory angles at each energy~\cite{SIE1994}. The outgoing
kaons were detected at small angles in the focal plane of the SPES4
spectrometer. Decay corrections were important because of the length of this
spectrometer.

Characteristic structures were seen at both the $\Lambda p$ and $\Sigma N$
thresholds and the first of these could be unambiguously associated with the
strong and attractive $\Lambda p$ final state interaction. These data were
therefore used to extract estimates for the scattering length and effective
range from the low energy $\Lambda p$ data. A major difficulty in the
determination of low energy $\Lambda p$ parameters from these data within a
final state interaction model was the resolution in the missing-mass, which
was typically about 4~MeV/$c^2$. An attempt was made by Laget~\cite{LAG1991}
to describe the whole data set, though one must recognize the ambiguities
inherent in such an inclusive measurement. It is also important to note that
the $S$-wave $\Lambda p$ system can be in either the spin-triplet or
spin-singlet state and the $pp\to K^+\Lambda p$ reaction produces some
mixture of these that need not follow a statistical population rule.

In addition to the $\Lambda$ and $\Sigma$ threshold phenomena, the SPES4
group found suspicions of a peak in the vicinity of 2097~MeV/$c^2$, though
its statistical significance was far from convincing~\cite{SIE1994}. The
experiment was therefore repeated at beam energies of 1.953 and 2.097~GeV by
the HIRES collaboration using the Big Karl spectrometer, where a missing-mass
resolution of $\sigma_M \approx 0.84$~MeV/$c^2$ was
achieved~\cite{BUD2010a,BUD2011}.

\begin{figure}[hbt]
\begin{center}
\includegraphics[width=0.9\columnwidth]{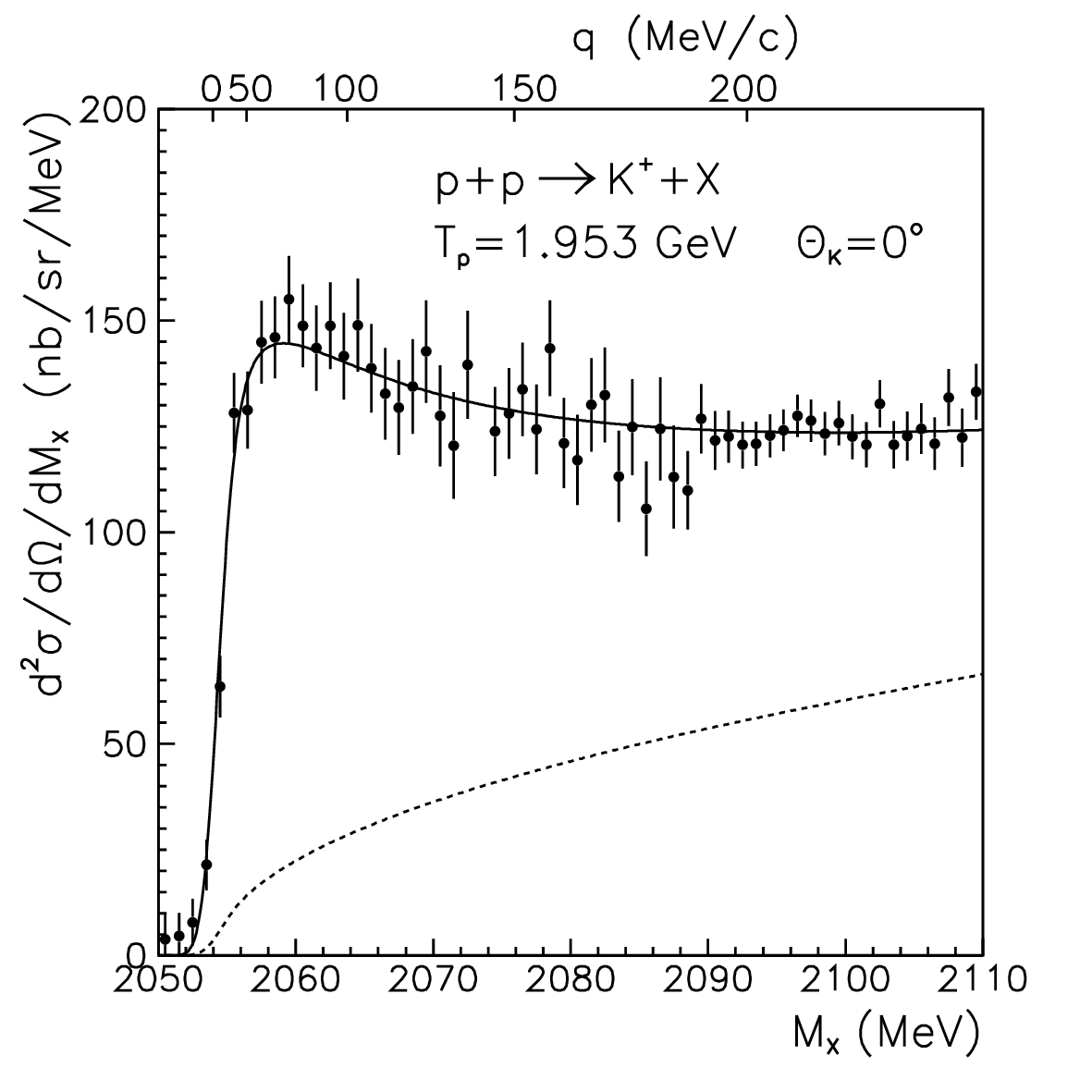}
\caption{\label{fig:HIRES4} Missing-mass spectrum of the reaction $pp\to
K^+\{\Lambda p\}$ measured at $T_p = 1.953$~GeV with the Big Karl
spectrometer placed at $\theta_K =0^{\circ}$~\cite{BUD2010a}. The upper axis
indicates the c.m.\ momentum $q$ of the $\Lambda p$ system. Solid line:
Fit including the $\Lambda p$ FSI. Dashed line: $pp\to K^+\Lambda p$ phase space
distribution.}
\end{center}
\end{figure}

In order to cover the range $2050 - 2110$~MeV/$c^2$ in missing mass, data
were taken using three overlapping settings of the spectrometer, with
enhanced luminosity in the highest mass interval. As shown in
Fig.~\ref{fig:HIRES4}, no structure was evident in the 2097~MeV/$c^2$ region
and upper limits were determined on the production cross sections of narrow
strange dibaryons over the whole missing-mass range~\cite{BUD2011}.

Also clearly seen in Fig.~\ref{fig:HIRES4} is the rapid rise from the
$\Lambda p$ threshold, which is very unlike the shape of the three-body phase
space that is also shown (with arbitrary normalization). The HIRES authors
tried to fit simultaneously this spectrum together with the limited data on
$\Lambda p$ elastic scattering assuming that the final state interaction
could be parameterized by the Jost function in the form of
$(q-i\alpha)/(q+i\beta)$ for both the singlet and triplet $\Lambda p$
states~\cite{BUD2010a}. However, it was difficult to reconcile the two data
sets and they found a best fit to the two sets where spin-singlet production
completely dominated the $pp\to K^+\Lambda p$ reaction. If the $\Lambda p$
system is indeed in a pure singlet state then the analyzing power measured
with polarized protons should be antisymmetric around $\theta_K=90^{\circ}$.
The COSY-TOF data discussed in sect.~\ref{Ayanda} do not support such a
conclusion.

By considering only the HIRES production data they found scattering length
and effective range of $\bar{a}=-2.43\pm0.16$~fm and $\bar{r}_0=2.21\pm
0.16$~fm, but these values represent some unknown averages for singlet and
triplet production~\cite{BUD2010a}. In terms of the Jost function parameters,
$\alpha = -0.31$~fm$^{-1}$ and $\beta =1.215$~fm$^{-1}$, so that there is a
virtual state of the $\Lambda p$ system with a ``binding energy'' of about
3.6~MeV. We will return later to attempts to the determine the scattering
length in the context of the exclusive COSY-TOF measurements.

\subsection[Hypernuclei lifetime measurements]{Hypernuclei lifetime measurements} 
\label{COSY13}

In free space the $\Lambda$ hyperon decays principally through the channels
$\Lambda \to p\pi^-$ or $\Lambda\to n\pi^0$ with a mean lifetime of
$\tau_{\rm free}=263\pm2$~ps~\cite{OLI2014}. The energy release in such a
decay is about 38~MeV so that, when a $\Lambda$ is bound deep inside a heavy
hypernucleus, these decays are strongly suppressed by the Pauli blocking of
the recoil neutron or proton. On the other hand, this reduction might be
compensated by the non-mesonic decays $\Lambda p\to np$ or $\Lambda n\to nn$.
The nucleons from such decays have typically energies of about 80~MeV, so
that they are largely unaffected by nuclear effects. It is therefore
suggested that the study of the lifetimes of heavy hypernuclei might be a
useful way of investigating non-mesonic decays.

\begin{figure}[hbt]
\begin{center}
\includegraphics[width=0.8\columnwidth]{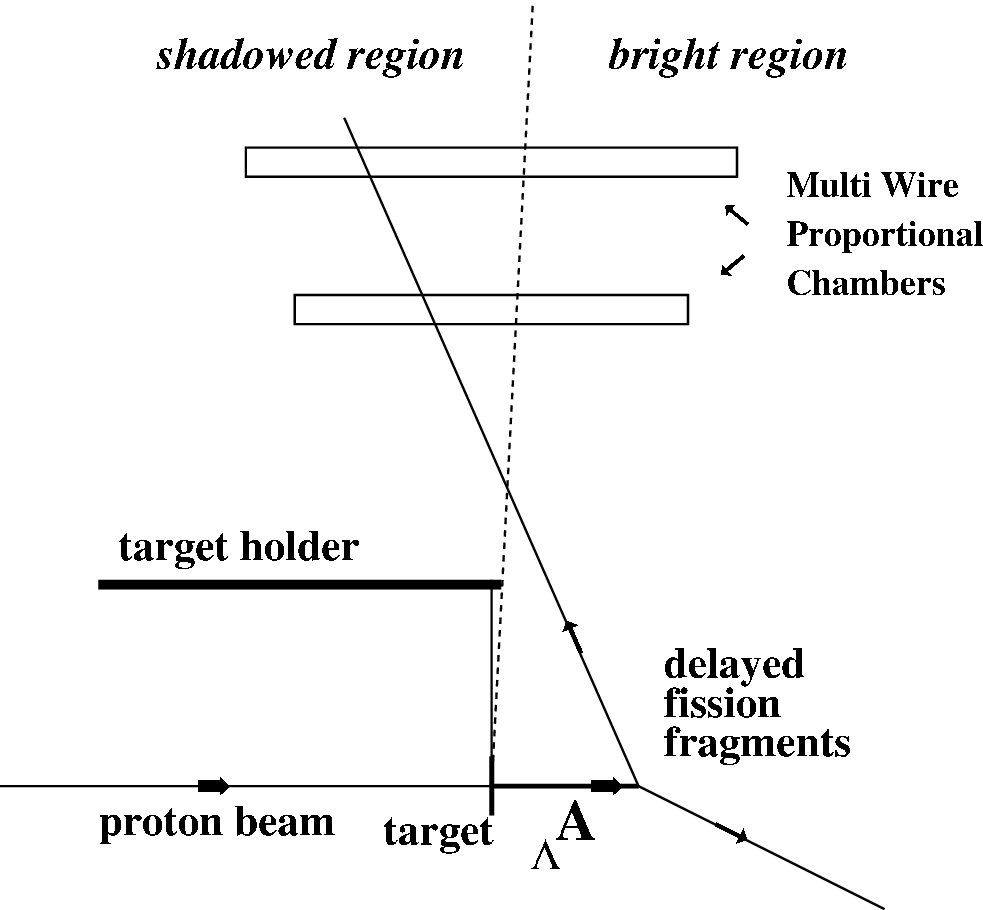}
\caption{\label{fig:CASSING4} Schematic view of the COSY-13 experimental
setup illustrating the recoil shadow method. An idealized trajectory of a
decay product from a produced hypernucleus  is shown arriving in the region
of the shadow. A particle coming directly from the target region is allowed
in the forward hemisphere but is blocked by the target holder in the backward
hemisphere. }
\end{center}
\end{figure}

The lifetimes of heavy hypernuclei have been investigated through the
interaction of antiprotons with Bismuth and Uranium, but the resulting error
bars are quite large~\cite{ARM1993}. The COSY-13 collaboration~\cite{CAS2003}
measured the decay of hypernuclei produced in the interaction of $\approx
1.9$~GeV protons with Bi, Au, and U targets using the recoil shadow method.
The principle of the technique is illustrated schematically in
Fig.~\ref{fig:CASSING4}.

After the proton beam hits a thin ribbon target, any particle emerging
directly from the intersection region is blocked by the target holder so that
it is not detected on the left hand side of the multiwire proportional
chambers. When a produced hypernucleus travels along the beam direction, the
products of its delayed decay may indeed reach the shadowed region, as
indicated by the idealized trajectory shown in the figure. The obvious
difficulty with this approach arises from the low production rate of
hypernuclei, such that the background from fragments of non-$\Lambda$ delayed
fission of recoil nuclei populates significantly the shadowed region. This
was estimated using data taken at a lower proton beam energy, where
hypernucleus formation is negligible.

The distribution of events in the shadowed region will reflect the
hypernucleus lifetime but significant modeling is required because the
recoils do not emerge from the target with a unique speed and, moreover, they
do not represent a unique hypernuclear species. Nevertheless, the lifetimes
deduced from all three targets are consistent and give a mean average of
$\tau = (145\pm 11)$~ps. This is also compatible with the value of
$(143\pm36)$~ps obtained from antiproton interactions in Bi and U
targets~\cite{ARM1993}.

In terms of the free $\Lambda$ decay time, the COSY-13 result may be written
as $\tau=(0.55\pm 0.04)\,\tau_{\rm free}$. Such a value is hard to explain
theoretically but could be understood if the $\Lambda N \to NN$ transition
were much stronger on neutrons than protons, but this would imply a violation
of the $\Delta I = \half$ rule~\cite{CAS2003}. However, this is in conflict
with the neutron/proton transition ratio of $0.51\pm0.14$ found for
$_{\phantom{,}\Lambda}^{12}$C~\cite{KIM2006}. Another alternative might be
that the systematic effects were underestimated in the COSY-13 experiment.
The COSY-13 result was criticized by the authors of an electroproduction
experiment~\cite{QIU2012} but that paper was subsequently withdrawn from the
arXiv!

It has been stressed that ``COSY-13 was a simple experiment while we were
waiting for ANKE to be ready''~\cite{SCH2015}.

\subsection[Inclusive $K^+$ production on nuclei]{Inclusive $\boldsymbol{K^+}$ production on nuclei} 
\label{sub-threshold}\setcounter{equation}{0}%

The threshold for producing a $K^+$ in proton-proton collisions is at
$T_p=1.58$~GeV but the meson might be produced at much lower energies in
collisions with a nuclear target due to a variety of effects, including Fermi
motion, two-step contributions, clustering, and kaon-nucleus potentials.
These are all phenomena that are exciting to investigate.

The inclusive momentum spectrum of $K^+$ emitted at laboratory angles
$\theta_K< 12^{\circ}$ was measured for 1.0~GeV protons hitting C, Cu, and Au
targets~\cite{KOP2001}. The experiment was carried out at ANKE using the
delayed-veto technique on the range telescopes~\cite{BUS2002}. Each telescope
covered a well-defined interval of $K^+$ momentum $p_K$ so that the 15
elements spanned the momentum range from 200 to 520~MeV/$c$~\cite{KOP2001}.
There were typically 100 $K^+$ counts per telescope from a four-day run with
a carbon target. Interpreted naively, the resulting cross section spectrum
would suggest that of the order of 5-6 target nucleons were involved in the
production process.

Many of the experimental uncertainties cancel when evaluating ratios of cross
sections for different nuclear targets and the average values measured were
$R(\textrm{Cu/C}) = 4.0\pm0.3$ and $R(\textrm{Au/C}) = 6.8\pm0.38$, but with
a slight indication that these values might increase with $K^+$ momentum.

\begin{figure}[hbt]
\begin{center}
\includegraphics[width=0.8\columnwidth]{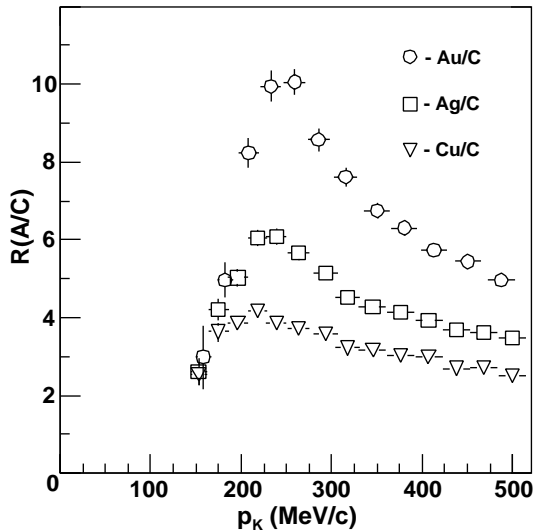}
\caption{\label{fig:Nekipelov} Ratios of the $K^+$ production cross sections
on Cu, Ag, and Au to C measured at a proton beam energy of $T_p=2.3$~GeV as a
function of the laboratory kaon  momentum~\cite{NEK2002}.}
\end{center}
\end{figure}

Though it is difficult to get a clear message from the 1.0~GeV data, the
group also made similar measurements with the same strip targets plus Ag
below and above the free $pp$ threshold, where the cross sections are
naturally much higher~\cite{NEK2002}. They measured the production cross
sections relative to C as a function of the $K^+$ momentum for 1.5, 1.75, and
2.3~GeV protons. What is striking is the rapid decrease in the ratios for all
targets and energies for $p_K\lesssim 200$~MeV/$c$. This is illustrated for
the 2.3~GeV data in Fig.~\ref{fig:Nekipelov}.

A large part of the suppression is due to the repulsive Coulomb potential
$V_C(r)$ between the $K^+$ and the residual nucleus. The situation has a
parallel in the well-known suppression of $\beta^+$ emission in heavy nuclei
at low positron momenta. Thus a $K^+$ produced at rest at some radius $R$ in
the nucleus would, in the absence of all other interactions, acquire a
momentum of $p_{\rm min} = \sqrt{2m_KV_C(R)}$. Taking $R$ to be the nuclear
edge, this purely classical argument leads to a minimum $K^+$ momentum for Au
of about 130~MeV/$c$. It is thought that, in addition, the strong interaction
$K^+$-nucleus potential is itself mildly repulsive. A fit to the data within
a transport calculation suggests that this is about $+20$~MeV at normal
nuclear matter density, $\rho_0\approx 0.16$~fm$^{-3}$~\cite{NEK2002}.

Later experiments by that group involved also a deuterium target with the aim
of investigating the production on the neutron~\cite{BUS2004}, but the
resulting uncertainties were very large.

%
%
\section{Hyperon production} 
\label{hyperon}\setcounter{equation}{0}%

\begin{figure}[htb]
\begin{center}
\includegraphics[width=1.0\columnwidth]{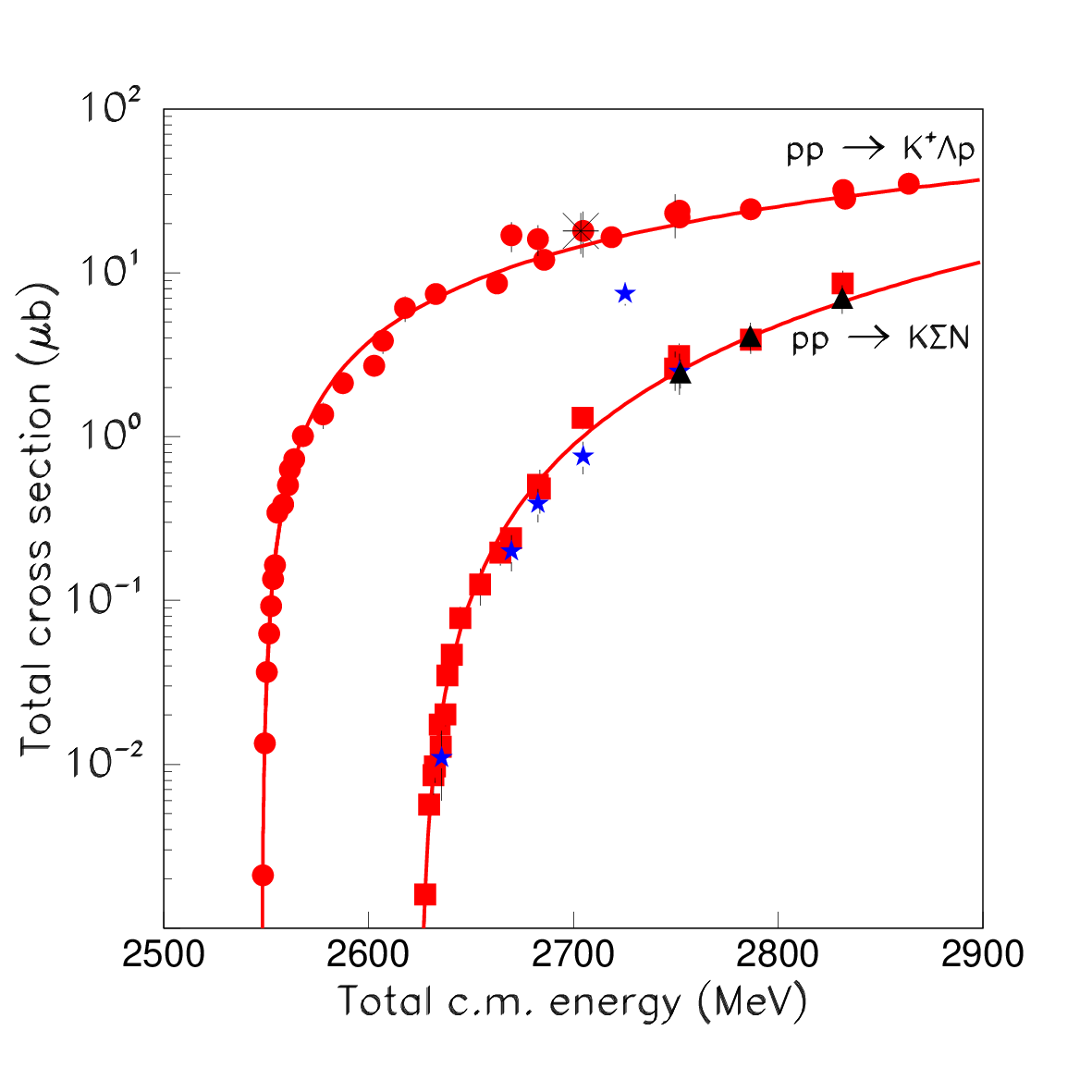}
\caption{\label{fig:Overview2} Total cross sections for $\Lambda$ and
$\Sigma$ production in proton-proton collisions near threshold as functions
of the total energy in the  c.m.\ system. Since the values for the various
$\Sigma$ channels overlap, the $pp\to K^+p\Sigma^0$ data are denoted by (red)
inverted triangles, the $pp\to K^+n\Sigma^+$ by (blue) stars, and the $pp \to
K^0p\Sigma^+$ by (black) triangles. All the data were obtained at COSY with
the exception of a bubble chamber measurement of $\Lambda$
production~\cite{FIC1962} that is shown by the (black) cross. The
phenomenological curves are discussed in the text.
 }
\end{center}
\end{figure}

Almost all measurements in hyperon production in proton-proton collisions
close to threshold were carried out at COSY. This dominance is well
illustrated by the summary presented in Fig.~\ref{fig:Overview2}, where the
only non-COSY point is derived from the 11 bubble chamber events
corresponding to the $pp\to K^+p\Lambda$ reaction~\cite{FIC1962}.

\subsection[The $pp\to K^+p\Lambda$ and $pp\to K^+p\Sigma^0$ reactions]{The $\boldsymbol{pp\to K^+p\Lambda}$ and $\boldsymbol{pp\to K^+p\Sigma^0}$ reactions} 
\label{Lambda}

Away from the threshold region the acceptance of the COSY-11 spectrometer
decreases rapidly and the extrapolation to the whole of phase space that is
necessary in order to evaluate a $pp\to K^+p\Lambda$ or $pp\to K^+p\Sigma^0$
total cross section becomes more model dependent. Nevertheless several
pioneering measurements of the total cross sections near threshold were made
at this facility~\cite{BAL1998,SEW1999,KOW2004}. Though the acceptance of the
ANKE spectrometer is somewhat larger than that of COSY-11, it has rather
similar limitations and this also restricted its use at the higher COSY
energies~\cite{VAL2007,VAL2010}.

In contrast, COSY-TOF has a much larger geometric acceptance and this enabled
reliable measurements to be made up to higher energies and also yielded
differential distributions that are so valuable for understanding the
underlying physics~\cite{BIL1998,SAM2006,SAM2010,ABD2010,SAM2013,ROD2013}.
Results on hyperon production were obtained at COSY-TOF at beam momenta of
2.5, 2.59, 2.68, 2.7, 2.75, 2.85, 2.95, 3.06, 3.2, and 3.3~GeV/$c$, though
the data in later years were much more detailed as the equipment was refined
through the addition of the straw tubes mentioned in sect.~\ref{TOF}.
However, some of the COSY-TOF results are only available in theses and should
be treated as preliminary until appearing in regular publications.

\begin{figure}[h!]
\begin{center}
\includegraphics[angle=-90,width=0.9\columnwidth]{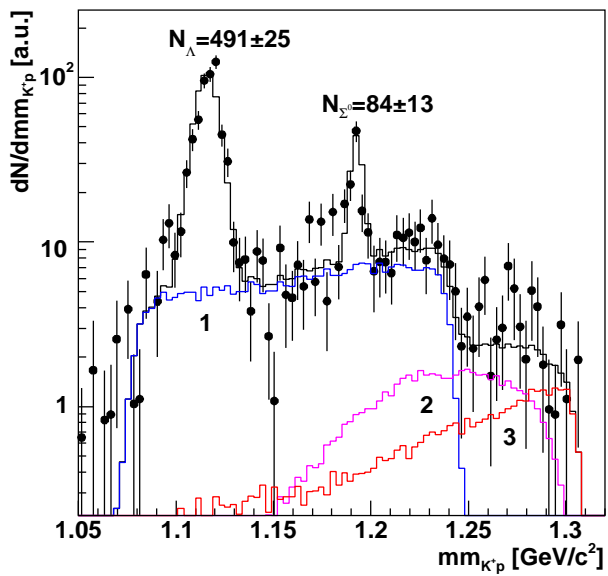}
\caption{\label{fig:Val1} Missing--mass distribution of $K^+p$ pairs from
$pp$ collisions at 2.16\,GeV~\cite{VAL2007}. The two peaks correspond to
direct protons from the $pp \to K^+p \Lambda/\Sigma^0$ reactions. The
continua, evaluated in Monte Carlo simulation, arise from secondary protons
from the $pp\to K^+p~( \Lambda \to \pi^-p)$ (histogram 1), $pp \to
K^+p~(\Sigma^0\to \gamma \Lambda \to \gamma\pi^-p)$ (2), and $pp\to
K^+n~(\Sigma^+ \to \pi^0 p)$ (3). The black line shows the sum of all
contributions. }
\end{center}
\end{figure}

It must be recognized that COSY-TOF is not well suited to the measurement of
small cross sections at low excess energies so that most of the data below
about $Q \approx 100$~MeV come from COSY-11 whereas at higher excess energies
COSY-TOF measurements are dominant.

The principle of the COSY-11 and ANKE experiments looks simple --- identify
and measure a $K^+$ and proton from a $pp \to K^+pX$ reaction and then
isolate the $\Lambda$ and $\Sigma^0$ from the missing-mass peaks. This is not
completely straightforward, especially at the higher energies. In the COSY-11
case there was a non-physical background from misidentified kaons, which was
estimated from sideband contributions around the $K^+$ mass. However, even if
the $K^+$ is unambiguously identified, there remains a physical source of
coincident protons coming from the decays $\Lambda\to p\pi^-$ and
$\Sigma^0\to \gamma \Lambda \to \gamma\pi^-p$. This is illustrated by the
data in Fig.~\ref{fig:Val1}, obtained at ANKE at 2.16~GeV with
well-identified $K^+$ mesons~\cite{VAL2007}.

\begin{figure}[hbt]
\begin{center}
\includegraphics[width=0.95\columnwidth]{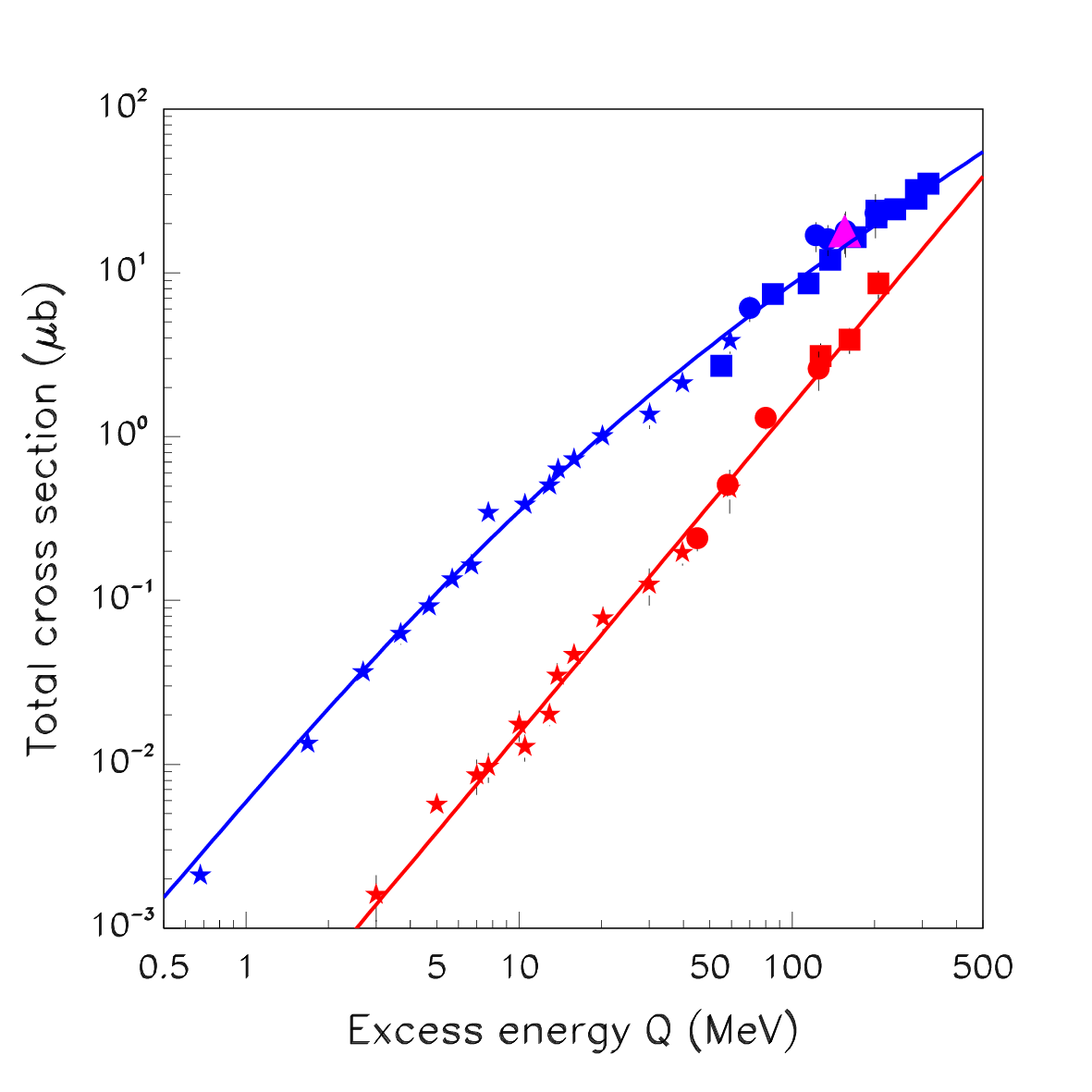}
\caption{\label{fig:lambda1} Upper (blue) points are experimental
measurements of the $pp\to K^+p\Lambda$ total cross section whereas the lower
(red) points represent data from the $pp\to K^+p\Sigma^0$ reaction. Stars are
COSY-11 values~\cite{BAL1998,SEW1999,KOW2004}, squares are from
COSY-TOF~\cite{BIL1998,SAM2006,SAM2010,ABD2010}, and circles from
ANKE~\cite{VAL2007,VAL2010}. Note that not all systematic uncertainties have
been included. The solitary bubble chamber point for $\Lambda$
production~\cite{FIC1962} is shown by the (magenta) triangle. $\Sigma^0$
production seems to follow the indicated $Q^2$ behaviour that is expected
from three-body phase space but there is evidence for a $\Lambda p$ final
state interaction and the (blue) curve is evaluated from Eq.~(\ref{simple2})
with $B_0=5.20$~MeV~\cite{ABD2010}.}
\end{center}
\end{figure}

The total cross sections for $\Lambda$ and $\Sigma^0$ production obtained at
the different COSY facilities are illustrated in Fig.~\ref{fig:lambda1}. It
is obvious from this presentation that there has been a truly impressive
amount of work done in this field at COSY. The only previous data came from
bubble chamber work, where 11 $K^+p\Lambda$ events were found at $Q\approx
156$~MeV~\cite{FIC1962}. These data did at least show that the cross section
was small! The total cross section for $\Sigma^0$ production seems to follow
closely the $Q^2$ behaviour expected from undistorted three-body phase space
and, as will be shown later in this section, such a behaviour is consistent
with data on other $\Sigma$ production reactions.

If one considers only the effects of an $S$-wave $\Lambda p$ final state
interaction then the expected energy variation is that given by
Eq.~(\ref{simple2})~\cite{FAL1996}. The $\Lambda$ data are well fit with the
position in energy of the antibound state $B_0=5.20$~MeV ($\alpha \approx
-0.37$~fm$^{-1}$)~\cite{ABD2010}, though it must be noted that this is an
effective parameter that will depend on the relative production of
spin-singlet and spin-triplet $S$-wave $\Lambda p$ states.

\begin{figure}[hbt]
\begin{center}
\includegraphics[width=0.9\columnwidth]{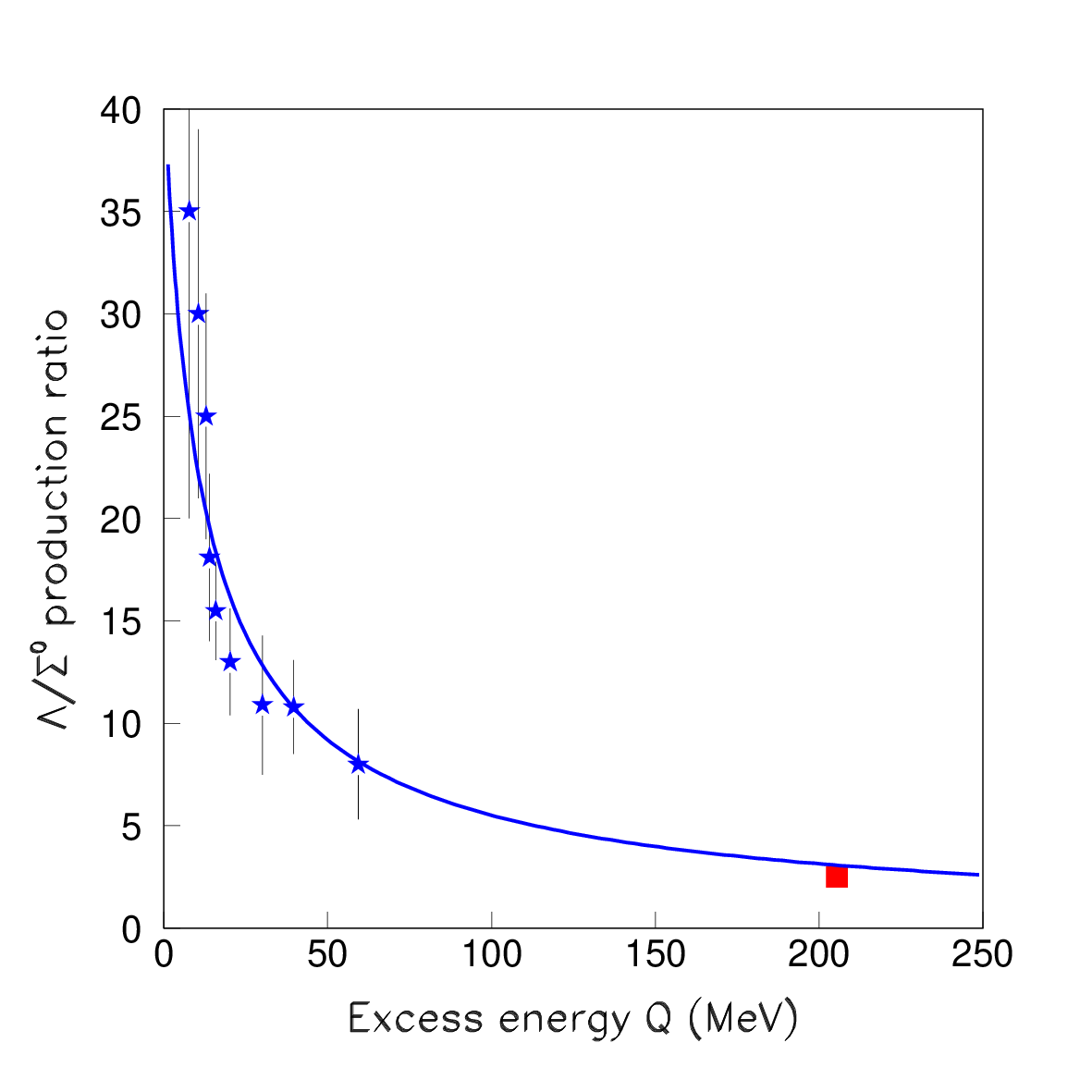}
\caption{\label{fig:lambda2} Ratio of the total cross sections for the $pp\to
K^+p\Lambda$ and $pp\to K^+p\Sigma^0$ reactions at the same values of excess
energy. The (blue) stars are from COSY-11 work~\cite{SEW1999,KOW2004} and the
(red) square from COSY-TOF~\cite{ABD2010}. Only data measured for the two
hyperons at similar values of $Q$ are shown here. The curve of
Eq.~(\ref{Lambda_FSI2}) assumes that there is a final state interaction
purely in the $\Lambda p$ system.}
\end{center}
\end{figure}

Deviations from Eq.~(\ref{simple2}) are, however, easier to see on the linear
scale of Fig.~\ref{fig:lambda2}. Here is shown the ratio $R$ of measured
$pp\to K^+p\Lambda$ and $pp\to K^+p\Sigma^0$ total cross sections compared to
the predictions that follow from Eq.~(\ref{simple2});
\begin{equation}
\label{Lambda_FSI2}
R = C^{\prime}\!\left/\left(1+\sqrt{1+Q/B_0}\right)^{\!2}\right.\!\!.
\end{equation}

\subsection{Differential distributions} 
\label{Differential}

Most phenomenological descriptions of the $pp\to K^+p\Lambda$ reaction are
based upon some form of a one-boson-exchange model and there has been
considerable controversy among theorists as to whether the data are dominated
by the exchange of strange (e.g.\ $\bar{K}$) or non-strange mesons (e.g.\
$\pi$). This problem was brought to the fore by the DISTO measurement of the
spin-transfer parameter $D_{NN}$ between the incident proton and final
$\Lambda$ in the $pp\to K^+p\Lambda$ reaction~\cite{BAL1999}. The negative
value of $D_{NN}$ found was taken as evidence for the dominance of kaon
compared to pion exchange~\cite{LAG1991}, but it is important to stress that
the possibility of $\rho$ exchange was not considered in this discussion.

In contrast, the early COSY-TOF differential cross section results came down
in favour of non-strange meson exchange by showing that the $pp\to
K^+p\Lambda$ data have evidence for the excitation of $N^*$ isobars in the
final $K^+\Lambda$ channel~\cite{SAM2006,SAM2010}. This approach has far less
model dependence than the $D_{NN}$ studies. The COSY-TOF collaboration found
that the $S_{11}(1650)$ plays a prominent role near threshold and there are
certainly similarities with the $pp\to pp\eta$ reaction, which is dominated
by the analogous $S_{11}(1535)$ near threshold. Away from the threshold
region the group also found evidence from the Dalitz plots for the importance
of the $P_{11}(1710)$ and/or the $P_{13}(1710)$ isobars in $\Lambda$
production~\cite{SAM2006,SAM2010}.

\begin{figure}[hbt]
\includegraphics[width=0.9\columnwidth]{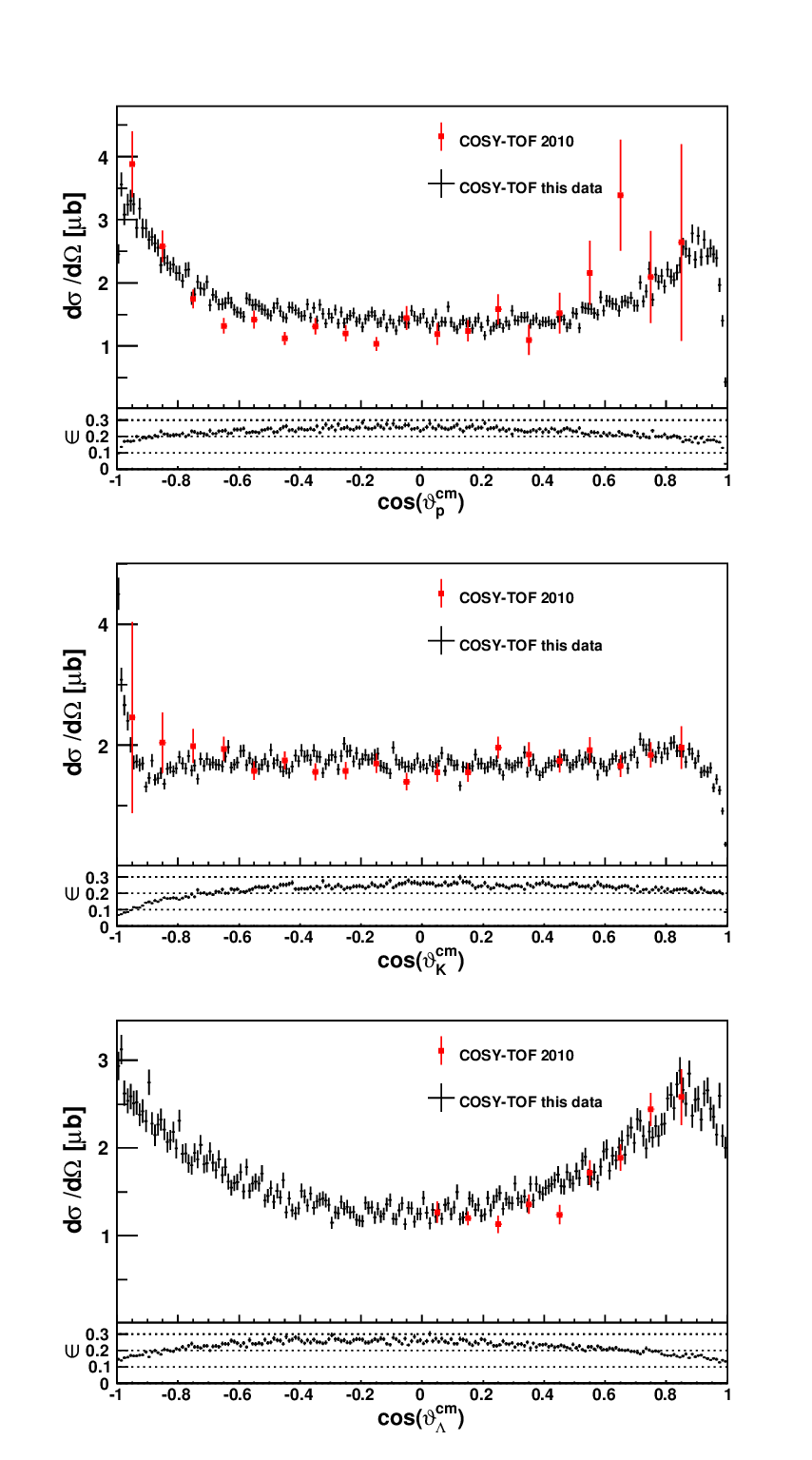}
\caption{\label{fig:sam2010a}
Differential cross section for the $pp\to K^+p\Lambda$ reaction at
2.95~GeV/$c$ in the overall c.m.\ frame. The (black) crosses are recent
COSY-TOF results~\cite{JOW2016} and these are to be compared to the
collaboration's earlier data shown as (red) circles~\cite{ABD2010}. The
distributions are for the proton (top panel), kaon (middle panel), and
$\Lambda$ (bottom panel). Beneath each distribution is shown the value of the
detector acceptance and efficiency.}
\end{figure}

\begin{figure}[hbt]
\includegraphics[width=0.9\columnwidth]{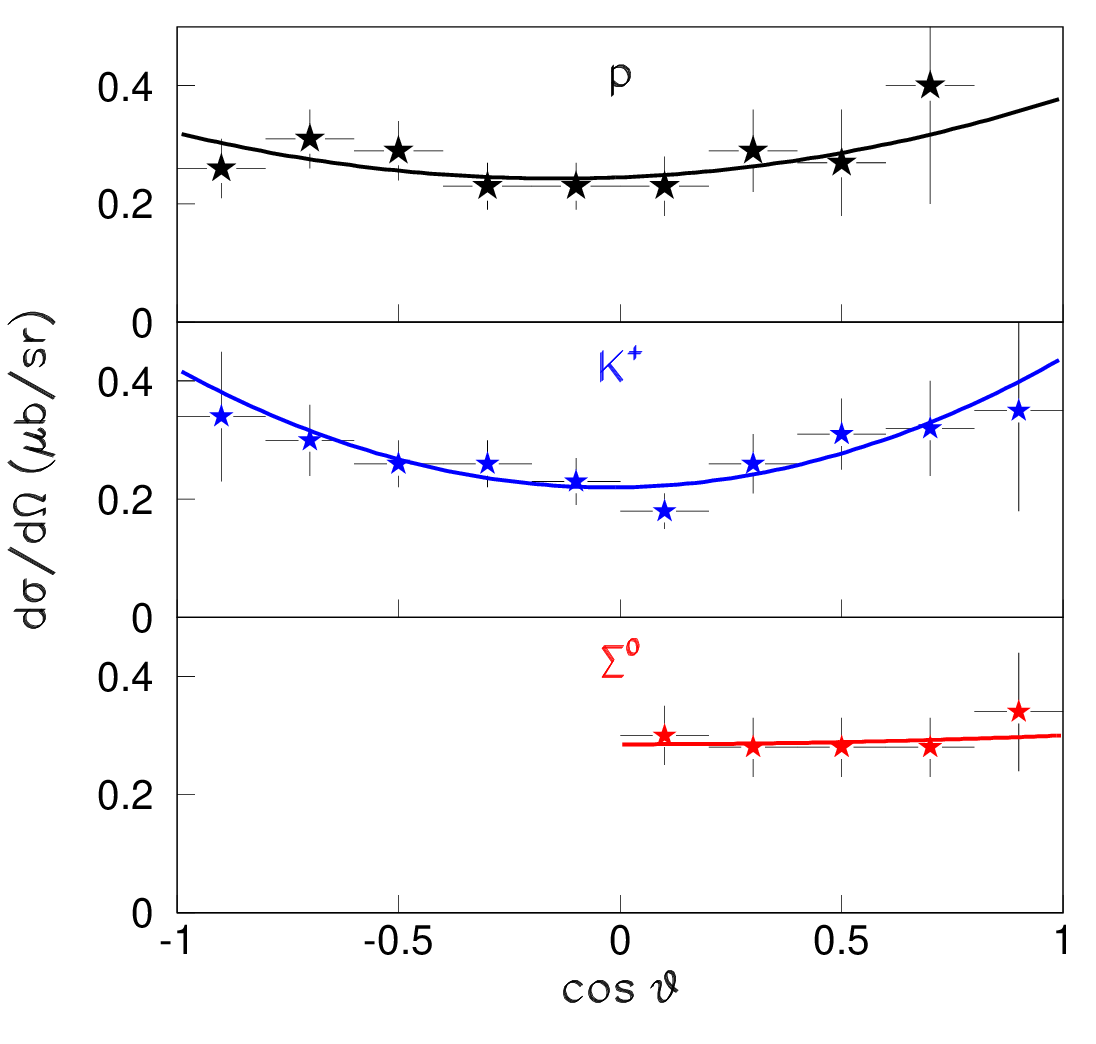}
\caption{\label{fig:sam2010b} Differential cross section for the $pp\to
K^+p\Sigma^0$ reaction at 3.06~GeV/$c$ in the overall c.m.\
frame~\cite{ABD2010}. The distributions are shown separately for the three
particles in the final state, though symmetry around $90^{\circ}$ has not
been imposed in the fits shown. Although the beam momentum is higher than
that shown in Fig.~\ref{fig:sam2010a}, the excess energy for $\Sigma^0$
production is somewhat less than that for the $\Lambda$.}
\end{figure}

Of great importance for theoretical modeling are the angular distributions
measured by the COSY-TOF collaboration, only a small fraction of which are
shown for $pp\to K^+p\Lambda$ at 2.95~GeV/$c$ in Fig.~\ref{fig:sam2010a} and
$pp\to K^+p\Sigma^0$ at 3.06~GeV/$c$ in Fig.~\ref{fig:sam2010b}. The
resulting differential cross sections in the overall c.m.\ frame with respect
to the incident proton direction should be symmetric about $90^{\circ}$
because of the identical particles in the initial state. In order to check
for instrumental bias, this has not been imposed for the $p$ and $K^+$ in the
fits shown but the coefficients of the terms that are odd in $\cos\theta$ are
small and often consistent with zero~\cite{ABD2010}. Tables of preliminary
values with finer binning are to be found in some COSY-TOF theses, e.g., at
2.7~GeV/$c$~\cite{HAU2014}.

There has been a remarkable advance in both the quantity and quality of the
COSY-TOF $pp\to K^+\Lambda p$ data in recent years and this is most evident
in Fig.~\ref{fig:sam2010a} which shows data on this reaction published in
2010~\cite{ABD2010} and 2015~\cite{JOW2016}. Though the two data sets are
clearly consistent, the newer one allows the fit parameters to be determined
much more precisely. On the other hand, the quality of the data shows more
clearly the limitations of COSY-TOF in the forward direction.

Apart from providing invaluable data for models, the results are also useful
in checking some of the assumptions made at other COSY facilities that do not
have the advantage of TOF's extensive angular coverage. It is, for example,
very helpful to see that the $K^+$ distribution in Fig.~\ref{fig:sam2010a} is
essentially consistent with isotropy, though the proton and $\Lambda$
distributions are forward-peaked.

In addition to the distributions in the c.m.\ angle shown in
Figs.~\ref{fig:sam2010a}, the collaboration also
evaluated distributions in the Jackson and helicity
angles~\cite{ABD2010,HAU2014,JOW2016}. By fitting simultaneously the angular
distributions in all three frames of reference, it was possible to confirm
the importance of $P_{11}(1710)$ and/or the $P_{13}(1720)$ isobars away from
threshold~\cite{ABD2010}.

\begin{figure}[hbt]
\includegraphics[width=0.9\columnwidth]{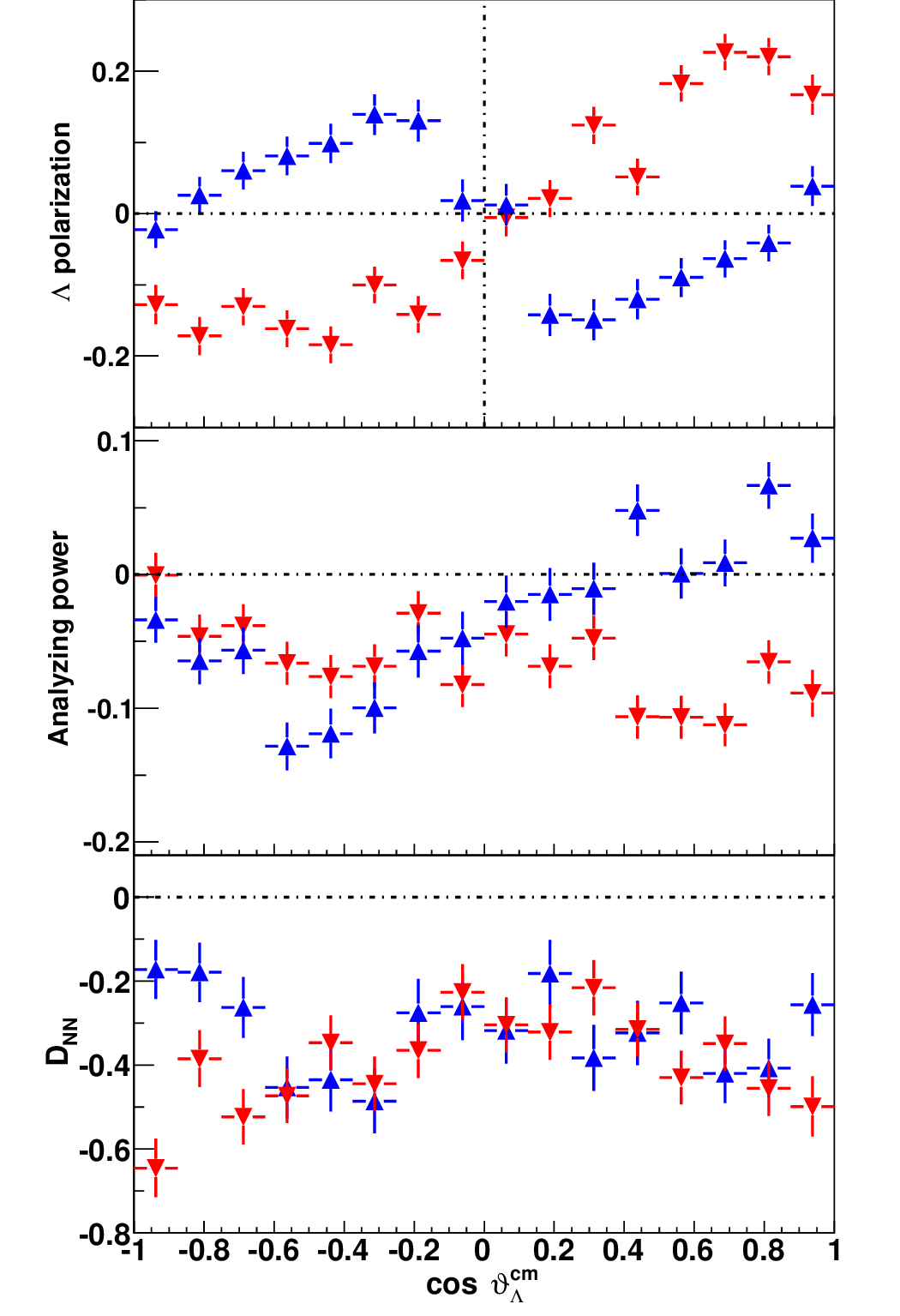}
\caption{\label{fig:TOFPol} COSY-TOF measurements of the $pp\to K^+\Lambda p$
reaction at a beam momentum of 2.7~GeV/$c$ (red downward-pointing triangles)
and 2.95~GeV/$c$ (blue upward-pointing triangles)~\cite{HAU2016}. Shown are
the $\Lambda$ polarization, the proton analyzing power with respect to the
$\Lambda$ direction, and the transverse spin-transfer coefficient $D_{NN}$
between the incident polarized proton and the produced $\Lambda$.}
\end{figure}

The azimuthal symmetry of the COST-TOF detector minimizes many systematic
uncertainties in measurements involving polarized particles.
Figure~\ref{fig:TOFPol} shows several measurements carried out at 2.7 and
2.95~GeV/$c$ in the $pp\to K^+\Lambda p$ reaction~\cite{HAU2016}. Though all
the events were fully reconstructed, only observables associated with the
direction of the $\Lambda$ are shown here. The first thing to notice is the
change of sign in the $\Lambda$ polarization as the beam momentum is
increased by 250~MeV/$c$. The proton analyzing power also changes, but not as
dramatically and only in the forward hemisphere. Both these quantities are
sensitive to interferences between partial waves but the transverse
spin-transfer parameter $D_{NN}$ is a much more robust observable that lends
itself to more direct interpretation. As already noted in connection with the
DISTO data, Laget~\cite{LAG1991} has shown that a positive value of $D_{NN}$
generally favours pion exchange whereas kaon exchange would generally lead to
negative values. This argument may have little real relevance since it is
believed that, even for $\eta$ production, $\rho$-meson exchange is more
important than pion~\cite{FAL2001} and this trend is likely to be reinforced
for the production of even heavier systems.

Although one sees from Fig.~\ref{fig:sam2010a} that the differential cross
section depends strongly on the $\Lambda$ c.m.\ angle, in most of the
analyzes it has nevertheless been assumed that one can expand the
polarization as a series in associated Legendre polynomials. This makes it
even harder to identify contributions from individual partial waves.

The corresponding cross section data for the $pp\to K^+p\Sigma^0$ reaction
are shown in Fig.~\ref{fig:sam2010b} at a beam momentum of 3.06~GeV/$c$.
Because of the much smaller cross sections the data have been put into wider
bins. Nevertheless one can see qualitative differences with $\Lambda$
production; the $K^+$ distribution is more bowed though the $\Sigma^0$ looks
flatter than the $\Lambda$. These differences might arise from the
possibilities of kaon exchange or $\Delta^*$ excitation in $\Sigma$
production~\cite{ABD2010}.

\subsection[Polarization and the $\Lambda p$ scattering length]{Polarization and the $\boldsymbol{\Lambda p}$ scattering length} 
\label{Ayanda}

\begin{figure}[hbt]
\includegraphics[width=0.9\columnwidth]{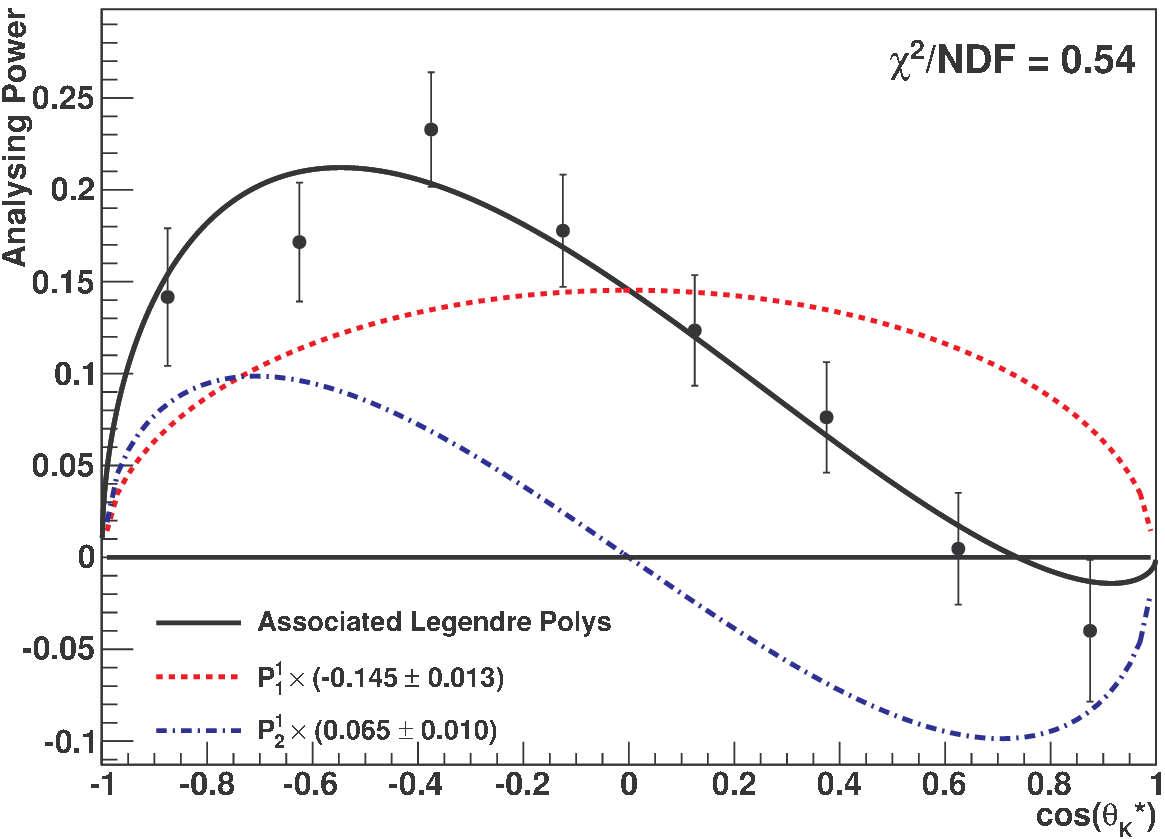}
\caption{\label{fig:Roder2} The proton analyzing power $A_y^K$ with respect
to the kaon direction for the full range of data taken on the $\pol{p}p\to
K^+p\Lambda$ reaction at 2.95~GeV/$c$~\cite{ROD2013}. Since it is seen in
Fig.~\ref{fig:sam2010a} that the $K^+$ angular distribution is essentially
isotropic, it is useful to expand $A_y^K$ in a series of associated Legendre
polynomials to give the fits shown.}
\end{figure}

Data with a polarized proton beam were taken by the COSY-TOF collaboration at
2.95~GeV/$c$~\cite{ROD2013} and at 2.7\ GeV/$c$~\cite{HAU2014}. Although one
should expand $A_y^p\dd\sigma/\dd\Omega_K$ in terms of associated Legendre
polynomials, the fact that the cross section in Fig.~\ref{fig:sam2010a} is
almost independent of the kaon angle means that a direct expansion of $A_y^p$
for the kaon asymmetry is not unreasonable. The fit
\[A_y^p=(-0.145\pm 0.013)P_1^1(\cos\theta)+(0.065\pm0.010)P_2^1(\cos\theta)\]
is shown in Fig.~\ref{fig:Roder2}. At least two terms are required in this
description, suggesting that kaon $d$-waves are important, despite there
being no sign of their presence in the differential cross section.

The results at 2.7~GeV/$c$ are broadly similar~\cite{HAU2014} and they both
show that the component that is symmetric about $\theta_K=90^{\circ}$ is as
important as the one that is antisymmetric. This proves that, in conflict
with the HIRES fit~\cite{BUD2010a}, there must be significant amount of
$\Lambda p$ spin-triplet production, the argument being very similar to that
for pion production in the $pp\to \{pp\}_{\!s}\pi^0$ reaction discussed in
sect.~\ref{pion}. Of course, since the analyzing power represents
interference between amplitudes, it is not possible from this picture to
determine the relative magnitudes of the singlet and triplet contributions,
though some weak limits might be established, depending upon the phase
assumptions.

Since the $\Lambda p$ spin-singlet contribution to the analyzing power with
respect to the kaon direction must vanish at $\theta_K=90^{\circ}$, the
product $\left.A_y^p\dd\sigma/\dd\Omega\right|_{\theta_K=90^{\circ}}$ is
sensitive to the spin-triplet scattering length. However, to study this
quantity in fine steps in the $\Lambda p$ invariant mass would require very
high statistics. The IKP theory group proposed an alternative procedure that
exploits more seriously the analyticity properties of the production
amplitudes~\cite{GAS2004,GAS2005}. This involves the evaluation of a dispersion
integral which, it is claimed, is less sensitive to the mass resolution and
can lead to a robust estimate of the error associated with the theory.

In the dispersion approach, $a=\lim_{{m}\to m_0}\{a(m)\}$, where
\begin{eqnarray}
\nonumber
a(m)&=&\frac1{2\pi}\sqrt{\frac{m_0}{m_{\rm red}}}\,
{\bf P} \int_{m_0^2}^{m_{\rm max}^2}\dd\mu^2
\sqrt{\frac{m_{\rm max}^2-{m}^2}{m_{\rm max}^2-\mu^2}}\times\\
& & \hspace{-0.5cm}
\frac1{\sqrt{\mu^2-m_0^2} \ (\mu^2-{m}^2)}
\log{\left\{\frac{1}{p}\left(\frac{\dd^2\sigma}{\dd\mu^2\dd t}\right)\right\}}
\cdot
\label{final}
\end{eqnarray}
Here $m_{\rm red}$ is the reduced $\Lambda$ and proton mass and
$m_0=m_{p}+m_{\Lambda}$. The bracket contains the double-differential cross
section for producing a $\Lambda p$ pair of invariant mass $\mu$, $p$ is the
relative momentum between that pair, and $t$ is the four-momentum transfer
between the incident proton and final kaon. The choice of the cut-off
parameter $m_{\rm max}$ is rather subjective. It should be as large as
possible, subject to the $\Lambda p$ system still being in an $S$-wave and,
in the ideal world, the theoretical corrections would be minimized if one
could let $m_{\rm max}\to \infty$. The authors argued that it would be
sufficient to take $m_{\rm max}=m_0 +40$~MeV/$c^2$~\cite{GAS2004,GAS2005}.

In the evaluation of Eq.~(\ref{final}) it is very convenient to parameterize
the production cross section in order to provide a simple estimate of the
principal value (${\bf P}$) integral. A particular choice of fit function
even allows the integral to be  evaluated
analytically~\cite{GAS2004,GAS2005}. The approach was tested on the Saclay
inclusive $K^+$ production data at 2.3~GeV. A spin-average scattering length
of $\bar{a} = (-1.5\pm 0.15\pm 0.3)$~fm was obtained, where the first error
corresponds to the uncertainty from the data and the second from the theory.
Already in this case it was found that the experimental error is smaller than
the theoretical one~\cite{GAS2004,GAS2005}.

Since the statistics are so much higher for the unpolarized distributions,
the dispersion integral method~\cite{GAS2004,GAS2005} was first used to
determine a spin-average scattering length from the COSY-TOF
data~\cite{HAU2016a}. This led to values for $\bar{a}$ of
\begin{eqnarray*}&&\hspace{-2mm}(-1.25\pm0.08\pm0.3)~\textrm{fm}\
\textrm{at~2.95~GeV/}c,\\
&&\hspace{-2mm}(-1.38_{-0.05}^{+0.04}{}_{\rm stat}\pm 0.22_{\rm syst}\pm0.3_{\rm theo})~\textrm{fm}\ \textrm{at~2.7~GeV/}c.
\end{eqnarray*}
for Refs.~\cite{ROD2013} and \cite{HAU2014}, respectively. The uncertainty in
the 2.95~GeV/$c$ value should be increased to include the systematic effects
of distortions due to the $N^*$ isobars~\cite{ROD2013}, which seem to be far
less important at 2.7~GeV/$c$, giving there perhaps an uncertainty of only
$\pm 0.1$~fm~\cite{HAU2014}.

The average scattering lengths from the COSY-TOF data~\cite{ROD2013,HAU2016a}
are not inconsistent with the value $(-1.5\pm 0.15\pm 0.3)$~fm obtained from
the SPES4 data~\cite{SIE1994} using the same analysis technique. However,
they are in conflict with the $(-2.43\pm0.16)$~fm quoted by the HIRES
collaboration~\cite{BUD2010a}. In this context one should note that the HIRES
data were obtained at a similar momentum (2.735~GeV/$c$) to those of
COSY-TOF~\cite{HAU2016a}, but with the Big Karl spectrometer being set to
take data around the forward direction. It seems likely that most of the
discrepancy in the scattering length determinations arises from relatively
small differences in the input in the logarithm of Eq.~(\ref{final}) near the
kinematic threshold of $m_p+m_{\Lambda}$.

The COSY-TOF data at 2.95~GeV/$c$ were not precise enough to extract a useful
value for the spin-triplet scattering length by weighting the data with the
$K^+$ analyzing power~\cite{ROD2013}. The conditions are far more favourable
at 2.7~GeV/$c$ and a value of
\[a_t = (-2.55_{-1.39}^{+0.72}{}_{\rm stat}\pm 0.6_{\rm syst}\pm0.3_{\rm theo})~\textrm{fm}\]
was obtained, where the error bar includes an estimate of the possible $N^*$
distortion~\cite{HAU2014}. The value found for the triplet scattering length
is not inconsistent with the spin-average result and, in view of the large
error bars, it is clearly going to be very hard to separate the singlet value
from the triplet with this method.

It has recently been pointed out that, although the scattering length changes
significantly when the maximum energy in the dispersion integral of
Ref.~\cite{GAS2004,GAS2005} is reduced, the position of the $\Lambda p$
virtual bound state at $k=i\alpha$ hardly moves at all~\cite{FAL2016}. In
fact, for the Jost parametrization used by the HIRES
collaboration~\cite{BUD2011}, the position of the virtual bound state is
completely independent of the cut-off energy in the dispersion relation. The
COSY-TOF spin-average value of $\alpha({\rm TOF})=-0.42$~fm$^{-1}$ should be
compared to the COSY-HIRES result of $\alpha({\rm HIRES})=-0.31$~fm$^{-1}$
and $\alpha(\sigma_T)=-0.37$~fm$^{-1}$ deduced from the energy dependence of
the $pp\to K^+\Lambda p$ total cross sections shown in
Fig.~\ref{fig:lambda1}. In order to assess the significance of the deviations
between these values, careful studies of the systematic and statistical
uncertainties in the different experiments are required. All the $\Lambda p$
potentials discussed in a recent review~\cite{GAL2016} generate virtual bound
states and these are typically at $\alpha({\rm singlet}) \approx
-0.28$~fm$^{-1}$ and $\alpha({\rm triplet}) \approx -0.38$~fm$^{-1}$. These
are not very different from the values derived from the HIRES and TOF data
but the COSY experimental results correspond to unknown spin averages.

\subsection[The $\Lambda:\Sigma$ cusp effect]{The $\boldsymbol{\Lambda:\Sigma}$ cusp effect} 
\label{cusp}

\begin{figure}[hbt]
\includegraphics[angle=-90,width=0.9\columnwidth]{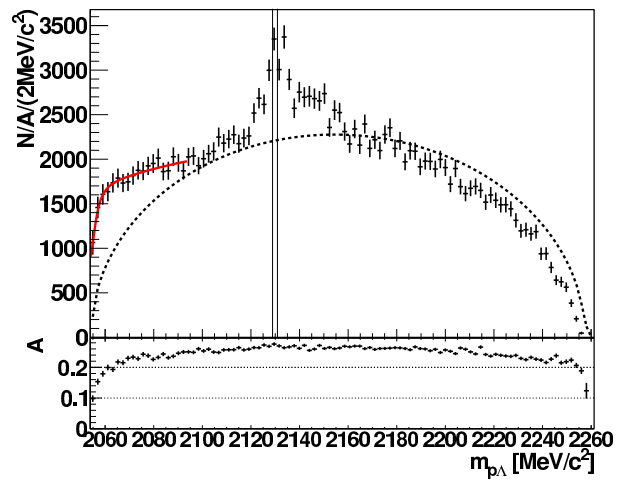}
\caption{\label{fig:Roder} The spectrum of $p\Lambda$ invariant mass taken
from the high resolution $pp\to K^+p\Lambda$ COSY-TOF data at 2.95~GeV/$c$ in
2~MeV/$c^2$ bins~\cite{ROD2013}. The data have been corrected for the
acceptance (A) that is shown at the bottom of the figure. The vertical lines
indicate the positions of the two $\Sigma N$ thresholds. An arbitrarily
scaled phase space distribution (dashed line) is shown to guide the eye. The
solid (red) line at low invariant masses represents the data used in the
scattering length determination discussed in the previous subsection.}
\end{figure}

It was already suspected from earlier COSY-TOF work that there was some kind
of anomaly in the differential distribution of the $pp\to K^+p\Lambda$
reaction at a $\Lambda p$ invariant mass corresponding to the $\Sigma N$
threshold~\cite{SAM2006,SAM2010}. However, by far the most detailed study of
this region is to be found in their Refs.~\cite{SAM2013,ROD2013}, where the
effect is ascribed to a cusp associated with the very strong $S$-wave
$\Lambda p \rightleftarrows \Sigma N$ transitions. As shown in
Fig.~\ref{fig:Roder}, there is a sharp but asymmetric peak in well-identified
$pp\to K^+p\Lambda$ events. These data resulted from the high resolution
COSY-TOF experiment at 2.95~GeV/$c$, where the invariant mass resolution of
2.6~MeV/$c^2$ was much narrower than the cusp peak.

There is a long history of cusps in nuclear and particle physics, especially
in connection with the production of strange particles~\cite{DAL1982}. In
lowest order of the transitions, the peak in Fig.~\ref{fig:Roder} can be
understood as being caused by an interference between a direct $pp \to
K^+\Lambda p$ amplitude and one arising from $pp \to K^+\Sigma N$ followed by
the conversion $\Sigma N \to \Lambda p$. The on-shell part of this second
term is proportional to the c.m.\ momentum in the $\Sigma N$ system,
$q_{\Sigma N}$. But this variable changes from being purely imaginary below
the $\Sigma N$ threshold to purely real above, which means that the
interference between the direct and conversion terms changes very abruptly at
the $\Sigma N$ threshold, often giving rise to a cusp shape.

The on-shell contribution, which is really a reflection of unitarity, could
be calculated using the physical $\Sigma N \to \Lambda p$ amplitudes (if we
knew them) but the off-shell ones would require a model for the coupled
channel $\Sigma N: \Lambda p$ potential. Since the cusp manifests itself
largely as an interference term, there is no reason for it to be symmetric or
to be Breit-Wigner in shape~\cite{SAM2013}. A full simulation of the effect
would depend on knowing both the $pp \to K^+\Lambda p$ and $pp \to K^+\Sigma
N$ amplitudes as well as the $\Sigma N:\Lambda p$ potential, or at the very
least the $\Sigma N \to \Lambda p$ amplitude. The situation is further
complicated by the fact that one has also to consider total spin-one and
-zero in the $\Sigma N: \Lambda p$ channels.

It is clear that it will take considerable effort to extract the full
information from the data shown in Fig.~\ref{fig:Roder}. The peak structure
in fact looks very similar to the very strong cusp effect observed with
stopping kaons in the $K^-d\to \pi^-\Lambda p$ reaction~\cite{TAN1969}. Just
as in Fig.~\ref{fig:Roder}, there is also an unexplained enhancement on the
high mass side of the $\Lambda p$ peak. However, there is no reason for the
cusp structure to look identical for the two reactions. Of course, if much
higher statistics were available, then one might expect the angular
distribution to change through the cusp region but it seems that in any case
the cusp peak is less than 20\% of the total events in the threshold region
of Fig.~\ref{fig:Roder} and this will limit the size of any angular change.

It could be argued that such a strong peak as that seen in
Fig.~\ref{fig:Roder} would only occur if the interaction had given rise to a
virtual state in the $\Sigma N$ system but there is no obvious evidence for a
significant final state interaction from the energy dependence of the $pp\to
K^+\Sigma^0 p$ reaction in Fig.~\ref{fig:lambda1}.

Since there is a mass splitting between the $\Sigma^0p$ and $\Sigma^+n$
thresholds, one might hope to see a double cusp~\cite{DAL1982} but this would
require mass resolutions that better than the 2~MeV/$c^2$ threshold
difference. Only the HIRES inclusive $K^+$ production data achieves this, but
even here it is hard to be sure if there is a two-peak structure in the
data~\cite{BUD2010a}.

\subsection[$pp\to K^+n\Sigma^+$]{$\boldsymbol{pp\to K^+n\Sigma^+}$} 
\label{Sigmap}

There are many reliable measurements of the $pp\to K^+p\Lambda$ and $pp\to
K^+p\Sigma^0$ total cross sections presented earlier in this section that
were achieved by detecting the $K^+$ and proton in coincidence and
identifying the neutral hyperons through peaks in the missing-mass
distributions. It is far harder to extend such a method to $\Sigma^+$
production.

The most direct approach to the study of the $pp\to K^+n\Sigma^+$ reaction
was attempted at $Q=13$ and 60~MeV at COSY-11~\cite{ROZ2006}, where the
momenta of the $K^+$ and neutron were measured and the corresponding missing
mass evaluated. For this purpose the standard COSY-11 facility was
supplemented through the introduction of a neutron detector. The total cross
sections thus obtained were extraordinarily large, being about two orders of
magnitude higher than those for $\Sigma^0$ production at similar values of
$Q$ and also much larger than the results reported from later COSY
experiments~\cite{VAL2007,VAL2010,BUD2010,ABD2012a}.

In order to avoid problems associated with the detection of neutrons, a
different three-prong approach was undertaken at ANKE at five beam energies,
corresponding to $Q=13$, 47, 60, 82, and 128~MeV~\cite{VAL2007,VAL2010}. All
three methods used the delayed-veto technique~\cite{BUS2002}, that was
already mentioned in connection with sub-threshold $K^+$ production in nuclei
in sect.~\ref{sub-threshold}. The approaches followed were:

\begin{enumerate}
\item[(a)]{A conservative upper limit on $\Sigma^+$ production was
    deduced through the study of inclusive $K^+$ production in the $pp\to
    K^+X$ reaction. This method suggested that the cross section for
    $\Sigma^+$ production was broadly similar to that for the $\Sigma^0$.
    This limit was already in severe conflict with the COSY-11
    claim~\cite{ROZ2006}.}
\item[(b)]{As discussed with respect to Fig.~\ref{fig:Val1}, there are
    two significant contributions to the $K^+p$ missing-mass distribution
    in $pp\to K^+pX$ near the highest values of $m_X$. These are $pp \to
    K^+p\,(\Sigma^0\to \gamma \Lambda \to \gamma\pi^-p)$ and $pp\to
    K^+n\,(\Sigma^+\to \pi^0 p)$. The first of these could be estimated
    from the measured $\Sigma^0$ production rate. After taking this into
    account, the measurement gave values for the $\Sigma^+$ cross section
    that were below the upper limits set in (a). However, it is hard to
    estimate quantitatively the systematic uncertainties involved.}
\item[(c)]{Although the statistics were not high, the cleanest signal for
    $\Sigma^+$ production was found from the $K^+\pi^+$ correlations that
    arise from the $pp\to K^+n\,(\Sigma^+{\to}\pi^+n)$ reaction. Even at
    the highest excess energy the background from the direct $pp\to
    K^+n\Lambda\pi^+$ is believed to be less than of the order of 2\%.}
\end{enumerate}

All three methods gave consistent results. The ratio $R(\Sigma^+/\Sigma^0)$
of $\Sigma^+$ to $\Sigma^0$ production was found to be $0.7\pm
0.1$~\cite{VAL2010}. As a cross check, the cross sections derived for the
$\Sigma^0$ and $\Lambda$ production cross sections from the missing-mass
peaks were found to be consistent with other results shown in
Fig.~\ref{fig:lambda1}, though there were more uncertainties in the $\Lambda$
case because of the higher excess energies and the limited ANKE acceptance.

\begin{figure}[hbt]
\begin{center}
\includegraphics[width=0.9\columnwidth]{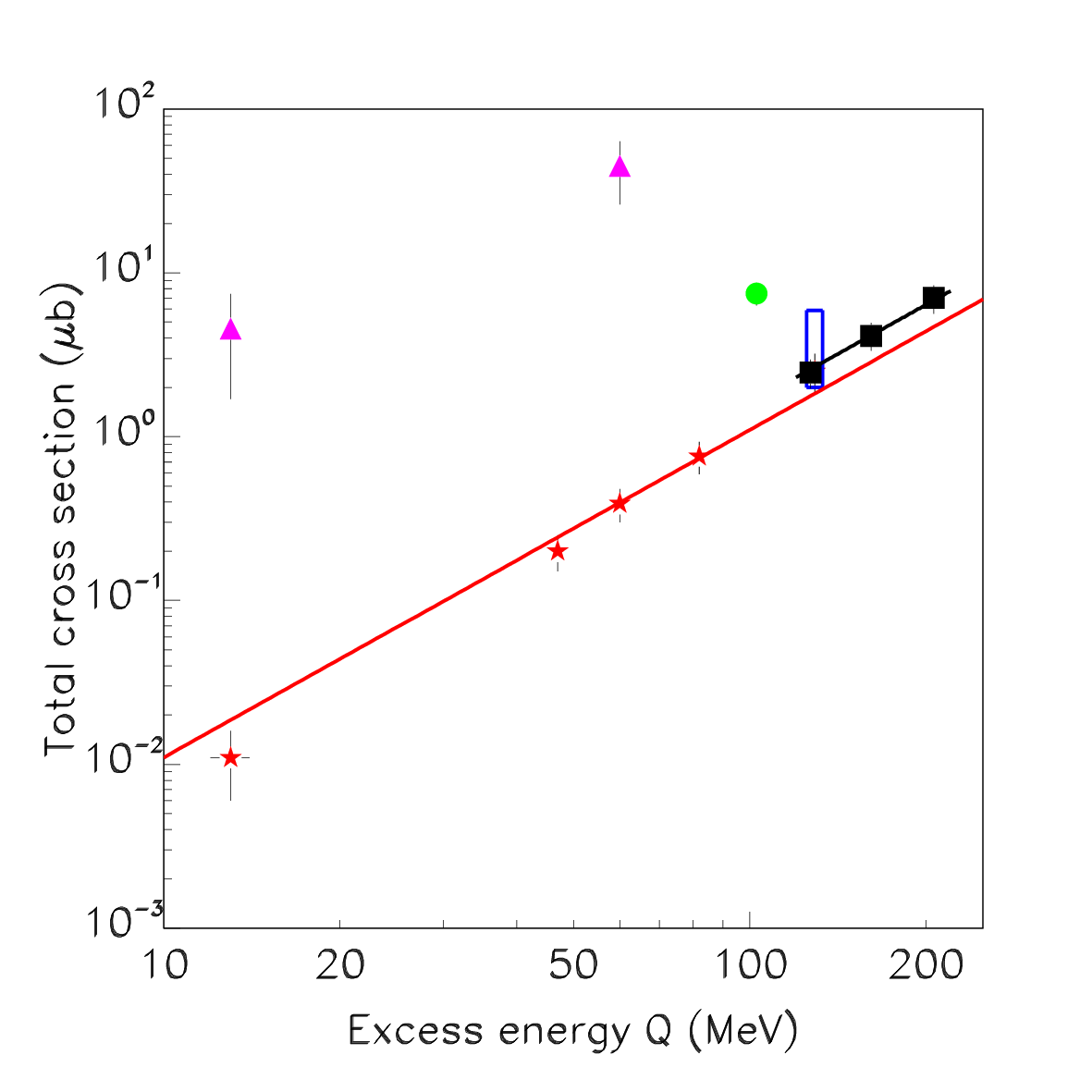}
\caption{\label{fig:yv} Values of the total cross section for the $pp\to
K^+\Sigma^+n$ reaction extracted from COSY experiments. The (magenta)
triangles are the original COSY-11 measurements~\cite{ROZ2006}; (red) stars
were obtained at ANKE~\cite{VAL2007,VAL2010}; the (green) circle was deduced
from the HIRES inclusive $K^+$ data~\cite{BUD2010a}; the (blue) open
rectangle is the range of values allowed by the TOF measurement at the 68\%
confidence level~\cite{ABD2012a}. In all cases the systematic and statistical
errors have been added in quadrature. Also shown by (black) squares are three
points corresponding to the $pp\to K^0p\Sigma^+$ total cross section measured
by the COSY-TOF collaboration~\cite{ABD2012b}. Note that one of these
partially obscures the highest ANKE point. The curves represent the $Q^2$
three-body phase-space behaviour normalized arbitrarily to the ANKE 82~MeV
point and the COSY-TOF 161~MeV point. }
\end{center}
\end{figure}

The COSY-11 total cross sections for $\Sigma^+$ production \cite{ROZ2006} are
compared with those obtained at ANKE~\cite{VAL2007,VAL2010} in
Fig.~\ref{fig:yv}. Also shown is the $Q^2$ dependence expected on the basis
of pure three-body phase space.

The HIRES collaboration~\cite{BUD2010a}, whose results were discussed
extensively in sect.~\ref{InclusiveK}, measured the inclusive $pp\to K^+X$
cross section in fine steps in kaon momentum using the high resolution Big
Karl spectrometer. The data taken at 2.87~GeV/$c$ show a big jump in the
missing-mass spectrum around the threshold for $\Sigma$ production and it was
assumed that this was mainly associated with the $pp\to K^+p\Sigma^0$ and
$pp\to K^+n\Sigma^+$ reactions, with some localized effect coming from
$\Lambda/\Sigma$ channel coupling. Using techniques developed
earlier~\cite{SIB2007}, and knowing the value of the $\Sigma^0$ production
cross section, they claimed that $R(\Sigma^+/\Sigma^0) \approx 5 \pm 1$ at
$Q=129$~MeV~\cite{BUD2010}.

Although the HIRES result casts significant doubts on the COSY-11 values in
Fig.~\ref{fig:yv}, there remains a very serious disagreement with the ANKE
data. It has been argued~\cite{VAL2011} that the HIRES analysis should really
be considered as an upper limit, principally because it underestimates the
significance of the coupling between the $pp\to K^+p\Lambda$ and $pp\to
K^+N\Sigma$ channels. The very robust measurements of the exclusive $pp\to
K^+p\Lambda$ reaction by the COSY-TOF collaboration displayed in
Fig.~\ref{fig:Roder} show a very strong cusp effect in the $\Sigma$ threshold
region~\cite{SAM2006,SAM2010} associated with $\Lambda p \rightleftarrows
\Sigma N$ transitions. As a consequence, some of what had been assumed by the
HIRES group to have been $\Sigma$ production~\cite{BUD2010} could, instead,
be due to $\Lambda$ production, where the $K^+$ missing-mass distribution is
far from smooth~\cite{SAM2006,SAM2010}.

The geometric acceptance of the COSY-TOF detector is much larger than that of
ANKE but, as discussed in sect.~\ref{TOF}, it was designed mainly for the
detection of charged particles. In order to study the $pp\to K^+n\Sigma^+$
reaction, the standard COSY-TOF design was extended through the addition of a
large neutron detector that was placed outside the TOF barrel, some 5.17~m
downstream of the target. An unambiguous signature of the $pp \to
K^+n\Sigma^+$ reaction was a primary track due to a charged kaon, a hit in
the neutron detector, and a decay of the $\Sigma^+$ hyperon, which resulted
in a kinked track~\cite{ABD2012a}.

Although the COSY-TOF events are very clean, the statistics achieved at
$Q=129$~MeV were extremely limited. Of the 9 identified events, it was
estimated that perhaps 2 were due to background and the rest to $\Sigma^+$
production. Applying Poisson statistics, the authors concluded that
$(2.0\pm0.8) < \sigma(pp\to K^+n\Sigma^+) < (5.9\pm 1.2)~\mu$b at the 68\%
confidence level~\cite{ABD2012a} and this limit is shown in Fig.~\ref{fig:yv}
by the open (blue) box. The ANKE value at 129~MeV~\cite{VAL2007} clearly
falls within this range but, if one assumes a reasonable energy dependence,
the HIRES point at 103~MeV seems very high.

Limits on the $pp\to K^+n\Sigma^+$ cross section can also be derived by
combining data on the $pp\to K^+\Sigma^+n$ and $pp\to K^0\Sigma^+p$ channels.
However, as will be discussed in the following subsection, these are
comparatively weak limits and merely suggest that the COSY-11 point at 60~MeV
is likely to be in error. There are also bubble chamber data that are shown
in some of the publications cited but these were taken at much higher excess
energies, where the underlying Physics might be significantly different

\subsection[$pp\to K^0p\Sigma^+$]{$\boldsymbol{pp\to K^0p\Sigma^+}$} 
\label{TOF_Sigmaplus}

The $pp\to K^0p\Sigma^+$ reaction is much more closely matched to the
capabilities of the COSY-TOF detector than $pp\to K^+\Sigma^+n$ because the
$K^0_s$ decay into two charged pions occurs mainly within the
barrel~\cite{ABD2012b}. Since the decay $\Sigma^+\to p\pi^0/n\pi^+$ largely
happens after the start counter, this means that there is the excellent
trigger of two charged tracks turning into four tracks within the volume of
the detector. Kinematic fitting procedures could then be applied with
confidence because there is relatively little background. This channel was
the basis of the COSY-TOF pentaquark search that is described in
sect.~\ref{pentaquark}.

Extensive angular and mass distributions were obtained at three excess
energies, $Q=126$, 161, and 206~MeV~\cite{ABD2012b} and the resulting total
cross sections are also shown in Fig.~\ref{fig:yv}. Over the small range in
excess energy, these behave like three-body phase space, i.e., $\sigma
\propto Q^2$, which is also indicated. This variation, which is similar to
that observed in $pp\to K^+p\Sigma^0$ and $pp\to K^+n\Sigma^+$, suggests that
any $\Sigma N$ FSI is quite weak. Just as for $\Lambda$
production~\cite{ABD2010}, the angular distributions were analyzed in the
three frames of reference and it was found that one nucleon isobar with a
mass $\approx 1720$~MeV/$c^2$ and width $\approx 150$~MeV/$c^2$ could
describe the bulk of these data.

Isospin invariance allows one to put limits on the $pp\to K^+n\Sigma^+$ cross
section in terms those for $pp\to K^+\Sigma^+n$ and $pp\to K^0\Sigma^+p$:
\begin{align}
\nonumber &\left[\sqrt{2\sigma(pp\to K^+ p \Sigma^0)}-\sqrt{\sigma(pp \to K^0 p\Sigma^+)}\,\right]^2\\
\nonumber &\hspace{2cm}\leq \sigma(pp \to K^+ n \Sigma^+)\\
&\leq \left[\sqrt{2\sigma(pp \to K^+ p \Sigma^0)}+\sqrt{\sigma(pp \to K^0 p \Sigma^+)}\,\right]^2.
\label{triangle}%
\end{align}
Using the COSY-TOF values~\cite{ABD2010,ABD2012b} as input. this results in
the rather wide limits~\cite{ABD2012a}
\begin{equation}
(0.9\pm0.8)~\mu\textrm{b} \leq \sigma(pp \to K^+ n \Sigma^+) \leq (16.8\pm 0.8)~\mu\textrm{b},
\end{equation}
which provide little constraint in Fig.~\ref{fig:yv}. The triangle constraint
of Eq.~(\ref{triangle}) is also valid for differential distributions but the
available data do not suggest that this would currently offer a very
profitable approach.

\subsection{The production of heavy hyperons} 
\label{heavy_hyperons}

Of the heavier hyperons, there is great interest in the production of the
$\Lambda(1405)$ because models based on unitary chiral perturbation theory
find two poles in the neighborhood of the $\Lambda(1405)$ that evolve from a
singlet and an octet in the exact SU(3) limit~\cite{OLL2001}. The existence
of two poles means that the lineshape measured in an experiment will depend
upon the particular reaction being studied~\cite{JID2003}. This unusual
situation is discussed clearly in Ref.~\cite{MEI2016}.

Naively one might hope to carry out the same type of missing-mass experiment
that was so successfully used for $\Lambda$ and $\Sigma^0$  production. This
is in fact not possible because of the presence of a nearby isospin-one
resonance, the $\Sigma(1385)$. Due to their finite widths, the two states
overlap and cannot be separated in a simple $pp \to K^+pX$ experiment.
Secondary protons are also more troublesome when the peaks are not narrow.
Extra particles have therefore to be detected and, in a spectrometer such as
ANKE, this leads to reduced acceptance and much greater ambiguity in the
evaluation of cross sections.

The suppression of the $\Sigma(1385)$ in the ANKE experiment~\cite{ZYC2008}
was achieved by looking for neutral decays because isospin forbids the
reaction $\Sigma^0(1385)\nrightarrow \Sigma^0\pi^0$. The basic principle of
the experiment was the search for the four--fold coincidence of two protons,
one positively charged kaon and one negatively charged pion, i.e., $pp
\rightarrow p K^+ p \pi^- X$. These could correspond to the reaction chains
\\ %
(\emph{i})\ $pp \rightarrow p K^+ \Sigma^0(1385) \rightarrow p K^+
\Lambda \pi^0 \rightarrow p K^+ p \pi^- \pi^0$ \\ %
(\emph{ii})\ $pp\rightarrow p K^+ \Lambda(1405) \rightarrow p K^+ \Sigma^0
\pi^0 \rightarrow p K^+ \Lambda \gamma \pi^0\\
\phantom{11111111111111111111111111111111} \rightarrow p K^+ p \pi^- \gamma
\pi^0$.

\begin{figure}[h!]
\begin{center}
\includegraphics[width=0.9\columnwidth]{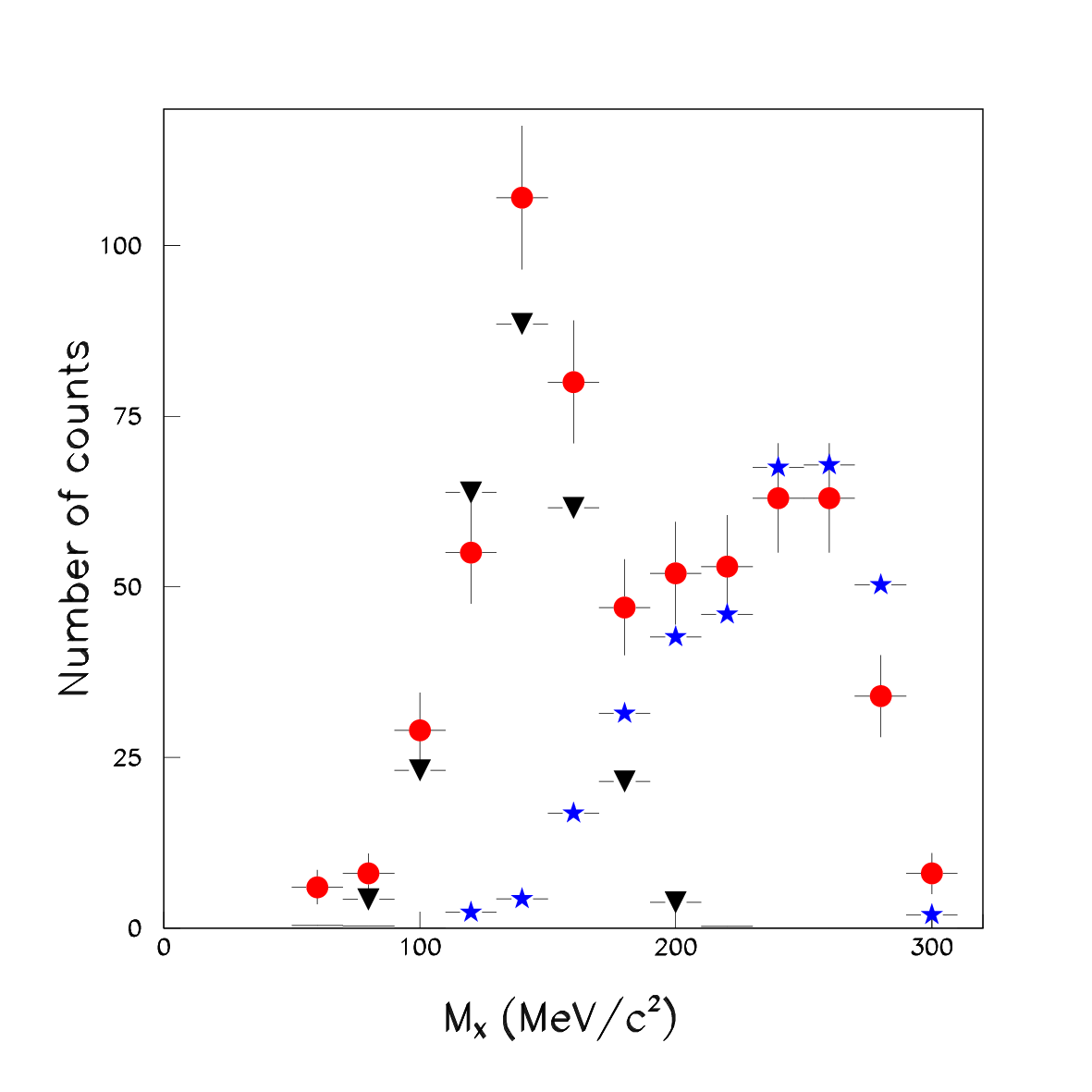}
\caption{\label{fig:iza} Missing-mass distribution from the $pp\to
K^+p\Lambda X$ reaction at 2.83~GeV~\cite{ZYC2008}. The (red) circles are
experimental data whereas the (blue) stars are simulations of the reaction
chain leading to $X=\pi^0\gamma$ normalized to the large $M_X$ data. After
subtracting this contribution, the (black) inverted triangles are a fitted
Gaussian representation of the $X=\pi^0$ signal.}
\end{center}
\end{figure}

Having identified the $\Lambda$ from its $\pi^-p$ decay, the final piece in
the jigsaw is to separate the missing masses corresponding to the $\pi^0$
(the $\Sigma(1385)$ case) and $\pi^0\gamma$ (the $\Lambda(1405)$ case). This
is illustrated in Fig.~\ref{fig:iza} for the ANKE data taken at
2.83~GeV~\cite{ZYC2008}. The experimental points are compared to the
simulation of the $\Lambda(1405)$ channel, where the $X=\pi^0\gamma$
distribution should be little affected by any uncertainties in the
experimental resolution. The remainder is a Gaussian fit to the $\pi^0$ peak,
which arises from the $\Sigma^0(1385)$ contribution. This fit shows that only
about 4 of the $\Sigma^0(1385)$ events lie above the experimental cut of
190~MeV/$c^2$ that was chosen to select the $\Lambda(1405)$
signal~\cite{ZYC2008}.

The biggest uncertainty in the quoted cross section,
\[\sigma_{\rm{tot}}(pp \rightarrow p K^+ \Lambda(1405)) = (4.5
\pm 0.6_{\rm{stat}}\pm 1.8_{\rm{syst}})\,\mu\textrm{b},\]%
comes from the extrapolation from the miniscule ANKE acceptance to the whole
of phase space, which is certainly a leap of faith. The rather complex
analysis reported in Ref.~\cite{ZYC2008} is typical of the approaches that
have to be undertaken in the study of heavy hyperon production at ANKE. The
acceptance at COSY-TOF is much larger but this spectrometer was not designed
for events with two neutral particles.

The $\Lambda(1405)$ mass spectrum produced in a single
reaction~\cite{ZYC2008} sheds little light on the two-pole
hypothesis~\cite{JID2003}, though it is informative regarding the
interpretation of kaon pair production presented in sect.~\ref{ppKK}.

A similar but more detailed experiment was carried out at the higher energy
of 3.5~GeV using the HADES spectrometer at GSI~\cite{AGA2013}. Within the
uncertainties of the two ANKE and HADES points, it seems that the $Q$
dependence of $\sigma_{\rm{tot}}(pp \rightarrow p K^+ \Lambda(1405))$ could
be similar to that for $\sigma_{\rm{tot}}(pp \rightarrow p K^+ \Lambda)$.

An earlier experiment at ANKE~\cite{ZYC2006} investigated the three-particle
correlations in $pp\to K^+p\pi^-X^+$ and $pp\to K^+p\pi^+X^-$, also at
2.83~GeV. For both reaction channels evidence was found for bumps that could
correspond to a heavy hyperon with mass $M(Y^{0*}) = (1480\pm 15)$~MeV/$c^2$
and width $\Gamma(Y^{0*}) = (60\pm 15)$~MeV/$c^2$. The isospin dependence of
this effect was not established and there is little supporting evidence for
either a $\Lambda$ or a $\Sigma$ state in this region in the current PDG
tables~\cite{OLI2014}.

\subsection{Pentaquarks} 
\label{pentaquark}

Following the introduction of the quark model, mesons were generally
categorized as $q\bar{q}$ pairs and baryons as $qqq$ triplets; any state that
did not fit into this scheme was classed as being \emph{exotic}. A prime
example of this was the $Z_0^*$ baryon, which has isospin-zero and
strangeness $+1$. Searches were made for such a state in the nineteen sixties
through the comparison of total cross sections of $K^+$ interactions with
hydrogen and deuterium but, because of decay losses in the kaon beams of low
momenta, the data were not sensitive to low mass
states~\cite{BUG1968,COO1970}.

The subject was opened again towards the end of the century with the proposal
for a narrow ($\Gamma < 15$~MeV/$c^2$) $Z_0^*$ baryon with a mass of $\approx
1530$~MeV/$c^2$~\cite{DIA1997}, which fell below the energy region of the
total cross section experiments~\cite{BUG1968,COO1970}. This suggestion
inspired numerous enthusiastic searches for such a state, which had been
renamed $\Theta^+$ by the theorists involved. Unlike some of the other
members of the proposed antidecuplet of states, there could be no doubt that
one in the $K^+n$ or $K^0p$ channel would be \emph{exotic} and so had, in the
quark model, to be of the form $q\bar{q}qqq$, i.e., be a \emph{pentaquark}.

In view of the mixed results achieved at other laboratories, the search
carried out at COSY-TOF looked very clean and promising because it was well
adapted to the unique characteristics of the spectrometer~\cite{ABD2004}. The
experiment consisted of a measurement of the $pp\to K^0p\Sigma^+$ reaction,
which was already described in sect.~\ref{TOF_Sigmaplus}. At a beam momentum
of 2.95~GeV/$c$, the excess energy was only $Q=126$~MeV, which limited the
$K^0p$ invariant mass to lie below about 1562~MeV/$c^2$.

Immediately after the TOF start counter there were two tracks from the proton
and the $\Sigma^+$ but, after a $K^0_s\to\pi^+\pi^-$ decay, these become four
tracks within the TOF barrel. Also within the barrel the $\Sigma^+$ decays
into either $\pi^0p$ or $\pi^+n$, and events were retained that showed the
resulting kink in the track. The decay kinematics and angular distributions
allow a clear suppression of the main background arising from the $pp \to
K^+p\Lambda$ reaction. Clear peaks were seen in the reconstruction
corresponding to the $\Sigma^+$ and $K^0$ and the resolution in these masses
was consistent with the simulations, which showed that the resolution in
$m(K^0p)$ should be $18\pm3$~MeV/$c^2$. There were some 939 events that
passed all the required cuts.

\begin{figure}[t]
\begin{center}
\includegraphics[width=1.0\columnwidth]{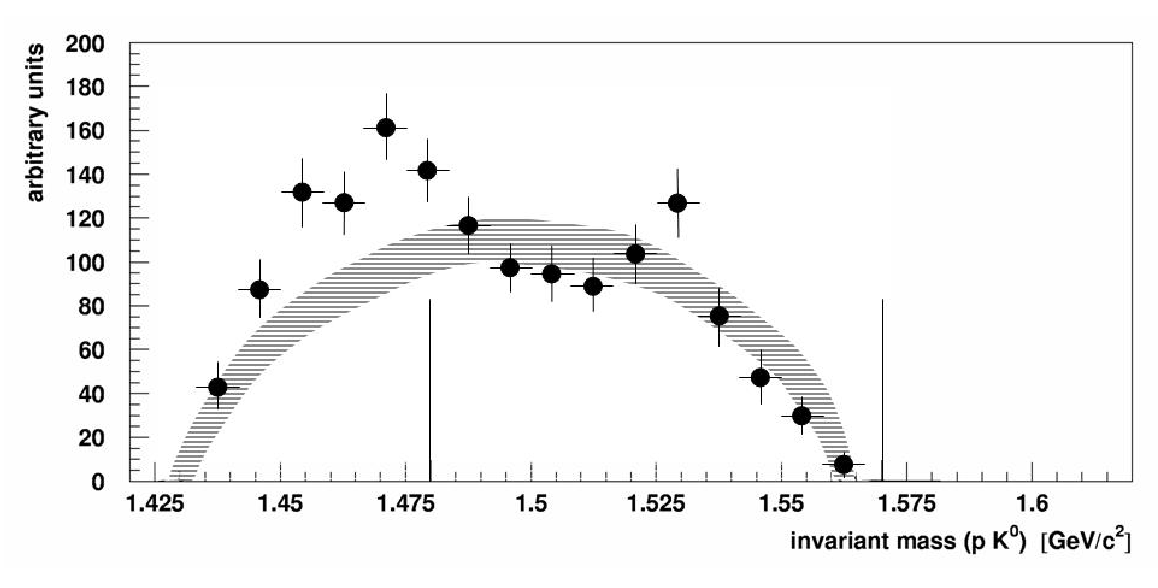}
\caption{\label{fig:penta1} Invariant $K^0p$ mass distribution from the
$pp\to K^0p\Sigma^+$ reaction. The points represent the results of the
initial COSY-TOF experiment at $Q=126$~MeV~\cite{ABD2004}. The shaded band
corresponds to the spectrum evaluated in the second experiment at
$Q=161$~MeV~\cite{ABD2007}. To account for the different values of $Q$, the
invariant mass axis has been scaled according to the kinematically allowed
range and the height of the band adjusted so that the averages agreed for
data between the two vertical lines, i.e., $m(K^0p)> 1480$~MeV/$c^2$.}
\end{center}
\end{figure}

The initial data shown in Fig.~\ref{fig:penta1} display a peak at
$1530\pm5$~MeV/$c^2$ with a width of $18 \pm 4$~MeV/$c^2$ (FWHM), which is
completely consistent with the expected mass resolution. This seemed to
present strong evidence for the existence of the $\Theta^+$ with the expected
properties.

The definitive COSY-TOF experiment was carried out at the slightly higher
momentum of 3.059~GeV/$c$~\cite{ABD2007}, so that the upper bound of the
$K^0p$ invariant mass was increased to 1.597~GeV/$c^2$, which is further away
from the suggested pentaquark peak. The experimental method was similar to
the previous one~\cite{ABD2004} but with the much higher statistics of more
than 12,000 independent events. In addition, three different approaches were
used in the analysis, depending largely on how the information regarding the
$\Sigma^+$ track was included. All three results agreed within statistical
uncertainties.

No evidence was found for any peak, especially one in the 1530~MeV/$c^2$
region, and an upper limit of the $pp \to \Sigma^+\Theta^+$ production cross
section of 150~nb was found at the 95\% confidence level. It should be noted
here that systematic studies of the instrumental background allowed
corrections to be made to the second data set~\cite{ABD2007} that were not
available for the earlier results~\cite{ABD2004}. An even lower limit was
found in a subsequent analysis~\cite{SCH2014}, depending upon the assumed
$\Theta^+$ mass.

To give a \emph{qualitative} illustration of the difference between the two
sets of COSY-TOF results, we show in Fig.~\ref{fig:penta1} the original COSY-TOF
points~\cite{ABD2004} together with a grey band corresponding to the newer
COSY-TOF results~\cite{ABD2007}. To account for the different beam energies,
the $K^0p$ invariant mass axis has been scaled according to the kinematically
allowed range.

A very different approach was undertaken at ANKE, where the four-body final
state in the $pp \to K^0p\pi^+\Lambda$ reaction was investigated at
2.83~GeV~\cite{NEK2007}. Here the $\Lambda$ was identified through its
$\pi^-p$ decay so that the actual final state was $K^0pp\pi^+\pi^-$. Rather
than measuring the $K^0$ through its $\pi^+\pi^-$ decay, as was done at
COSY-TOF~\cite{ABD2004,ABD2007}, it was deduced from the missing mass of the
other four particles, viz.\ $pp\pi^+\pi^-$. With this selection procedure,
1041 events were found corresponding to the $pp \to K^0p\pi^+\Lambda$
reaction. Assuming a phase-space dependence this led to a total cross section
of $\sigma(pp \to K^0p\pi^+\Lambda) = 1.41 \pm 0.05 \pm 0.33~\mu$b.

No peak was found that could correspond to the $\Theta^+$ and the upper limit
for producing this state
\[\sigma\left(pp\to \Lambda\pi^+\Theta^+\right) < 58~\textrm{nb}\]
at the 95\% confidence level depended weakly on the assumed position and
width of the state. Without a reaction model for the production of this
non-existent state, it is not useful to try to compare this limit with that
obtained at COSY-TOF for the $pp\to \Theta^+\Sigma^+$
reaction~\cite{ABD2007}, though the order of magnitude larger statistics in
the COSY-TOF case should certainly be noted.

Despite the negative results of the COSY-TOF experiment, pentaquarks are far
from being dead since the LHCb collaboration at CERN have recently claimed
two peaks in the $J/\Psi p$ system with high statistical significance but
with masses over 4~GeV/$c^2$~\cite{AAI2015}.

\subsection{Hyperon production in proton-neutron collisions} 
\label{pn_strange}

There is relatively little information available on hyperon production in
proton-neutron collisions and what does exist is rarely very systematic. The
first indications of the strength of production on the neutron came from a
comparison of inclusive $K^+$ production by protons on hydrogen and deuterium
targets~\cite{VAL2011a}. Below the threshold for $\Sigma$ production, the
$K^+$ rates are dominated by quasi-free $pp\to K^+p\Lambda$ and $pn\to
K^+n\Lambda$ reactions but, at higher energies, $\Sigma$ production and even
the formation of kaon pairs must also be considered.

The obvious disadvantage of this approach is that the centre-of-mass energy
is not well determined, due to the Fermi motion in the deuteron target, and
some modeling is required. After doing this, it was found that the weighted
average of the production ratio over the three lowest beam energies was
$\sigma^{K^+}_{pd}/\sigma^{K^+}_{pp}=1.4\pm0.2$ which, after taking shadowing
into account, means that the ratio of $K^+$ production cross sections in $pn$
and $pp$ collisions
\begin{equation}
\label{ratty}
R=\sigma_{pn}^{K^+}/\sigma_{pp}^{K^+}=0.5\pm0.2\,.
\end{equation}

At low energies, where only $\Lambda$ production is possible, the $0.5\pm0.2$
is comfortably above the isospin lower bound $R \geqq 0.25$. The result in
this region for the ratio of cross sections of definite isospin becomes
\begin{equation}
\sigma^{I=0}(NN\to KN\Lambda)/\sigma^{I=1}(NN\to KN\Lambda) = 1.0\pm0.8\,.
\end{equation}%
It should, of course, be noted in this context that, for a neutron target,
half of the $I=0$ signal would be associated with $K^0$ production.

The inclusive approach was later refined by detecting a \emph{spectator}
proton in an STT in coincidence with the $K^+$ measured in
ANKE~\cite{DZY2010}. Apart from being more selective, the spectator proton
allowed the reconstruction of the c.m.\ energy on an event-by-event basis.
Although extensive data were taken below the $\Sigma$ threshold, definitive
results on the quasi-free $pn\to K^+\Lambda n$ cross section are still not
available.

\begin{figure}[hbt]
\begin{center}
\includegraphics[width=0.9\columnwidth]{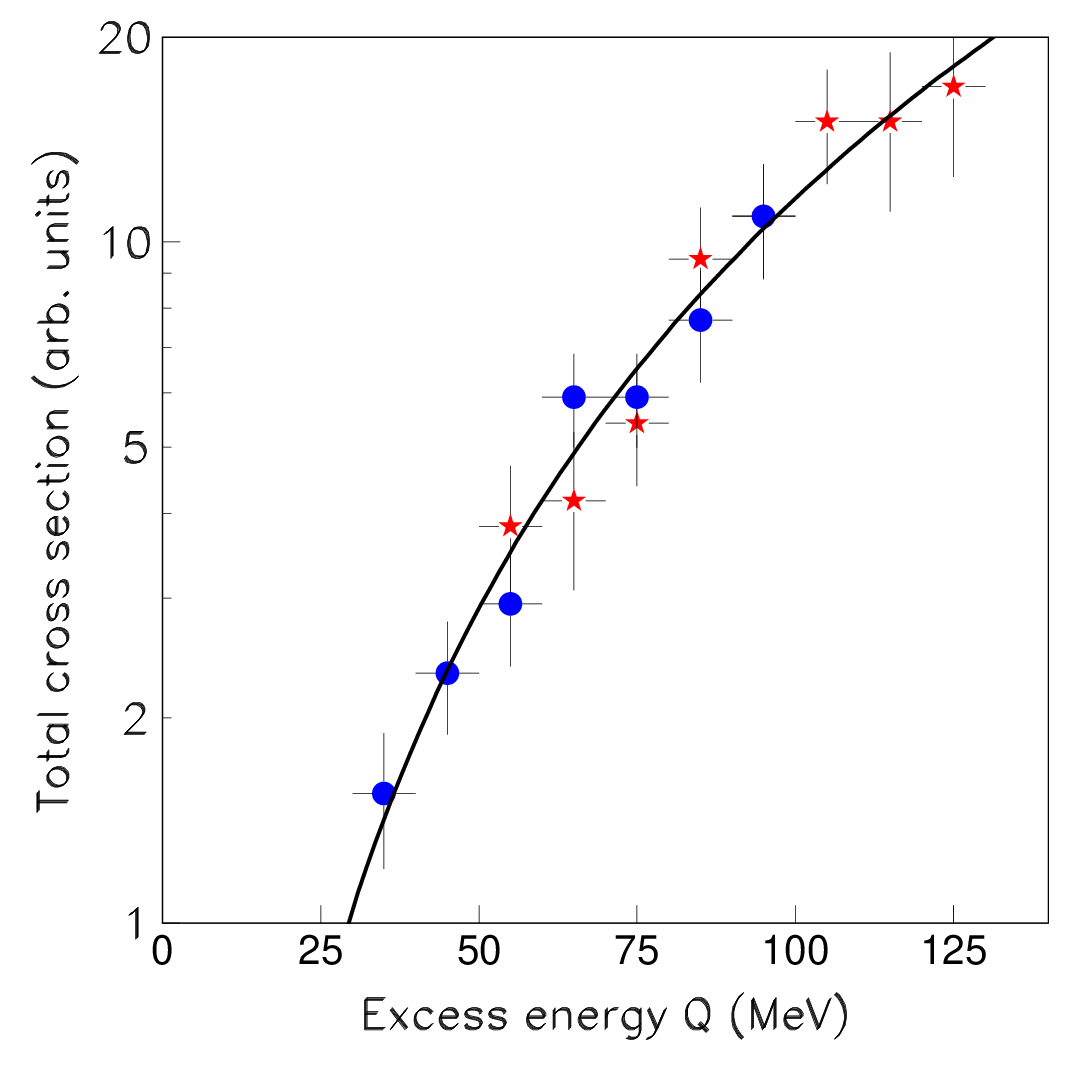}
\caption{\label{fig:Shikov} The total cross section for the quasi-free $pn\to
K^+p\Sigma^-$ reaction extracted from proton-deuteron collisions at
2.915~GeV/$c$ (blue circles) and 3.015~GeV/$c$ (red stars)~\cite{SHI2013}.
Due to difficulties in evaluating the acceptance involving the STT, these
preliminary values are not normalized and the 3.015~GeV/$c$ results have been
multiplied by a factor of 1.3 compared to Ref.~\cite{SHI2013} in order to
ensure good relative normalization between the two data sets. The curve,
$\sigma \propto Q^2$, represents undistorted three-body phase space. }
\end{center}
\end{figure}

A similar use of the STT was made in the study of the quasi-free measurement
of the $pd\to p_{\rm sp}K^+p\Sigma^-$ reaction at ANKE~\cite{SHI2013}. By
measuring the spectator proton $p_{\rm sp}$ in a silicon tracking telescope,
a scan over a wide range of centre-of-mass energies was achieved while
keeping the beam energy fixed. The $K^+$ and primary proton could then be
detected in ANKE and the $\Sigma^-$ identified through the missing-mass peak.
The experiments were carried out at beam momenta of 2.915 and 3.015~GeV/$c$.
If these had been undertaken on a free neutron target they would have
corresponded to excess energies of about 110 and 140~MeV, respectively. These
values are reduced by the deuteron binding energy and the energy taken by the
spectator proton in the STT. However, the biggest effect comes from the
placing of the STT towards the backward hemisphere so that the useful
coverage shown for the total cross sections of Fig.~\ref{fig:Shikov} is
$30\leq Q \leq 130$~MeV.

Due to complications in evaluating the acceptance involving the STT, the
total quasi-free $pn\to K^+p\Sigma^-$ cross sections of Fig.~\ref{fig:Shikov}
have not yet been reliably normalized and they are displayed in arbitrary
units. Nevertheless, over the range of energies shown, the cross section
seems to follow the $Q^2$ behaviour expected from three-body phase space.
This is seen in other $\Sigma$ production reactions but, of course, the
effects of a strong $\Sigma N$ final state interaction might well only show
up closer to threshold.

By charge symmetry, the cross sections for $pn\to K^0\Lambda p$ should be
identical to that for $pn\to K^+\Lambda n$. The reaction $pd\to p_{\rm
sp}K^0p\Lambda$ was studied in the COSY-TOF detector~\cite{KRA2012}.
Immediately after the target, only the track of the fast proton is seen, the
\emph{spectator} proton $p_{\rm sp}$ being undetectable. However, after
$\Lambda\to p\pi^-$ and $K^0\to\pi^+\pi^-$ decays, five tracks are present,
which is an excellent signal for a good event. The main difficulty is
identifying the two decay vertices from the four tracks with their
measurement errors -- the combinatorial background. In practice, this
combinatorial background can be much more of a problem than the physical
background.

In the data taken at 2.261~GeV, fewer than one in $10^5$ events corresponded
to the $pd\to p_{\rm sp}K^0p\Lambda$ reaction and so the robust trigger was
crucial~\cite{KRA2012}. Nevertheless, the preliminary cross section ratio of
Eq.~(\ref{ratty}) seems to be only just above the isospin limit of $R=0.25$,
though this value depends sensitively upon the detector efficiencies. In
stark contrast to the ANKE result~\cite{VAL2011a}, this would imply very
little isospin $I=0$ production. However, a more detailed COSY-TOF data set
is currently under analysis~\cite{ROD2016}.

%
%

\section{Kaon pair production} 
\setcounter{equation}{0}%

Since strangeness is conserved in strong interactions, the production of a
$K^+$ must be associated with that of particles with a net strangeness $-1$.
In the previous section these were hyperons but now we turn to $K^+K^-$
production. However, the specific case of $pd \to{}^3\textrm{He}\,K^+K^-$
will be deferred until sect.~\ref{pd3hephi}, where it will be treated along
with other $pd \to{}^3\textrm{He}\,X$ reactions.

\subsection[Kaon pair production in nucleon-nucleon collisions]{Kaon pair production in nucleon-nucleon collisions} 
\label{ppKK}

Well above the threshold for hyperon production in proton-proton collisions
it is possible to produce kaon pairs through the $pp\to ppK^+K^-$ reaction.
The early COSY experiments in this area were carried out by the COSY-11
collaboration~\cite{WOL1997,QUE2001,WIN2006,SIL2013} and these results are
crucial in the determination of the energy dependence of the total cross
section near threshold. The larger acceptance available at the ANKE
spectrometer allowed experiments to be carried out at higher excess energies,
including above the $\phi$ threshold. The higher statistics also meant that
more differential observables could be usefully
measured~\cite{HAR2006,MAE2008,YEa2012,YEa2013}.

In the COSY-11 experiments the two protons were first identified and measured
using the time-of-flight information in combination with the momentum
analysis~\cite{WOL1997,QUE2001,WIN2006}. Due to the decay of the $K^+$, the
probability that it reached the second stop counter is of the order of a few
percent and so its four-momentum was evaluated using the time difference from
the target to the first stop counter, using start information derived from
the proton measurements. Though $K^-$ were also detected by a combination of
scintillator and silicon pads, the reaction was confirmed through the
evaluation of the missing mass with respect to the two protons and the $K^+$
candidate, which showed a peak at the $K^-$ mass. The value of the overall
luminosity was reliably derived by measuring proton-proton elastic scattering
in parallel. More problematic was the $\approx 10\%$ uncertainty in the
efficiency, which was estimated by including only the $pp$ final state
interaction in the four-body phase space. Distortions due to the $K^-p$ and
$K^+K^-$ final state interactions were not considered for this purpose.

The experimental procedure was somewhat different at
ANKE~\cite{HAR2006,MAE2008,YEa2012,YEa2013}, where the $K^+$ candidate was
first selected using time-of-flight information\footnote{A use of the
delayed-veto trigger for the $K^+$ would have reduced the acceptance by
almost an order of magnitude.}. The signal from the $K^+$ stop counter was
then used in the determination of the momenta of the $K^-$ and one of the
protons. The reaction was finally identified by looking for the proton peak
in the missing mass with respect to the $pK^+K^-$ recoiling system. The
uncertainties in the overall acceptance are slightly less than those at
COSY-11 but, for data taken above the $\phi$ threshold, there is an
additional uncertainty associated with the separation of $\phi$ and
non-$\phi$ events. The resulting $\phi$ data will be reviewed in the next
subsection. Proton-proton elastic scattering was also used as the basis for
the luminosity determination, though there was greater ambiguity in the $pp$
database at the small angles used at ANKE, which has only recently been
clarified~\cite{MCH2016}.

\begin{figure}[hbt]
\begin{center}
\includegraphics[width=0.9\columnwidth]{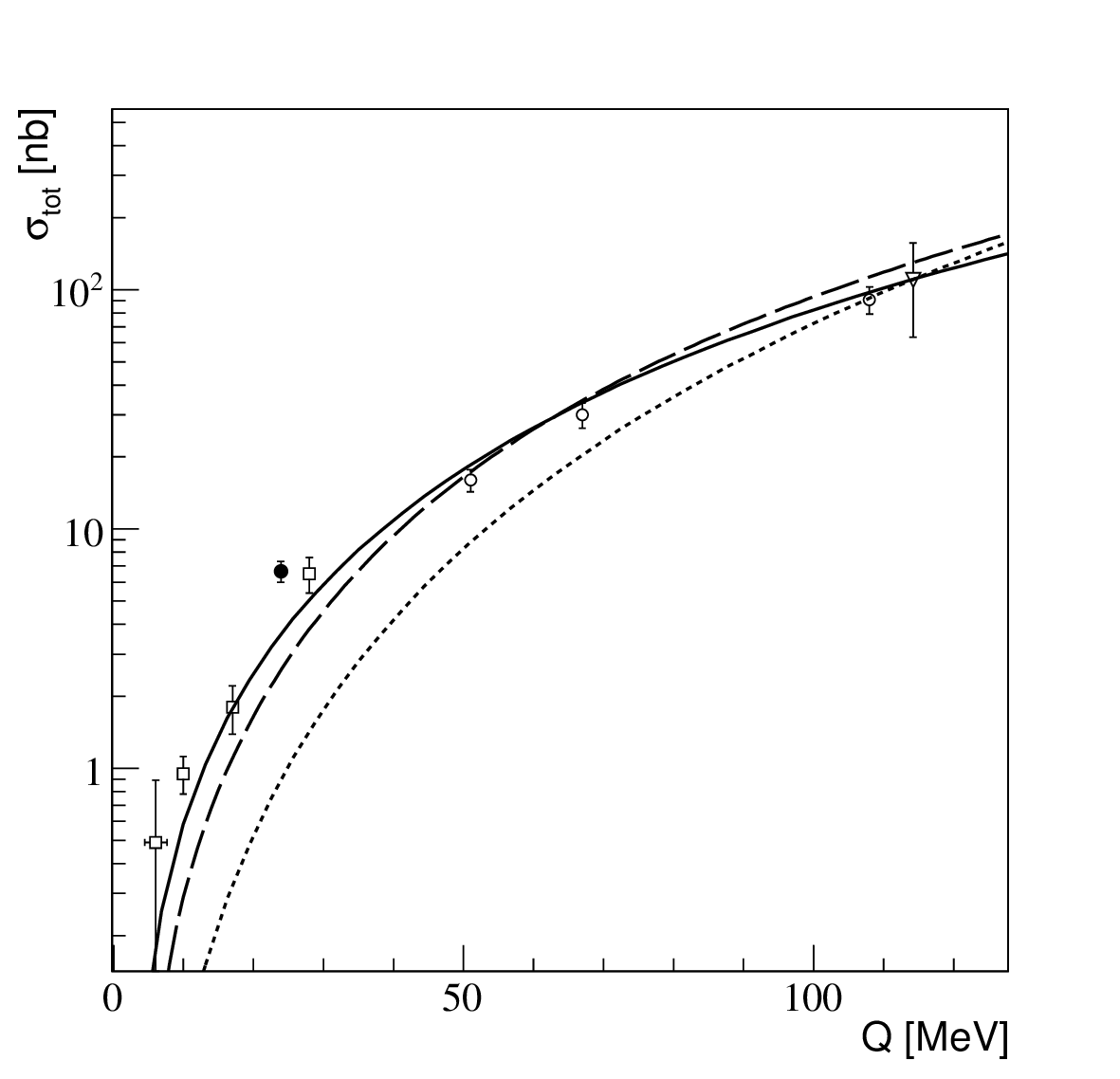}
\caption{\label{fig:Ye8} Total cross section for the non-$\phi$ contribution
to the $pp\to ppK^+K^-$ reaction as a function of the excess energy $Q$. The
data are taken from DISTO (triangle)~\cite{BAL2001}, COSY-11
(squares)~\cite{WOL1997,QUE2001,WIN2006}, and ANKE (open and closed
circles)~\cite{HAR2006,MAE2008,YEa2012,YEa2013}. The dotted line shows the
four-body phase space simulation, whereas the solid line represents the
simulations of Eq.~(\ref{assume}) with $a_{K^-p}=1.5i$~fm. The predictions of
a one-boson exchange model are shown by the dashed line~\cite{SIB1997}. }
\end{center}
\end{figure}

The total cross sections measured at COSY for the non-$\phi$ contribution to
the $pp\to ppK^+K^-$ reaction are shown in Fig.~\ref{fig:Ye8} along with one
point measured earlier by the DISTO collaboration~\cite{BAL2001}. These show
a steady rise with excess energy $Q$ but the four-body phase space normalized
in the 100~MeV region seriously underestimates the low energy
data\footnote{Unpublished COSY-11 data suggest that the cross section is
below $0.1$~nb at $Q\approx 4.5$~MeV~\cite{GIL2015}, which is more
restrictive than the previous COSY-11 upper limit of 0.16~nb at
$Q=3$~MeV~\cite{QUE2001a}. This departure from the trend shown by the higher
energy data might be due to the Coulomb repulsion between the $K^+$ and the
two protons, which must be critical so close to threshold.}. Part of this can
be compensated through the introduction of a $pp$ final state interaction
but, in order to get a reasonable description near threshold, some attraction
is required between the $K^-$ and each of the final protons. Clear evidence
for this is to be found in the differential distributions, to which we now
turn.

\begin{figure}[hbt]
\begin{center}
\includegraphics[width=0.9\columnwidth]{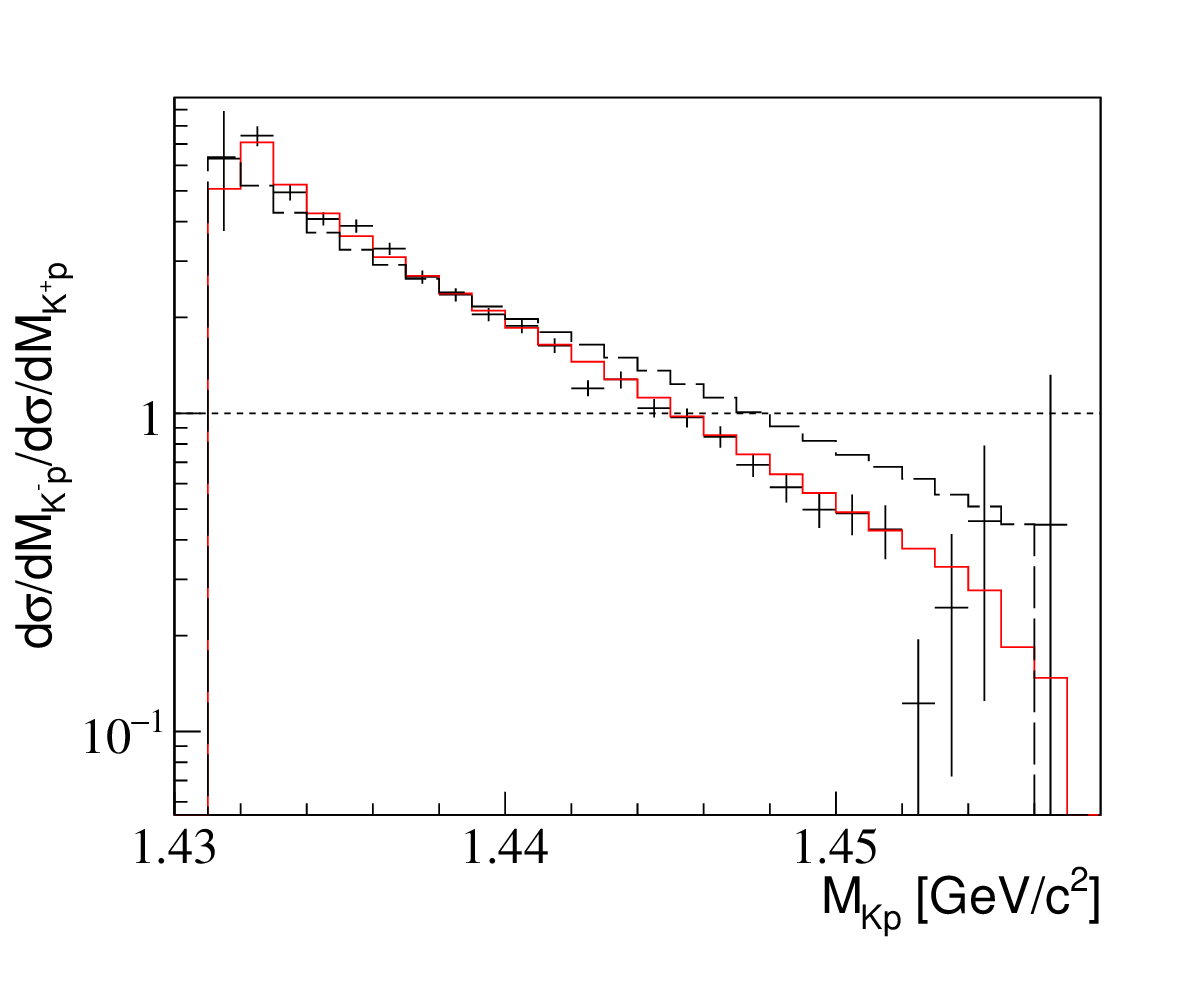}
\includegraphics[width=0.9\columnwidth]{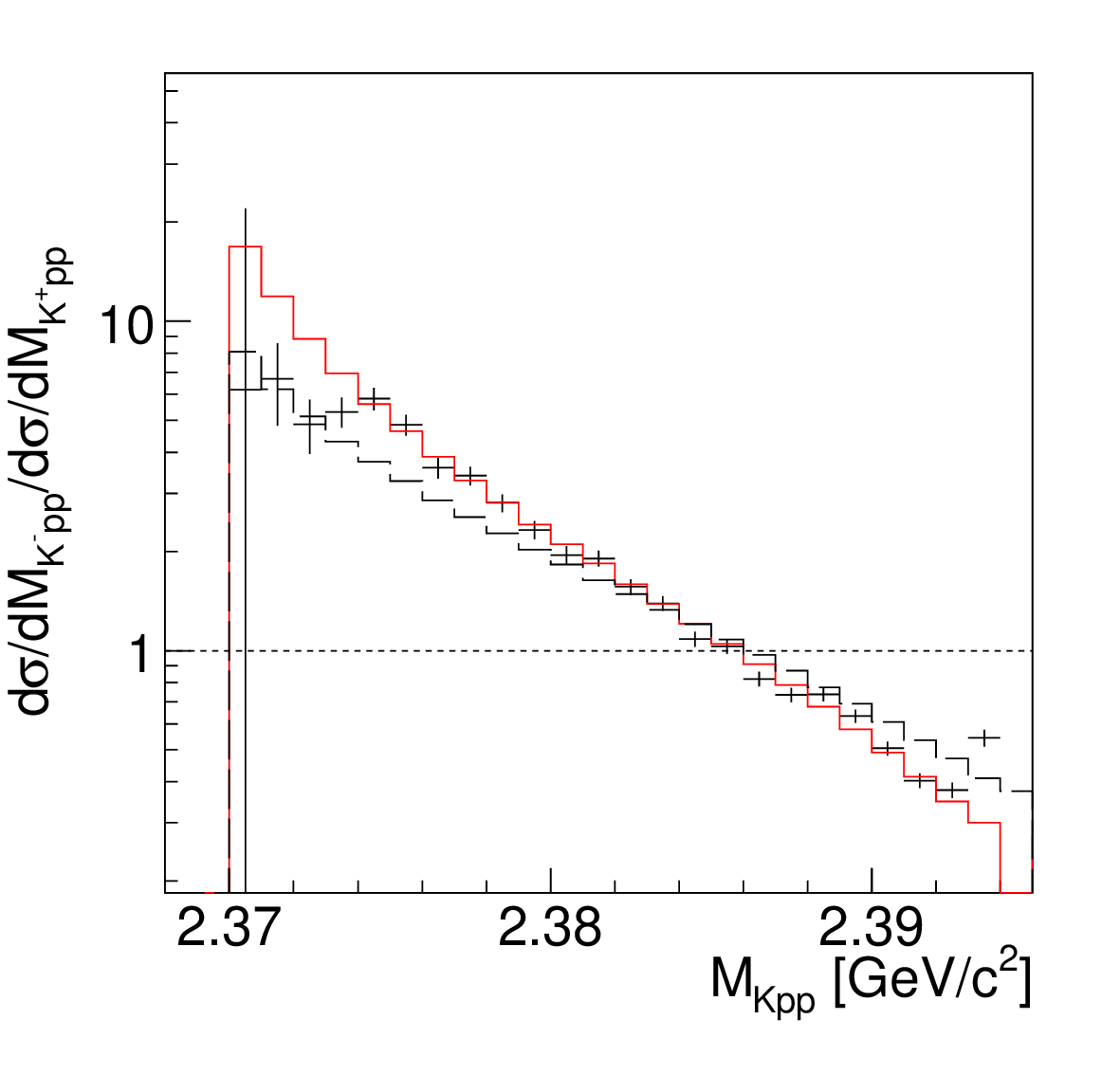}
\caption{\label{fig:Nick}  The ratios $R_{Kp}$ and $R_{Kpp}$ of
Eq.~(\ref{IMratio}) at $Q=24$~MeV~\cite{YEa2013}. The red solid and broken
black histograms represent estimations based on Eq.~(\ref{assume}) that take
into account $K^-p$, $pp$ and $K^+K^-$ final state interactions with
$a_{K^-p}=2.45i$~fm and $a_{K^-p}=1.5i$~fm, respectively. }
\end{center}
\end{figure}

It was first apparent in the COSY-11 data at $Q=10$ and 28~MeV~\cite{WIN2006}
that the $K^-p$ interaction was strongly attractive because the measured
events clustered around low $K^-p$ or even low $K^-pp$ invariant masses. This
observation was taken up by the ANKE
collaboration~\cite{MAE2008,YEa2012,YEa2013} which, following COSY-11,
constructed the ratios of cross sections with respect to the $K^{\pm}p$ and
$K^{\pm}pp$ invariant masses:
\begin{equation}
\label{IMratio}
R_{Kp} = \frac{\dd\sigma/\dd M_{K^-p}}{\dd\sigma/\dd M_{K^+p}}, \hspace{5mm}
R_{Kpp} = \frac{\dd\sigma/\dd M_{K^-pp}}{\dd\sigma/\dd M_{K^+pp}}\cdot
\end{equation}
The $R_{Kp}$ and $R_{Kpp}$ ratios at $Q=24$~MeV are shown in
Fig.~\ref{fig:Nick}~\cite{YEa2013}.

It is well known that there can be no rigorous estimation of an enhancement
factor $F$ when three or more particles interact in the final state.
Nevertheless, in the case of $pp\to pp\eta$, where there are strong
interactions between all three pairs of particles, the data can be well
described by taking the overall enhancement as the product of the three
two-body enhancements~\cite{BER1999}. This ansatz was also adopted in the
analysis of the ANKE data, where it was assumed that
\begin{equation}
\label{assume}
F = F_{pp}(q_{pp}) \times F_{Kp}(q_{Kp_{1}}) \times F_{Kp}(q_{Kp_{2}}) \times F_{KK}(q_{KK}),
\end{equation}
where $q_{pp}$, $q_{Kp_{1}}$, $q_{Kp_{2}}$, and $q_{KK}$ are the magnitudes
of the relative momenta in the $pp$, the two $K^-p$, and the $K^+K^-$
systems, respectively. It is believed that the $K^+p$ interaction is weakly
repulsive and may be neglected compared to the uncertainties in the other
effects.

The critical interaction in the $R_{Kp}$ ratio of Fig.~\ref{fig:Nick} is that
between the $K^-$ and each of the protons. A good description of the data is
achieved by assuming a simple $K^-p$ scattering length formula with
$a_{K^-p}=2.45i$~fm~\cite{YEa2013}. What is more surprising is that the
ansatz of Eq.~(\ref{assume}) gives an equally good description of the $Kpp$
data at $Q=24$~MeV. However, it must be admitted that $a_{K^-p}$ is an
effective parameter and should not necessarily be equated to the free $K^-p$
scattering length because the factorization assumption clearly does not
contain all of the relevant physics. In fact the ANKE data above the $\phi$
threshold are better fit with $a_{K^-p}=1.5i$~fm, though it should be noted
that the data are not very sensitive to the phase of
$a_{K^-p}$~\cite{MAE2008,YEa2012}.

\begin{figure}[hbt]
\begin{center}
\includegraphics[width=0.9\columnwidth]{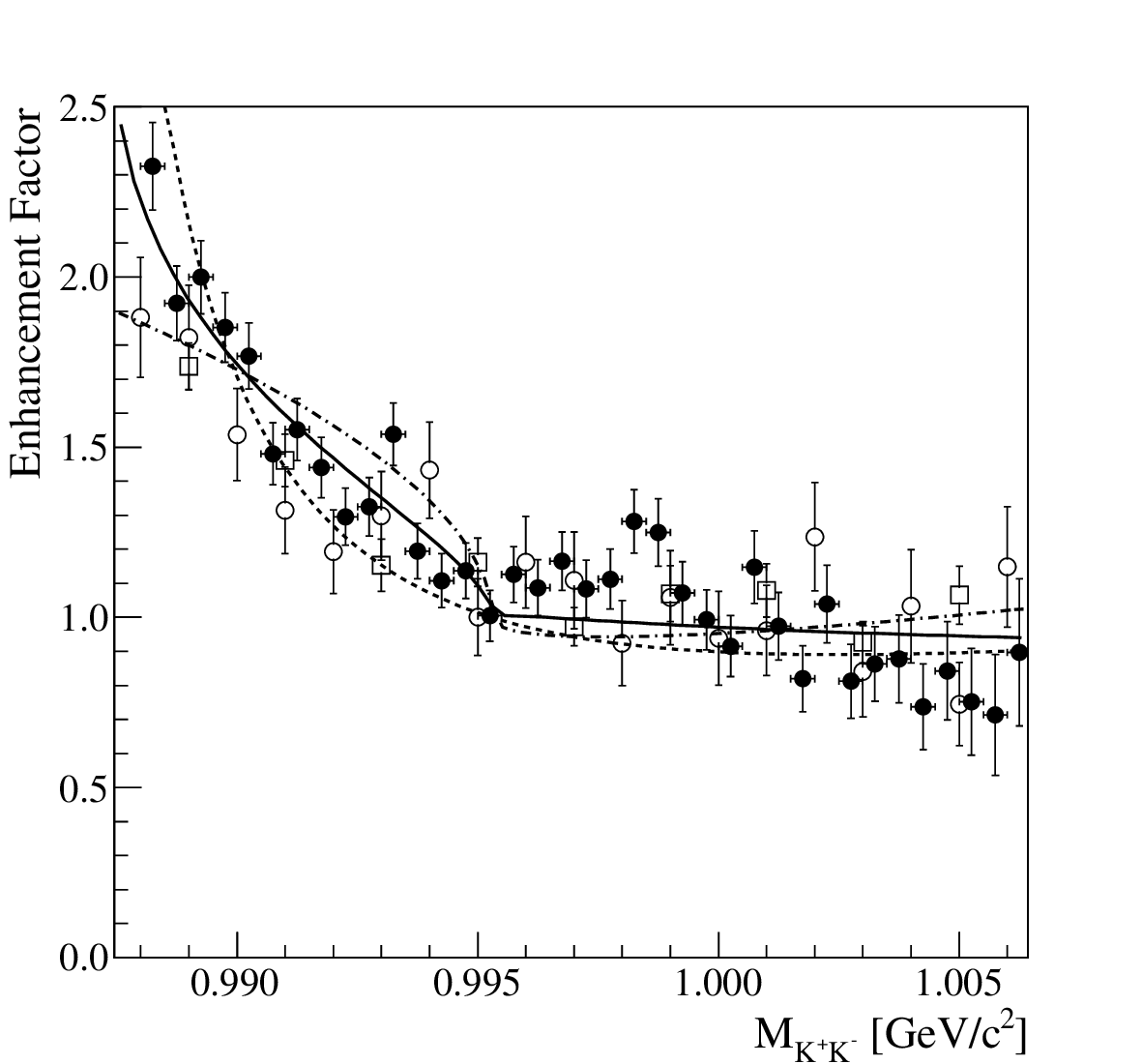}
\caption{\label{fig:Ye4} Ratio of $K^+K^-$ invariant mass distributions
measured at different energies at ANKE in the $pp\to ppK^+K^-$
reaction~\cite{MAE2008,YEa2012,YEa2013} to a simulation that includes only
$K^-p$ and $pp$ final state interactions. The solid curve represents the best
fit in a model that includes elastic $K^+K^-$ FSI and $K^0\bar{K}^0
\rightleftharpoons K^+K^-$ charge-exchange~\cite{DZY2008}. The best fits
neglecting charge exchange or neglecting the elastic $K^+K^-$ FSI are shown by
the dashed and dot-dashed curves, respectively. }
\end{center}
\end{figure}

Interesting effects also arise from mass differences. For example,
$2m_{K^0}-2m_{K^{\pm}} \approx 8$~MeV/$c^2$ and a change in behaviour might
be seen at the $K^0\bar{K}^0$ threshold in the $pp\to ppK^+K^-$ data as a
function of $m_{K^+K^-}$ shown in Fig.~\ref{fig:Ye4}. The inclusion of final
state interactions through $K^0\bar{K}^0 \rightleftharpoons K^+K^-$ in a
coupled-channel formalism was explored in Ref.~\cite{DZY2008} though even
these high statistics data were not sufficient to identify unambiguously a
cusp at the $K^0\bar{K}^0$ threshold. On the other hand, there is no evidence
for the production of the $a_0/f_0$ scalar mesons in this reaction, which was
one of the motivations for measuring $pp\to ppK^+K^-$.

Another mass difference that might be significant is $m_{K^0}+m_n-m_{K^-}-m_p
\approx 5.3$~MeV/$c^2$, which means that there might be some kind of anomaly
in the $K^-p$ mass distribution of Fig.~\ref{fig:Nick} at $m_{K^-p}\approx
1437$~MeV/$c^2$, though there is little sign of this in the data.

The inclusion of the $pp$, $K^-p$, and $K^+K^-$ final state interactions
improves significantly the prediction of the energy dependence of the $pp\to
ppK^+K^-$ total cross section shown in Fig.~\ref{fig:Ye8}.

\begin{figure}[hbt]
\begin{center}
\includegraphics[width=0.49\columnwidth]{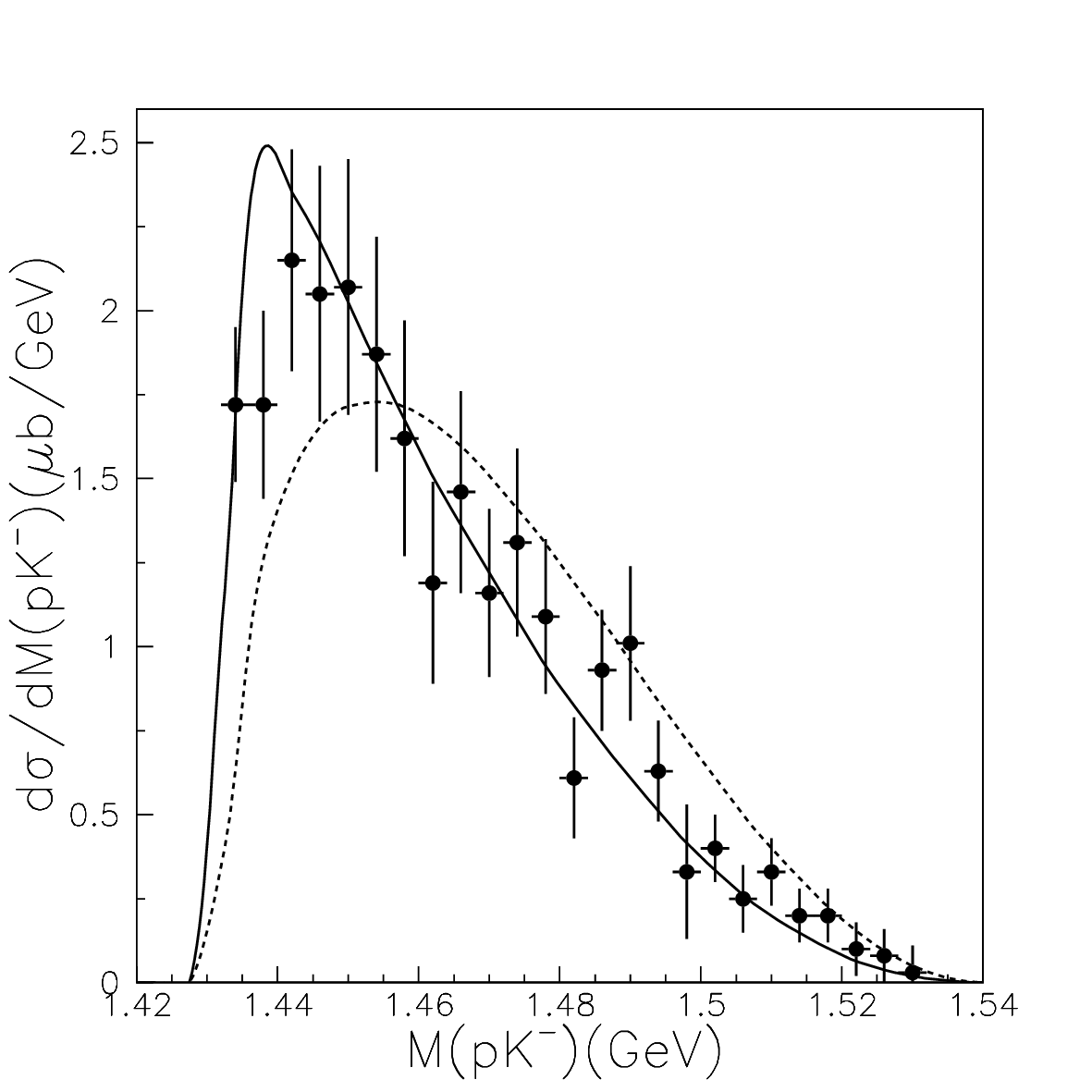}
\includegraphics[width=0.49\columnwidth]{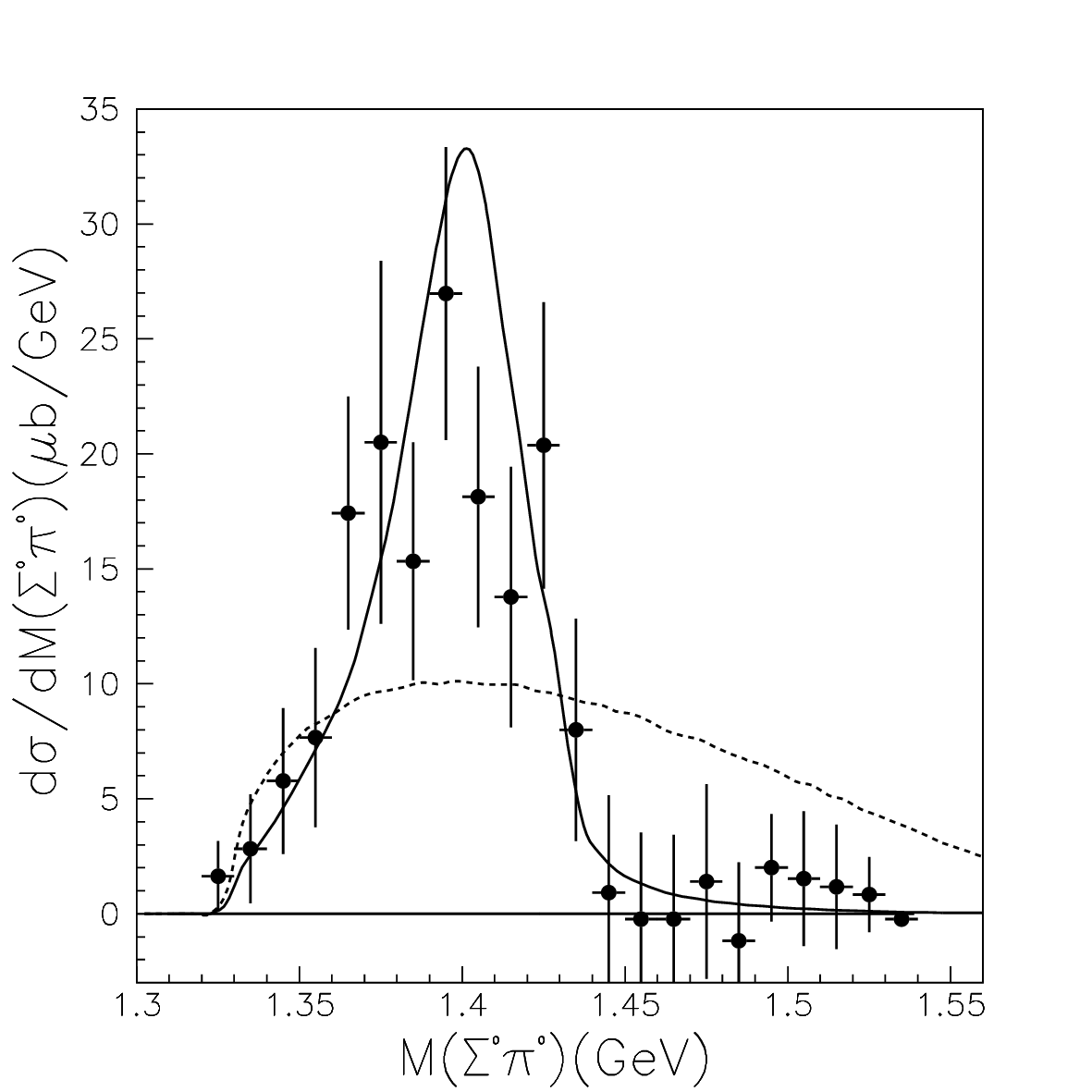}
\caption{\label{fig:Xie} Left: Differential cross section for the $pp \to
pK^+ K^-p$ reaction at $T_p=2.83$~GeV as a function of the $K^-p$ invariant
mass~\cite{MAE2008}. The predictions of the Lagrangian model are shown by
the solid line~\cite{XIE2010}. Right: Differential cross section for the $pp
\to pK^+ \pi^0 \Sigma^0$ reaction also at 2.83~GeV. The predictions of the
Lagrangian model (solid line) have been scaled by a factor of $0.4$ before
being compared to the ANKE data~\cite{ZYC2008}. In both cases the dashed line
represents normalized four-body phase space~\cite{XIE2010}.}
\end{center}
\end{figure}

The ANKE data on $\Lambda(1405)$ production in the reaction $pp\to
K^+p\,(\Lambda(1405)\to \Sigma^0\pi^0)$ were discussed in
sect.~\ref{heavy_hyperons}. Although the centre of this hyperon lies just
below the $K^-p$ threshold, it has a finite width and the high mass tail can
decay into the $K^-p$ channel. There is therefore a strong possibility that some of
the kaon pair production observed at COSY might be proceeding through this
doorway state, i.e., $pp\to K^+p\,(\Lambda(1405)\to K^-p)$. This would
account for the strong enhancements that are observed for low $K^-p$
invariant masses. This idea has been tested in a specific Lagrangian
model~\cite{XIE2010} and the results are shown in Fig.~\ref{fig:Xie}.

The shapes of both mass distributions shown in Fig.~\ref{fig:Xie} are very
encouraging and the normalization factor of $0.4$ may be within the combined
experimental and theoretical uncertainties. Though the predictions were based
upon a specific model whose normalization was tuned to fit the $pp \to pK^+
K^-p$ data, it is quite likely that the results and normalization for the $pp
\to pK^+ \pi^0 \Sigma^0$ reaction are more general than the particular model
used in Ref.~\cite{XIE2010}. It is clear that, if the $\Lambda(1405)$ plays
an important role in kaon pair production, one will find it very hard to
identify a signal for the production of the scalar bosons
$a_0(980)/f_0(980)$.

The background for the $a_0(980)$ could be smaller in the $K^+\bar{K}^0$
charged channel and searches were undertaken at ANKE for scalar meson
production in the $pp\to K^+\bar{K}^0d$ reaction at proton beam energies of
2.65~GeV~\cite{KLE2003} and 2.83~GeV~\cite{DZY2006}. Selection rules play a
very important role here because the combination of the Pauli principle and
angular momentum and parity conservation do not allow all three final
particles to be in relative $s$-waves. An analysis of the two data sets
suggests that it is the $\bar{K}^0d$ system that is dominantly in the $s$
wave, being driven by the attraction of the antikaon to nucleons and
nuclei~\cite{DZY2008a}. There is only very weak evidence for the possible
production of the $a_0^+(980)$ but the two final state interactions together
do reproduce a little better the energy dependence of the total cross
section.

\begin{figure}[hbt]
\begin{center}
\includegraphics[width=0.9\columnwidth]{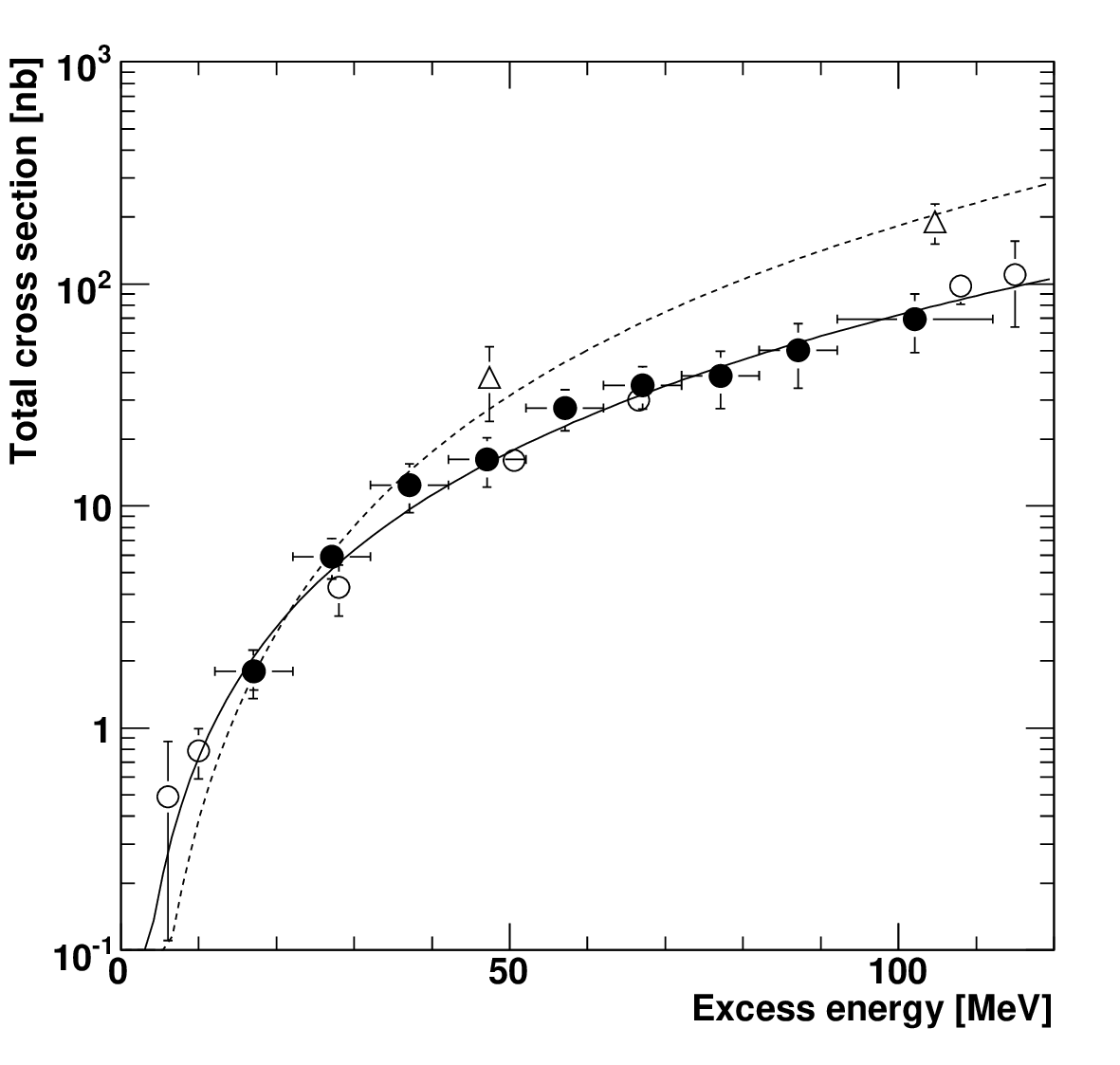}
\caption{\label{fig:Yoshi4} Total cross section for non-$\phi$ $K\bar{K}$
production in nucleon-nucleon collisions near threshold. The closed circles
denote $pn\to dK^+K^-$ data~\cite{MAE2009} and the open triangles $pp\to
dK^+\bar{K}^0$~\cite{KLE2003,DZY2006}, whereas the open circles show the
results for $pp\to ppK^+K^-$ up to
2008~\cite{WOL1997,QUE2001,WIN2006,MAE2008,BAL2001}. The dotted curve is the
best fit to the $pp\to dK^+\bar{K}^0$ data within a simple final state
interaction model~\cite{FAL1996}, whereas the solid curve includes also the
isospin-zero contribution needed to describe the energy dependence of the
$pn\to dK^+K^-$ total cross section.  }
\end{center}
\end{figure}

A closely allied reaction is $pn\to K^+K^-d$, which was studied at ANKE in
quasi-free kinematics using a deuterium target~\cite{MAE2009}. The effective
luminosity was determined using the Schottky technique that was discussed in
sect.~\ref{technical}~\cite{STE2008}. Though the beam energy was fixed at
$2.65$~GeV, the reconstruction of the $K^+K^-d$ centre-of-mass energy allowed
the reaction to be studied up to excess energies of around 100~MeV. Above
$Q=32.1$~MeV there was also the problem of separating direct $K\bar{K}$
production from that of the $\phi$, whose results are discussed in
sect.~\ref{NNphi}. The resulting total cross sections are shown in
Fig.~\ref{fig:Yoshi4} along with those for $pp\to
dK^+\bar{K}^0$~\cite{KLE2003,DZY2006}.

The relationship between the two reactions is seen more clearly in the
isospin basis of Eq.~(\ref{isospin}):
\begin{equation}
\sigma(pp\to dK^+\bar{K}^0)=\sigma_1,\hspace{2mm}
\sigma(pn\to dK^+K^-)=\fmn{1}{4}(\sigma_1+\sigma_0)
\label{isospin}
\end{equation}
An interpolation of the $pn$ results to energies where $pp\to dK^+\bar{K}^0$
was measured~\cite{KLE2003,DZY2006} gives isospin ratios of
$\sigma_0/\sigma_1 = 0.9\pm 0.9$ at $Q=47$~MeV and $0.5\pm0.5$ at 105~MeV,
where the large error bars arise from the subtraction implicit in
Eq.~(\ref{isospin}). All that one can reasonably conclude from this is that
$\sigma_0$ cannot be much larger than $\sigma_1$, despite the necessity for
having a $p$-wave in the final state in the $I=1$ case.

Also shown in Fig.~\ref{fig:Yoshi4} are values of the $pp\to ppK^+K^-$ total
cross sections~\cite{WOL1997,QUE2001,WIN2006,MAE2008,BAL2001}, which are very
similar in magnitude to those for $pn\to dK^+K^-$~\cite{MAE2009}. However,
some account must be taken of the difference between the 3-body and 4-body
final states but, when this is done, one sees that
\begin{equation}
\sigma(pp\to ppK^+K^-)/\sigma(pn\to \{pn\}_{I=0}K^+K^-) \approx 1.5\,,
\end{equation}
though this estimate is rather model-dependent.

The $K^-d/K^+d$ cross section ratio, i.e., the analogue of
Eq.~(\ref{IMratio}), shows the usual preference for the $K^-$ to be attracted
to the deuteron. A reasonable agreement with the data was achieved with a
scattering length of $a_{K^-d} = (-1.0 + i1.2)$~fm, which would be in line
with theoretical expectations~\cite{MEI2006}.

%
%

\subsection[$pp\to pp\,\phi$ and $pn\to d\,\phi$ reactions]{$\boldsymbol{pp\to pp\,\phi}$ and $\boldsymbol{pn\to d\,\phi}$ reactions} 
\label{NNphi}

\begin{figure}[hbt]
\begin{center}
\includegraphics[width=0.9\columnwidth]{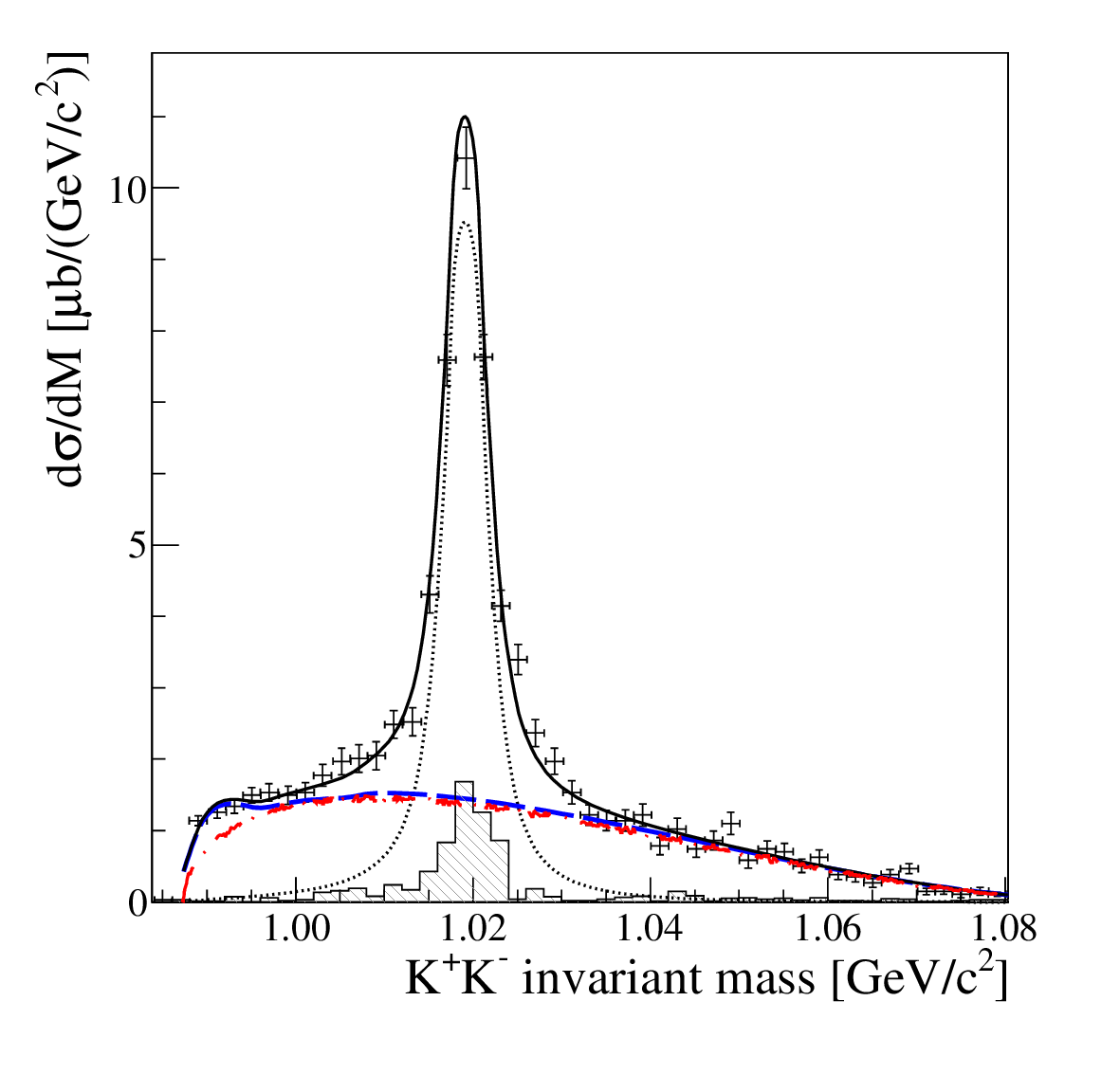}
\caption{\label{fig:Nick8} The $pp\to ppK^+K^-$ differential cross section at
2.83~GeV as a function of the $K^+K^-$ invariant mass~\cite{YEa2012}. The
error bars are statistical; systematic uncertainties are shown by the hatched
histogram. The dashed-dotted (red) curve represents four-body phase-space
that is distorted slightly by the FSI to give the dashed (blue) curve. The
dotted curve is the fit to the $\phi$ contribution, whereas the solid line is
the incoherent sum of the $\phi$ and non-$\phi$ contributions. }
\end{center}
\end{figure}

The ANKE experiments on $K^+K^-$ production in both
$pp$~\cite{HAR2006,MAE2008,YEa2012} and $np$~\cite{MAE2006} collisions were
primarily motivated by the study of $\phi$ production, where the meson was
detected through its decay $\phi\to K^+K^-$. Since the cross sections are
low, and the multipion backgrounds are high, it is hard to isolate the $\phi$
in $pp$ collisions just by detecting the final protons. Even by reducing the
background by several orders of magnitude by demanding the presence of a
$K^+K^-$ pair in the final state, the separation of the $\phi$ from direct
$K^+K^-$ production is non-trivial, as illustrated in Fig.~\ref{fig:Nick8}.

Of immediate interest here is the ratio $R_{\phi/\omega}$ of the production
of the $\phi$ and $\omega$ vector mesons in reactions where there are no
strange particles in the initial state. According to the Okubo-Zweig-Iizuka
(OZI) rule~\cite{OZI1963}, one expects $R_{\phi/\omega}$ to be on the order
of $R_{\rm OZI}=4\times 10^{-3}$. Using $\omega$ production data from
Refs.~\cite{HIB1999,SAM2001} it was found that in $pp$ collisions
$R_{\phi/\omega} \approx 6-8\times R_{\rm OZI}$~\cite{HAR2006} and a similar
conclusion was reached using later COSY-TOF data~\cite{ABD2007a}. This is
consistent with the result obtained in the $pn$ case~\cite{MAE2006} on the
basis of $pn\to d\omega$ data taken at COSY~\cite{BAR2004}. These values are
to be contrasted with the $1-2.4\times R_{\rm OZI}$ found in high energy
$pp\to ppV$ data~\cite{BAL1998b}.

Close to threshold, the two final-state protons in the $pp\to pp\phi$
reaction must be in the $^{1\!}S_0$ wave, and the $\phi$ in an $s$ wave
relative to this pair. The $\phi$ is then aligned with polarizations $m=\pm
1$ along the beam direction. The polar angular distribution of the decay
kaons in the $\phi$ meson rest frame must then display a
$\sin^2\theta_{\phi}^K$ shape, which is consistent with the $Q=18.5$~MeV
data~\cite{HAR2006}. On the other hand, at $Q=76$~MeV the kaon angular
distribution is almost isotropic, which means that the $\phi$ is essentially
unpolarized~\cite{YEa2013}. This is an unambiguous proof of the importance of
higher partial waves in $\phi$ production at 76~MeV.

A similar conclusion can be drawn from the study of the $pn\to d\phi$
data~\cite{MAE2006}. At threshold a pure $\sin^{2}\theta^{K}_{\phi}$ is
required but, if the data are parameterized as
\begin{equation}
\dd\sigma/\dd\Omega_{\phi}^{K} = 3\,(a\sin^{2}\theta^{K}_{\phi} +
2 b \cos^{2}\theta^{K}_{\phi})/8\pi, %
\end{equation}%
then the best fit gives $b/a \approx 0.012\,(Q/\textrm{MeV})$. The $\phi$ are
therefore produced unpolarized for $Q\approx 40$~MeV.

\begin{figure}[hbt]
\begin{center}
\includegraphics[width=0.9\columnwidth]{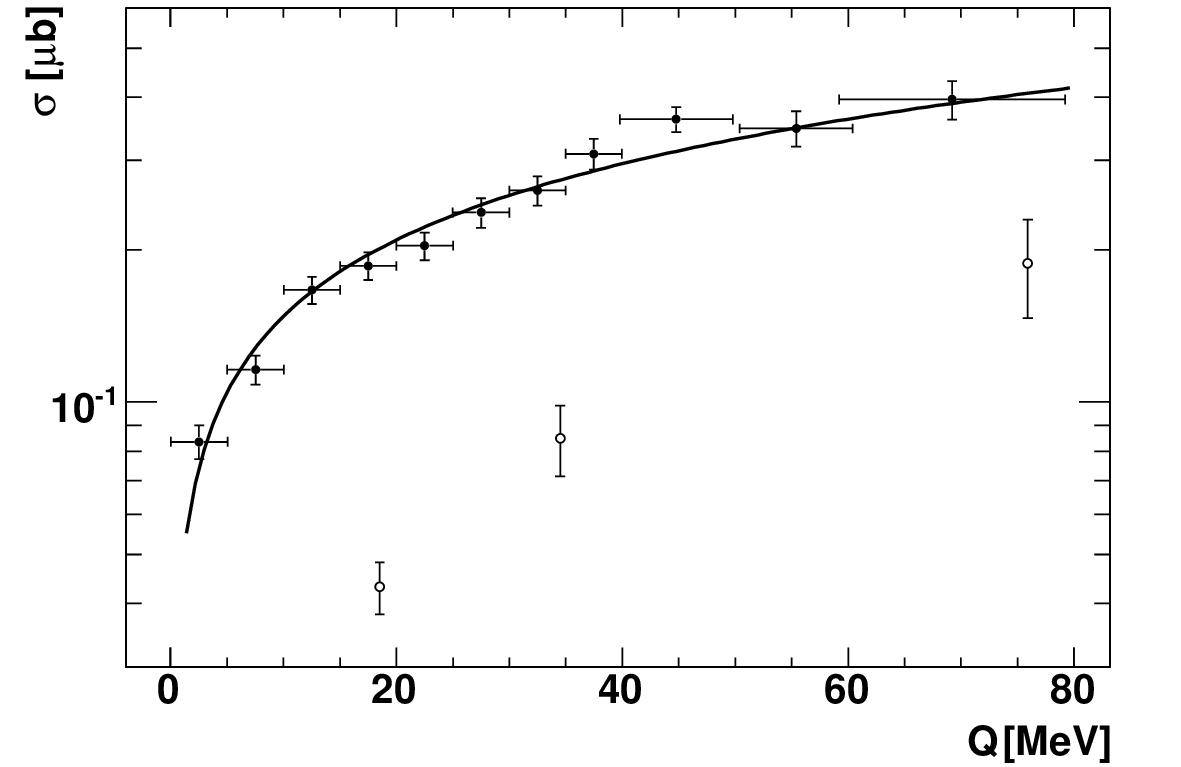}
\caption{\label{fig:Yoshi3} Total cross section for the quasi--free $pn\to
d\phi$ reaction as a function of the excess energy (filled
circles)~\cite{MAE2006}. The curve represents a phase--space $\sqrt{Q}$
behaviour. For comparison, also shown by open circles are the values obtained
for $pp\to pp\phi$ in the first ANKE experiment~\cite{HAR2006}.}
\end{center}
\end{figure}

Despite the evidence for higher partial waves coming from the angular
distributions, there is no sign of their effect in the total cross section.
Thus the total cross section for the quasi-free $pn\to d\phi$ reaction shown
in Fig.~\ref{fig:Yoshi3} is well described by the curve representing
$48\sqrt{Q/\textrm{MeV}}$. The figure also shows that, over the range in $Q$
measured, the $pn\to d\phi$ cross section is much bigger than that of $pp\to
pp\phi$, though one has to bear in mind the difference between a two-body and
a three-body final state.

%
%

\subsection[$pA\to K^+K^-X$ and $pA\to \phi X$]{$\boldsymbol{pA\to K^+K^-X}$ and $\boldsymbol{pA\to \phi X}$} 

In the experiments described so far, and those that will be discussed in
sec~\ref{pd3hephi}, only exclusive $\phi$ production was studied. In such
hard processes the interaction of the $\phi$ with nucleons plays only a minor
role compared to the uncertainty of the reaction mechanism itself. In order
to investigate how the $\phi$ interacts with the nuclear medium, its
inclusive production was studied with 2.83~GeV protons incident on C, Cu, Ag,
and Au nuclear targets~\cite{POL2011,HAR2012}. Only the $K^+K^-$ pair was
detected in the ANKE facility but, unlike the $pp\to ppK^+K^-$ experiments
that were the subject of sects.~\ref{ppKK} and \ref{NNphi}, the delayed-veto
technique~\cite{BUS2002} was employed to identify the $K^+$ unambiguously.

\begin{figure}[hbt]
\begin{center}
\includegraphics[width=0.9\columnwidth]{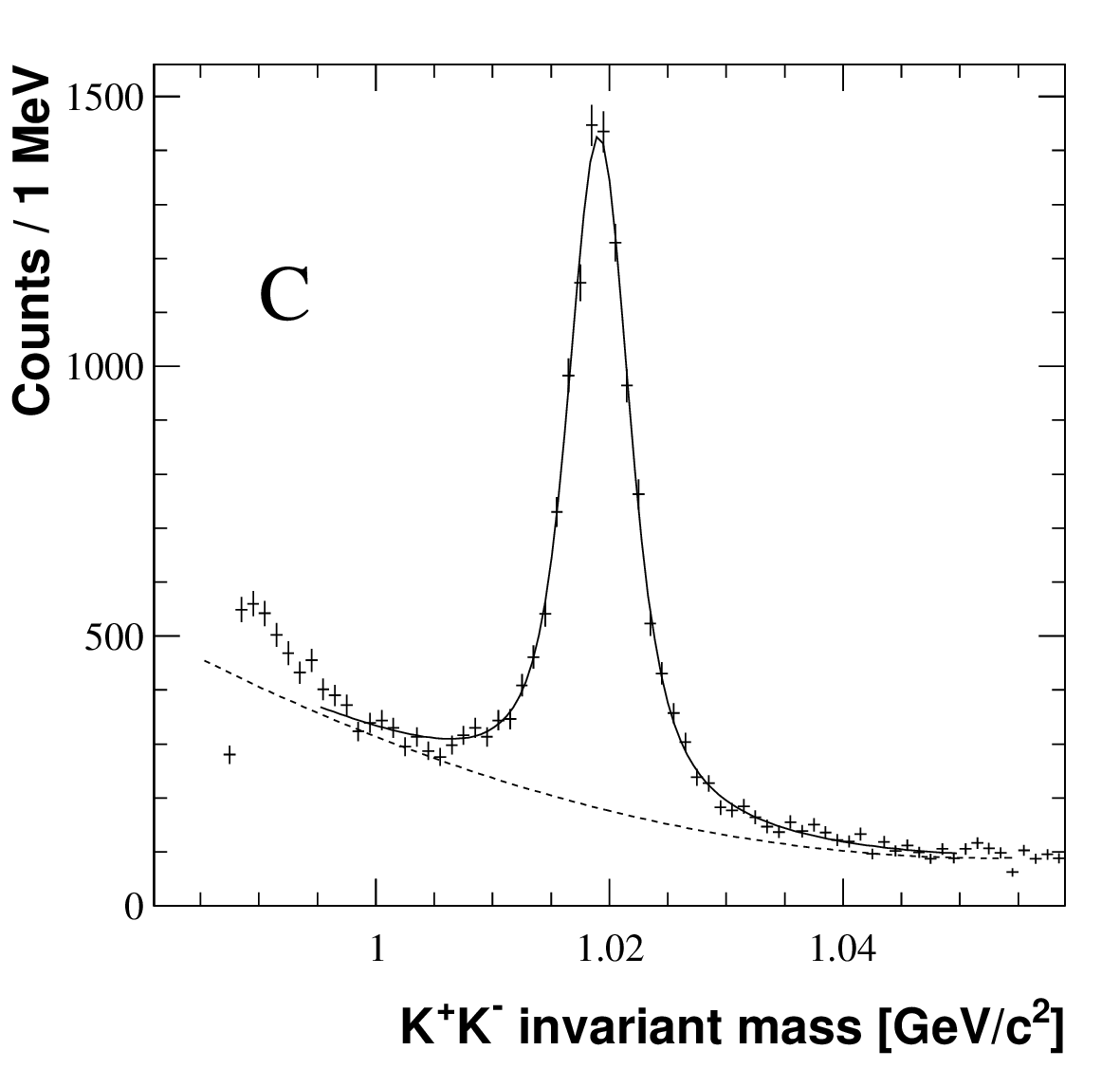}
\caption{\label{fig:Pol1a} Invariant mass distributions for $K^{+}K^{-}$
pairs produced in $p$C collisions at 2.83~GeV~\cite{POL2011}. The
experimental data are not acceptance-corrected. The dashed line is a
second-order polynomial representation of the background in the region of
the $\phi$ peak.}
\end{center}
\end{figure}

The $K^+K^-$ peak corresponding to the $\phi$ meson was clearly seen for all
four nuclear targets, the carbon example being shown in Fig.~\ref{fig:Pol1a}.
The background, which was dominantly due to direct pair production, was
parameterized by a quadratic function in $m_{K^+K^-}$ in order to make
subtractions under the invariant mass peak. The resulting number of
reconstructed $\phi$ mesons for each target was between 7000 and 10000. It
has, of course, to be recognized that these numbers depended sensitively upon
the ANKE acceptance for the positive and negative kaons.

The standard way of describing such data is through the evaluation of the
so-called transparency ratios
\begin{equation}
\label{transparency}%
R = \frac{ 12~\sigma_{pA \to \phi X'}}{A~\sigma_{p{\rm C} \to \phi X}}
\end{equation}
normalized to carbon. Here $\sigma_{pA \to \phi X'}$ and $\sigma_{p{\rm C}
\to \phi X}$ are inclusive cross sections for $\phi$ production in $pA$ and
$p\,$C collisions, respectively. By dividing by the carbon data one takes
into account production on neutrons as well as protons, though it must be
noted that there is a significant neutron excess in the heavier targets.

\begin{figure}[hbt]
\begin{center}
\includegraphics[width=0.9\columnwidth]{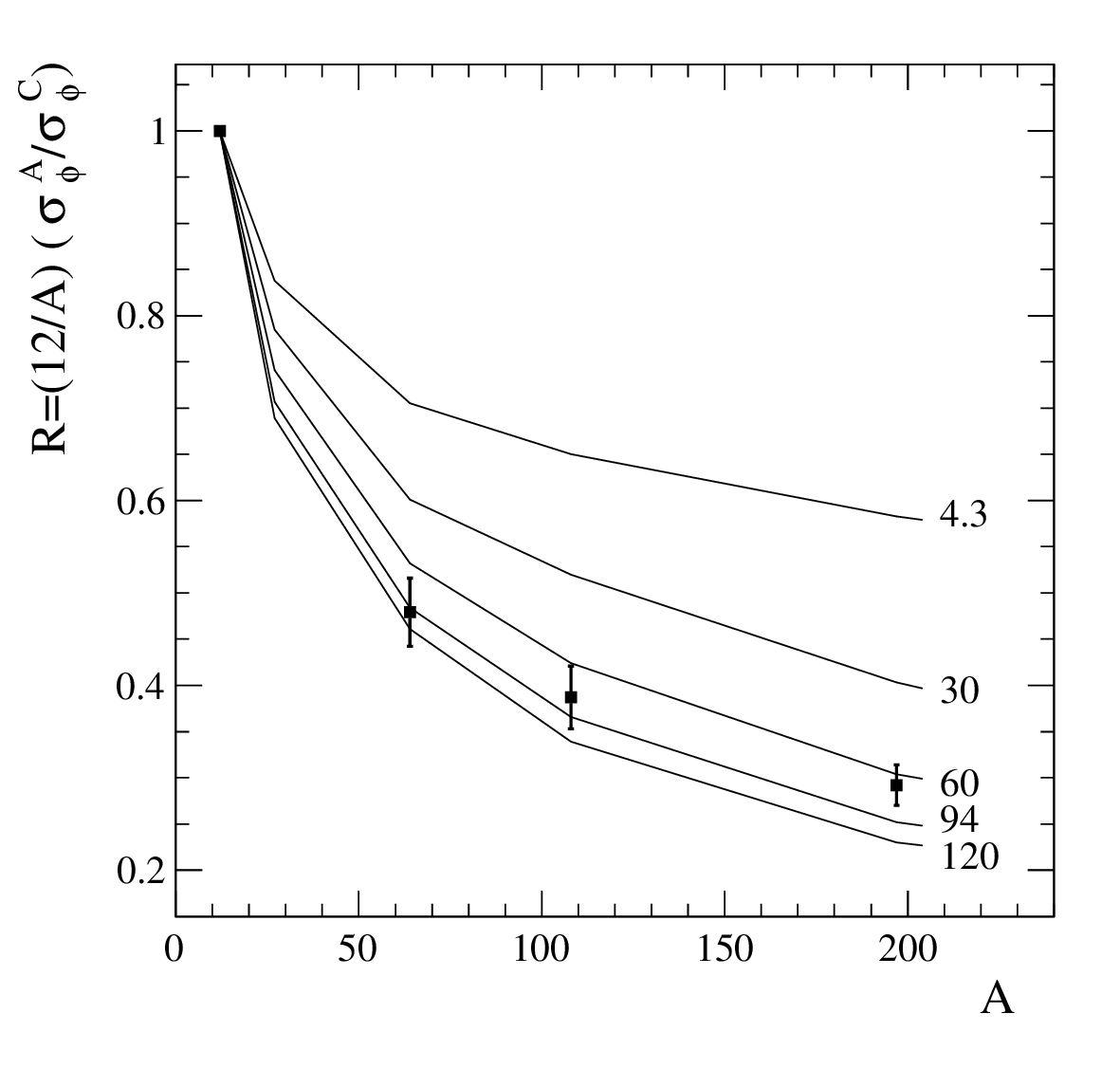}
\caption{\label{fig:Pol2b} Comparison of the measured transparency ratio $R$
with the predictions of the Paryev model~\cite{PAR2008} for various values of
the $\phi$ width in the nuclear medium (in MeV/$c^2$)~\cite{POL2011}.}
\end{center}
\end{figure}

The values obtained for the transparency ratios are shown in
Fig.~\ref{fig:Pol2b}. Any interpretation of these data has to rely on a
detailed theoretical treatment and the curves shown in the figure are the
predictions of one specific model~\cite{PAR2008} for various values of the
$\phi$ width $\Gamma_{\phi}$ in nuclear matter, taking into account the
effects of the ANKE cuts on the distributions in the laboratory $\phi$
momenta and production angles. It should be noted that in the low density
limit the medium contribution to $\Gamma_{\phi}$ is proportional to the
$\phi$-nucleon total cross section.

The best fit for the in-medium width within this model is achieved with
$\Gamma_{\phi} = {73}^{+14}_{-10}$~MeV/$c^2$. Since the average $\phi$
momentum at ANKE is around 1.1~GeV/$c$, this corresponds to $\Gamma_{\phi}
\approx 50$~MeV/$c^2$ in the nuclear rest frame. However, there is
significant model dependence, especially regarding the relative production
rates on protons and neutrons, but also on the importance of two-step
processes induced, e.g., by intermediate pions. Independent of any particular
model, the results suggest a substantial increase in the total $\phi$ width
in the nuclear environment.

In a subsequent analysis~\cite{HAR2012}, the dependence of the transparency
ratio on the $\phi$ momentum was investigated over the range $0.6 < p_{\phi}
< 1.6$~GeV/$c$. In all the models tested, the experimental results show
evidence for an increase of $\Gamma_{\phi}$ with the $\phi$ momentum.
However, these models do not reproduce satisfactorily the large numbers of
$\phi$ mesons produced with momenta below 1~GeV/$c$.

Figure~\ref{fig:Pol1a} shows, in addition to the $\phi$ peak, a large number
of $K^+K^-$ pairs produced ``directly'' in the $pA\to K^+K^-X$ reaction.
These have been used to investigate the interaction of the $K^-$ with nuclear
matter~\cite{KIS2015}. In the framework of a specific reaction
model~\cite{PAR2015}, the data are best described by a moderately attractive
$K^-$-nucleus potential of depth $\approx 60$~MeV for an average kaon
momentum $\approx 0.5$~GeV/$c$.

%
%

\section[The $pd\to{}^3$He$\,X(^3$H$\,X')$ family of reactions]{The $\boldsymbol{pd\to{}^3}$He$\boldsymbol{\,X}(^3$H$\boldsymbol{\,X'})$ family of reactions} 
\setcounter{equation}{0}%

Despite the $pd\to{}^3$He$\,X(^3$H$\,X')$ reactions generally involving large
momentum transfers, they have the big experimental advantage of requiring the
measurement of only one charged particle, $^3$He or $^3$H, in the final state
in order to reconstruct $X$ through the missing mass in the reaction. For
this reason there have been numerous studies of such reactions over the
years. At COSY there were important measurements of single and two-pion
production as well as the formation of $K^+K^-$ pairs, including $\phi$
production. However, the most fascinating data are associated with $\eta$
production, which has proved to be a very rich field to explore at COSY. Data
taken at WASA on the combined measurement of the production of $\eta$ mesons
and single and multiple pions above the $\eta$ threshold are currently being
analyzed but definitive results with absolute normalizations are not yet
available~\cite{KHO2016}.

\subsection[$pd\to{}^3$He$\,\pi^0$ and $pd\to{}^3$H$\,\pi^+$]{$\boldsymbol{pd\to{}^3}$He$\,\boldsymbol{\pi^0}$ and 
$\boldsymbol{pd\to{}^3}$H$\,\boldsymbol{\pi^+}$}
\label{pd3hepi}

The most extensive published measurements of the differential cross section
of the $pd\to{}^3$He$\,\pi^0$ and $pd\to{}^3$H$\,\pi^+$ reactions was
undertaken by the GEM collaboration, where the final $^3$He or $^3$H was
detected~\cite{BET2001,ABD2003}\footnote{Data on the $pd\to{}^3$He$\,\pi^0$
reaction taken by the WASA collaboration at a variety of energies are
currently under analysis~\cite{KHO2016}}. Isospin invariance predicts that
there should be a factor of two difference in the cross sections, though the
spin dependence of the observables should be identical in the two channels.

\begin{figure}[h!]
\begin{center}
\includegraphics[width=0.9\columnwidth]{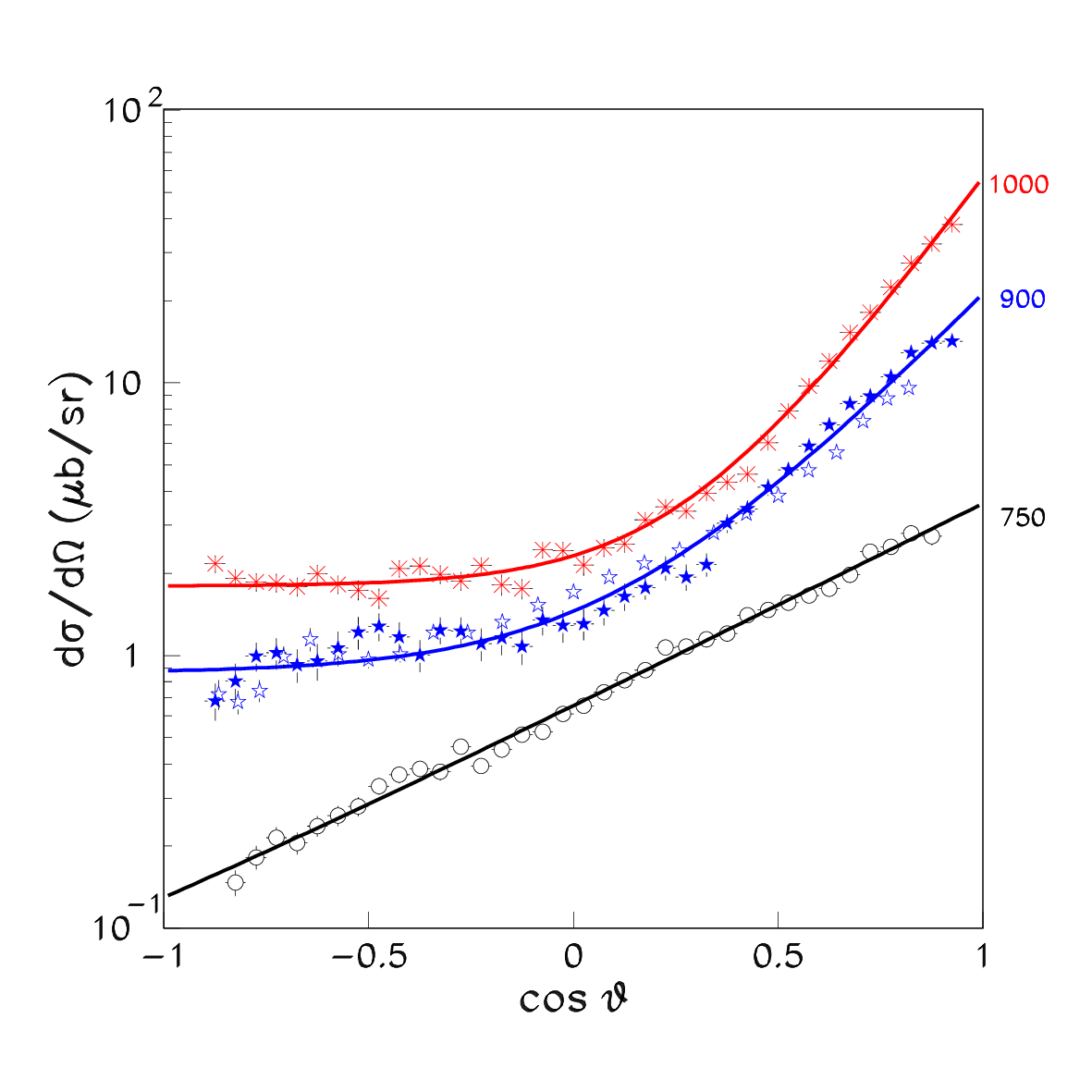}
\caption{\label{fig:gem_dcs2} Differential cross sections for the
\mbox{$pd\to{}^{3}\textrm{He}\,\pi^0$} reaction  measured by the GEM
collaboration at the three beam momenta indicated~\cite{BET2001,ABD2003}. For
presentational purposes, the data at 900~MeV/$c$ have been scaled by a factor
of 2.25 and at 1000~MeV/$c$ by (2.25)$^2$. The (blue) open stars represent the
results of the TRIUMF measurement at 883~MeV/$c$~\cite{CAM1987} using the
900~MeV/$c$ scaling factor. Also shown are the fits based upon
Eq.~(\ref{GEM_par}).}
\end{center}
\end{figure}

Data were taken over the whole angular range for both $\pi^0$ and $\pi^+$
production for laboratory momenta between 750 and 1000 or 1050~MeV/$c$.
Samples of the $\pi^0$ data are plotted in Fig.~\ref{fig:gem_dcs2}. Also
shown are data obtained at TRIUMF at a beam momentum of
883~MeV/$c$~\cite{CAM1987}, which are broadly in line with the GEM results at
900~MeV/$c$.

\begin{figure}[h!]\centering
\includegraphics[width=0.8\columnwidth,clip=]{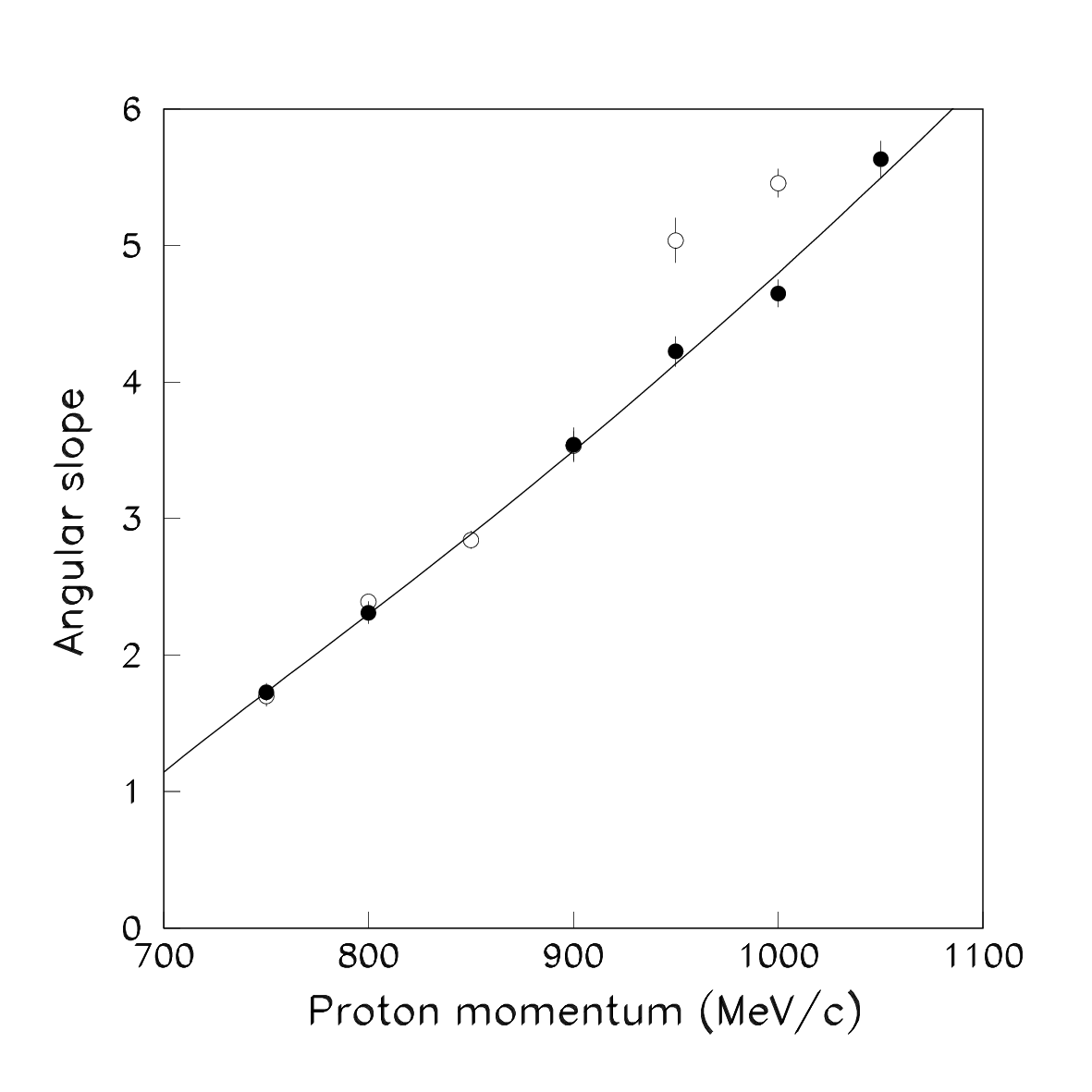}
\caption{Slope parameter $c$ of Eq.~(\ref{GEM_par}) deduced from fits to the
GEM data on \mbox{$pd\to{}^{3}\textrm{He}\,\pi^0$} (closed circles) and
\mbox{$pd\to{}^{3}\textrm{H}\,\pi^+$} (open circles) differential cross
sections~\cite{BET2001,ABD2003}. Since the slope parameters should be similar
for the two channels, the values of $c$ for the $\pi^+$ case may be
questioned at the two highest momenta. Also shown is the arbitrarily
normalized curve $1.31pk$, where the proton ($p$) and pion ($k$) c.m.\
momenta are measured in fm$^{-1}$. \label{c_slope}}
\end{figure}

The shape of the data changes with beam momentum and the results are well fit
with:
\begin{equation}
\label{GEM_par}
\frac{\dd\sigma}{\dd\Omega} = a + \exp\{b + c(\cos\theta -1)\}.
\end{equation}
The GEM authors~\cite{BET2001,ABD2003} argue that the change in the slope
parameter $c$ is mainly kinematic; at larger beam momenta the momentum
transfer between the deuteron and $^3$He increases faster with angle. As a
consequence, the slope parameter should vary like $pk$, where $p$ and $k$ are
the proton and pion momenta in the c.m.\ frame. This behaviour is clearly
seen in the values of the fit parameters shown in Fig.~\ref{c_slope}.

%
%

Equation~(\ref{GEM_par}) also suggests that there are at least two mechanisms
that play important roles here. At small angles the reaction might involve a
spectator nucleon but at large angles all nucleons seem to be
involved~\cite{GER1990}.

\begin{figure}[hbt]
\begin{center}
\includegraphics[width=1.0\columnwidth]{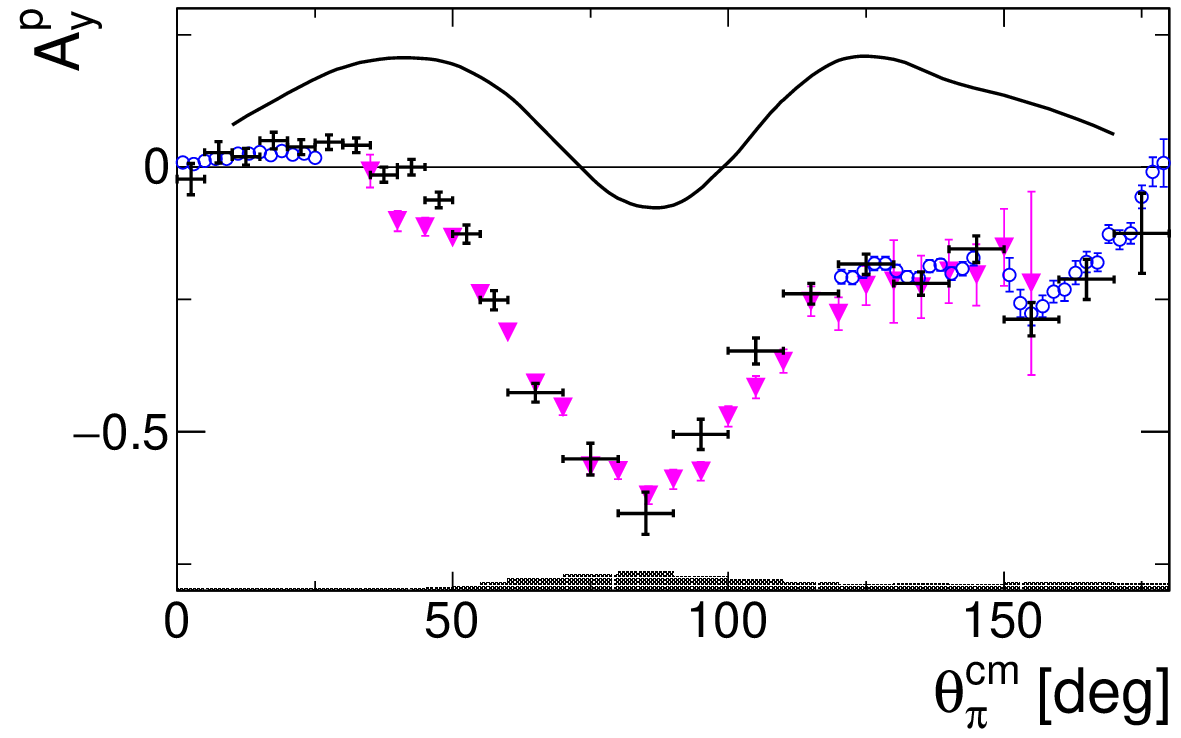}
\end{center}
\caption{TRIUMF data on the proton analysing power $A_y^p$ in the
$\pol{p}d\to{}^{3}\textrm{He}\,\pi^0$ reaction at 350~MeV~\cite{CAM1987}
(magenta triangles) are compared to ANKE results obtained at 353~MeV (blue
open circles) and at 363~MeV per nucleon with a polarized hydrogen target
(black crosses)~\cite{DYM2016}. The curve corresponds to the predictions by
Falk in a cluster-model approach~\cite{FAL2000a}.} \label{fig:ayp}
\end{figure}

As by-products of studying other reactions at ANKE, data were taken on the
proton analyzing powers in $pd\to{}^{3}\textrm{He}\,\pi^0$ and
$pd\to{}^{3}\textrm{H}\,\pi^+$ and also the deuteron and proton analyzing
powers in $\pol{d}\pol{p}\to{}^{3}\textrm{He}\,\pi^0$ and
$\pol{d}\pol{p}\to{}^{3}\textrm{H}\,\pi^+$~\cite{DYM2016}. The proton
analyzing power data shown in Fig.~\ref{fig:ayp} complement the earlier
TRIUMF measurements at 350~MeV~\cite{CAM1987} and show much structure from
the higher partial waves than is apparent in the differential cross section
shown in Fig.~\ref{fig:gem_dcs2}.

\begin{figure}[htb]
\centering
\includegraphics[width=1.0\columnwidth]{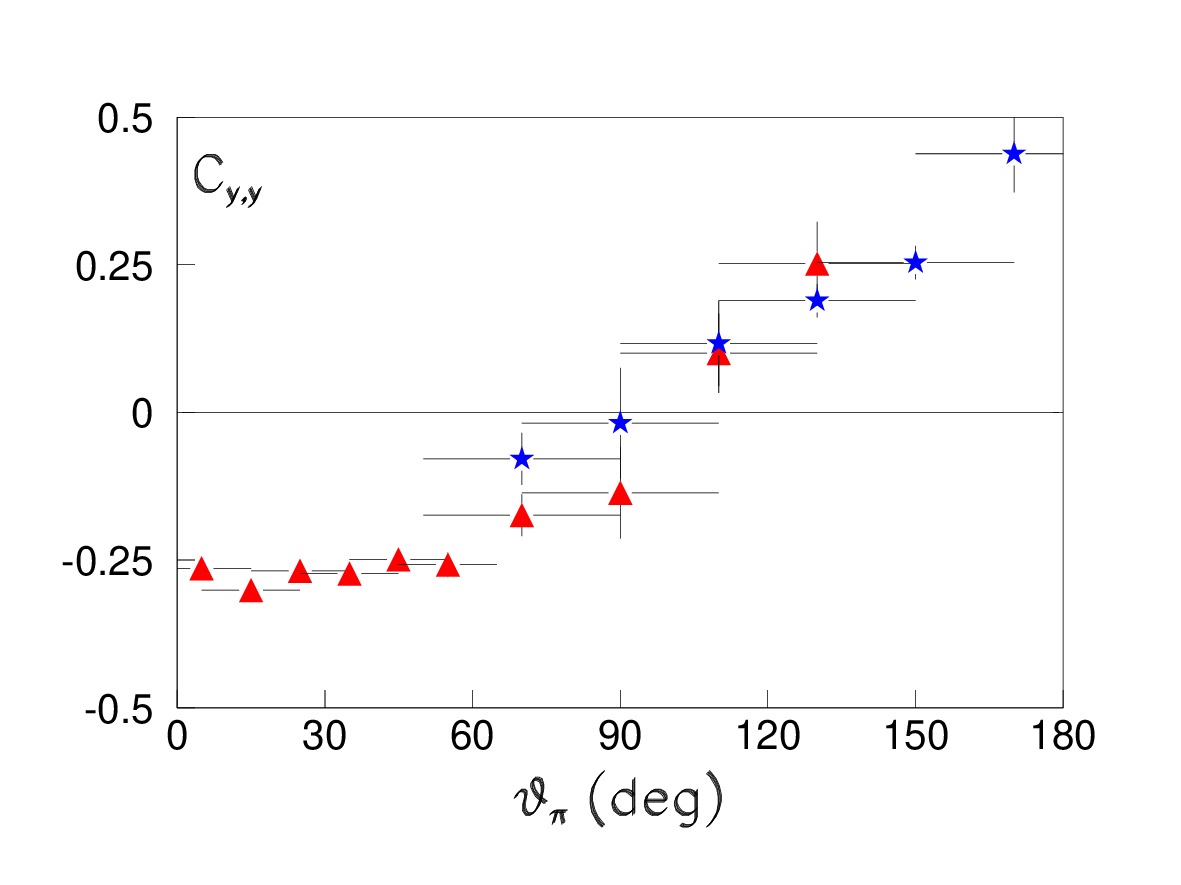}
\caption{Transverse spin correlation coefficient $C_{y,y}$ measured in the
$\pol{d}\pol{p}\to {}^3\rm{He}\,\pi^0$  and $\pol{d}\pol{p}\to
{}^3\rm{H}\,\pi^+$  reactions at 363~MeV per nucleon~\cite{DYM2016}. The
(red) triangles were obtained through $^3$He detection and the (blue) stars
through $^3$H detection. Only statistical errors are shown. The systematic
ones are about 11\% for $^3$He detection and a little bit larger in the $^3$H
case.} \label{fig:CxxCyy}
\end{figure}

The deuteron beam measurements also allowed transverse spin correlations to
be studied for the first time and values obtained for $C_{y,y}$ and $C_{x,x}$
at 363~MeV per nucleon are shown in Fig.~\ref{fig:CxxCyy}~\cite{DYM2016}. In
the forward and backward directions the the number of independent amplitudes
reduces from six to two, $A$ and $B$, and their values can be determined
through measurements of the cross section, spin correlation, and deuteron
tensor analyzing power, $T_{20}$, which was measured at
Saclay~\cite{KER1986}:
\begin{eqnarray}
\nonumber
\frac{\dd\sigma}{\dd\Omega} & = & \frac{kp}{3}(|A|^2+2|B|^2),\\
\nonumber
T_{20} & =& \sqrt{2}\frac{|B|^2-|A|^2}{|A|^2+2|B|^2},\\
C_{y,y}=C_{x,x}& = & -\frac{2Re(A^*B)}{|A|^2+2|B|^2},\label{eq:collinear}
\end{eqnarray}
where $k$ and $p$ are the pion and proton c.m.\ momenta. Thus the ANKE
measurements of $C_{x,x} = C_{y,y}$ in collinear kinematics determines the
relative phase of the $A$ and $B$ amplitudes. Whereas at 363~MeV per nucleon
there is strong interference between $A$ and $B$, at 600~MeV per nucleon the
two amplitudes are almost out of phase in the forward
direction~\cite{DYM2016}.

\subsection[$pd\to{}^3$He$\,\eta$]{$\boldsymbol{pd\to{}^3}$He$\,\boldsymbol{\eta}$} 
\label{pd3heeta}

\begin{figure}[h!]
\begin{center}
\includegraphics[width=0.9\columnwidth]{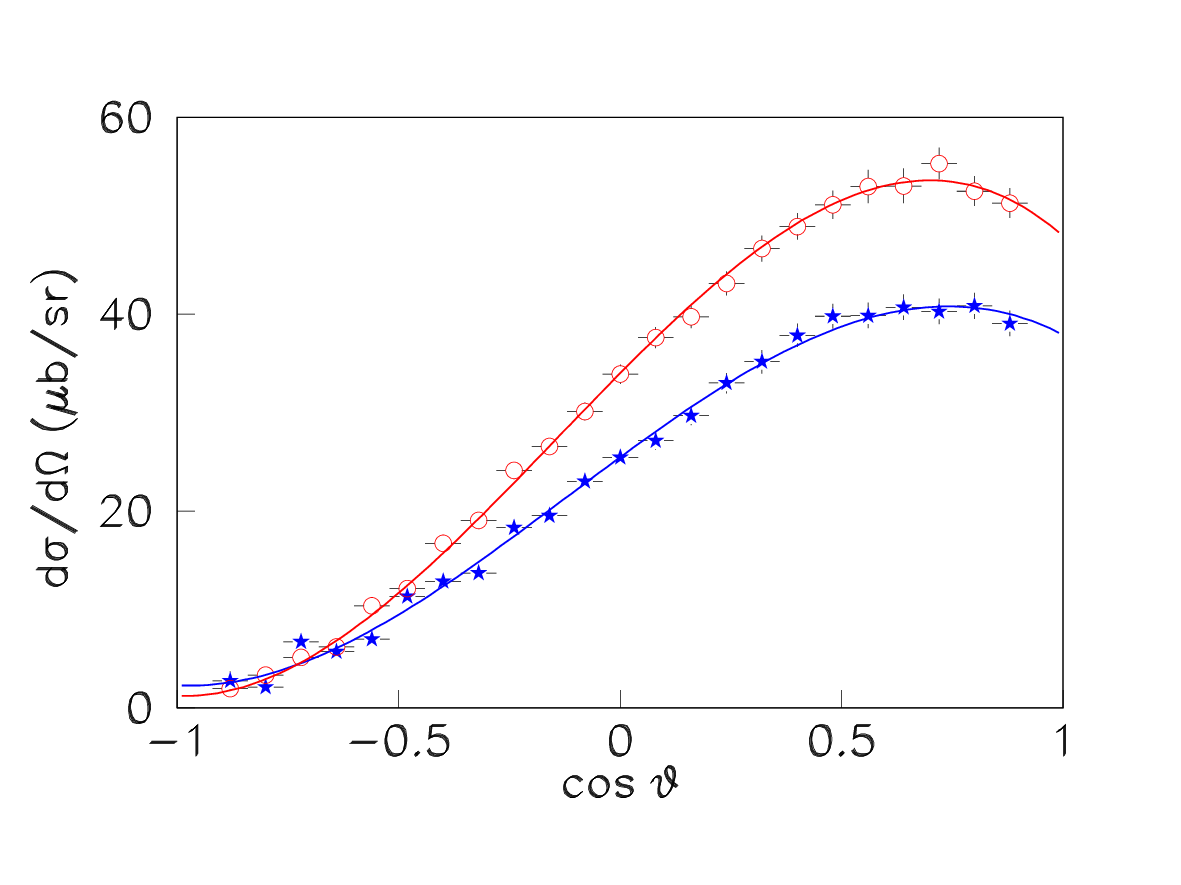}
\caption{\label{fig:Alfons1} Differential cross section for the unpolarized
$pd\to {}^{3}\textrm{He}\,\eta$ reaction obtained at
WASA-at-COSY~\cite{ADL2014c} at $Q=49$~MeV (blue filled stars) and 60~MeV
(red open circles), respectively. The curves are cubic fits in $\cos\theta$,
where $\theta$ is the c.m.\ angle between the initial proton and final
$\eta$. The relative normalization between the two data sets was established
through a comparison with the $pd\to {}^{3}\textrm{He}\,\pi^0$ results. The
reliability of this procedure is currently being checked on a larger data
sample~\cite{KHO2016}.}
\end{center}
\end{figure}

There have been numerous missing-mass measurements of the differential cross
section for the $pd\to{}^3$He$\,\eta$ reaction away from threshold at
COSY~\cite{BET2000,ADA2007,RAU2009,ADL2014c} and these have confirmed the
striking angular dependence illustrated at two energies by the WASA data
shown in Fig.~\ref{fig:Alfons1}.

Although the cross section in the backward hemisphere, where the momentum
transfer between the deuteron and the $^3$He is very large, is strongly
suppressed, the data seem to turn over well before reaching
$\theta=0^{\circ}$. This is in sharp contrast to the corresponding pion
production data discussed in sect.~\ref{pd3hepi} and is an indication that
the impulse approximation might not be the dominant driving mechanism for
$\eta$ production, even at small angles. Three-nucleon mechanisms, involving
intermediate pions, have been suggested to describe these large momentum
transfer reactions. Though classical~\cite{KIL1990} and quantum
mechanical~\cite{FAL1995} calculations have had some success near threshold, they
have not provided any real insight away from the small $Q$ region.

The values of the total cross sections, which are summarized in
Ref.~\cite{ADL2014c}, have a wide scatter, due in part to the different
techniques used to obtain the absolute normalizations. An observable that is
independent of such uncertainties is the logarithmic slope at the mid-point,
defined by
\begin{equation}
\label{alpha}
\alpha=\left.\frac{\dd\phantom{x}}{\dd(\cos\theta_{\eta})}
\ln\left(\frac{\dd\sigma}{\dd\Omega}\right)\right|_{\cos\theta_{\eta}=0}\:.
\end{equation}

\begin{figure}[h!]
\begin{center}
\includegraphics[width=0.9\columnwidth]{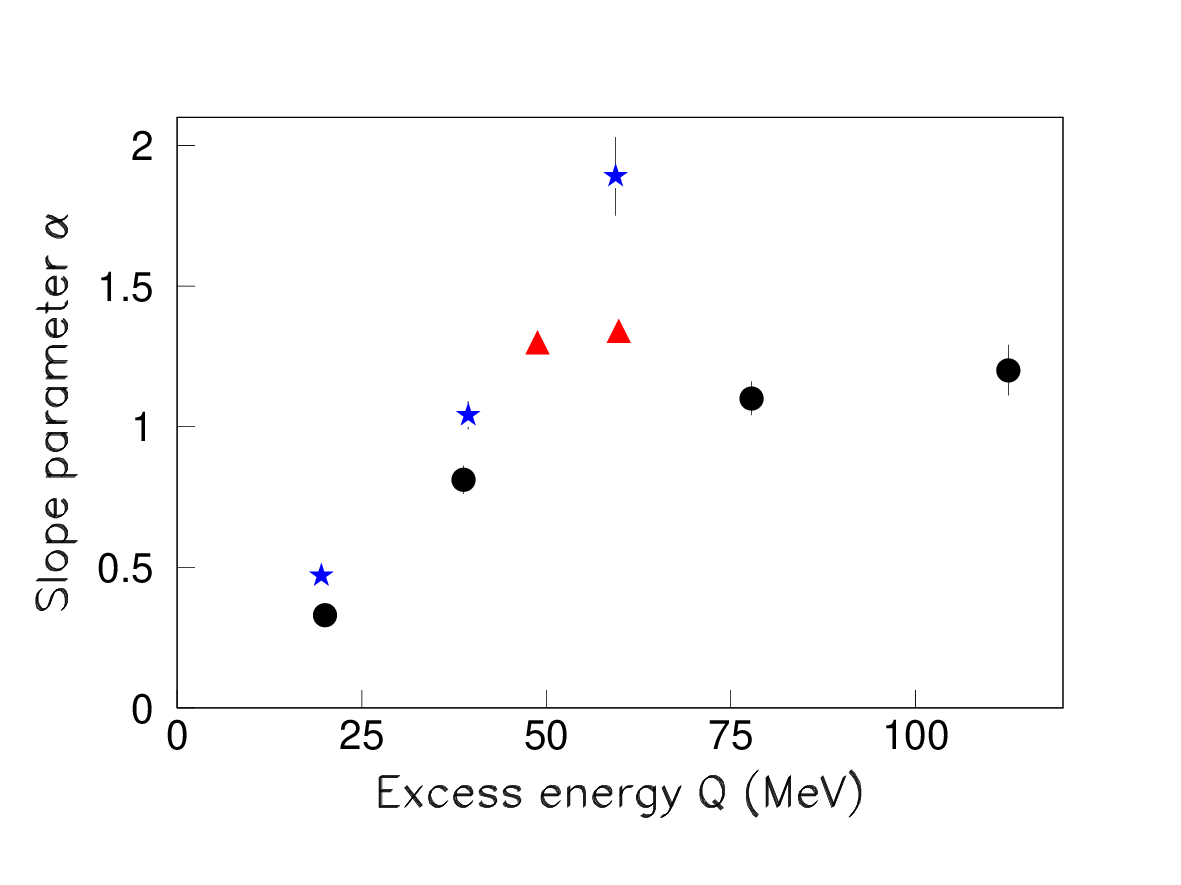}
\caption{\label{fig:alpha} Values of the symmetry parameter $\alpha$, defined
by Eq.~(\ref{alpha}), deduced from measurements of the $pd\to{}^3$He$\,\eta$
differential cross section. These are shown as (black) circles~\cite{BIL2002},
(blue) stars~\cite{RAU2009}, and (red) triangles~\cite{ADL2014c}. Only
statistical errors are shown.}
\end{center}
\end{figure}

Apart from one exceptional point at $Q=59.4$~MeV~\cite{RAU2009}, the values
of $\alpha$ shown in Fig.~\ref{fig:alpha} seem to display a steady rise with
$Q$. However, it is clear from the disparity between the different
experimental results, which are much larger than the statistical errors, that
there must be significant systematic uncertainties. These could originate
from the understanding of the acceptance of the various spectrometers used.
The conflicts are likely to be underscored when the new WASA data are
published~\cite{KHO2016}.

The general physics interest is much greater in the two near-threshold
measurements, carried out simultaneously at COSY-11~\cite{SMY2007} and
ANKE~\cite{MER2007}, where values of the $dp\to{}^3$He$\,\eta$ cross section
were extracted at many energies with robust relative normalizations. In both
experiments, only the recoiling $^3$He was detected and the $\eta$-meson
identified through the peak in the missing-mass distribution.

The first point to note is that the missing-mass $m_X$ resolution gets better
as the threshold is approached. The reason for this is most easily seen by
using non-relativistic kinematics in the c.m.\ frame, where
\begin{equation}
\label{mm}
m_X = W - m_{\tau} - k^2/2m_{\rm red}.
\end{equation}
Here $W$ is the total c.m.\ energy, $m_{\tau}$ the mass of $^3$He, $k$ the
momentum of the $^3$He, and $m_{\rm red}$ the reduced mass of $m_{\tau}$ and
$m_X$. The $\eta$ threshold corresponds to $m_X=m_{\eta}$ and $k=0$, so that
\begin{equation}
\label{mm2}
\left|\partial{m_X}/\partial{k}\right| = k/m_{\rm red} \to 0\ \ \textrm{as}\ \ k\to 0.
\end{equation}
The value of $m_X$ is therefore stationary at threshold in the c.m.\ frame
and this is also true for small changes of the $^3$He momentum from threshold
in the laboratory frame.

The improvement in resolution near threshold is, of course, more general than
this  particular reaction and a similar improvement in the missing-mass
resolution is to be expected in, for example, the $pp\to pp\eta^{\prime}$
reaction near its threshold. The power of this result is, however, diluted by
the smearing arising from the finite momentum bite of the COSY beam.

The early Saclay experiments~\cite{BER1988,MAY1996} showed that the $dp(pd)
\to {}^{3}\textrm{He}\,\eta$ total cross section jumped very rapidly in the
$\eta$ threshold region and it was suggested that this is likely to be the
consequence of a strong $\eta{}^3$He $s$-wave final state interaction, that
might even lead to the $\eta$ being quasi-bound to the
nucleus~\cite{WIL1993}. Such states had been predicted
previously~\cite{HAI1986,HAI2002}, but their positions were rather ambiguous
due to uncertainties in the parameters of the $\eta$-nucleon interaction.
More detailed experiments were needed to investigate the possible existence
of $_{\eta}^3$He.

\begin{figure}[h!]
\begin{center}
\includegraphics[width=0.9\columnwidth]{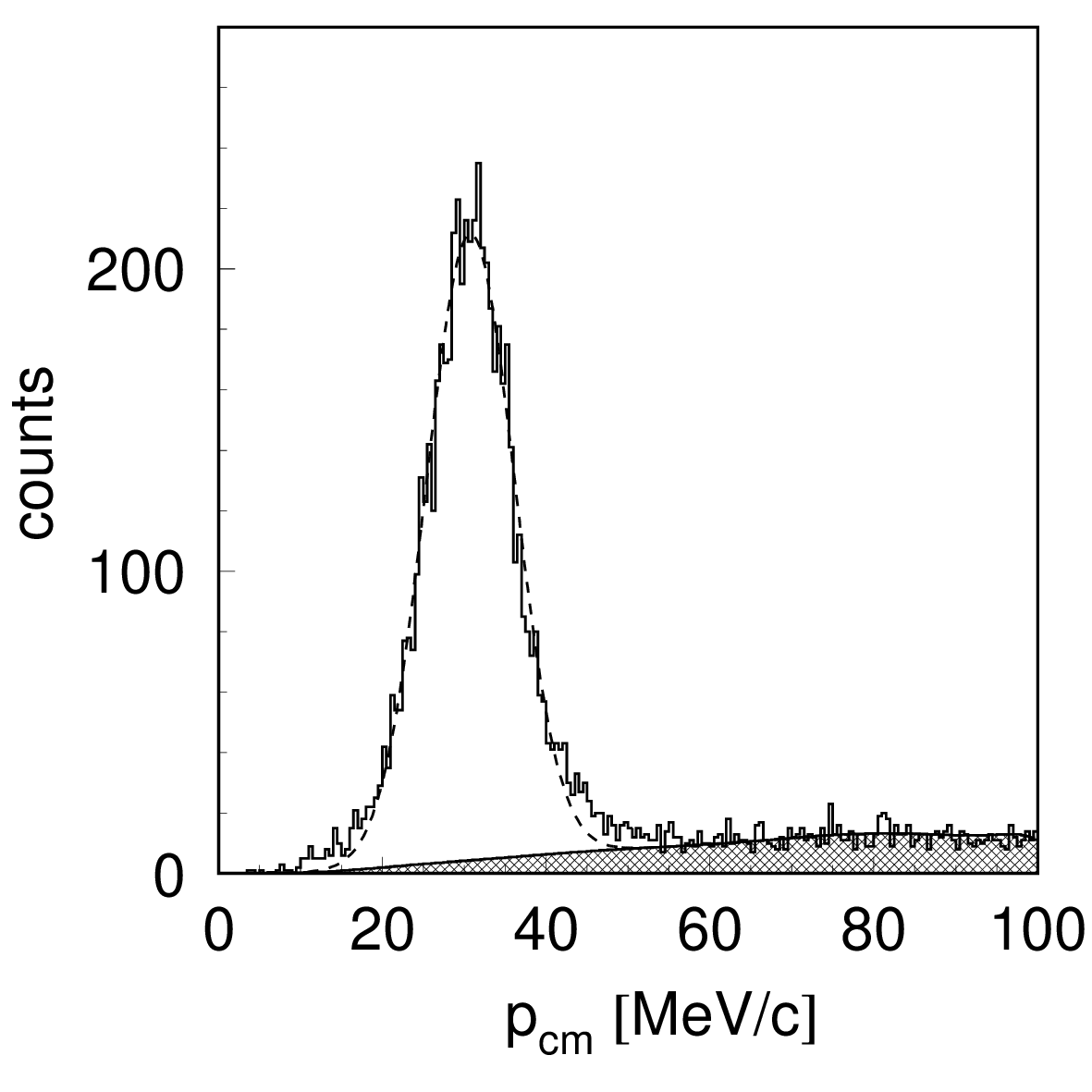}
\caption{\label{fig:smy1} Distribution of $^3$He momenta in the c.m.\ frame
at $Q\approx 1.88$~MeV~\cite{SMY2007}. The dashed line represents a Gaussian
fit to the $\eta$ peak and the shaded region corresponds to the background
estimated from below-threshold measurements. }
\end{center}
\end{figure}

The COSY-11 and ANKE experiments were carried out in very similar ways, using
a deuteron beam that was steadily accelerated through each cycle. The binning
in beam energy of events taken in this ramping mode could be chosen at will
at the analysis stage and, as will be shown shortly, the two groups did make
very different choices here. The background to the $dp\to{}^3$He$\,\eta$ data
was, in both cases, estimated from data taken below the $\eta$ threshold.
These were shifted so that the kinematic limits coincided and scaled to give
consistency outside the $\eta$-peak region. As can be seen from the COSY-11
data shown in Fig.~\ref{fig:smy1}, this procedure gives a very plausible
description of the data.

\begin{figure}[hbt]
\begin{center}
\includegraphics[width=01.0\columnwidth]{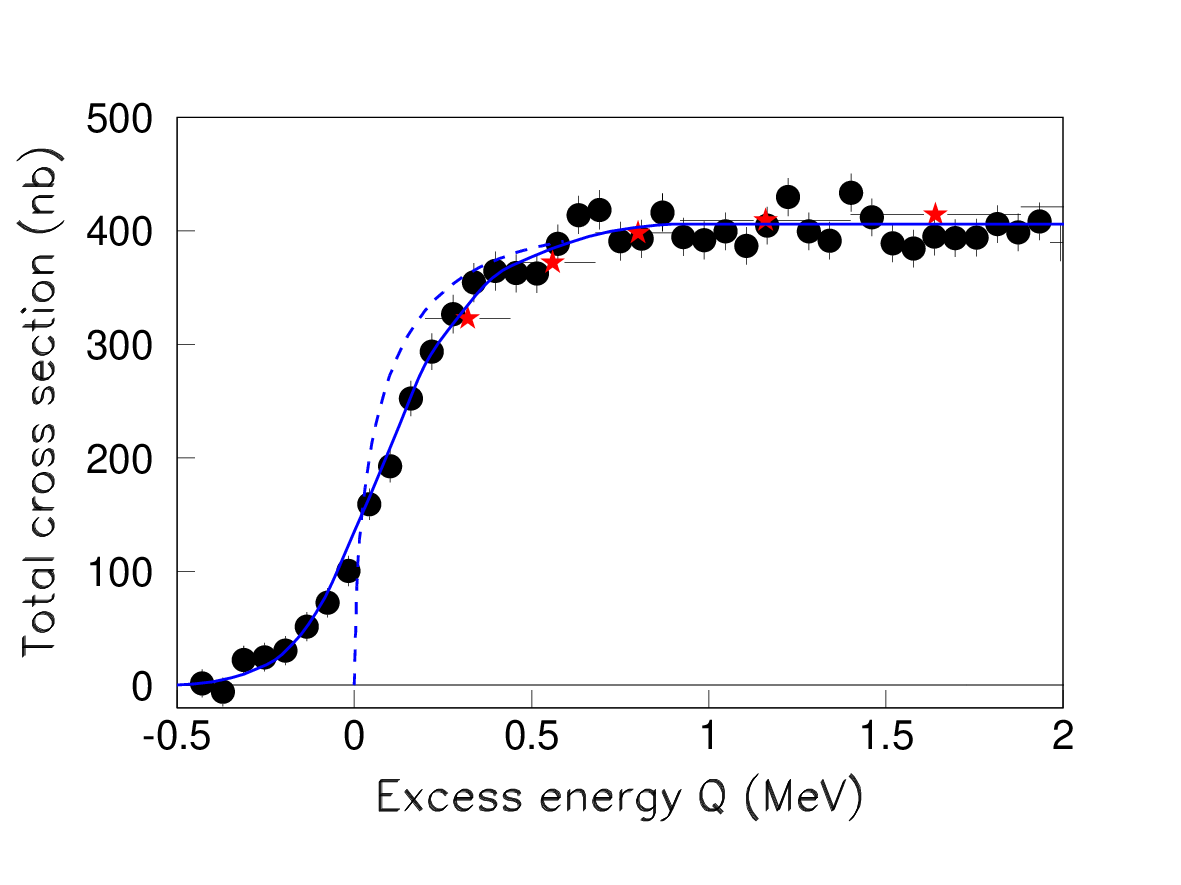}
\caption{\label{fig:timo} Near-threshold $dp\to{}^3$He$\,\eta$ total cross
sections measured by ANKE (black circles)~\cite{MER2007} and COSY-11 (red
stars)~\cite{SMY2007}. Only statistical errors are shown. The solid curve is
the fit of Eq.~(\ref{eq:fsi2}) to the ANKE data, where a 171~keV smearing in
$Q$, arising mainly from the deuteron beam profile, is taken into account.
The dashed curve shows what the data should look like if the deuteron beam
were truly monochromatic and other sources of smearing were neglected. }
\end{center}
\end{figure}

The absolute normalizations of the cross sections were achieved by measuring
deuteron-proton elastic scattering in parallel, though in different kinematic
regions in the two experiments. This gave overall systematic errors of
$\approx 15\%$~\cite{MER2007} and $\approx 10\%$~\cite{SMY2007} and these
uncertainties are not included in the results presented in
Fig.~\ref{fig:timo}. Only the small $Q$ region is shown because the total
cross sections at higher energies are almost constant, up to the limits of
the two experiments~\cite{SMY2007,MER2007}. The larger acceptance for single
particles at ANKE allowed a somewhat larger range in $Q$ to be studied but
this is not crucial here because the important Physics is contained in the
first few MeV of excess energy.

It is important to note that the values of the COSY-11 and ANKE total cross
sections shown in Fig.~\ref{fig:timo} are completely consistent, especially
in view of the overall luminosity uncertainties. The COSY-11 group omitted
effects associated with smearing in $Q$ and, by fitting their data with an
$\eta{}^3$He scattering length formula, obtained a relatively modest value of
the magnitude of the scattering length. This can easily be understood because
the energy smearing softens the jump to the cross section plateau.

In the alternative approach, it was noted in the ANKE data of
Fig.~\ref{fig:timo} that there were many $\eta$ events produced below the
nominal threshold and these must have arisen from the spread in momentum of
the COSY deuteron beam. The existence of such a spread of the right order of
magnitude was confirmed independently from the spin depolarizing measurements
described in sect.~\ref{Paul1}~\cite{GOS2010}.

After defining an average production amplitude in terms of the cross section
by dividing by the ratio of the $\eta$ and deuteron momenta in the c.m.\
frame, %
\begin{equation}
\label{amps}
|f|^2= \frac{p_d}{p_{\eta}}\frac{\dd\sigma}{\dd\Omega},
\end{equation}
the ANKE total cross section data were fitted with a two-pole final state
interaction factor,
\begin{equation}
f = \frac{f_B}{(1-p_\eta/p_1)(1-p_\eta/p_2)}\,,
\label{eq:fsi2}
\end{equation}
smeared with a Gaussian distribution in $Q$ and put into finite energy bins.
Here $f_B$ is assumed to be constant over the energy range where the FSI is
important. The fit parameters obtained were~\cite{MER2007}
\begin{eqnarray}
\label{mom}
p_1 &=& [(-5.2\pm7.0^{+1.5}_{-0.8})\pm i(18.7\pm2.4^{+0.7}_{-0.8})]\,\textrm{MeV}/c\phantom{111}\\
\nonumber
p_2 &=& [(106.3\pm4.5^{+0.2}_{-0.3})\pm
i(75.6\pm12.5^{+0.7}_{-1.8})]\,\textrm{MeV}/c
\end{eqnarray}
and a smearing that could arise from a beam spread of $\delta p_d/p_d \approx
2.2\times 10^{-4}$. Here the first errors are statistical and the second
systematic. It is important to note here that the error bars quoted refer to
the specific form of Eq.(\ref{eq:fsi2}) and the nearby pole may move by more
than these if one assumed a different fit function.

It is possible to deduce scattering length and effective range parameters
from the numbers given in Eq.~(\ref{mom}), but it must be realized that $p_2$
is really an effective parameter that might be hiding some of the energy
dependence of $f_B$. The real physics is contained in the value of $p_1$,
which shows that there is a pole in the $\eta{}^3$He scattering amplitude at
$Q_{\rm pole} = [(-0.30\pm0.15\pm0.04)\pm i(0.21\pm0.29\pm0.06)]$~MeV. The
sign of the imaginary part of $Q_{\rm pole}$, i.e., whether the state is
``bound'' or ``antibound'' cannot be determined from above-threshold
measurements but, nevertheless, the proximity of the pole to the origin and
the associated large scattering length have excited a lot of interest in the
mesic nuclei community.

It is important to check that the pole really is due to an interaction in the
$s$-wave $\eta{}^3$He final state. This state can be accessed from either the
total spin $S=\frac{3}{2}$ or the $S=\frac{1}{2}$ deuteron-proton initial
system and the differences will influence the deuteron tensor analyzing power
$T_{20}$. The pure $s$-wave FSI hypothesis requires that $T_{20}$ should
remain constant, despite the strange behaviour of the unpolarized cross
section.

The tensor analyzing power of the $\pol{d}p \to {}^{3}\textrm{He}\,\eta$
total cross section was measured at ANKE for $Q\lesssim
11$~MeV~\cite{PAP2014} using a similar system to that employed earlier for
the unpolarized cross section~\cite{MER2007}. The results are indeed
consistent with a constant value of $T_{20}$, which offers strong support to
the FSI interpretation of the near-threshold energy dependence. It is
important to note here that the detection system was independent of the
deuteron beam polarization so that many of the systematic effects cancel. The
result is not totally unexpected because, if the poles in the two threshold
amplitudes had been significantly separated, the single pole fit of
Eq.~(\ref{mom}) would not have resulted in such a small value of the
imaginary part of $Q_{\rm pole}$.

Further evidence in support of the FSI hypothesis is to be found from
studying the $Q$ dependence of the slope parameter $\alpha$ of
Eq.~(\ref{alpha}) in the near-threshold region. At high $Q$ the cross
sections are forward peaked and the $\alpha$ of Fig.~\ref{fig:alpha} are all
positive. However, there were already suspicions from the Saclay
data~\cite{MAY1996} that $\alpha$ might be slightly negative near threshold
and this was confirmed by the ANKE data~\cite{MER2007}. It was argued in
Ref.~\cite{WIL2007} that this behaviour could only occur if the interference
between the $s$- and $p$-wave production amplitudes changed significantly
near threshold. This is precisely what would be expected if there were a
complex pole in the $\eta{}^3$He amplitude.

\begin{figure}[htb]
\includegraphics[width=01.0\columnwidth]{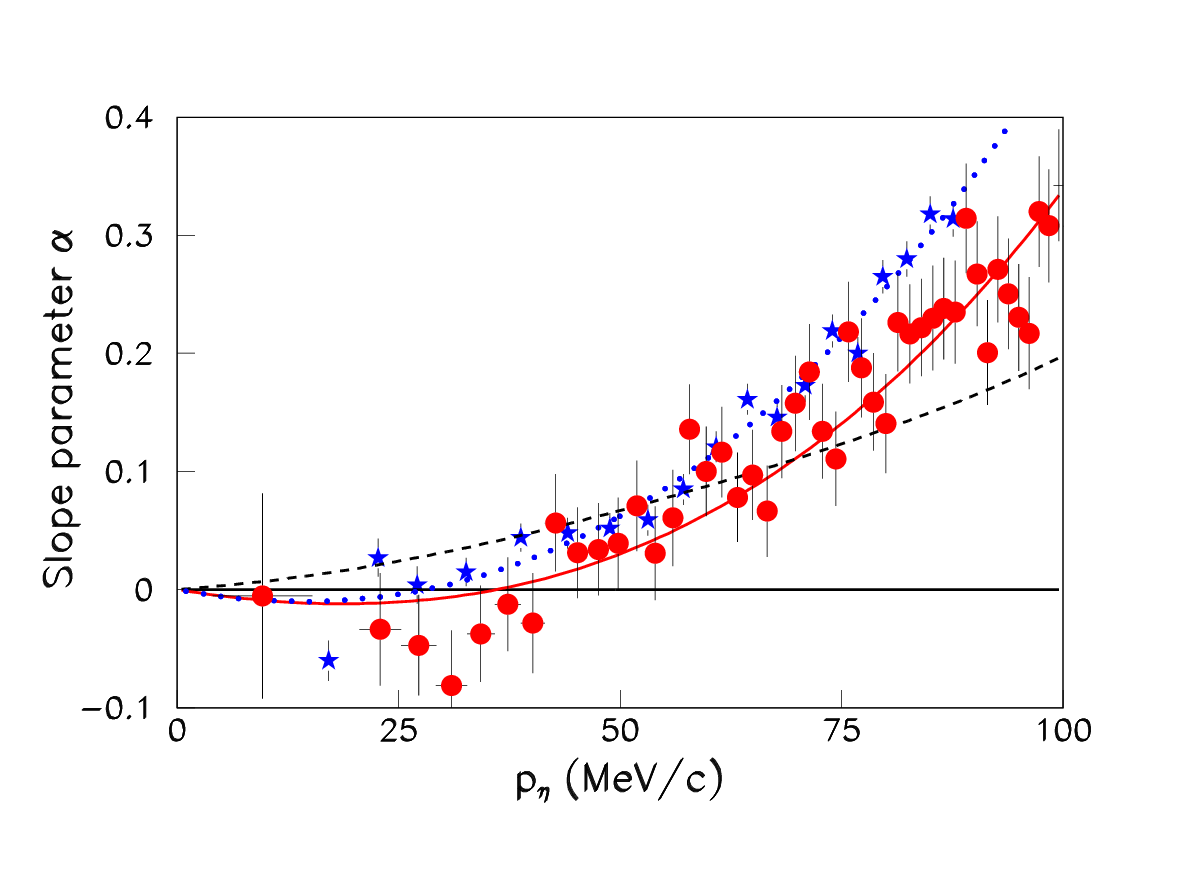}
\caption{ \label{alpha_mom} Slope parameter $\alpha$ of the $dp\to
{}^{3}\textrm{He}\,\eta$ reaction as a function of the $\eta$ c.m.\
momentum. The experimental data from COSY--ANKE (red closed
circles)~\cite{MER2007} and COSY--11 (blue stars)~\cite{SMY2007} are
compared to fits (solid red curve and blue dots) where the phase variation
of the $s$--wave amplitude is taken into account. If the phase
variation is neglected, the best fit (black dashed curve) fails to describe
the data.}
\end{figure}

Figure~\ref{alpha_mom} shows the results for $\alpha$ obtained by the
ANKE~\cite{MER2007} and COSY-11~\cite{SMY2007} collaborations and the best
fits to these data achieved when taking the phase variation from the $s$-wave
pole into account~\cite{WIL2007}. For comparison, the best fit obtained when
the phase variation is neglected is also shown. It seems clear that these
data are better described by including the phase variation. It is interesting
to note that an analogous reversal of the slope parameter near threshold is
also observed in the $\gamma{}^3\textrm{He}\to \eta{}^3\textrm{He}$
photoproduction data~\cite{PHE2012}, though the details will depend on the
phase of the $p$-wave amplitude in any particular reaction.

However, one must be cautious because data or their interpretation may
change. Thus it could well be argued that, instead of using
Eq.~(\ref{eq:fsi2}), one would be on slightly firmer ground by assuming that
\begin{equation}
f = f_B\times(1-p_\eta/p_2)/(1-p_\eta/p_1).
\label{eq:fsi4}
\end{equation}
As expected, by fitting the ANKE $dp\to{}^3$He$\,\eta$ total cross section
data~\cite{MER2007} with Eq.~(\ref{eq:fsi4}) it is seen that the nearby pole
is still in the region where $|Q|< 1$~MeV, but it is now on an unphysical
sheet with $p_1 = -28\pm2$~MeV/$c$, with a relatively small imaginary
part~\cite{MER2016}. Since the extracted value of $p_2$ is also large and
real, the phase of the $s$-wave amplitude changes little with momentum and so
the value of the slope parameter $\alpha$ could never change sign and so the
data shown in Fig.~\ref{alpha_mom} would not be reproduced in this model.

However, very precise $dp\to{}^3$He$\,\eta$ differential cross section data
were taken in connection with the measurement of the mass of the $\eta$
meson~\cite{GOS2012} that is discussed in the next section. Unlike the
ramping mode used in the initial ANKE experiment~\cite{MER2007}, these
consisted of a series of 14 flat tops with $p_{\eta}< 100$~MeV/$c$. In
contrast to the data shown in Fig.~\ref{alpha_mom}, the \emph{preliminary}
results from this analysis show little indication of $\alpha$ going negative
close to threshold~\cite{FRI2016} and this is precisely what one would expect
if the $s$-wave amplitude did not show a rapid phase variation.

\subsubsection{Measurements of the mass of the $\eta$ meson} 
\label{eta_mass}

The first measurement of the mass at the $\eta$ meson at COSY by the GEM
collaboration~\cite{ABD2005} yielded a value that was about 0.5~MeV/$c^2$
lower than the results of other modern determinations that were reported by
the PDG group~\cite{OLI2014}. In contrast, the later experiment, carried out
with a circulating deuteron beam using the ANKE spectrometer~\cite{GOS2012},
is completely consistent with the PDG recommended value and the error bars
are among (or possibly are) the best in the World.

The GEM experiment used a proton beam that was electron-cooled at injection
energy and then stochastically extracted. This was incident on a thin liquid
target, with the charged particles produced being detected in the Big Karl
spectrometer, described in sect.~\ref{BK}. Big Karl was calibrated by
measuring separately the proton and positive pion from the $pp\to d\pi^+$
reaction~\cite{ABD2005}.

In order to extract a value for the $\eta$ mass, the beam momentum must be
well measured and the kinematics of a reaction where the $\eta$ meson is
produced fully determined. These requirements were met in the Big Karl
experiment by measuring simultaneously the $\pi^+$ and $^3$H from one branch
of the $pd\to{}^3$H$\,\pi^+$ reaction and the $^3$He from one branch of
$pd\to{}^3$He$\,X$ at an excess energy $Q\approx 34$~MeV with respect to the
threshold for $\eta$ production. At this energy the relevant $\pi^+$, $^3$H,
and $^3$He all have similar rigidities and can be detected in parallel in the
Big Karl focal plane. Though different in detail, there are some similarities
with the SATURNE experiment~\cite{PLO1992}, where the same $\eta$-production
reaction was studied and pion production was also used to determine the beam
momentum.

\begin{figure}[h!]
\begin{center}
\includegraphics[width=0.8\columnwidth]{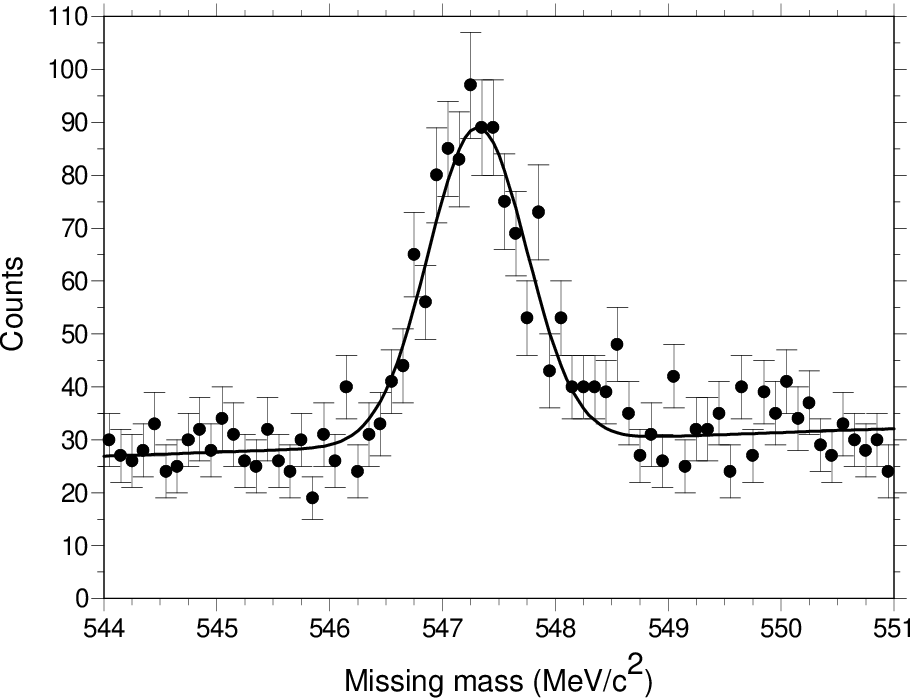}
\caption{\label{fig:HMmass} Missing-mass distribution from the
$pd\to{}^3$He$\,X$ distribution at Big Karl, where the beam momentum was determined
through the simultaneous study of the $pd\to{}^3$H$\,\pi^+$
reaction~\cite{ABD2005}.}
\end{center}
\end{figure}

It is clear from the missing-mass distribution shown in Fig.~\ref{fig:HMmass}
that there is little difficulty in separating the $\eta$ peak (with
$\textrm{FWHM}\approx 1$~MeV/$c^2$) from the slowly varying background. The
mass scale shown here was fixed using the measurements in parallel of the
$pd\to{}^3$H$\,\pi^+$ reaction. The problem is that the central value of
$m_{\eta}=547.31$~MeV/$c^2$ is lower than the PDG recommended mean by of the
order of $0.5$~MeV/$c^2$. This disagreement is very large compared to the
$\pm 0.03(\textrm{stat})\pm0.03(\textrm{syst})$~MeV/$c^2$ errors quoted in
the GEM paper~\cite{ABD2005}. To put the deviation into some kind of context,
the $0.5$~MeV/$c^2$ off-set would correspond to about twice the energy loss
of the $^3$He in the $1~\mu$m mylar window.

Rather than repeating the Big Karl experiment and analysis, it was decided to
carry out an $\eta$ mass measurement using the circulating deuteron beam in
COSY~\cite{GOS2012}, with the aim of reducing the error bars to below those
quoted in the literature~\cite{OLI2014}. Such a $dp\to{}^3$He$\,\eta$
experiment offered distinct advantages over the Big Karl measurement. As
shown in sect.~\ref{Paul1}, the momentum of the deuteron beam could be
determined to better than $3\times 10^{-5}$ by inducing an artificial
depolarizing resonance~\cite{GOS2010}. The energy loss of the $^3$He in the
hydrogen cluster-jet target is negligible and, by working very close to
threshold, good missing-mass resolution and low backgrounds could be
achieved~\cite{GOS2012}. Furthermore, as will be shown later, this allowed
the data to be extrapolated to threshold, which reduced the systematic
uncertainties.

As discussed earlier, the background under the $\eta$ peak in the
missing-mass distribution could be reliably estimated using data taken a
little below the $\eta$ threshold. The $^3$He were measured in the Forward
Detector of the ANKE spectrometer and this has full geometric acceptance for
the $dp\to{}^3$He$\,\eta$ reaction for $Q\lesssim 11$~MeV. This was of
crucial importance because it allowed the study of the effects of the finite
momentum resolution in the three different directions in space. If this had
not been done, the value obtained for $m_{\eta}$ would have depended on the
production angle, with differences of up to $0.5$~MeV/$c^2$ between
$\cos\theta_{\tau} =\pm1$ and $\cos\theta_{\tau}=0$. It may be interesting to
note that the Big Karl data were taken only in the forward
direction~\cite{ABD2005}. The careful corrections to the measurements of the
$^3$He momenta to compensate for the spectrometer resolution are thoroughly
described in Ref.~\cite{GOS2012}.

\begin{figure}[h!]
\begin{center}
\includegraphics[width=0.8\columnwidth]{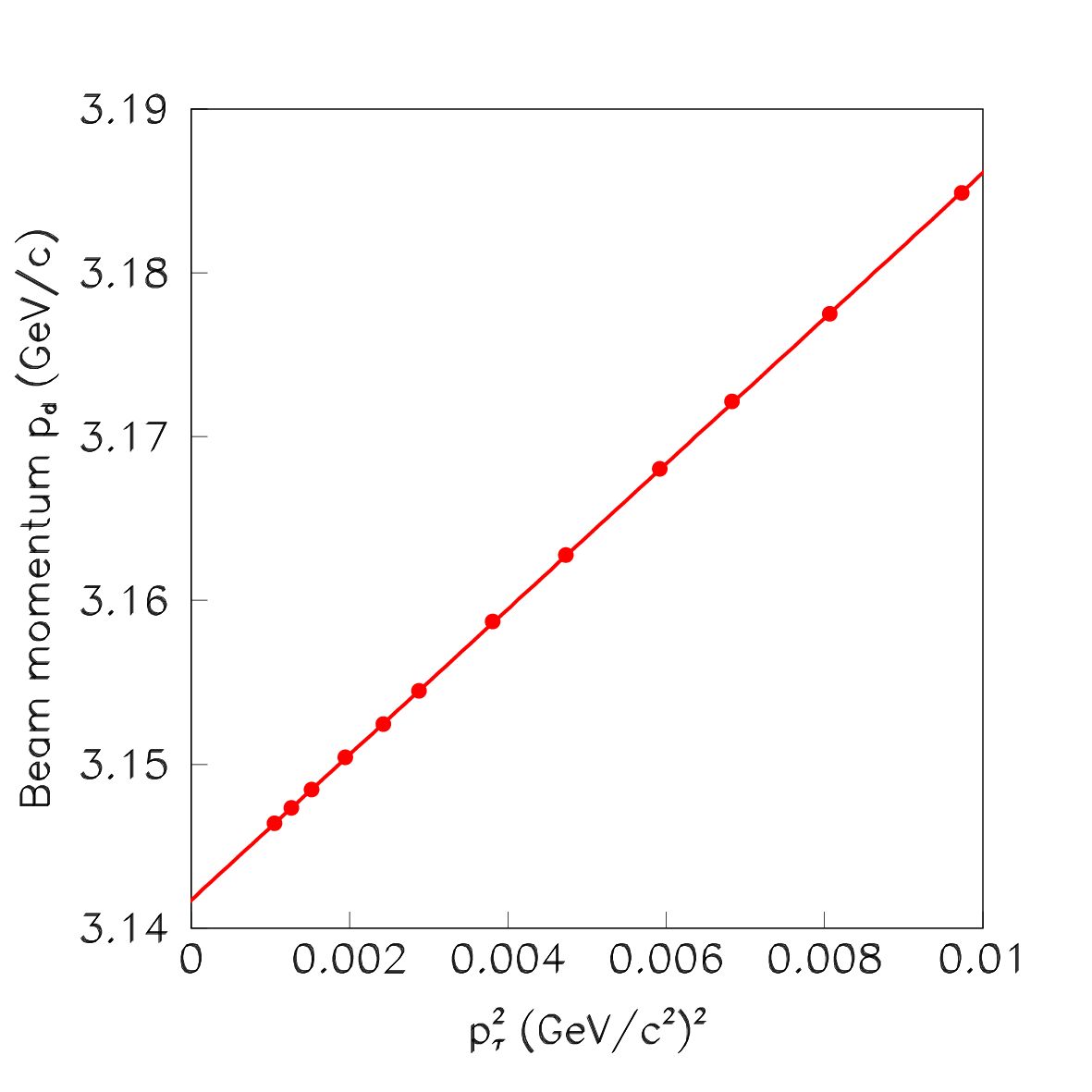}
\caption{\label{fig:paul3} The momentum of the circulating deuteron beam
$p_d$ compared to the square of the corresponding $^3$He momentum $p_{\tau}$
for the 12 points $dp\to{}^3$He$\,\eta$ points measured at
COSY~\cite{GOS2012}. The shape of the function $p_d=f(p_{\tau})$ and its
value at $p_{\tau}=0$ are governed by the single parameter $m_{\eta}$ but to a
good approximation the data are well described by the straight line shown,
viz.\ $p_d=3.14171+4.44\,p_{\tau}^{\,2}$.}
\end{center}
\end{figure}

Figure~\ref{fig:paul3} shows the squares of the resolution-corrected $^3$He
momentum $p_{\tau}$, as measured in the ANKE Forward Detector at twelve
incident deuteron momenta $p_d$. It is important to note that, for the
kinematics of the $dp\to{}^3$He$\,\eta$ reaction, the only free parameter is
$m_{\eta}$ so that, once the intercept is fixed, the form of the fit function
in Fig.~\ref{fig:paul3} is completely determined \emph{a priori}. These
considerations show that the fit should deviate slightly from a straight line
and this must be taken into account when extracting the best value for
$m_{\eta}$.

It is, however, sufficient for the discussion here to consider the free
linear fit shown in the figure, $p_d=3.14171+4.44\,p_{\tau}^{\,2}$. With a
perfectly
tuned spectrometer, the slope of this line should be%
\begin{equation}
\label{slope}
\textrm{slope} = \frac{(m_{\tau}+m_{\eta})^2}{m_{\tau}m_{\eta}} \frac{E_d}{2m_pp_d} = 4.54~(\textrm{GeV}/c)^{-1},
\end{equation}
where $p_d$ and $E_d$ are, respectively, the deuteron momentum and total
energy evaluated at the $\eta$ threshold. The 1.1\% correction to the
momentum that this exhibits, compared to the 0.8\% in the refined
fit~\cite{GOS2012}, is of no real importance because it does not affect the
extrapolation to threshold. On the other hand it would be relevant if the
experiment were conducted at an isolated energy above threshold. Thus the
extrapolation to $p_{\tau}=0$ reduces the systematic uncertainties in the
mass determination.

The value of the $\eta$ mass given in Ref.~\cite{GOS2012} is
\begin{equation}
\label{eq:massfinal}
m_{\eta}=(547.873\pm0.005_{\textrm{stat}}\pm0.023_{\textrm{syst}})~\textrm{MeV}/c^2,
\end{equation}
where the systematic error is dominated by that associated with the
determination of the beam momentum through the induced depolarizing resonance
technique. This might be improved but there seems to be currently no pressing
need to know the value of the $\eta$ mass to better than 23~keV/$c^2$. The
statistical and systematic errors are both marginally better than those
of other modern measurements quoted in the PDG compilation~\cite{OLI2014}.
With a relative uncertainty of about $4\times 10^{-5}$, this is certainly the
most precise measurement carried out within the hadron physics programme at
COSY.

\subsection[The $pd\to{}^3$He$\,\pi^+\pi^-$ reaction]{The $\boldsymbol{pd\to{}^3}$He$\,\boldsymbol{\pi^+\pi^-}$ reaction} 
\label{pd3hepipi}

At the height of the meson hunt, Abashian, Booth, and Crowe~\cite{ABA1963}
measured the inclusive cross sections for $pd \to{}^3\textrm{He}\,X^0$ and
$pd \to{}^3\textrm{H}\,X^+$ at a beam energy of $T_p=743$~MeV. This
corresponds to an excess energy with respect to the $\pi^+\pi^-$ threshold of
$Q=W-M_{^3\rm{He}}-2M_{\pi^+}=184$~MeV, where $W$ is the total energy in the
centre-of-mass system. In addition to the expected single-pion peaks, a
striking enhancement was seen in the $^3$He case at a missing mass of about
310~MeV/$c^2$, with a width $\approx 50$~MeV/$c^2$. Being so close to the
$\pi^+\pi^-$ threshold, it could be assumed that the pions were in a relative
$s$-wave and hence had an overall isospin of $I=0$. This is consistent with
the lack of a similar signal in the $pd \to{}^3\textrm{H}\,X^+$ data. This
behaviour has since become known as the ABC effect or enhancement. However,
the ABC parameters change with the experimental conditions and it is believed
that the ABC is a kinematic effect, related to the presence of nucleons,
rather than the hoped-for $s$-wave isoscalar $\pi\pi$
resonance~\cite{AMS2012}.

A similar inclusive measurement could have been carried out using the high
resolution Big Karl spectrometer described in sect.~\ref{BK}. However, to
investigate the $pd \to{}^3\textrm{He}\,X^+X^-$ reaction in greater depth,
one needs more information on the distributions of the mesons $X$ produced.
For this purpose Big Karl was used in conjunction with the MOMO vertex
detector that was described in sect.~\ref{MOMO}. In a low energy run, an
event with two charged particles in the vertex detector and a $^3$He in Big
Karl was considered to be a candidate for the $pd\to{}^{3}\mbox{\rm
He}\,\pi^+\pi^-$ reaction. Its identification and complete reconstruction
involved a two-constraint kinematic fit. About 15,000 unambiguous events were
obtained at a beam energy of 546~MeV ($Q=70$~MeV).

\begin{figure}[h]
    \includegraphics[width=0.49\linewidth]{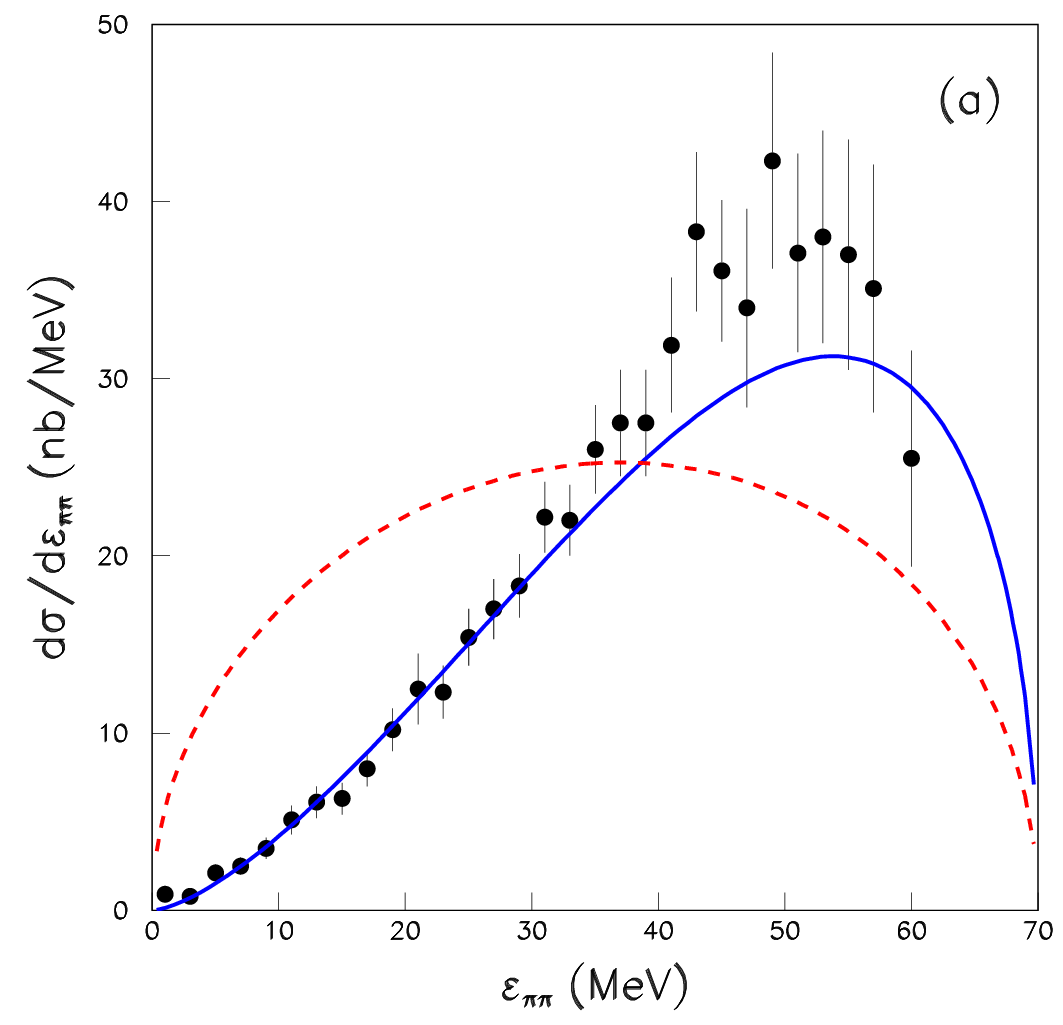}
	\includegraphics[width=0.49\linewidth]{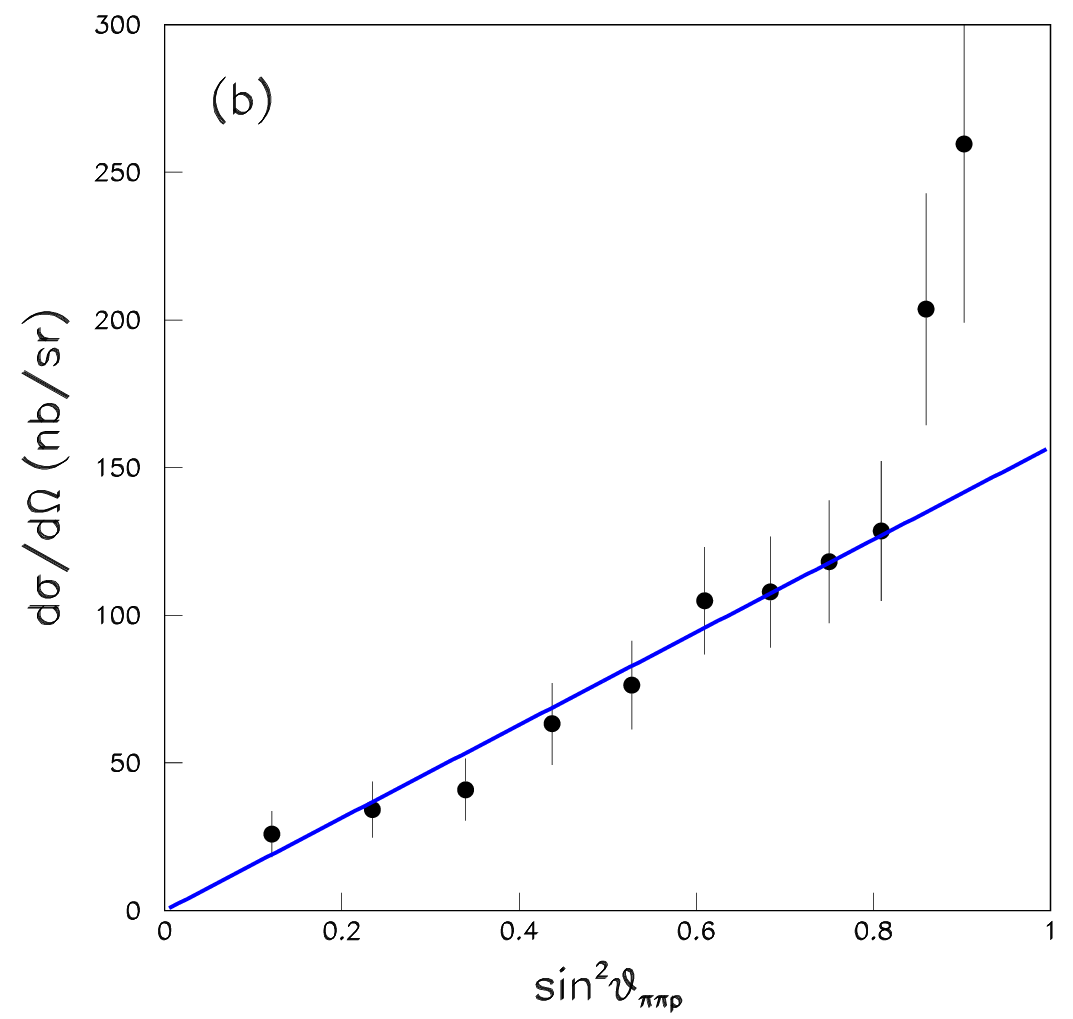}
\caption{Differential cross sections for the $pd\to{}^{3}\mbox{\rm
He}\,\pi^+\pi^-$ reaction at $T_p=546$~MeV~\cite{BEL1999} as a function of
(a) the pion-pion excitation energy $\varepsilon_{\pi\pi}$ and (b) the angle
$\theta_{\pi\pi p}$ between the two-pion relative momentum and the beam axis
in the $\pi^+\pi^-$ rest frame. The dashed curve represents phase space
normalized to the data whereas the solid ones are predictions assuming that
the pion pair emerges in a relative $p$-wave with spin projection $m=\pm 1$.}	
\label{Momo_pipi}
\end{figure}

The only variable accessible in a single-arm experiment~\cite{ABA1963} is the
pion-pion excitation energy
$\varepsilon_{\pi\pi}=(m_{\pi\pi}-2m_{\pi})\,c^2$, where $m_{\pi\pi}$ is the
two-pion invariant mass. This distribution, which is shown in
Fig.~\ref{Momo_pipi}a, could have been studied using just the Big Karl
measurement, but it would then have summed the charged and neutral pion data.
In marked contrast to the original ABC experiments~\cite{ABA1963}, which
showed an enhancement over phase space in the region of
$\varepsilon_{\pi\pi}\approx 30$~MeV, the MOMO data were pushed towards
maximum $\varepsilon_{\pi\pi}$. The MOMO authors suggested that this
distortion might be due to the $\pi^+\pi^-$ pair emerging in a relative
$p$-wave and the solid curve in Fig.~\ref{Momo_pipi}a represents phase space
multiplied by a kinematic factor of $\varepsilon_{\pi\pi}$.

Further evidence in support of the $p$-wave \emph{ansatz} is given in
Fig.~\ref{Momo_pipi}b, which shows the distribution in the angle between the
$\pi^+\pi^-$ relative momentum in the dipion rest frame relative to the beam
direction. If the $\pi^+\pi^-$ pair had been in an $s$-wave, as was expected
for the ABC, the distribution would be isotropic. This is far from being the
case and the MOMO authors argued that the observed $\sin^2\theta_{\pi\pi p}$
behaviour was consistent with the production of a $p$-wave dipion with
angular momentum projection $m=\pm 1$ along the beam direction. Other
distributions presented by the MOMO collaboration~\cite{BEL1999} did not
disagree with this $p$-wave hypothesis.

Although the angular distribution in Fig.~\ref{Momo_pipi}b clearly
demonstrates the presence of higher partial waves in the $\pi^+\pi^-$ system,
the MOMO interpretation is not unambiguous. Similar effects could arise from
$s$-$d$ interference in the $I=0$ channel and, in a two-step
model~\cite{FAL2000}, the shape of the $\varepsilon_{\pi\pi}$ distribution
was described in terms of $\pi^-p\to \pi^0\pi^0p$ amplitudes. However, the
uncertainty in the normalization in such a model means that a significant
$p$-wave component cannot be excluded.

\begin{figure}[h!]
\begin{center}
\includegraphics[width=0.9\columnwidth]{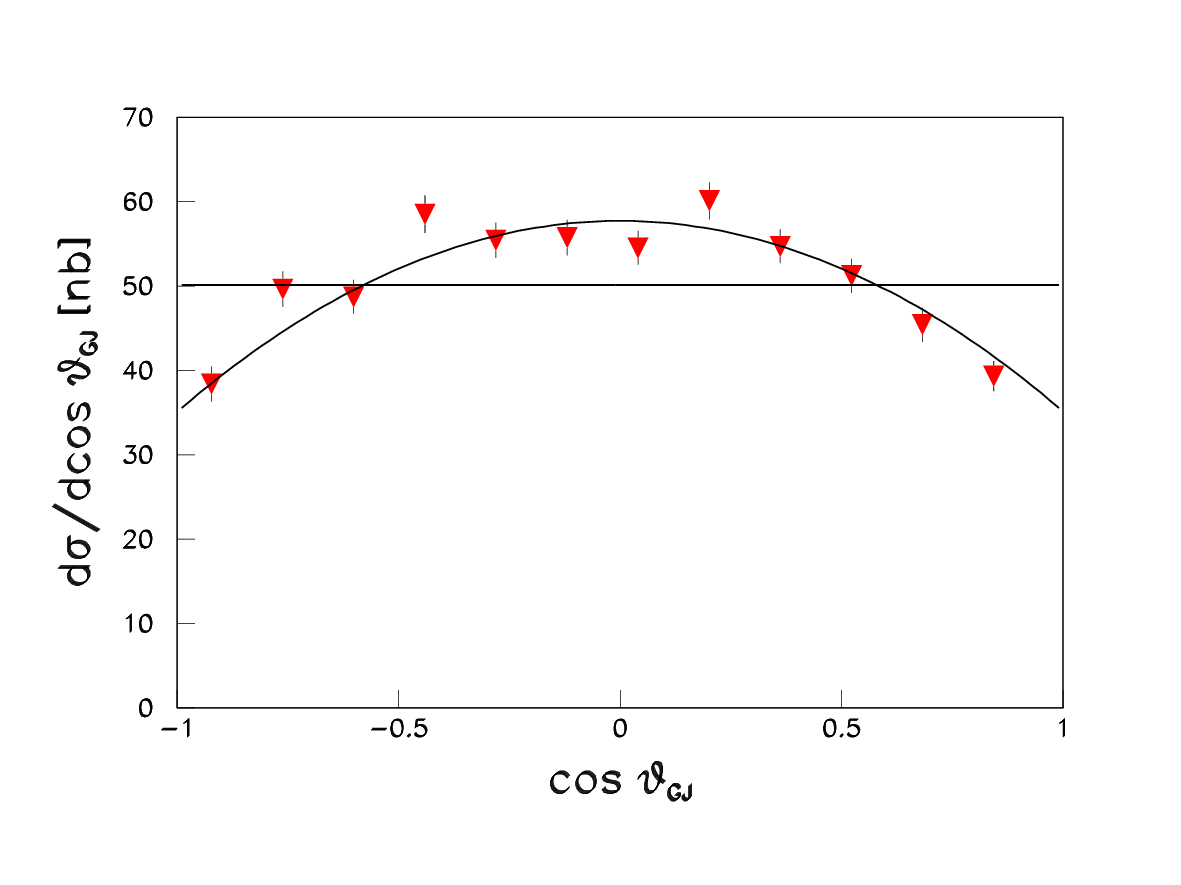}
\caption{Distribution of the MOMO $dp \to{} ^3\textrm{He}\,\pi^+\pi^-$ data
at $Q=28$~MeV in the Gottfried-Jackson angle~\cite{BEL2016}. The data are
symmetric about 90$^{\circ}$ because the sign of the charges on the pions was
not measured. The solid curve is a linear fit to the data in
$\cos^2\theta_{GJ}$. } \label{Fig:28c}
\end{center}
\end{figure}

The anti-ABC behaviour near threshold was also seen in subsequent MOMO data
taken with proton and deuteron beams at $Q=92$, 28, and 8~MeV~\cite{BEL2016}.
The clearest proof for the importance of higher partial waves in the
$\pi^+\pi^-$ system produced in the $dp \to{} ^3\textrm{He}\,\pi^+\pi^-$
reaction at $Q=28$~MeV is provided by the distribution in the
Gottfried-Jackson angle $\theta_{GJ}$. This is the angle between the relative
momentum between the two pions and the direction of the deuteron beam,
evaluated in the dipion rest frame. The MOMO data shown in Fig.~\ref{Fig:28c}
are symmetric about 90$^{\circ}$ because the $\pi^+$ and $\pi^-$ are not
distinguished in this detector. The deviation from isotropy could be a signal
for a superposition of $s$- and $p$-wave pion pairs but even higher partial
waves are not definitively excluded.

A different behaviour for the $pd\to{}^{3}\mbox{\rm He}\,\pi^+\pi^-$ reaction
at low energies is also suggested by the simple $\Delta(1232)$ decay model
used to evaluate the acceptance for this process at ANKE at $Q=265$~MeV, and
which is discussed later in this section. Though this does not predict the
anti-ABC shape shown by the MOMO data in Fig.~\ref{Momo_pipi}a, it does
suggest that the ABC effect might have largely vanished when the energy is
reduced to $Q\approx 70$~MeV.

Values of the $pd\to{}^{3}\mbox{\rm He}\,\pi^0\pi^0$ cross section could be
derived by subtracting the exclusive $pd\to{}^{3}\mbox{\rm He}\,\pi^+\pi^-$
cross section measured using MOMO plus Big Karl from the inclusive
$pd\to\,^{3}\mbox{\rm He}\,X^0$ data obtained using Big Karl
alone~\cite{BEL2016}. However, it is hard to quantify the systematic errors
associated with this procedure.

A first fully exclusive measurement of the $pd \to{}^3\textrm{He}\,\pi\pi$
reaction well away from threshold was undertaken by the WASA collaboration
working close to the $\eta$ threshold at CELSIUS~\cite{BAS2006}. In addition
to seeing ABC peaks in both the $\pi^0\pi^0$ and $\pi^+\pi^-$ invariant mass
distributions, the fully reconstructed events and the large WASA acceptance
allowed the group to study also the individual $^3$He$\,\pi^+$ and
$^3$He$\,\pi^-$ mass distributions. The measurements of $pd
\to{}^3\textrm{He}\,\pi^0\pi^0$ were continued by the group working at COSY
at a higher energy, corresponding to $Q=338$~MeV~\cite{ADL2015b}.

Of greater interest for the low energy discussion is the fact that data were
also taken with a 1.7~GeV deuteron beam and the results analyzed in terms of
quasi-free $dd \to{}^3\textrm{He}\,\pi^0\pi^0n_{\rm sp}$. For this purpose
the photons from the $\pi^0$ decays were detected in coincidence with the
$^3$He and the neutron spectator $n_{\rm sp}$ identified from the
missing-mass peak. This allowed the $pd \to{}^3\textrm{He}\,\pi^0\pi^0$
reaction to be studied simultaneously over a range of energies. It was seen
there that even as low as $Q=172$~MeV there was evidence for some ABC
enhancement at low $\pi\pi$ masses~\cite{ADL2015b}. Of course it must be
realized that pion-pion $p$-waves are excluded in this channel and so it
would really be most interesting to get data also in the $pd
\to{}^3\textrm{He}\,\pi^+\pi^-$ channel above the highest MOMO energy of
92~MeV.

The use of the ANKE magnetic spectrometer improved the resolution relative to
that achieved with WASA. Furthermore, the higher beam energies available at
COSY meant that the $dp \to{}^3\textrm{He} \,\pi^+\pi^-$ differential cross
section could be measured with a deuteron beam incident on a hydrogen target
which, in the absence of a beam pipe hole, increases significantly the
acceptance~\cite{MIE2014}. It is important to realize that, in contrast to
experiments using the MOMO device, the charges on each pion could be
determined. At least one pion had to be detected in coincidence with the
$^3$He, the other being identified through the missing-mass peak, though in
some cases all three particles were measured.

The ANKE $dp \to{}^3\textrm{He}\,\pi^+\pi^-$ data were also taken close to
the $\eta$ threshold as background measurements in the experiment to
determine the mass of the $\eta$ meson~\cite{GOS2012}. Since the ANKE
acceptance is very limited, a model was needed to estimate the necessary
corrections. The one used was based loosely on the idea of the Roper
resonance emitting a $p$-wave pion and decaying into the $\Delta(1232)$
resonance, which also emits a $p$-wave pion when it decays. They therefore
assumed that
\begin{equation}
\sigma \propto \left|[M_{\pi}^{\,2} + B\,\mathbf{k}_1\cdot\mathbf{k}_2](3\Delta^{++}
+ \Delta^{0}) \right|^2,
\label{acccorrformula1}
\end{equation}
where the $\mathbf{k}_{i}$ are the pion momenta in centre-of-mass frame, the
factors 3 and 1 result from the isospin couplings of the $\Delta$
propagators, and $B$ is a complex fit parameter. Note that this \emph{ansatz}
neglects any dependence on the direction of the beam. Other models gave
rather similar correction factors, as did that of the multidimensional matrix
approach.

\begin{figure}[h]
    \includegraphics[width=1.0\linewidth]{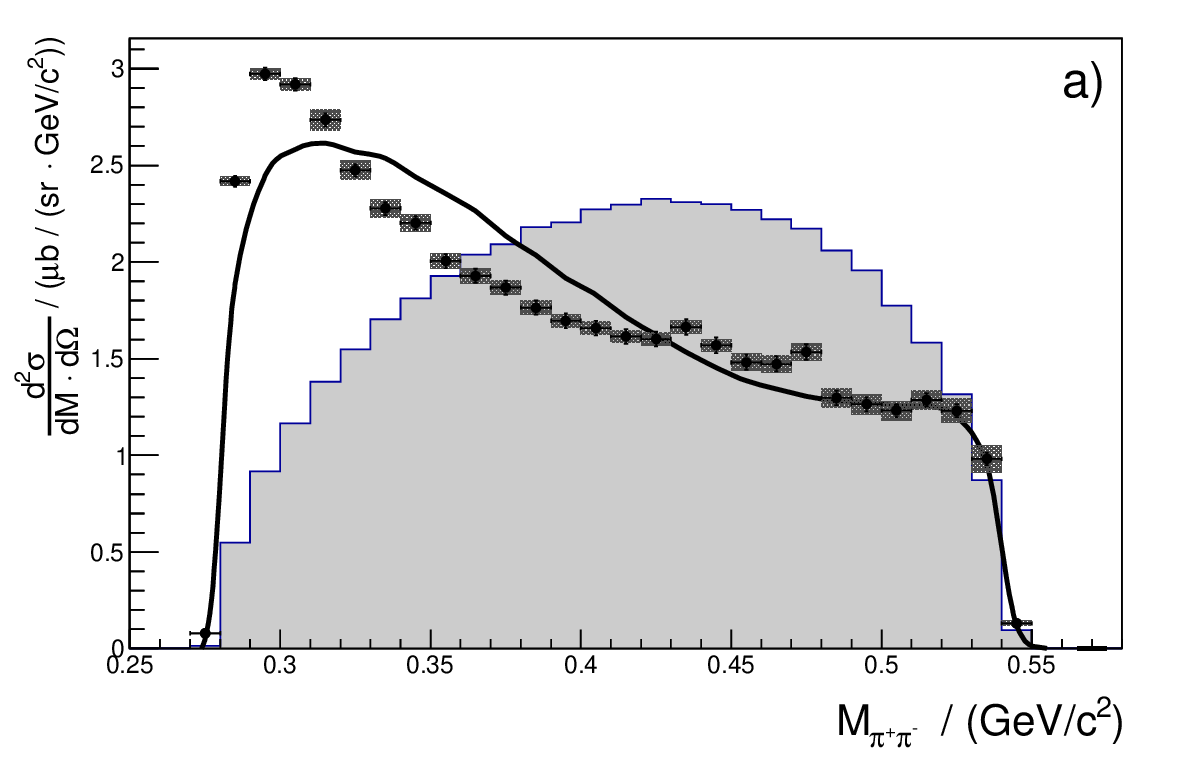}
	\includegraphics[width=1.0\linewidth]{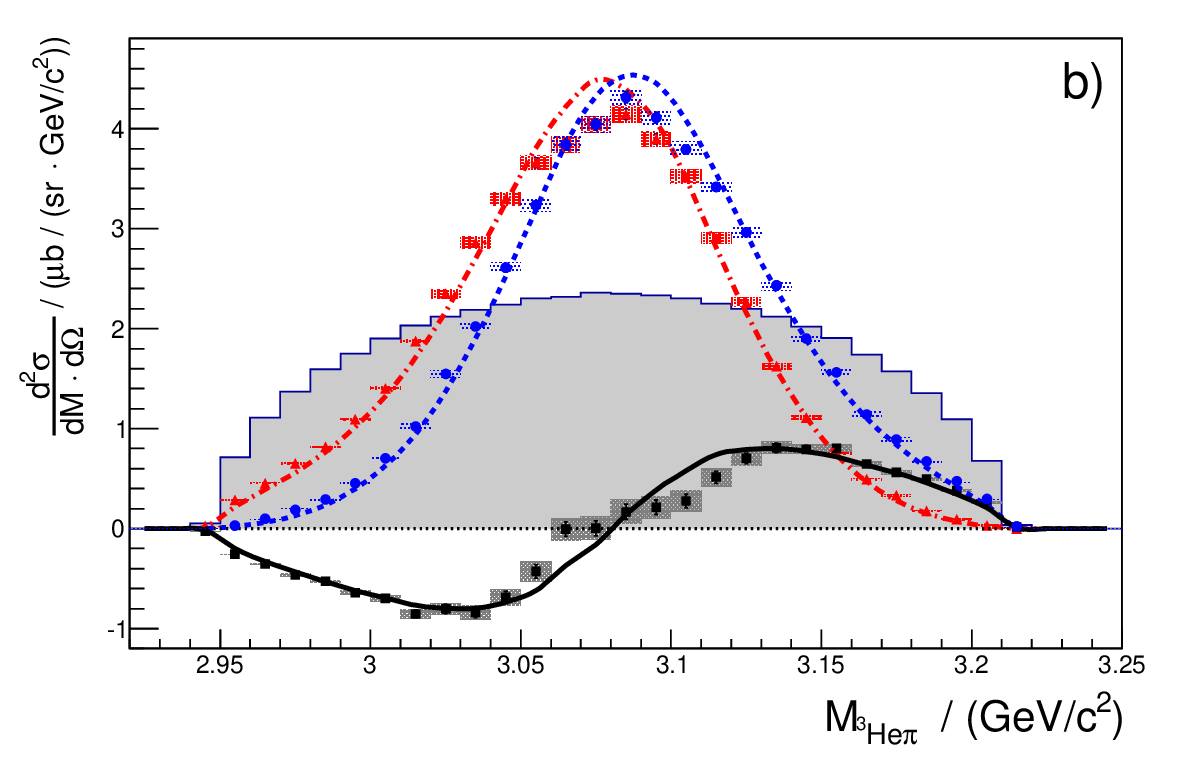}
\caption{\label{invariantmasses} Centre-of-mass double differential cross
sections for the $dp \to{}^3\textrm{He}\,\pi^+\pi^-$ reaction averaged over
$143^{\circ} < \vartheta^{CMS}_{^3\rm{He}} < 173^{\circ}$~\cite{MIE2014} in
terms of (a) $M_{\pi^+\pi^-}$ and (b) $M_{^3\rm{He}\,\pi^+}$ (blue circles)
and $M_{^3\rm{He}\,\pi^-}$ (red triangles). The differences between the two
$M_{^3\rm{He}\,\pi}$ distributions are plotted as black squares. There is, in
addition, an overall normalization uncertainty of 6\%. The curves correspond
to Eq.~(\ref{acccorrformula1}) and the shaded areas are phase-space
distributions normalized to the integrated cross section.}	
\end{figure}

The model of Eq.~(\ref{acccorrformula1}) describe well the two
$M_{^3\rm{He}\,\pi}$ spectra of Fig.~\ref{invariantmasses} and their
difference but the ABC peak in the $\pi^+\pi^-$ mass distribution is not
quite sharp enough, though this might be adjusted through the introduction of
a modest $\pi^+\pi^-$ form factor. If the $\pi^+\pi^-$ spectrum is purely
isoscalar then the $\pi^{\pm}{}^3$He distributions should be identical but,
as is clearly shown in Fig.~\ref{invariantmasses}b, the peak in the
$\pi^-{}^3$He distribution is at a lower mass than that of $\pi^+{}^3$He.
This is indicative of an $I_{\pi\pi}=1$ amplitude interfering with one for
$I_{\pi\pi}=0$. It is therefore a much more sensitive test of isovector pion
pairs than, say, a comparison of $\pi^+\pi^-$ and $\pi^0\pi^0$ production
rates and, moreover, there is no ambiguity accounting for the pion mass
differences. The broad features of the difference spectrum in
Fig.~\ref{invariantmasses}b are reproduced by the simple model of
Eq.~(\ref{acccorrformula1}) and this should contribute to the understanding
of the double-pion-production reaction. It further suggests that there must
have been also some isovector two-pion production on the original ABC
experiment~\cite{ABA1963}. It is unfortunate that there are no similar fully
exclusive measurements closer to the MOMO domain.

\subsection[The $pd\to{}^3$He$K^+K^-(\phi)$ reactions]{The $\boldsymbol{pd\to{}^3}$He$\,\boldsymbol{K^+K^-(\phi)}$ reactions} 
\label{pd3hephi}

The MOMO/Big Karl combination was also used to measure the $pd
\to{}^3\textrm{He}\,K^+K^-$ differential cross section~\cite{BEL2007}, though
the multipion background is here much larger than for two--pion
production~\cite{BEL1999}. In view of this, and in order to identify the
produced particles unambiguously as kaons, the detector was supplemented by a
hodoscope of 16 wedge--shaped scintillators. Charged kaons could thus be
detected and their production vertex measured with full azimuthal acceptance
within a polar angular range of $8^{\circ}<\theta_{\rm lab}<45^{\circ}$.

\begin{figure}[h!]
\begin{center}
\includegraphics[width=0.9\columnwidth]{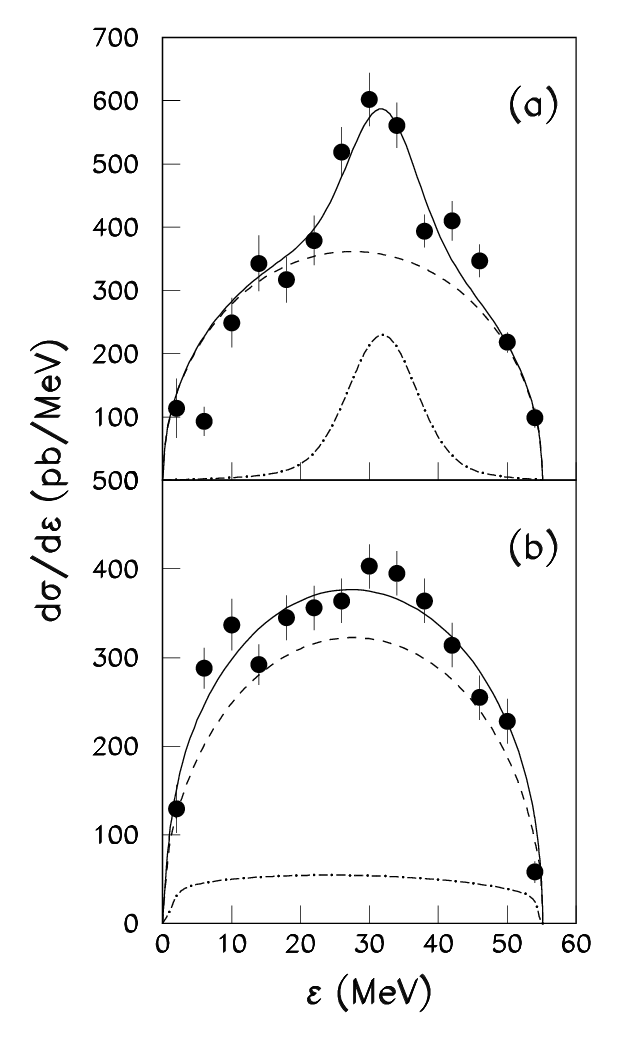}
\caption{\label{fig:MOMOkk1}
Differential cross section for the $pd \to{}^3\textrm{He}\,K^+K^-$  reaction
at an overall excess energy of $Q=55.2$~MeV~\cite{BEL2007}. The curves are
fits to the distributions in terms of phase space coming from prompt $K^+K^-$
production (dashed line), proceeding via $\phi$-meson formation (chain), and
their sum (solid line). The distributions are shown in terms of (a) the
excitation energy $\epsilon$ in the $K^+K^-$ system and (b) that in the
$K{}^3$He system. }
\end{center}
\end{figure}

The experiments were carried out at three beam momenta, corresponding to
excess energies of $Q=35.1$, 40.6, and 55.2~MeV with respect to the
$^3\textrm{He}\,K^+K^-$ threshold (i.e., 3.0, 8.5, and 23.1~MeV with respect
to the nominal $^3\textrm{He}\,\phi$ threshold). The separation of $\phi$
production from direct $K^+K^-$ production was done on the basis of the
$K^+K^-$ invariant mass distribution, an example of which is shown in
Fig.~\ref{fig:MOMOkk1}a at $Q=55.2$~MeV. This is in fact the most challenging
energy because, in addition to the resolution getting worse kinematically as
$Q$ increases, the beam conditions were also less favourable.

The $\phi$ contribution to the differential cross section was modeled in
terms of a peak with a natural width of $\Gamma = 4.2$~MeV, smeared with the
expected energy resolution. In contrast, the direct $K^+K^-$ component was
taken to be proportional to three-body phase space, which is very different
to the distributions found by MOMO for $\pi^+\pi^-$
production~\cite{BEL1999}. The sum of these two elements reproduces very well
the $K^+K^-$ data at all three values of $Q$, an example of which is shown in
Fig.~\ref{fig:MOMOkk1}a. This therefore gave confidence in using such a model
in correcting the overall acceptance. However, the large background under the
$\phi$ made it difficult to extract separate angular distributions at this
energy, though the conditions are much more favourable close to the $\phi$
threshold.

The distribution in the $K\,{}^3$He invariant mass shown in
Fig.~\ref{fig:MOMOkk1}b is little sensitive to the $\phi/K^+K^-$ separation.
There is no sign of any $K^-p$ enhancement that was so evident in the $pp\to
ppK^+K^-$ case of sect.~\ref{ppKK} because it was not possible to distinguish
between $K^-$ and $K^+$ with the MOMO detector.

By measuring all three final particles it was possible to construct several
angular distributions, the most interesting of which is the angle between the
relative $K^+K^-$ direction in its rest frame and the initial beam direction.
The $Q= 35.1$~MeV data in a region where the $\phi$ cannot contribute
($\epsilon < 28$~MeV) are flat. On the other hand, the $\phi$-rich region
($\epsilon > 28$~MeV) has a very strong angular dependence that could only be
explained by the $\phi$ being produced almost exclusively with polarization
projection along the beam direction of $m=0$. This clear effect is in
complete contrast to the behaviour in $pd \to{}^3\textrm{He}\,\omega$, where
the $\omega$ mesons are produced effectively unpolarized~\cite{SCH2008}. This
is the clearest signal for a violation of the OZI rule that relates $\omega$
and $\phi$ production rates~\cite{OZI1963}, since there is much less
ambiguity in accounting for the effects of the meson mass difference.

\begin{figure}[h!]
\begin{center}
\includegraphics[width=0.9\columnwidth]{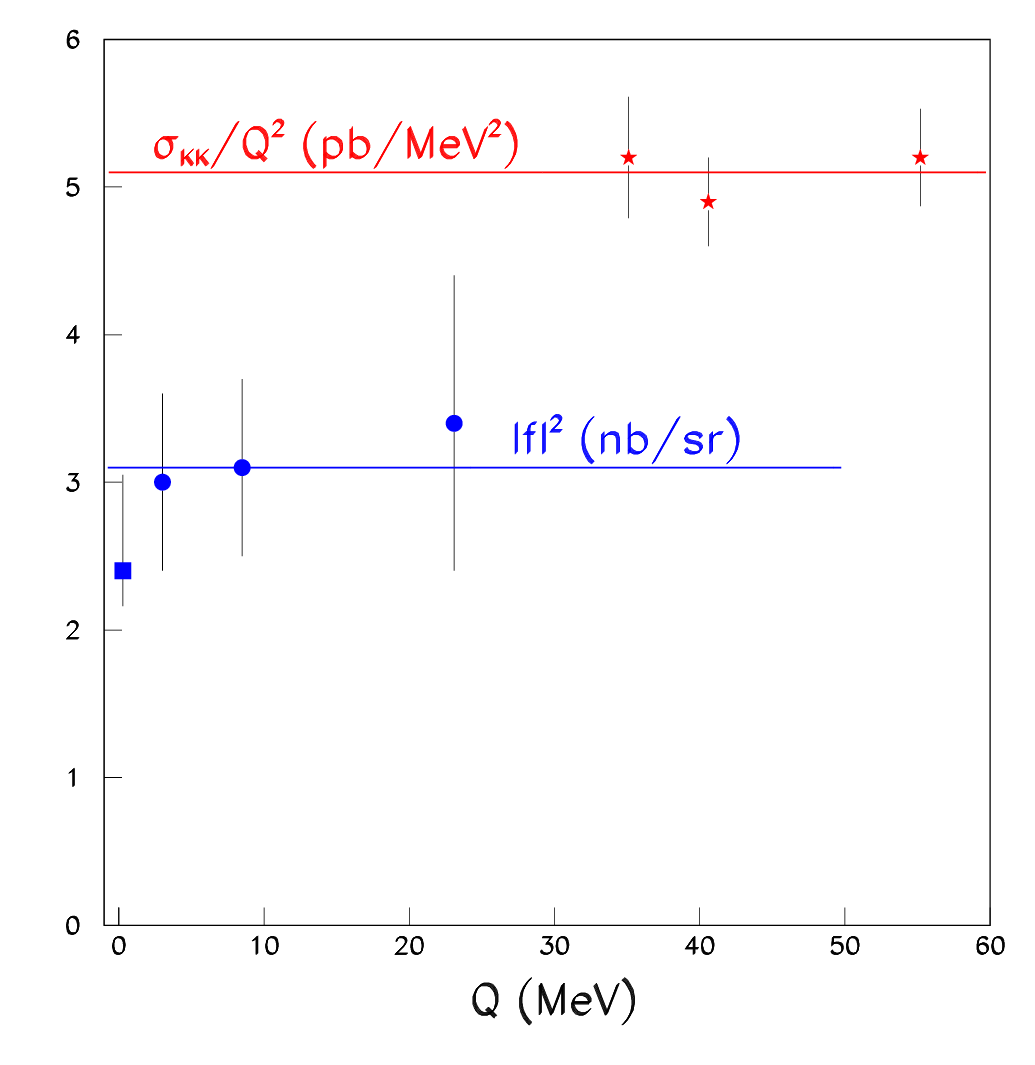}
\caption{\label{fig:MOMOkk2} Circles (blue) denote MOMO values of the
amplitude-square for $\phi$ production~\cite{BEL2007}, as defined by
Eq.~(\ref{amps}); the SPES4 result at $Q = 0.3$~MeV~\cite{WUR1996} is shown
by a square. The data are consistent with $|f|^2$ being constant, as
indicated. The MOMO data for the total cross section for direct $K^+K^-$
production divided by the phase space factor of $Q^2$ are shown by the (red)
stars~\cite{BEL2007}.}
\end{center}
\end{figure}

Though the shapes of the distributions and the angular dependence are largely
unaffected by the overall normalizations, these are of course important for
the cross section evaluations, which have systematic uncertainties of the
order of 10\%. The resulting energy dependence is illustrated in
Fig.~\ref{fig:MOMOkk2}. For $\phi$ production, this is shown in the form of
the angular average of the square of the production amplitude, as defined by
Eq.~(\ref{amps}). Within the experimental uncertainties, $|f(pd
\to{}^3\textrm{He}\,\phi)|^2$ is consistent with being constant, and this
value agrees with the Saclay near-threshold missing-mass
measurement~\cite{WUR1996}.

On the other hand, it is seen in Fig.~\ref{fig:MOMOkk1}a that the cross
section for direct $K^+K^-$ production looks very much like phase space. This
is confirmed in Fig.~\ref{fig:MOMOkk2}, where it is shown that the total
cross section divided by a phase space factor of $Q^2$ is effectively
constant.

Whereas in the vicinity of the $\phi$ threshold the resolution and the
signal-to-background ratio allow the $pd \to{}^3\textrm{He}\,\phi$ cross
section to be measured in a missing-mass experiment, at higher excess
energies this is no longer possible and the detection of a $K^+K^-$ pair in
coincidence is necessary. It is unfortunate that the MOMO detector could not
determine the charges on the mesons. Otherwise it could have investigated the
$K^-{}^3$He final state interaction in $pd \to{}^3\textrm{He}\,K^+K^-$.

%
%

\section[The $dd\to{}^4$He$X^0$ family of reactions]{The $\boldsymbol{dd\to{}^4}$He$\boldsymbol{X^0}$ family of reactions} 
\label{dd4HeX}\setcounter{equation}{0}%

Meson production rates in the $dd\to{}^4$He$\,X^0$ reaction are expected to
be very low because two deuterons, with mean diameters of about 4~fm, have to
be squeezed to form the much smaller $\alpha$-particle. An exception is the
$dd\to{}^4$He$\,\pi^+\pi^-$ reaction~\cite{PAD2012}, but there may be special
reasons for this~\cite{GAR1999}. However, the rates are often disappointingly
low as, for example, in the search for evidence for the production of the
$f_0$ meson in the $dd\to{}^4$He$\,K^+K^-$ reaction~\cite{YUA2009}.
Nevertheless, some channels have to be studied in depth because of their
importance in Physics and two such examples are outlined in this section.

Evidence from COSY-11 and ANKE was presented in sect.~\ref{pd3heeta} for the
possible existence of the $_{\eta}^3$He $\eta$-mesic nucleus. The
interpretation of these results suggest that the $\eta$ might bind to
$^4$He but that signals could be harder to detect for heavier nuclei. Given
that the final $s$-wave is forbidden by spin-parity constraints in the
$\gamma{}^4$He$\to \eta{}^4$He reaction, and that the use of a tritium target
in the $p{}^3$H$\to\eta^4$He reaction presents its own special problems, the
attention naturally turns to $dd\to\eta^4$He, which is the subject of
sect.~\ref{dd4heeta}.

Symmetry properties have been one of the cornerstones of particle and nuclear
physics and, as such, have been extensively investigated for over sixty
years. In modern parlance, charge symmetry corresponds to invariance under
the interchange of the $u$ and $d$ quarks. It has even been dismissed as an
\emph{accidental} symmetry that arises because of the near equality of the
$u$ and $d$ masses. Charge symmetry violation associated with isospin mixing
within multiplets is responsible for the well-known mixing of the $\rho^0$
and $\omega$ mesons but there has been far less direct evidence for such
violations in nuclear reactions.

Under the charge symmetry operation, the $\pi^0$ changes sign whereas the
deuteron and $\alpha$-particle are unaffected. As a consequence, a
non-vanishing rate for the $dd\to \alpha\pi^0$ reaction, which is discussed
in sect.~\ref{CSB}, is a clear example of
charge symmetry breaking (CSB). It is also the most convincing example in
nuclear reactions because it is proportional to the square of a CSB
amplitude, with no contribution from interference terms.

\subsection[The $dd\to{}^4$He$\,\eta$ reaction]{The $\boldsymbol{dd\to{}^4}$He$\,\boldsymbol{\eta}$ reaction} 
\label{dd4heeta}

\begin{figure}[h!]
\begin{center}
\includegraphics[width=1.0\columnwidth]{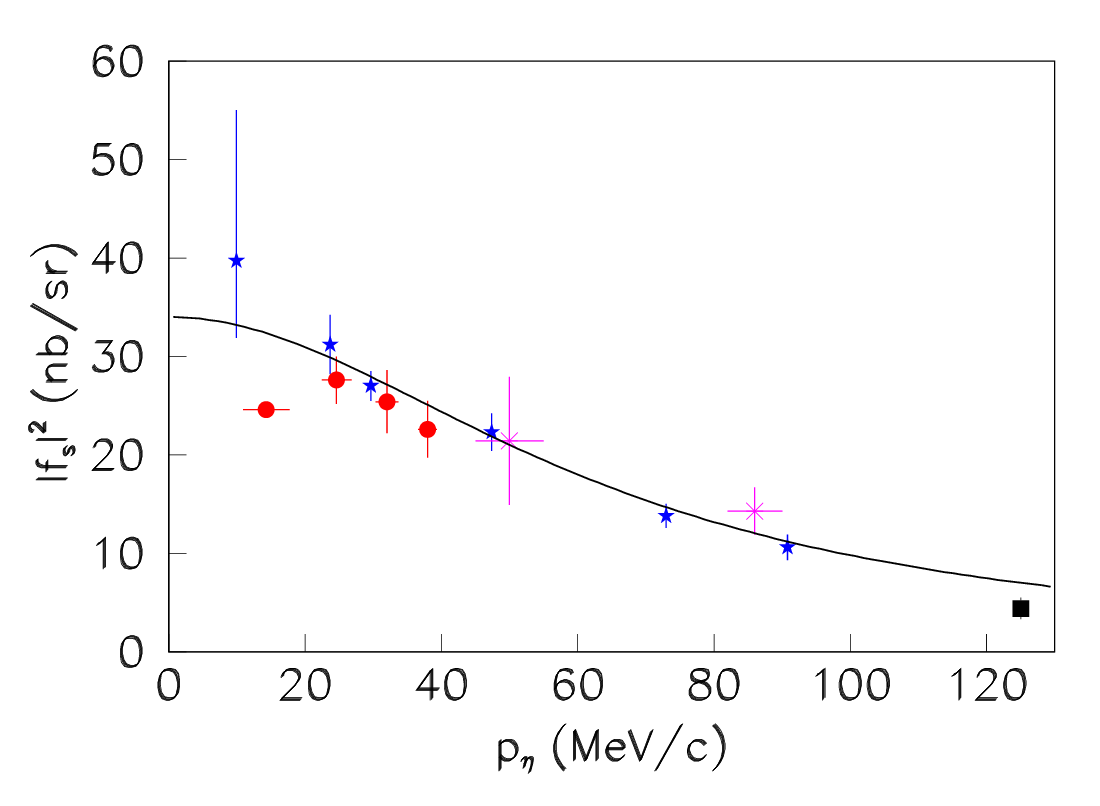}
\caption{\label{fig:dd2ae} Experimental values of the square of the $dd \to
{}^{4\!}\textrm{He}\,\eta$ $s$-wave amplitude. The data are taken from
Refs.~\cite{FRA1994} (red circles), \cite{WIL1997} (blue stars),
\cite{WRO2005} (magenta crosses), and \cite{BUD2009} (black closed
square). The curve is a scattering-length fit: $|f_s|^2 = 34/[1+
(p_{\eta}/64)^2]$~nb/sr, where $p_{\eta}$ is measured in MeV/$c$.
}
\end{center}
\end{figure}

The first measurements of the $dd\to {}^4$He$\,\eta$ total cross sections
were carried out using the SPES4~\cite{FRA1994} and the SPES3~\cite{WIL1997}
spectrometers at SATURNE. Figure~\ref{fig:dd2ae} shows the data converted
into averaged-squared-amplitudes $|f_s|^2$ on the basis of the analogue of
Eq.~(\ref{amps}) used for the $pd\to{}^3$He$\,\eta$ reaction. Corrections,
which will be discussed later, have been made to eliminate effects of higher
partial waves. The first point to notice is that the production rate at small
$p_{\eta}$ is about a factor of 50 lower for $^4$He$\,\eta$ than for
$^3$He$\,\eta$.

Though the acceptance of SPES3 was much larger than that of SPES4, its
missing-mass resolution was certainly poorer and it was therefore difficult
to overcome the multipion background. However, the SPES3 experiment was
carried out with a polarized deuteron beam. Since $s$-wave $\eta$ production
is forbidden by Bose symmetry for deuterons with spin projection $m=0$, the
tensor analyzing power near threshold must be $A_{xx}=-\half$. Any deviations
from this would be a signal for $p$ or higher $\eta$ partial waves. The SPES3
analysis assumed that this constraint was also valid for all the energies so
that they studied the shape of the multipion background by forming the $m=0$
combination $(1+2A_{xx})\,\dd\sigma/\dd\Omega$. Thus, if the analyzing power
were small away from threshold, the SPES3 results for the total cross section
would be too low by a factor of $1.5$.

The unpolarized differential cross section was studied later at two excess energies
$Q$ with the ANKE spectrometer~\cite{WRO2005}. Whereas the results were
consistent with isotropy at $Q=2.6$~MeV, there was clearly a need for a linear
term in $\cos^2\theta_{\eta}$ at $Q=7.7$~MeV. However, there is no way from
these data to discover if this effect is due to the square of a large
$p$-wave amplitude or arises from a much smaller $d$-wave interfering with the
dominant $s$-wave amplitude. For that one needs well identified events
obtained with a tensor polarized deuteron beam (or target).

The GEM collaboration used the high resolution Big Karl spectrometer to study
the $dd\to {}^4$He$\,\eta$ reaction with both polarized and unpolarized
deuteron beams at an excess energy of $Q=16.6$~MeV~\cite{BUD2009}. By
expanding the unpolarized differential cross section in terms of Legendre
polynomials it was found that terms up to at least $P_4(\cos\theta_{\eta})$
were required to describe the data and the corresponding coefficient $a_4$
was in fact the largest in the series. This demonstrated that the production
involved significant contributions from $d$ or higher waves.

In the second stage, data were taken with a polarized deuteron beam but, due
to the side yoke in the first magnetic dipole of Big Karl, the acceptance had
severe cuts in the azimuthal angle and the results were only sensitive to the
$A_{xx}$ deuteron analyzing power~\cite{BUD2009}. Nevertheless this is
sufficient to separate the contributions from even and odd partial waves. An
analysis of the data shows that at this energy the reaction is dominated by
the $d$-wave in the  $\eta ^4$He system. By assuming that each of the partial
waves varied with the threshold factor $(p_{\eta})^{\ell_{\eta}}$, it was
possible to correct all the previous data in order to extract values for the
$s$-wave contributions, and this is precisely what is shown in
Fig.~\ref{fig:dd2ae}.

It is clear that $|f_s|^2$ falls far less rapidly than the corresponding
quantity for $dp\to {}^3$He$\,\eta$ discussed in sect.~\ref{pd3heeta}. Even
if one restricts a parametrization of the data to one in terms of a
scattering length, there would still be strong coupling between the real and
imaginary parts. The curve shown in the figure, $|f_s|^2 = 34/[1+
(p_{\eta}/64)^2]$~nb/sr, corresponds to a purely imaginary scattering length.
Here $p_{\eta}$ is measured in MeV/$c$ and the pole is at $Q\sim -4$~MeV,
though there is no way of knowing from $\eta$ production data whether this
would correspond to a bound or anti-bound state.

Limits on the production of $_{\eta}^{4}$He were also obtained through the
COSY-WASA measurements of the cross sections for the
$dd\to{}^{3}$He$\,p\pi^-$ and $dd\to{}^{3}$He$\,n\pi^0$ reactions, where the
exotic nucleus is expected to decay via $\eta n\to \pi^-p$ or $\eta n\to
\pi^0n$~\cite{ADL2016}. These limits are, however, comparable to theoretical
estimates, which themselves have a large degree of
uncertainty~\cite{WYC2014}.

The use of a polarized deuteron beam helped significantly in the
understanding of the $dd \to {}^{4}$He$\,\eta$ reaction and constrained the
position of the $s$-wave pole by reducing the influence of higher partial
waves. Unfortunately this has not proved possible for the much rarer $dd \to
{}^4$He$\,\pi^0$ reaction, to which we turn in the next section.

However, before we leave entirely the topic of $\eta$-mesic nuclei, mention
should be made of one COSY experiment that investigated the possibility of
the formation of heavier nuclei~\cite{BUD2009b,MAC2015}. Here the ENSTAR
detector, described in sect.~\ref{ENSTAR}, was used in combination with the
Big Karl spectrometer. ENSTAR detected back-to-back $\pi^-p$ pairs from the
hoped-for $\eta N$ decays, with Big Karl measuring the $^3$He in the
$p{}^{27\!}\textrm{Al}\to{}^3\textrm{He}\, p\,\pi^-X$ reaction. The
kinematics were cunningly chosen such that, for a weakly bound $\eta$-nucleus
state, the meson was produced almost at rest so that it had a higher chance
of \textit{sticking} to the residual nucleus.

\begin{figure}[h!]
\begin{center}
\includegraphics[width=1.0\columnwidth]{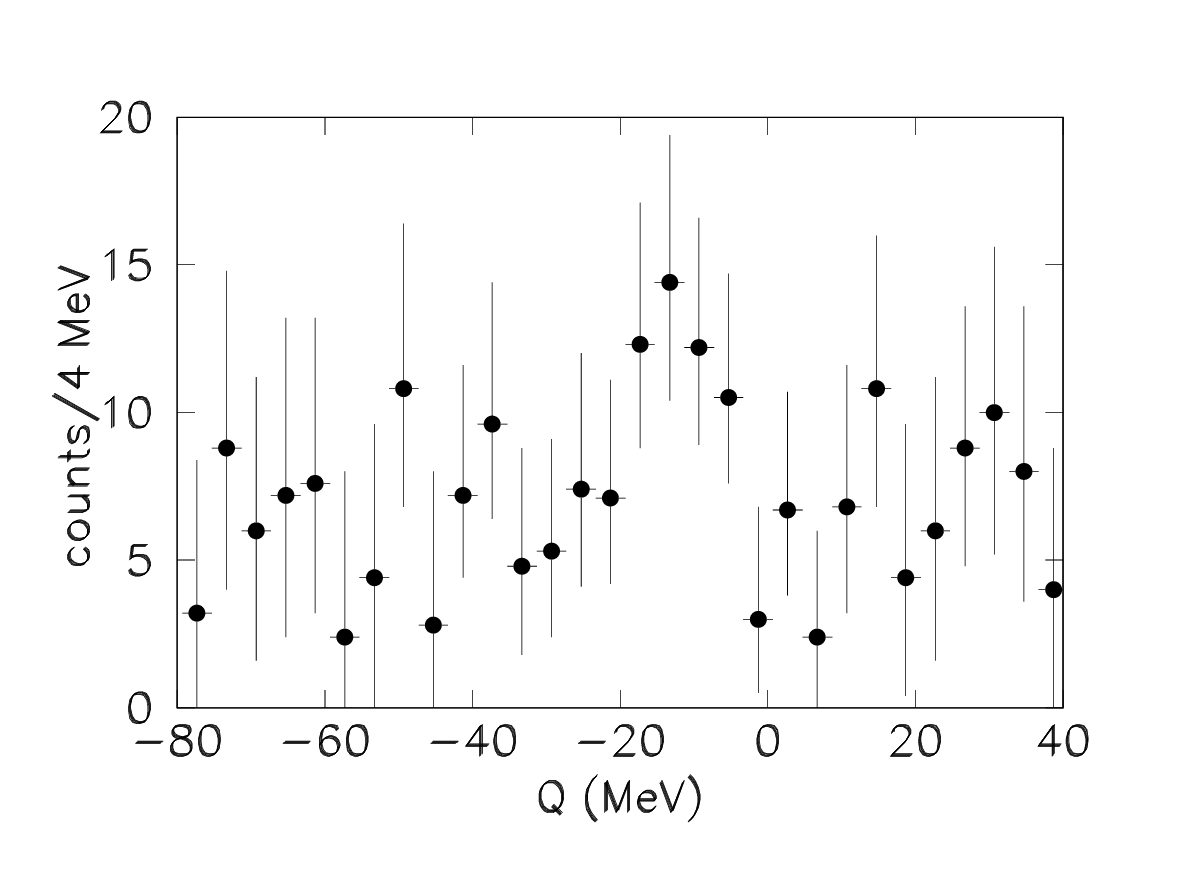}
\caption{\label{fig:ENSTAR2} Counts for the $p{}^{27\!}\textrm{Al} \to
{}^3\textrm{He}\, p\, \pi^- X$ reaction as a function of the excess energy in
the $\eta^{25}$Mg system~\cite{BUD2009b}.
}
\end{center}
\end{figure}

It was suggested that the excess of events for $Q\approx -13$~MeV with a FWHM
of $\approx 10$~MeV in Fig.~\ref{fig:ENSTAR2} might be a signal for a
$_{\,\,\eta}^{25}$Mg bound state. If this were indeed the case, then the
production cross section for this state is estimated to be $0.46\pm 0.16({\rm
stat}) \pm 0.06 ({\rm syst})$~nb. Haider and Liu~\cite{LIU2015} predicted a
state in $_{\,\,\eta}^{25}$Mg at $Q\approx -7$~MeV but offered an explanation
for the difference of 6~MeV. However it is still possible that the structure
is a statistical fluctuation and so further investigations are required.

\subsection[The $dd\to{}^4$He$\,\pi^0$ reaction and charge symmetry]{The $\boldsymbol{dd\to{}^4}$He$\,\boldsymbol{\pi^0}$ reaction and charge symmetry} 
\label{CSB}

\begin{figure}[h!]
\begin{center}
\includegraphics[width=0.9\columnwidth]{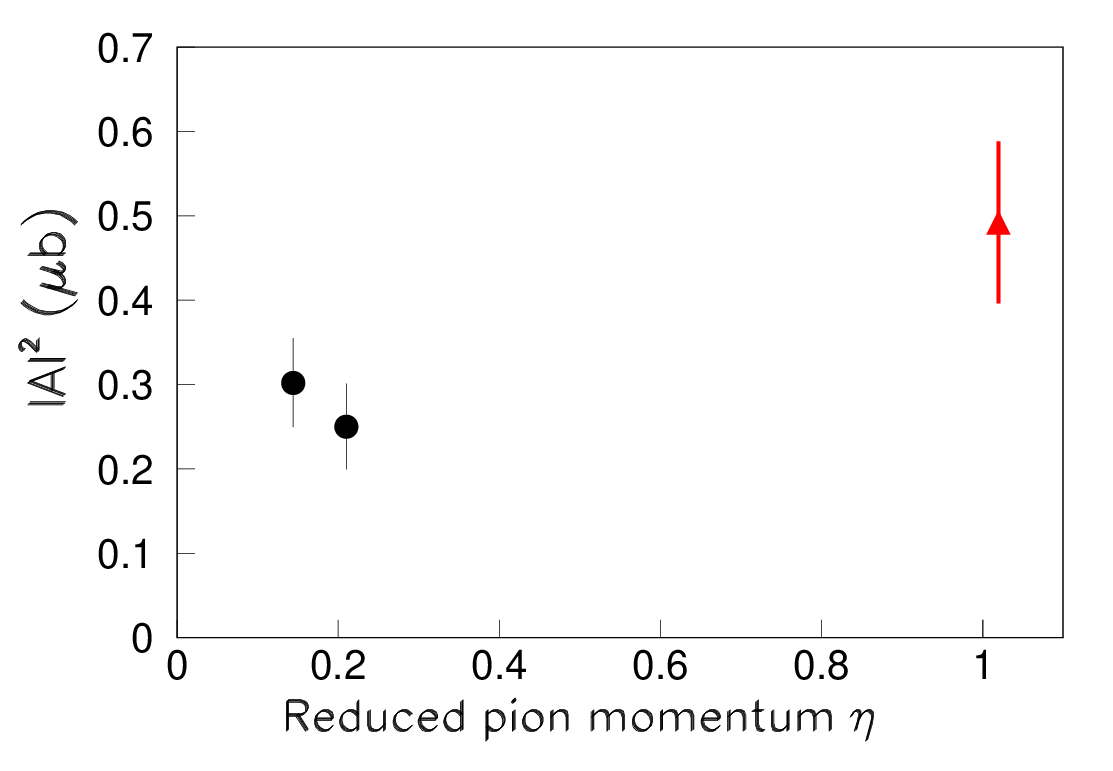}
\caption{\label{fig:Volker2} Values of the total cross sections measured for
the $dd\to \alpha\,\pi^0$ reaction measured at IUCF (black)
circle~\cite{STE2003} and WASA-at-COSY (red) triangle~\cite{ADL2014b}. The
data are multiplied by the phase-space factor of the ratio of the initial to
the final c.m.\ momenta and plotted in terms the reduced pion c.m.\
momentum $\eta=p_{\pi}/m_{\pi}$.}
\end{center}
\end{figure}

There have been many attempts over the years to detect the $dd\to
\alpha\,\pi^0$ reaction but the first positive claim~\cite{GOL1991} was
possibly a misidentification of the non-pionic three-body
$\alpha\gamma\gamma$ final state~\cite{DOB1999}. The first unambiguous
measurement of the $dd\to \alpha\,\pi^0$ reaction was made at IUCF at two
energies close to threshold~\cite{STE2003}. The momentum of the
$\alpha$-particle was determined by magnetic analysis and time-of-flight
measurements, which allowed the missing mass in the $dd\to \alpha X$ reaction
to be evaluated reliably. Photons from the $\pi^0$ decay were detected in
lead glass arrays placed to the left and right of the gas jet target. The
resulting missing-mass distribution was therefore that of $dd\to
\alpha\gamma\gamma$ with the $\pi^0$ in the final state being recognized
through the peak in the distribution with $\textrm{FWHM}\approx
600~\textrm{keV}/c^2$. The three-body $\alpha\gamma\gamma$ background was
predicted to vary smoothly with $m_{\gamma\gamma}$, though the rates
extracted were about twice those estimated in an \emph{ab initio}
model~\cite{DOB1999}. The IUCF total cross sections of $\approx 14$~pb are
shown in Fig.~\ref{fig:Volker2} divided by the phase-space factor and plotted
in terms of the pion reduced momentum $\eta = p_{\pi}^{cm}/m_{\pi}$.

The two IUCF points shown in Fig.~\ref{fig:Volker2} are consistent with
$s$-wave production but, in order to put extra constraints on the theoretical
modeling, data are required away from the threshold region where higher
partial waves may play a role. The decay photons are, of course, very well
measured in the WASA central detector but the $^4$He identification and
measurement in the WASA forward detector was less effective than in the IUCF
setup and some $^3$He may have been falsely identified as $^4$He. In
addition, in common with many meson production experiments, the missing-mass
peaks become less pronounced as one moves away from threshold.

\begin{figure}[h!]
\begin{center}
\includegraphics[width=0.9\columnwidth]{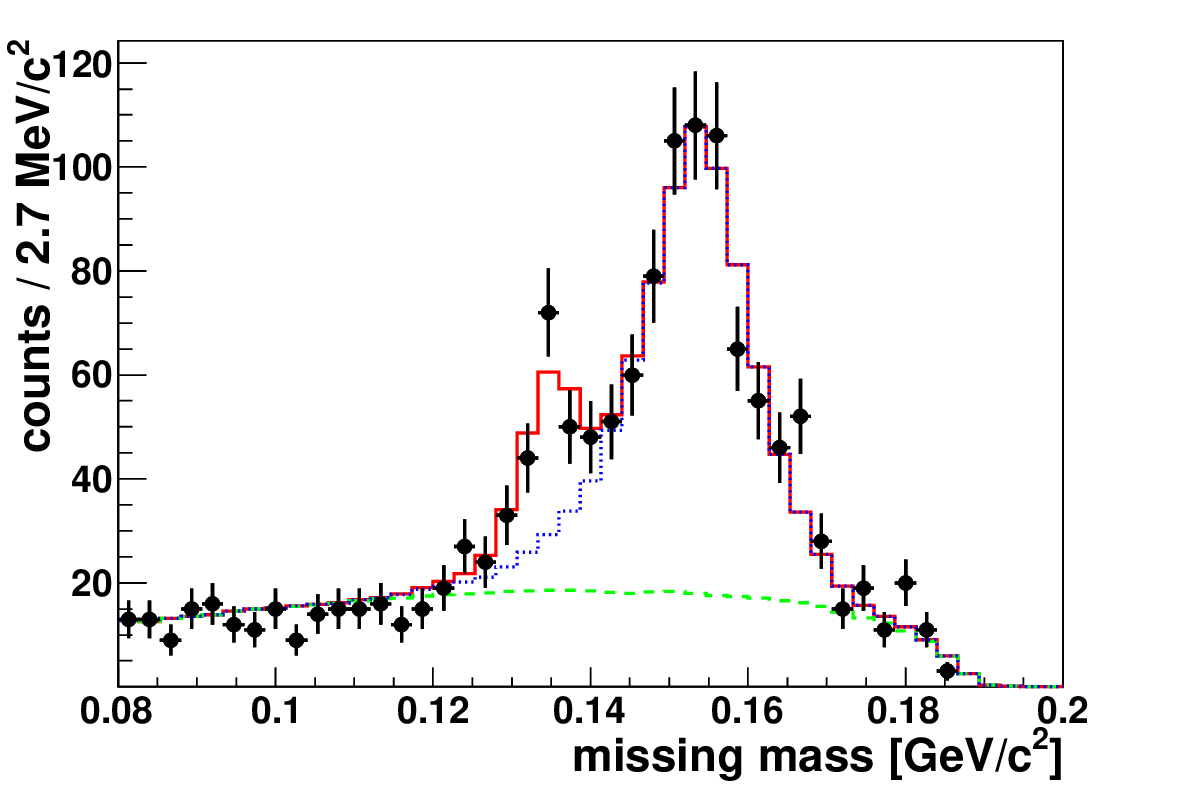}
\caption{\label{fig:Volker1} Missing-mass distribution for the $dd \to
\mathrm{{^4}He}\,X$ reaction measured by the WASA-at-COSY
collaboration~\cite{ADL2014b}. Fits were made that included the three-body
$dd \to \mathrm{{^4}He}\,\gamma\gamma$ contribution (green dashed), plus the
$dd \to \mathrm{{^3}He}\,n\pi^{0}$ reaction (blue dotted), plus the desired
signal (red solid). }
\end{center}
\end{figure}

The WASA-at-COSY experiment was carried out at an excess energy of $Q=60$~MeV
with respect to the $\pi^0$ production threshold. The missing-mass
distribution from the initial publication is shown in Fig.~\ref{fig:Volker1},
which was evaluated on the assumption that all the remaining recoiling helium
ions were $^4$He~\cite{ADL2014b}. Though much suppressed by the kinematics,
the most prominent peak is that due to the charge-symmetry-allowed $dd\to
\mathrm{{^3}He}\,n\pi^{0}$ reaction, which was actually used to establish the
normalization, using data previously obtained by the group~\cite{ADL2013}.
There is, in addition, the rather featureless $dd \to
\mathrm{{^4}He}\,\gamma\gamma$  background, though there may also be a small
contamination here from the $^3\textrm{He}\,n\gamma\gamma$ final state.
Nevertheless, the evidence for the $dd \to \mathrm{{^4}He}\,\pi^0$ reaction
in Fig.~\ref{fig:Volker1} is quite clear.

Extrapolating to the whole of phase space, including events lost down the
beam pipe, a total cross section $\sigma_{\rm tot}=(118\pm24)$~pb was
obtained, which is about $5\times 10^{-5}$ times smaller than the allowed $dd
\to \mathrm{{^3}He}\,n\pi^{0}$ cross section at the same beam
energy~\cite{ADL2013}. The value obtained for the total $dd\to
\mathrm{{^4}He}\,\gamma\gamma$ cross section ($\approx 1$~nb) contained
significant model dependence and was not compared to theoretical estimates.

The phase-space-modified $dd\to \alpha\pi^0$ total cross section is shown
along with the IUCF points in Fig.~\ref{fig:Volker2}. The increase in
$p_d^{cm}\sigma_{\rm tot}/p_{\pi}^{cm}$ with $\eta$ might be taken as
evidence for higher partial waves but is not conclusive. Limited information
was also obtained on the angular distribution, which suggests some
anisotropy, which was confirmed in a more extensive later run with higher
statistics~\cite{ZUR2016}. An angular dependence could arise from either
$s$-$d$ interference or from the square of a $p$-wave amplitude. These
possibilities could only be separated through a measurement of the deuteron
tensor analyzing power, as was done for the analogous but allowed $dd \to
\mathrm{{^4}He}\,\eta$ reaction~\cite{BUD2009}. Such information would be
very valuable for theorists trying to understand the origin of the symmetry
breaking.

%
%
\section[Rare decays of $\eta$ and $\pi^0$ mesons]{Rare decays of $\boldsymbol{\eta}$ and $\boldsymbol{\pi^0}$ mesons} 
\label{WASA}
\setcounter{equation}{0}%

The initial motivation for the study of $\eta$ decays at hadronic machines
came from the SATURNE measurement of the $dp \to {}^{3}\textrm{He}\,\eta$
total cross section~\cite{BER1988}. This showed that there was a very strong
$\eta$ signal even within a few MeV of threshold but that the multipion
background under the $\eta$ missing-mass peak was very low. Some $\eta$
decays were studied at SATURNE using this facility but the chance to create
an $\eta$-meson factory~\cite{ETA1994} was lost when the CsI crystals, which
were the basis of the detector design, could not be delivered on time and in
budget before the closure of the SATURNE accelerator. Much of the physics
programme was taken up at the WASA detector installed at the CELSIUS storage
ring but the full impact was only felt after the transfer of WASA to COSY.

It is important to realize that, although the tagging of the $\eta$ is very
good indeed in the $pd \to {}^{3}\textrm{He}\,\eta$ reaction, the counting
rates are limited and so this source cannot be used for the very rare decays.
Higher counting rates can be achieved in proton-proton collisions using the
$pp\to pp\eta$ reaction, though the backgrounds may then be larger than those
found at electron machines. Some WASA $\eta$-decay data obtained with the
$pp\to pp\eta$ reaction are still at the analysis stage~\cite{ADL2015}.

The situation is more challenging for the study of the $\eta^{\prime}$, where
the production rates with proton beams are low but the multipion background
high~\cite{WOL2013}.

\subsection[$\eta$ decays]{$\boldsymbol{\eta}$ decays} 
\label{eta_decays}

The WASA detector, as installed at COSY, was introduced in
sect.~\ref{WASA_detector}. Its first production run at COSY used $\eta$
mesons generated in $pp$ collisions at 1.4~GeV~\cite{ADO2009}. The $1.2\times
10^5$ fully reconstructed events of the $\eta\to\pi^0\pi^0\pi^0$ decay were a
sub-sample of the $8\times 10^5$ identified $pp\to pp6\gamma$ events. In
order to select $\pi^0$ candidates from the six reconstructed photons, all
fifteen possible combinations of the photon pairs were considered and only
solutions with reasonable probabilities retained.

Since the final $\pi^0$ are identical particles, the $\eta\to\pi^0\pi^0\pi^0$
Dalitz plot must be symmetric under the exchange of any two of their kinetic
energies $T_i$. For a constant matrix element the population of the Dalitz
plot should be uniform but this may be distorted by, among other things,
pion-pion scattering in the final state. Because of the identity of the
$\pi^0$, one would expect the population to vary as $1+2\alpha z$, where the
variable $z$, defined by $z=[(T_1-T_2)^2+3(T_3-\langle T\rangle)^2]/3\langle
T\rangle^2$, with $\langle T\rangle =(T_1+T_2+T_3)/3$, is indeed symmetric
under particle interchange.

The asymmetry found, $\alpha=-0.027 \pm 0.008({\rm stat}) \pm 0.005({\rm
syst})$, is consistent with the other modern measurements in the PDG
tabulation~\cite{OLI2014}, though the KLOE and Crystal Barrel collaborations
quote slightly smaller error bars. The statistical precision of the WASA data
is unfortunately insufficient to investigate the cusp effect that must be
present when the two-pion invariant mass is around twice the mass of the
charged pion.

\begin{figure}[hbt]
\begin{center}
\includegraphics[width=0.9\columnwidth]{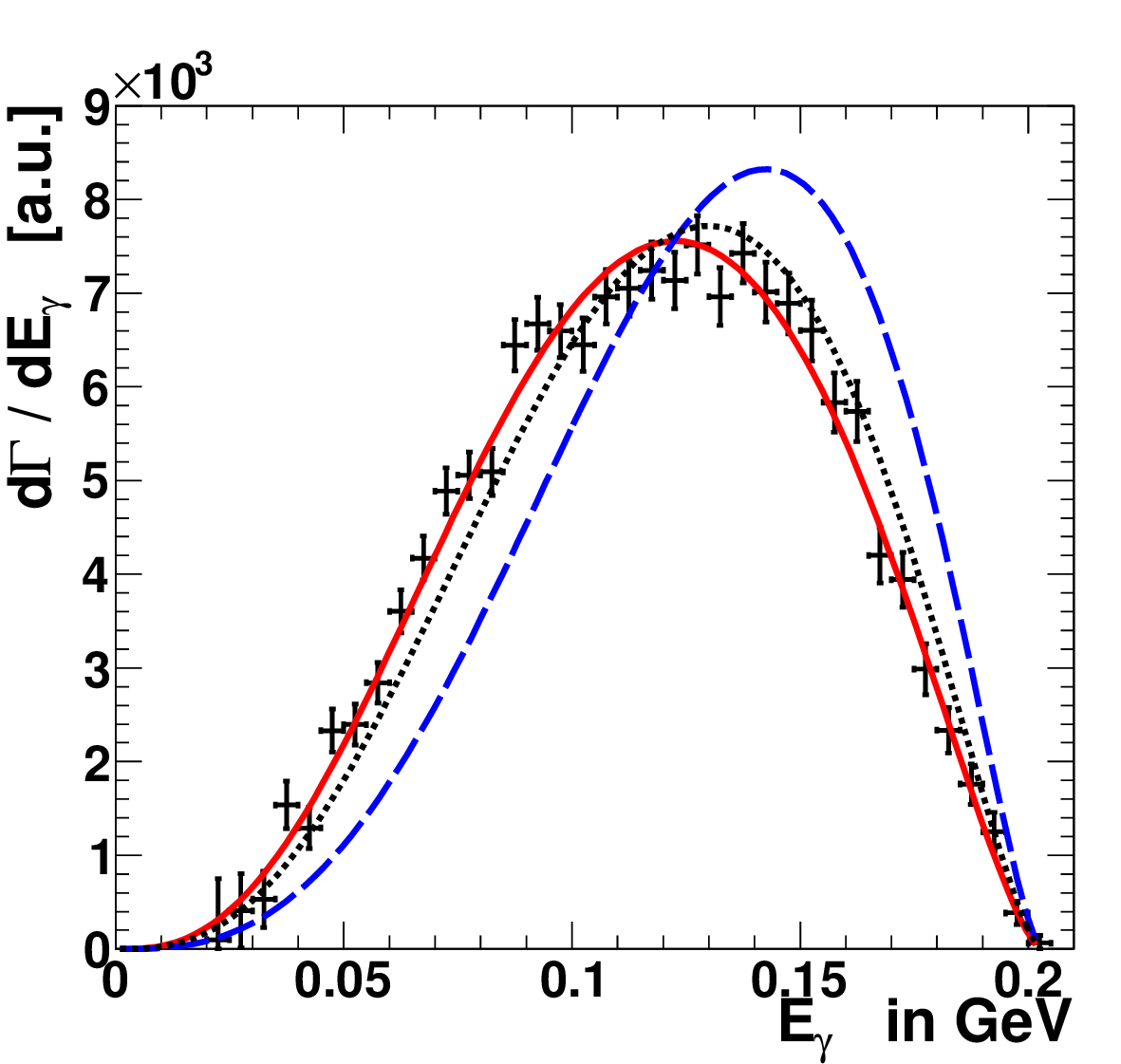}
\caption{\label{fig:ADL2012} The photon energy distribution measured in the
$\eta$ rest frame for the decay $\eta\to\pi^+\pi^-\gamma$~\cite{ADL2012}. The
dashed (blue) curve represents the shape expected from a point interaction,
$E_{\gamma}^2q^2$, with $q$ being the relative momentum in the pion-pion rest
frame. Multiplying this by the square of the pion vector form factor leads to
the dotted (black) curve. This has to be modified slightly to give the solid
(red) curve that fits the WASA data. All curves are normalized to the same
total number of events.}
\end{center}
\end{figure}

Later experiments at WASA used the $pd \to {}^{3}\textrm{He}\,\eta$ reaction
as the source of $\eta$ mesons but a compromise had to be made between the
better signal-to-background ratio close to threshold and the larger $\eta$
counting rates achievable at a slightly higher energy. Thus, in the
measurement of the decay $\eta\to \pi^+\pi^-\gamma$, the data were taken at a
beam momentum of 1.7~GeV/$c$, which corresponds to an excess energy of
$Q=60$~MeV~\cite{ADL2012}. Cuts were required to remove unwanted backgrounds,
some of which were reflections of other $\eta$ decay modes or direct meson
production. The group reconstructed $1.4\times 10^4$
$\eta\to\pi^+\pi^-\gamma$ events out of a total of $1.2\times 10^7$
candidates that contained an $\eta$ and the resulting distribution in photon
energy in the $\eta$ rest frame is shown in Fig.~\ref{fig:ADL2012}.

Just as for the $\eta\to\pi^0\pi^0\pi^0$ mode, the differential distributions
are strongly affected by $\pi\pi$ final state interactions. The big
difference in this case is that the $\pi^+\pi^-$ spectrum is governed by the
$\rho$-channel isovector $p$-wave form factor, which tends to favour higher
$\pi\pi$ masses. Though this explains most of the distortion apparent in
Fig.~\ref{fig:ADL2012}, it needs a little ``help'' to give a perfect fit to
the WASA data.

No attempt was made in Ref.~\cite{ADL2012} to extract an overall
$\eta\to\pi^+\pi^-\gamma$ branching ratio but the decay rates for four
charged modes were extracted in a later analysis from an extended data set
that started from $3\times 10^7$ $\eta$ events obtained from twelve weeks of
runs~\cite{ADL2015}. No absolute values of the decay probabilities were
extracted but the rates were measured relative to the well established
$\eta\rightarrow\pi^{+}\pi^{-}(\pi^{0}\to \gamma\gamma)$ decay, which means
that many of the systematic effects cancel in the ratios.

The results for the $\eta\rightarrow\pi^{+}\pi^{-}\gamma$, $\eta\rightarrow
\pi^{+}\pi^{-}e^{+}e^{-}$, $\eta\rightarrow e^{+}e^{-}\gamma$, and
$\eta\rightarrow e^{+}e^{-}e^{+}e^{-}$ are shown in
Table~\ref{t_BRsRel}~\cite{ADL2012}. The value extracted for the
$\eta\rightarrow\pi^{+}\pi^{-}\gamma$ channel is about a factor of 1.1 larger
than the PDG average~\cite{OLI2014}, which is based upon the results of other
modern measurements. The difference would correspond to 2.4 times the WASA
systematic error but, as yet, there is no explanation for the discrepancy.
The other three branching ratios reported in Table~\ref{t_BRsRel} are
consistent with those reported by PDG~\cite{OLI2014}.

\begin{table}[h]
\begin{center}
\begin{tabular}{ |l|l| }
\hline
Channel  &  Branching ratio with respect to\\
          & $\eta\rightarrow\pi^{+}\pi^{-}(\pi^{0}\to \gamma\gamma)$\\
\hline
$\eta\rightarrow\pi^{+}\pi^{-}\gamma$ & $0.206\pm0.003_{\rm stat/fit}\pm0.008_{\rm sys}$ \\
$\eta\rightarrow \pi^{+}\pi^{-}e^{+}e^{-}$ & $(1.2\pm0.1_{\rm stat}\pm0.1_{\rm sys})\times 10^{-3}$\\
$\eta\rightarrow e^{+}e^{-}\gamma$ & $(2.97\pm0.03_{\rm stat/fit}\pm0.13_{\rm sys})\times 10^{-2}$ \\
$\eta\rightarrow e^{+}e^{-}e^{+}e^{-}$ & $(1.4 \pm 0.4_{\rm stat} \pm 0.2_{\rm sys}) \times 10^{-4}$ \\
\hline
\end{tabular}
\caption{Summary of WASA measurements~\cite{ADL2015} of the branching ratios
for charged $\eta$ decays relative to the
$\eta\rightarrow\pi^{+}\pi^{-}(\pi^{0}\to \gamma\gamma)$ normalization
channel.} \label{t_BRsRel}
\end{center}
\end{table}

One should note, however, the large statistical error bar on the double
Dalitz decay $\eta\rightarrow e^{+}e^{-}e^{+}e^{-}$. With such a small
branching ratio the counting rate was only about 1.5 per week and this shows
the limitations of the $pd \to {}^{3}\textrm{He}\,\eta$ reaction to access
rare decays, such as those associated with $CP$ violation. Nevertheless, a
search for possible $CP$ violation was attempted in the study of the angular
distribution between the $\pi^+\pi^-$ and the $e^+e^-$ decay planes in the
rest frame of the $\eta$. No significant asymmetry was found but, once again,
the limitation came from the statistical precision.

The Dalitz plot for the $\eta\to\pi^+\pi^-\pi^0$ decay is clearly much richer
than that for the $\eta\to\pi^0\pi^0\pi^0$ because the pions are no longer
identical particles and at least five parameters may be relevant. A study of
the Dalitz plot for this decay was carried out with the initial $1.2\times
10^7$ events obtained using the $pd \to {}^{3}\textrm{He}\,\eta$ reaction at
1.7~GeV/$c$~\cite{ADL2014d}. The basic limitation compared to existing KLOE
data~\cite{AMB2008} is the restricted statistics, which might eventually be
expanded using $pp\to pp\eta$ as the source of $\eta$ mesons. Although it is
claimed that the results are generally compatible with those of KLOE, there
are deviations of more than $2\sigma$ in a few of the parameters, though one
has to realize that some of these were strongly correlated in the fits.

Superficially the decay $\omega\rightarrow\pi^{+}\pi^{-}\pi^{0}$ looks very
similar to the analogous $\eta$ decay discussed earlier in this section, but
there are two very important differences. The natural width of the $\omega$
means that in any production experiment there will be a significant
background, as illustrated in Fig.~\ref{fig:HF5}. This can be modeled in
terms of explicit multipion production~\cite{ABD2010a} or fitted empirically
using information from data on either side of the $\omega$ peak.

The big theoretical difference is that the
$\omega\rightarrow\pi^{+}\pi^{-}\pi^{0}$ decay is \emph{allowed} by the
strong interaction conservation laws and these show that each of the pion
pairs must be in a relative $p$-wave and this has to be taken into account
when modeling the Dalitz plot of the decay.

The three-pion decay of the $\omega$ was studied in three runs by the WASA
collaboration working at COSY~\cite{ADL2016c}. The $pd\to{}^3$He$\,\omega$
reaction was measured at proton beam energies of $T_p=1.45$ and 1.50~GeV and
$pp\to pp\omega$ at 2.063~GeV. In all three cases, combined fits were made
for an $\omega$ peak sitting on an empirical polynomial background.

Shortly after the $\omega$ discovery it was pointed out that the simple
$p$-wave form of the Dalitz plot would be distorted by the strong $p$-wave
attraction between pion pairs, caused by the low mass tail of the $\rho$
meson~\cite{GEL1962}. From the combined study of 44,000 $\omega\to
\pi+\pi^-\pi^0$ decays, the WASA collaboration has found the first clear
evidence of this effect at the $4.1\sigma$ level and determined deviations
from the simple $p$-wave Dalitz plot that are consistent with the
expectations of a $\rho$-meson-type final-state interaction~\cite{ADL2016c}.

\subsection{Dark photons} 
\label{dark}

One of the early extensions to the Standard Model of particle physics that
could accommodate some aspects of dark matter was the suggestion of an extra
Abelian symmetry. This would give rise to a $U(1)$ vector boson that could
mix with the normal photon to form new eigenstates. This idea has gained
traction in recent years because of the suggestion that such a \emph{dark
photon} might be the origin of the 3.6 standard deviations of the results of
the muon $g$--2 experiment from theoretical explanations. In view of the
importance of finding physics beyond the Standard Model, searches for such
\emph{dark photons} have been initiated at many laboratories throughout the
World.

The \emph{dark photon} search at COSY involved the study of the Dalitz decay
of the $\pi^0$ meson~\cite{ADL2013a} using the WASA detector that was
described in sect.~\ref{WASA_detector}. The pions were produced in the $pp\to
pp\pi^0$ reaction at 550~MeV, which is below the two-pion threshold. However,
the large excess energy of $Q=122$~MeV meant that the geometric acceptance
for both protons to enter the WASA forward detector illustrated in
Fig.~\ref{fig:wasadet} was only 19\%.

The $pp \to pp\,(\pi^0\to e^+e^-\gamma)$ reaction was clearly identified by
requesting, in addition to the two protons in the forward detector, two
oppositely curved tracks in the MDC in the central detector of
Fig.~\ref{fig:wasadet} with scattering angles between $40^{\circ}$  and
$140^{\circ}$.  A photon hit cluster in the calorimeter with an energy
deposit above 20 MeV was also demanded. Although the requirement of the mere
presence of electron tracks improved significantly the background in the
missing-mass spectrum, the most accurate identification of the $\pi^0$ signal
came from the $e^+e^-\gamma$ invariant mass, where the $\pi^0$ peak had a
FWHM $\approx 30$~MeV/$c^2$ with almost no background. The peak was very well
described by a Monte Carlo simulation where, in addition to the Dalitz decay,
there was also a contribution from $\pi^0\to \gamma\gamma$ where one of the
photons underwent an external conversion in the beryllium beam pipe. This
contribution could be largely suppressed by identifying the origin of the
$e^+e^-$ pair and discarding events that lay outside the target region.

\begin{figure}[h!]
\begin{center}
\includegraphics[width=1.0\columnwidth]{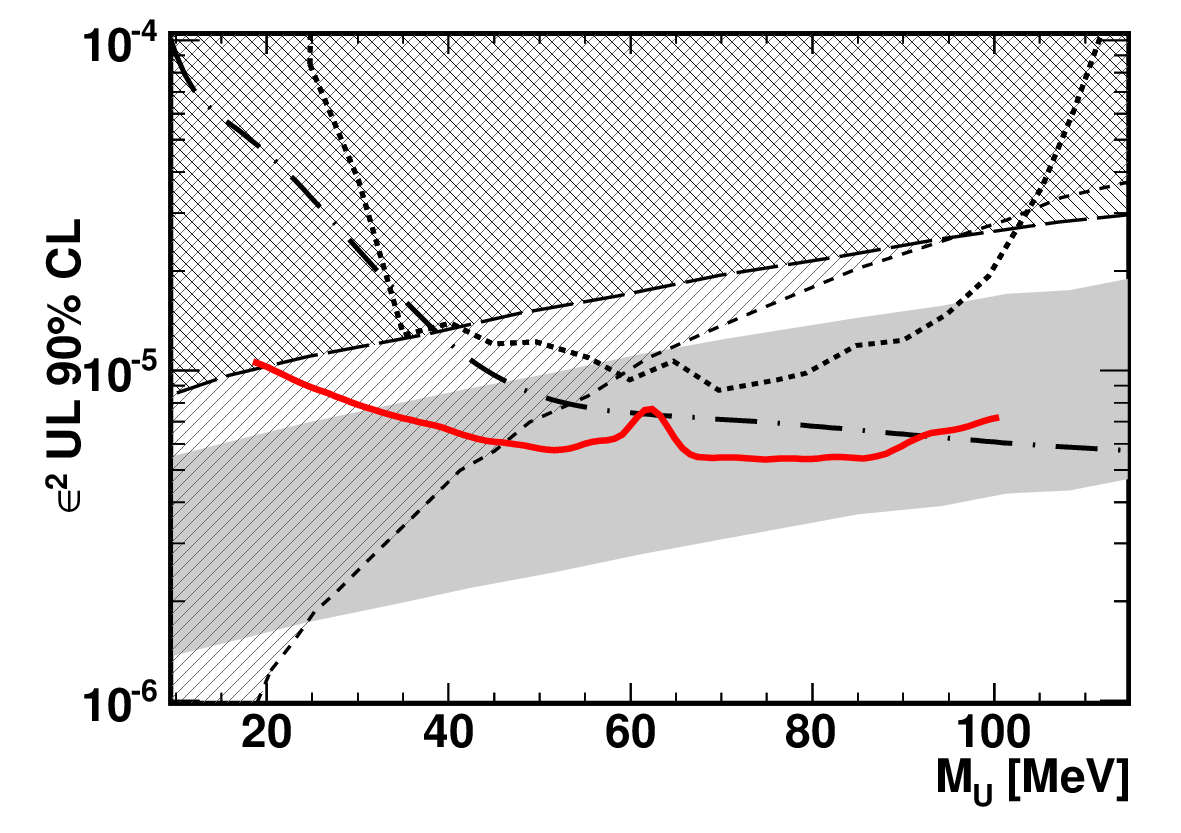}
\caption{\label{fig:dark1} Summary of the 90\% confidential upper limits for
the mixing parameter $\epsilon^2$ as a function of the \emph{dark photon}
mass from the WASA experiment~\cite{ADL2013a} (red solid line). Other lines
represent similar upper limits from earlier experiments but the grey area
represents the $\pm 2\sigma$  preferred band around the present value of the
muon ($g$--2) measurement. }
\end{center}
\end{figure}

The invariant mass distribution of the $e^+e^-$ pair from the Dalitz decay of
the $\pi^0$ can be predicted quite reliably, the only significant ambiguity
being associated with a form factor, whose effects are small and slowly
varying. The $5\times 10^5$ events collected agree with the prediction up to
at least 0.1~GeV/$c^2$, after which point random coincidences become
relatively much more important. If there were a contribution from a dark
photon in the decay $\pi^0\to\gamma\,(U\to e^+e^-)$ it should show up as a
peak in this distribution, depending upon the $(\gamma,U)$ mixing parameter
$\epsilon$ and the mass $m_U$ of the state. No sign of such a peak was seen
in the data and this allowed the group to put the upper limit on the dark
photon that is shown in Fig.~\ref{fig:dark1}. Also shown in the figure is a
grey area corresponding to the $\pm 2\sigma$ band of values that could
explain the $g$--$2$ experimental data. From this it is seen that the WASA
data would not rule out this possibility in certain regions in the plane.

Since dark matter is such an exciting area of current research, the WASA
limit could not be the last word on the subject. In the year that followed
there were measurements by collaborations at HADES~\cite{AGA2014},
BaBar~\cite{LEE2014}, MAMI~\cite{MER2014}, PHENIX~\cite{ADA2015}, and
NA48/2~\cite{BAT2015} and the much stricter upper limits found by summing all
these later results seem to exclude almost completely dark photons as being
the origin of the $g$--$2$ discrepancy. However, in general these were
inclusive measurements and the WASA data still represents the best exclusive
measurement of the Dalitz decay of the $\pi^0$.

%
%
\section{Conclusions} 
\setcounter{equation}{0}%

Though the hadron physics programme has terminated, COSY has now embarked on
a new career of precision measurements and we should here like to sketch what
the future holds in this domain. Currently there are two main themes that are
in different stages of preparation. The principal one involves the search for
an Electric Dipole Moment (EDM) of an elementary particle in an accelerator
environment. The second is the search for the breaking of Time Reversal
Invariance in COSY (TRIC) by measuring a particular spin correlation in
proton-deuteron collisions. We give below outlines of these two important
programmes.

The theory of the Big Bang postulates that, in the initial stages of the
development of the Universe, matter and antimatter were produced in equal
amounts. However, the Cosmic Microwave Background Probe has measured the
ratio of the difference in the numbers of baryons and antibaryons compared to
the number of relic photons in the visible part of the Universe and found a
deviation of eight orders of magnitude compared to the predictions of the
Standard Model of particle physics. This effect, the so-called Baryon
Asymmetry of the Universe, represents one of the most serious challenges to
the Standard Model. There is clearly a need for new sources of $CP$ violation
beyond the Standard Model to allow baryons to be generated at the expense of
antibaryons.

Sakharov~\cite{SAK1967} set down three conditions that are necessary in order
that baryonic matter should dominate the observable Universe:
\begin{itemize}
\item[$\bullet$] Baryon number conservation must violated sufficiently
    strongly,
\item[$\bullet$] $C$ and $CP$ must be violated so that baryons and
    antibaryons are produced with different rates,
\item[$\bullet$] The Universe must have evolved outside a realm of
    thermal equilibrium.
\end{itemize}
The COSY aim is therefore to try to find signals for the violation of $CP$
conservation that are bigger than the well-known ones that are found, for
example, in kaon decay. A permanent electric dipole moment of an elementary
particle would simultaneously violate both parity $P$ and time reversal
symmetry $T$. Assuming that the $CPT$ theorem holds, it means that an EDM
would also violate the combined symmetry $CP$~\cite{LEN2016}.

An EDM can arise due to the separation between the charges in a particle so
that the natural scale for an EDM in atomic physics is $10^{-8}$~e.cm whereas
the much smaller sizes in hadronic physics bring this down to
$10^{-13}$~e.cm. A water molecule has an EDM of about $2\times 10^{-9}$~e.cm
and this is close to the natural scale because such objects have degenerate
ground states of opposite parity and so parity violation does not lead to a
big suppression. In contrast, the neutron does not have a partner with
opposite parity and so it is not surprising that, despite steady improvements
over the years, there is currently only an upper limit on its EDM of $3\times
10^{-26}$~e.cm~\cite{PEN2015,BAK2014}.

For a neutron at rest, there is no Coulomb force associated with an electric
field which thus acts only through the neutron's EDM. This leads to
precession frequency shifts that depend on the orientation of the neutron
spin and hence its EDM. It might seem to be much more difficult to
investigate an EDM of a charged particle under clean conditions but that is
the goal of the JEDI collaboration at COSY~\cite{JEDI2014}. In the short term
an EDM measurement of a proton or deuteron will be undertaken at COSY but
with a limited sensitivity. On a longer time scale, the design of a dedicated
storage ring will be undertaken. A highly sensitive accelerator-based
experiment would allow the EDM of a charged particle to be inferred from its
very slow spin precession in the presence of large electric fields, and could
reach a limit of $10^{-29}$~e.cm. This huge improvement over current neutron
values is due mainly to the larger number of particles available in a stored
beam, compared with the number of ultra-cold neutrons usually found in trap
experiments, and also the potentially longer observation time that is
possible because such experiments are not limited by the particle decay time.
It must also be stressed that, knowing the EDM of just one particle, would
not be sufficient to identify unambiguously the $CP$-violating source. For
this reason the new design must allow for its use with a variety of light
ions.

The basic ideas behind the COSY measurements are as follows. The
proton/deuteron spin precesses in the horizontal plane at a rate determined
by its magnetic dipole moment. If the particle has an EDM, the spin vector
experiences an additional torque that will creates a vertical spin component
proportional to the size of the EDM. The main challenge of such kind of
experiment is the very small expected vertical component of the spin excited
by the EDM and the relatively large contributions from false spin rotations
due to field and misalignments errors of accelerator elements.

The coherent buildup of the vertical polarization only takes place on a time
scale where the spins of the particle ensemble stay aligned. Since the spin
tune is a function of the betatron and synchrotron amplitudes of the
particles in the six-dimensional phase space, spin decoherence, which is
caused by beam emittance and momentum spread of the beam, leads to a gradual
decrease of the polarization buildup rate in the vertical direction. To reach
the anticipated statistical sensitivity of $10^{-29}$~e.cm, a spin coherence
time of 1000~s must be reached. This has been achieved for deuterons at COSY
by a combination of beam bunching, electron cooling, sextupole field
corrections, and the suppression of collective effects through beam current
limitations~\cite{EVE2015,GUI2016}.

There are many technical problems to overcome, even with the precursor
experiment that aims to put limits on the deuteron EDM. For example, half the
time a longitudinally polarized beam would have its polarization parallel to
its momentum and half the time antiparallel, which could lead to no net EDM
effect. This could be overcome by making the spin precession in the machine
resonate with the orbital motion which might be done through the action of a
judiciously tuned \emph{rf} Wien filter, a possibility that is now being
actively pursued. After other upgrades to COSY through, for example, the
introduction of precise beam position monitors, it should be possible to make
the first direct measurement of the deuteron EDM using the COSY ring. This
will provide a proof-of-principle measurement for EDM searches of charged
particles (and a first direct measurement of EDM limits for protons and
deuterons) in storage rings. All the problems and challenges that have to be
faced in turning this programme into a reality will surely stimulate
developments in storage ring technology.

\begin{figure}[h!]
\begin{center}
\includegraphics[width=1.0\columnwidth]{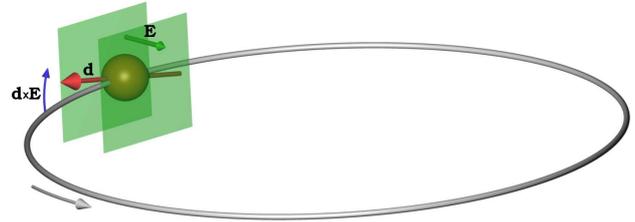}
\caption{\label{fig:CERN}
In a dedicated experiment to measure the EDM of a charged particle, a radial
electric field would be applied to an ensemble of particles circulating in a
storage ring with polarisation vector aligned to their momentum. The
existence of an EDM would generate a torque that would slowly rotate the spin
out of the plane of the ring and into the vertical direction. }
\end{center}
\end{figure}

However, the use of the COSY ring, where the orbits are governed purely by
magnetic fields, may not allow the EDM measurements of charged particles to
reach the limits that are suggested to be theoretically \emph{interesting}.
For this purpose a new dedicated electric storage ring, as illustrated
schematically in Fig.~\ref{fig:CERN}, might be required and its design will
be the major tasks for the JEDI collaboration in the upcoming years. Such a
precision storage ring for polarized light ions (proton, deuteron, and
$^3$He), possibly with two counter-rotating beams, would unquestionably
present a multitude of new technical features. Preliminary estimates suggest
that it might be possible to get down to the $10^{-29}$~e.cm level for these
light ions with such a facility.

In contrast to the EDM searches, the TRIC experiment seeks to find evidence
for an interaction that breaks time reversal but conserves
parity~\cite{EVE2013}. There is in fact an observable $A_{y,xz}$ in
proton-deuteron elastic scattering with polarized beam and target that would
vanish in the absence of some $T$-odd $P$-even interaction. By the
generalized optical theorem, the imaginary part of this observable in the
forward direction is linked to the dependence of the proton-deuteron total
cross section on these spin orientations. At COSY this could be studied by
arranging the proton beam polarization along the perpendicular to the
horizontal COSY plane while keeping the tensor polarization of the deuteron
in the horizontal plane, with an angle of $45^{\circ}$ between the
polarization vector and the beam direction.

\begin{figure}[h!]
\begin{center}
\includegraphics[width=1.0\columnwidth]{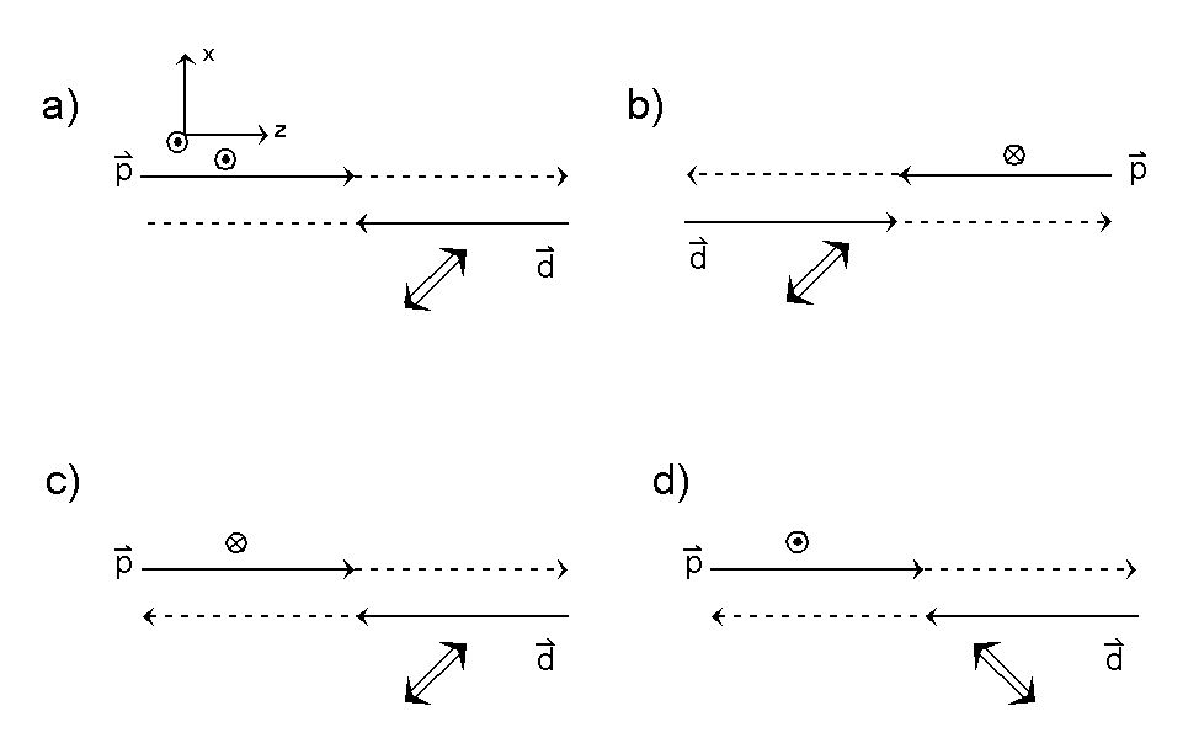}
\caption{\label{fig:tric1} Graphical illustration in the c.m.\ frame that a
time-reversed situation can be prepared in double-polarized proton-deuteron
total cross sections by flipping the spin of either the proton or
deuteron~\cite{EVE2013}. a) The basic system is shown. b) The time-reversal
operator is applied. In order to have a direct comparison between situation
a) and b), rotations through 180$^{\circ}$ about either the $y$- or $x$-axis
lead to configurations c) and d), respectively. \tiny$\bigotimes$\normalsize\
proton spin up (in the $y$-direction), \tiny$\bigodot$\normalsize\ proton
spin down, $\Leftarrow\!\Rightarrow$ deuteron tensor polarization.}
\end{center}
\end{figure}

The principle of the experiment is illustrated in the c.m.\ frame in
Fig.~\ref{fig:tric1}, where it is shown that the combined application of the
time-reversal and rotation operators leads to the same initial configuration
but with signs of the polarizations of either the proton or deuteron being
reversed. Time reversal therefore means that the total cross section measured
with configuration c) or d) should be identical to that obtained using a).
The ability to invert either of these polarizations allows a useful check on
the systematics of the experiment.

The measurement of such a $\pol{p}\pol{d}$ total cross section in a
transmission experiment with an external beam is problematic because of the
difficulties in preparing a suitable polarized deuterium target. In COSY one
can measure a total cross section by studying the lifetime of the beam as it
passes through the target with a revolution frequency of up to 1.6~MHz.

The first test of using the COSY beam lifetime to study the spin dependence
of a total cross section was carried out several years ago at
COSY~\cite{EVE2009}. The total cross section asymmetry $A_{y,y}$ in
transversally polarized $\pol{p}\,\pol{p}$ scattering has been measured at
1690~MeV/$c$ with stochastic cooling and polarizations of the beam and target
above 80\%. This test measurement gave a total cross section difference of
$-3.2\pm 9.6$~mb compared to that deduced from direct total cross section
measurements of $-3.0$~mb at this momentum~\cite{PER1986}. The experimental
conditions were rather poor and the error bar could have been reduced by a
factor of 20 and even higher precision could be achieved now with the current
equipment available at COSY. This will be further improved with the dedicated
high precision beam current measurement system that is being specially built
for the TRIC experiment.

In principle the TRIC measurement of $A_{y,xz}$ could be undertaken at any
proton beam energy, but there are good arguments for choosing $T_p=135$~MeV.
It was predicted~\cite{BEY1989} that the sensitivity to $T$-violating forces
should be maximal for $T_p$ of the order of 150~MeV and this has been checked
independently~\cite{UZI2015}. On the practical side, high quality polarimetry
data for proton-deuteron elastic scattering are available at
135~MeV~\cite{PRZ2006}. At this energy the electron cooler can continuously
cool the COSY beam over the complete cycle of measurements. It should also be
noted that only one depolarizing resonance has to be crossed to arrive at
this energy.

The TRIC experiment could be carried out at the PAX internal target station,
which was discussed in sec.~\ref{PAXsection}. This facility is located in one
of the straight low-$\beta$ sections of COSY, which leads to a reduction in
the beam emittance in the centre of the target. PAX is equipped with a high
intensity Atomic Beam Source, storage cell, multi-purpose detector, holding
field, and Breit-Rabi polarimeter. Using the new high precision beam current
measurement system, COSY will serve as accelerator, storage ring, and ideal
zero-degree spectrometer and detector for the TRIC experiment. However, the
timescale is a little uncertain because TRIC is in competition for resources
with the EDM project, which has perhaps more worldwide resonance.

As well as looking forward to the future, we must also look back at the past.
Since the aim of this review is to emphasize some of the legacy left by the
programme of hadron physics carried out at COSY over twenty years, it seems
appropriate to finish by trying to identify some elements that will leave a
lasting impression on the field. The ordering given below has no objective
significance!
\begin{enumerate}
\item The systematic EDDA
    studies~\cite{ALT2005,ALB1997,ALB2004,ALT2000,BAU2003,BAU2005} of the
    differential cross section and a wide variety of spin observables in
    elastic proton-proton scattering over almost all the COSY energy
    range have led to radical changes in the extraction of isospin $I=1$
    phase shifts. These data have been supplemented by measurements at
    small angles at ANKE but only of the differential cross section and
    analyzing power and at discrete energies~\cite{BAG2014,MCH2016}.
    Differential cross sections at even smaller angles have been measured
    by the PANDA collaboration with the KOALA detector~\cite{HUX2016}
    and, when good data become available in the Coulomb peak, this would
    lead to yet another independent means to normalize the cross
    sections.
\item The WASA measurements of two-pion production in the $np\to
    d\pi^0\pi^0$ reaction~\cite{BAS2009,PAD2011,PAD2016,PAD2012} have had
    an electrifying influence on the whole field because of the evidence
    for the production of an isoscalar dibaryon resonance.
\item Extra evidence in support of the dibaryon hypothesis was found from
    the WASA measurements of the analyzing power in neutron-proton
    elastic scattering~\cite{ADL2014,ADL2014a}, which led to a revision
    of the SAID~\cite{ARN2000} $I=0$ amplitude analysis. There had
    already been indications from $\pol{d}p\to \{pp\}_{\!s}n$
    measurements at ANKE~\cite{MCH2013a} that the existing SAID $np$
    solution was defective at high energies. Far more measurements are
    needed in neutron-proton elastic scattering above 1~GeV to give
    definitive solutions.
\item The most ambitious programme of meson production was carried out at
    ANKE~\cite{TSI2012,DYM2012,DYM2013} where the aim was to perform a
    full amplitude analysis of the $Np \to\{pp\}_{\!s}\pi$ reaction.
    Although there remain some discrete ambiguities, this is the most
    complete data set in the World to test theoretical models, be they
    phenomenological or more fundamental, e.g., effective field theory.
\item In order to constrain models in hyperon production in
    nucleon-nucleon collisions one needs fully exclusive measurements and
    the only facility for producing such data at COSY is the
    Time-of-Flight spectrometer. Of the many successful experiments
    carried out with COSY-TOF, the one that stands out is the measurement
    of $pp\to K^+p\Lambda$~\cite{SAM2013,ROD2013}, which shows a
    prominent cusp at the $\Sigma N$ thresholds due to the coupling to
    the $pp\to K^+\Sigma N$ channels. Though the resolution in the $K^+$
    momentum was better in the HIRES inclusive $pp\to K^+X$
    measurement~\cite{BUD2010}, it was not possible there to separate
    cleanly $\Lambda$ from $\Sigma$ production.
\item Although close to threshold the $pp\to K^+p\Lambda$ reaction is
    dominated by the excitation of the $N^*$ isobar $S_{11}(1650)$, it is
    only through the study of the different angular distributions away
    from threshold that one can confirm the importance of the higher
    $N^*$. Such angular distributions could only be measured at COSY with
    the help of the TOF detector~\cite{ABD2010,HAU2014,JOW2016}, and this
    is also true for the associated spin dependence~\cite{HAU2016}.
\item The COSY-11 and ANKE measurements of the
    $pd\to{}^{\,3}\textrm{He}\,\eta$ cross section near
    threshold~\cite{SMY2007,MER2007} have proved crucial in the search
    for $\eta$-mesic nuclei. After taking the beam momentum spread into
    account, the data show that there is a pole in the $\eta{}^{\,3}$He
    system within 1~MeV of threshold~\cite{MER2007}. This is the best
    signal of an $\eta$-mesic nucleus in the literature.
\item The COSY-11 measurements completely revolutionized the database on
    the total cross sections for single meson production in proton-proton
    collisions near threshold. This is best illustrated with the study of
    $pp\to pp\eta^{\prime}$, where the precision was such that an
    absolute value of the $\eta^{\prime}$ width could be
    obtained~\cite{CZE2010} and bounds deduced on the $\eta^{\prime}p$
    scattering length~\cite{CZE2014}.
\item COSY-11 data showed for the first time that the $K^-$ in kaon pair
    production $pp \to ppK^+K^-$ is strongly attracted to the
    proton~\cite{WIN2006} and this was made more quantitative in later
    ANKE experiments~\cite{YEa2013}. The interaction of $K^-$ with
    protons and light nuclei is likely to remain an intriguing field for
    years to come.
\item The mass of the $\eta$ meson was measured with unparalleled
    precision at ANKE~\cite{GOS2010} but this was only possible because a
    technique was found to determine the momentum of the deuteron beam to
    about $10^{-5}$. The $\eta$ mass will probably stay with little
    impact at the head of the PDG tables for years to come but more
    important for the future at COSY is the determination of the beam
    momentum with such precision.
\end{enumerate}

The choice shown in the list above is clearly subjective, but only time will
tell whether some of the items selected will have sunk without trace or
whether others, that have been overlooked, will flourish.

\section*{Acknowledgements}
This review was commissioned by the Directors of the Institut f\"ur
Kernphysik working at the COSY laboratory and the author is grateful for the
material and physics support offered by Professors Mei{\ss}ner, Ritman, and
Str\"{o}her. It would be invidious to list all the COSY colleagues who have
supplied me with relevant information but I have to take personal
responsibility for the inevitable oversights and misunderstandings that have
crept into this paper. One of the colleagues insisted that the review was
biased because it has been heavily influenced by the elements that I find
``interesting''. I apologize --- but this may indeed be the case!

%
%

\end{document}